\documentclass[12pt]{article}
\usepackage[margin=0.9in]{geometry}
\usepackage[utf8]{inputenc}
\title{Survival Analysis via Ordinary Differential Equations}
\author{Weijing Tang\thanks{Department of Statistics, University of Michigan, Ann Arbor, Michigan.} \and Kevin He\thanks{Department of Biostatistics, School of Public Health, University of Michigan, Ann Arbor, Michigan.} \and Gongjun Xu\footnotemark[1] \and Ji Zhu\footnotemark[1]}
\date{}

\usepackage{amsfonts,amsmath,amsthm,amssymb,color,graphicx, bm, bbm}
\usepackage{graphicx}
\usepackage{enumerate}
\usepackage{physics}
\usepackage[ruled,linesnumbered]{algorithm2e}
\usepackage{enumitem}
\usepackage{caption}
\usepackage{multirow}
\usepackage{enumitem}
\usepackage{natbib}
\usepackage{mathtools}
\usepackage{setspace}
\usepackage[flushleft]{threeparttable} % http://ctan.org/pkg/threeparttable
\usepackage{booktabs,caption}
\DeclarePairedDelimiter{\ceil}{\lceil}{\rceil}
\usepackage{soul}

\newcommand{\dev}{\textit{d}}
\newcommand{\mc}[1]{\mathcal{#1}}
\newcommand{\mb}[1]{\mbox{\textbf{#1}}}

\newcommand{\reals}{{\mbox{\textbf{R}}}}

\newtheorem{theorem}{Theorem}
\newtheorem*{theorem*}{Theorem}
\newtheorem{lemma}{Lemma}
\newtheorem{remark}{Remark}
\newtheorem{proposition}{Proposition}

\newcommand{\new}[1]{\textcolor{black}{#1}}
\newcommand{\newnew}[1]{\textcolor{black}{#1}}

\providecommand{\keywords}[1]
{
  \small	
  \textbf{\textit{Keywords---}} #1
}

\begin{document}

\maketitle
\begin{abstract}
\new{This paper introduces an Ordinary Differential Equation (ODE) notion for survival analysis. The ODE notion not only provides a unified modeling framework, but more importantly, also enables the development of a widely applicable, scalable, and easy-to-implement procedure for estimation and inference. Specifically, the ODE modeling framework unifies many existing survival models, such as the proportional hazards model, the linear transformation model, the accelerated failure time model, and the time-varying coefficient model as special cases. The generality of the proposed framework serves as the foundation of a widely applicable estimation procedure. As an illustrative example, we develop a sieve maximum likelihood estimator for a general semi-parametric class of ODE models. In comparison to existing estimation methods, the proposed procedure has advantages in terms of computational scalability and numerical stability. Moreover, to address unique theoretical challenges induced by the ODE notion, we establish a new}
%This paper introduces a general framework for survival analysis based on ordinary differential equations (ODE). 
%Specifically, this framework unifies many existing survival models, including proportional hazards models, linear transformation models, accelerated failure time models, and time-varying coefficient models as special cases. Such a unified framework provides a novel perspective on modeling censored data and offers opportunities for designing new and more flexible survival model structures. 
%Further, the aforementioned existing survival models are traditionally estimated by procedures that suffer from lack of scalability, statistical inefficiency, or implementation difficulty. Based on well-established numerical solvers and sensitivity analysis tools for ODEs, we propose a novel, scalable, and easy-to-implement general estimation procedure that is applicable to a wide range of models. 
%In particular, we develop a sieve maximum likelihood estimator for a general semi-parametric class of ODE models as an illustrative example. 
%We also establish a 
general sieve M-theorem for bundled parameters and show that the proposed sieve estimator is consistent and asymptotically normal, and achieves the semi-parametric efficiency bound. The finite sample performance of the proposed estimator is examined in simulation studies and a real-world data example.
 \end{abstract}


\section{Introduction}
\label{s:intro}

Survival analysis is an important branch of statistical modeling, where the primary outcome of interest is the time to a certain event. In practice, event times may not be observed due to a limited observation time window or missing follow-up during the study, which is referred to as censored data. Many statistical models have been developed to deal with censored data in the literature. For example, the Cox proportional hazard model is probably the most classical semi-parametric model for handling censored data~\citep{cox1975partial}, and it assumes that the covariates have a constant multiplicative effect on the hazard function. Although easy to interpret, the constant hazard ratio assumption is often considered as overly strong for real-world applications. As a result, many other semi-parametric models have been proposed as attractive alternatives, such as accelerated failure time (AFT) models, transformation models, and additive hazards models.  See \citet{aalen1980model}, \citet{10.2307/2335161}, \citet{Gray_1994}, \citet{doi:10.1002/sim.4780020223}, \citet{10.1093/biomet/82.4.835}, \citet{10.1093/biomet/85.4.980}, and \citet{chen2002semiparametric} for a sample of references. Given different assumptions made in these semi-parametric models, different estimation and inference procedures have also been developed accordingly, such as maximum partial likelihood based estimators (MPLE)~\citep{Zucker_1990, Gray_1994, bagdonavicius2001accelerated, chen2002semiparametric}, least square and rank-based methods~\citep{10.2307/2335161, lai1991large, tsiatis1990estimating, jin2003rank, jin2006least}, non-parametric maximum likelihood estimators (NPMLE)~\citep{murphy1997maximum, zeng2007maximum}, and sieve maximum likelihood estimators (MLE)~\citep{huang1999efficient, shen1994, ding2011, Zhao_2017}. % Zeng_2005, Chen_2007, 

In this paper, we introduce a novel Ordinary Differential Equation (ODE) notion and show that it provides a unified view of aforementioned survival models and, more importantly, facilitates the development of a scalable and easy-to-implement estimation and inference procedure, which can be applied to a wide range of ODE survival models.  We note that the proposed approach is founded upon well-established numerical solvers and sensitivity analysis tools for ODEs, and it overcomes various practical limitations 
of existing estimation methods when applied to different survival models for large-scale studies.

Specifically, the proposed framework models the dynamic change of the cumulative hazard function through an ODE. Let  $T$ be the event time 
and $X$ be covariates. Denote the conditional cumulative hazard function of $T$ given $X=x$ as $\Lambda_{x}(t)$. Then $\Lambda_x(t)$ is characterized by the following ODE with a fixed initial value
\begin{align}
\label{general ode}
\left\{
\begin{array}{lr}
\Lambda'_{x}(t) = f(t, \Lambda_{x}(t), x) \\
\Lambda_{x}(t_0) = c(x)
\end{array}
\right.,
\end{align}
where the derivative is with respect to $t$, $f(\cdot)$ and $c(\cdot)$ are functions to be specified, and $t_0$ is a predefined initial time point. In particular, function $c(\cdot)$ determines the probability of an event occurring after $t_0$; for instance, $\Lambda_x(0)=0$ corresponds to the case when no event occurs before time $0$. Further, function $f(\cdot)$ determines how covariates $x$ affect the hazard function at time $t$ given an individual's own cumulative hazard. Thus, different specifications of the function $f(\cdot)$ lead to different ODE models.

\new{Next, we comment on both benefits of the ODE approach in terms of modeling and computation and new theoretical challenges induced by the ODE notion.}
\begin{itemize}
	\item \new{Firstly, the ODE modeling framework is general enough to unify many aforementioned existing survival models through different specifications of the function $f(\cdot)$, which serves as the foundation of a widely applicable estimation procedure that will be developed later}. For example, the ODE~(\ref{general ode}) is equivalent to the Cox model when $f(\cdot)$ takes the form $\alpha(t)\exp(x^T\beta)$ for some function $\alpha(\cdot)$, and it is equivalent to the AFT model when $f(\cdot)$ takes the form $q(\Lambda_x(t))\exp(x^T\beta)$ for some function $q(\cdot)$. Similarly, we can obtain many more models such as the time-varying variants of the Cox model, the linear transformation model, and the additive hazards model to name a few (see Section \ref{s: examples} for details). \new{We note that the ODE notion can provide new and sometimes more explicit interpretations in terms of the hazard by re-writing the existing models in the ODE form. In addition, the generality of the proposed framework offers an opportunity for designing more flexible model structures and model diagnostics}.
	\item \new{Secondly, and also more importantly, introducing the ODE notion facilitates the development of a general and easy-to-implement procedure for estimation and inference in large-scale survival analysis.} 
		In this paper, we illustrate the proposed procedure by using a general class of ODE models as an example. In particular, this general class includes 
		the most flexible linear transformation model, where both the transformation function and the error distribution are unspecified. Since the $f(\cdot)$ function for the general model contains both finite-dimensional and infinite-dimensional parameters, we propose a spline-based sieve MLE that directly maximizes the likelihood in a sieve space. 
		\new{We provide an easy-to-implement gradient-based optimization algorithm founded upon \textit{local sensitivity analysis} tools for ODEs~\citep{Dickinson_1976}, where numerical ODE solvers are used to compute the log-likelihood function and its gradients. Since efficient implementations of both ODE solvers and splines are available in many software, the resultant algorithm is easy to carry out in practice. It is worth noting that, in comparison to existing estimation methods, the proposed procedure has advantages in various aspects, such as scalability against MPLE for the time-varying Cox model, optimization-parameter efficiency against NPMLE, statistical efficiency and numerical stability against rank-based methods for the linear transformation model. We demonstrate these advantages through extensive simulation studies. For example, when the sample size is $8,000$, it takes the proposed ODE approach about $6$ seconds to estimate the semi-parametric ODE-AFT model while the rank-based method needs $350$ seconds.} 
	\item \new{Finally, we note that the ODE notion brings new challenges to asymptotic distributional theory.} While many asymptotic distributional theories for M-estimation in semi-parametric models have been developed (see \citet{huang1999efficient}, \citet{Shen_1997}, \citet{ai2003efficient}, \citet{wellner2007}, \citet{ZHANG_2010}, \citet{He_2010}, \citet{ding2011} for a sample of references), they cannot be directly applied to our setting. Among them, the proposed theory in \citet{ding2011} considers bundled parameters where the infinite-dimensional parameter is an unknown function of the finite-dimensional Euclidean parameter and has been applied to the AFT model, and recently, to the accelerated hazards model in \citet{Zhao_2017}. \new{However, for the general class of ODE models, the estimation criterion is parameterized with more general bundled parameters where the nuisance parameter is an unknown function of not only finite-dimensional regression parameters of interest but also other infinite-dimensional nuisance parameters. To accommodate this different and challenging scenario induced by the ODE notion, we develop a new sieve M-theorem for more general bundled parameters.} By applying it to the general class of ODE models along with ODE related methodologies\new{~\citep{Walter1998}}, we show consistency, asymptotic normality, and semi-parametric efficiency for the estimated regression parameters. The proposed theory can also be extended to develop the asymptotic normality of estimators for other ODE models.
\end{itemize}

The rest of the paper is organized as follows. We introduce the ODE framework and present a general class of ODE models as special cases in Section \ref{s: ode framework}. We provide the estimation procedure in Section \ref{s: mle} and establish theoretical properties in Section \ref{s: asymptotics}. Simulation studies and a real-world data example are presented in Sections \ref{s: simulation} and \ref{s: real data} respectively.

%%%%%%%%%%%%%%%%%%%%%%%%%%%%%%%%%%%%%%%%%%%%%%%%
\section{The ODE Framework}
\label{s: ode framework}
To characterize the conditional distribution of $T$ given $X$, the conditional hazard function, denoted as $\lambda_x(t)=\Lambda'_x(t)$, 
provides a popular modeling target as it describes the instantaneous rate at which the event occurs given survival. In this paper, we view the hazard function as the dynamic change of the cumulative hazard function and quantify them using an ODE.

In our ODE framework, the hazard function depends not only on the time and covariates but also on the cumulative hazard as shown in (\ref{general ode}), where function $f(\cdot)$ specifies the dynamic change of $\Lambda_{x}(t)$ and covariates $x$ serve as additional parameters in terms of the ODE. 
The initial value in (\ref{general ode}) implies that, for an individual with covariates $x$, the probability for an event to occur after $t_0$ is controlled by $\exp(-c(x))$. For example, it is often the case that time $0$ is defined prior to the occurrence of events, which implies that an event always occurs after time $0$, i.e. the survival function $S_x(0)=1$, and it follows that $\Lambda_x(0)=0$. We use this initial value in the ODE framework hereafter for simplicity, while the estimation method and the theoretical properties established later can be extended to the general case where $c(x)$ can be a function of covariates.
Under certain smoothness conditions~\citep[page 108]{Walter1998}, the initial value problem~(\ref{general ode}) has exactly one solution, which uniquely characterizes the conditional distribution of the event time.

Next, we present a general class of ODE models \newnew{as an instantiation of the ODE framework}. 
Suppose there are two groups of covariates denoted by $X \in \reals^{d_1}$ and $Z\in \reals^{d_2}$ respectively. 
We consider ODE models in the form of  
\begin{equation}
\label{model: ltm + cox}
    \Lambda'_{x, z}(t) = \alpha(t)\exp(x^T\beta +z^T \boldsymbol{\eta}(t))q(\Lambda_{x, z}(t)),
\end{equation}
where $\alpha(\cdot)$ and $q(\cdot)$ are two unknown positive functions, and 
given an individual's own cumulative hazard, both covariates $x$ and $z$ have multiplicative effects on the hazard, one with time-independent coefficients $\beta \in \reals^{d_1}$ and the other with time-varying coefficients $\boldsymbol{\eta}(t) \in \reals^{d_2}$. Here $\boldsymbol{\eta}(\cdot)=(\eta_1(\cdot), \dots, \eta_{d_2}(\cdot))^T. $\footnote{Throughout this paper, we bold vectors only when each element is a function.}
\newnew{We note that this general class of ODE models is a specific example; other examples beyond this class are included in Remark~\ref{rm: other examples} to further illustrate the flexibility of the proposed ODE framework.}
In particular, this general class covers many existing models as special cases. As shown below, model~(\ref{model: ltm + cox}) reduces to the time-varying Cox model when $q(\cdot)=1$, to the linear transformation model when covariates $z$ are not considered, 
and further reduces to the AFT model if $\alpha(\cdot)=1$. 
In the following subsections, we will also show that by rewriting many existing models under the format (\ref{general ode}), the ODE framework brings them new interpretations in terms of the hazard function.

%%%%%%%%%%%%%%%%%%%%%%%%%%%%%%%%%%%%%%%%%%%
\label{s: examples}

%%%%%%%%%%%%%%%%%%%%%%%%%%%%%%%%%%
\subsection{Cox model and time-varying Cox model}
\label{subs: cox}
The Cox proportional hazard model assumes that the covariates have a multiplicative effect on the hazard function, i.e. $\lambda_{x}(t)=\alpha(t)\exp(x^T \beta)$,  
where $\alpha(t)$ is a baseline hazard function and $\exp(x^T \beta)$ is the relative risk, and extensions of the Cox model allow for time-varying coefficients~\citep{Zucker_1990, Gray_1994}. Here we write the Cox model with both time-independent and time-varying effects as a simple ODE, whose right-hand side does not depend on the cumulative function, i.e.
\begin{equation}
\label{cox ode}
	 \Lambda'_{x, z}(t) = \alpha(t) \exp(x^T \beta + z^T \boldsymbol{\eta}(t)),
\end{equation}
which allows covariates $x$ to have time-independent effects and covariates $z$ to have time-varying effects on the hazard function. The baseline hazard function $\alpha(t)$ and time-varying effects $\boldsymbol{\eta}(t)$ can be specified in a parametric model or left unspecified in a semi-parametric model.

%%%%%%%%%%%%%%%%%%%%%%%%%%%%%%%%%%
\subsection{Accelerated failure time model}
\label{subs: aft}
The AFT model assumes that the log transformation of $T$ is linearly correlated with covariates, i.e. $\log T = - X^T \beta + \epsilon$. 
In the proposed ODE framework, the AFT model can be written as 
\begin{equation}
\label{eq: aft ode}
	 \Lambda'_{x}(t) = q(\Lambda_{x}(t)) \exp(x^T \beta),
\end{equation}
where the function $q(\cdot)$ uniquely determines the distribution of error $\epsilon$ in the following way. Let $H_q(u) = \int_0^{-\ln u} {q^{-1}(v)}\dev v$ and $ G_{q}(u) = H_q^{-1}(u)$, then $G_{q}$ is the survival function of $\delta =\exp(\epsilon)$ as shown in \citet{bagdonavicius2001accelerated}. For example, if $q(t)=v k^{\frac{1}{v}} t^{1-\frac{1}{v}}$, then $\delta$ follows a Weibull distribution with $G_q(t)=\exp(-kt^v)$. When the error distribution is unknown (as in a semi-parametric AFT model), we can leave the function $q(\cdot)$ unspecified.

The ODE (\ref{eq: aft ode}) provides a new and clear interpretation on how covariates affect the hazard for the AFT model. Specifically, it implies that given an individual's own cumulative hazard, covariates $x$ have a multiplicative constant effect on the hazard function. Further, besides the direct effects of covariates, if $q(\cdot)$ is a monotonic increasing function, then an individual with a higher cumulative hazard at a particular time would have a higher ``baseline" hazard. Note that although we can also present the hazard directly as a function of covariates and time, i.e. $\lambda_x(t) = \lambda_{\delta}(t \exp(x^T\beta))\exp(x^T\beta)$, the covariate effects are entangled with the baseline hazard $\lambda_\delta$ in this representation, which is more difficult to interpret. 

%%%%%%%%%%%%%%%%%%%%%%%%%%%%%%%%%%
\subsection{Linear transformation model}
\label{subs: ltm}
As an extension of the AFT model, the linear transformation model assumes that, after a monotonic increasing transformation $\varphi(\cdot)$, the event time $T$ is linearly correlated with covariates, i.e. $\varphi(T) = -X^T\beta+\epsilon$. In the proposed ODE framework, it can be written as 
\begin{equation}
    \label{transformation ode}
   \Lambda'_{x}(t)  = q(\Lambda_{x}(t)) \exp(x^T \beta) \alpha(t),
\end{equation}
where $q(\cdot)$ corresponds to the distribution of $\epsilon$ in the same way as in the AFT model, and $\alpha(\cdot)$ is uniquely determined by the equation $\varphi(t) = \log \int_0^t \alpha(s)\dev s$.
In comparison to model (\ref{eq: aft ode}), the hazard function at time $t$ depends not only on the current cumulative hazard and covariates, but also on the current time $t$ directly.

Different specifications of $\varphi(\cdot)$ and $\epsilon$ have been proposed in the literature for the linear transformation model. We consider the case where both the transformation and the error distribution are unknown. This specification is especially preferred when parametric assumptions on the transformation function or the error distribution cannot be properly justified. However, when both $q(\cdot)$ and $\alpha(\cdot)$ are unknown, they may not be identifiable. The equivalent linear regression representation, $\varphi(T) = - x^T \beta+\epsilon$, allows us to see the identifiability issue clearly. \new{Note that, when no covariate is associated with survival, i.e.,  $\beta=0$, non-identifiability issue arises because parameters $(\varphi, \epsilon)$ and $(f(\varphi), f(\epsilon))$ give the same event time distribution for any arbitrary function $f$. Therefore, we consider $\beta \neq 0$, in which case \citet{horowitz1996semiparametric} showed that the model parameters are identifiable up to a scale and a location normalization under certain regularity conditions.  
 Following that result, we have developed Proposition \ref{identi} that characterizes the identifiability of parameters in~ (\ref{transformation ode}), while Proposition \ref{degeneration} provides necessary and sufficient degeneration conditions for AFT and Cox models.  The proofs are given in the Supplemental Material.} 

\begin{proposition}
\label{identi}
Suppose \newnew{at least one of the covariates in $x$ is continuous and this covariate has a non-zero $\beta$ coefficient}, which without loss of generality is assumed to be positive. Let $(q(\cdot), \beta, \alpha(\cdot))$ specify the survival distribution through (\ref{transformation ode}). Then for any other $(\tilde{q}(\cdot), \tilde{\beta}, \tilde{\alpha}(\cdot))$ that gives the same survival distribution, if and only if there exist positive constants $c_1$ and $c_2$ such that $\tilde{\beta} = c_1\beta,$ $\int_{0}^t\tilde{\alpha}(s)\dev s= c_2 (\int_{0}^t \alpha(s)\dev s)^{c_1}$,
and $\int_{0}^t {\tilde{q}^{-1}(s)}\dev s =  c_2 (\int_{0}^t {q^{-1}(s)}\dev s)^{c_1}$
for any $t>0$.
\end{proposition}

\begin{proposition}
\label{degeneration}
Suppose the conditions in Proposition~\ref{identi} hold, then the linear transformation model in~(\ref{transformation ode}) coincides with the Cox model if and only if there exist positive constants $c_1$ and $c_2$ such that $q(u) = c_2u^{1-c_1}$, 
and it coincides with the AFT model if and only if there exist positive constants $c_1$ and $c_2$ such that $\alpha(t)=c_2 t^{c_1-1}$ for $t>0$. 
\end{proposition}

\begin{remark}
	\new{Note that the original forms of the AFT model and the linear transformation model do not directly take time-varying coefficients. Existing works on the linear transformation model that consider varying coefficients choose to model them as a function of certain covariates rather than a function of time \citep{10.2307/29777150, QIU2015144}. In contrast, the equivalent ODE forms of the AFT model in (\ref{eq: aft ode}) and the linear transformation model in (\ref{transformation ode}) can naturally accommodate time-varying coefficients. For example, we can consider the generalization in (\ref{model: ltm + cox}), where given an individual's own cumulative hazard covariates $z$ have time-varying multiplicative effects $\boldsymbol{\eta}(t)$ on the hazard. In particular, this generalization is equivalent to a covariate-dependent transformation model
	\[\varphi_Z(T) = -X^T\beta+\epsilon,\]
	where $\varphi_z(t) = \log  \int_0^t \alpha(s)\exp(z^\top \boldsymbol{\eta}(s))\dev s$, i.e., covariates $z$ have multiplicative time-varying effect $\boldsymbol{\eta}(t)$ on the gradient of $\exp(\varphi_z(t))$. }
\end{remark}

\begin{remark}
\label{rm: other examples}
The proposed ODE framework is general enough to cover other existing models as well. For example, both the additive hazard model  \citep{aalen1980model, MCKEAGUE_1994} and the additive-multiplicative hazard model \citep{lin1995semiparametric} can be viewed as a specific ODE model, i.e. $\Lambda'_{x, z}(t)=r_1 (x^T \beta) + \alpha (t) r_2(z^T \eta)$, where $r_1(\cdot)$ and $r_2(\cdot)$ are some known link functions. Subsequently, the generalized additive hazards model and the generalized additive-multiplicative hazards model \citep{bagdonavicius2001accelerated} can be written as $\Lambda'_{x}(t)=q(\Lambda_{x}(t))(r_1 (x) + \alpha (t) r_2(x))$. The generalized Sedyakin's model \citep{bagdonavicius2001accelerated}, which was proposed as an extension of the AFT model, can also be viewed as a special case of (\ref{general ode}) with $\Lambda'_{x}(t) = f(\Lambda_{x}(t), x)$.
\end{remark}

\begin{remark}
Further, the proposed ODE framework and the estimation method in Section \ref{s: mle} can also be extended to deal with time-varying covariates. Suppose the covariate is a stochastic process $X(t), t\geq 0$ and $T_{X(\cdot)}$ is the failure time under $X(\cdot)$. Denote the conditional survival, the hazard function, and the cumulative function by
$S_{x(\cdot)}(t) = P\{T_{X(\cdot)}\geq t|X(s) = x(s), 0\leq s\leq t\}$,
$\lambda_{x(\cdot)}(t) =- \frac{S_{x(\cdot)}'(t)}{S_{x(\cdot)}(t)}$, and $\Lambda_{x(\cdot)}(t) =-\log(S_{x(\cdot)}(t))$, respectively.  Then the ODE (\ref{general ode}) can be extended to $\Lambda'_{x(\cdot)}(t) = f(t, \Lambda_{x(\cdot)}(t), x(t))$. \newnew{This extension also covers many existing models as special cases. For example, the linear transformation model with time-varying covariates \citep{10.1093/biomet/93.3.627} can be written as $\Lambda_{x(\cdot)}'(t)=q(\Lambda_{x(\cdot)}(t))\exp(x(t)^T \beta )\alpha(t)$, and the Cox model with time-varying covariates can be viewed as a special case with $q(\cdot) \equiv 1$.} 
For presentation simplicity, we focus on models in the form of (\ref{model: ltm + cox}) in this paper.
\end{remark}

\subsection{Related estimation methods and their limitations}
\label{s:estimation}

The maximum partial likelihood estimator (MPLE) \citep{cox1975partial} was first proposed for the Cox model, and the asymptotic property of MPLE was established by \citet{Andersen_1982} via the counting process martingale theory. For time-varying Cox models, many different estimation methods have been developed while relying on maximizing the partial likelihood \citep{Zucker_1990, Gray_1994}. 
However, evaluating the partial likelihood for an uncensored individual requires access to all other observations who were in its risk set. This prevents parallel computing for partial likelihood-based methods, which is a drawback when analyzing large scale data.

For the linear transformation model, different specifications of the transformation and the error distribution along with different estimation methods have been proposed. For example, \cite{10.1093/biomet/82.4.835}, \cite{10.1093/biomet/85.4.980}, \cite{10.1093/biomet/85.1.165}, \cite{chen2002semiparametric}, and \cite{bagdonavicius1999generalized} have considered an unknown transformation with a known error distribution, which includes the Cox model and the proportional odds model \citep{doi:10.1002/sim.4780020223} as special cases.  The corresponding
modified MPLE~\citep{chen2002semiparametric, bagdonavicius1999generalized}, sieve MLE~\citep{10.1093/biomet/85.1.165}, and NPMLE~\citep{murphy1997maximum, zeng2007maximum} have also been developed.  
However, due to the large number of nuisance parameters, it is difficult to obtain NPMLE in practice, especially in large-scale applications.  Alternatively, \citet{10.1093/biomet/92.3.619} considered a parametric Box-Cox transformation with an unknown error distribution, which includes the semi-parametric AFT model as a special case, and least square and rank-based methods have been proposed to estimate the regression parameters \citep{10.2307/2335161, lai1991large, tsiatis1990estimating, jin2003rank, jin2006least}. 
Nevertheless, they are not asymptotically efficient and may suffer additional numerical errors resulting from discrete objective functions. Subsequently, under the AFT model, \citet{zeng2007efficient} and \citet{Lin_2012} proposed efficient estimators based on a kernel-smoothed profile likelihood, and \citet{ding2011} developed an efficient sieve MLE. 
When both the transformation function and the error distribution are unknown, a partial rank-based method has been proposed  \citep{khan2007partial, Song_2006}, and its computation is analogous to that of the partial likelihood, where the rank of an uncensored individual is determined by all other individuals in its risk set, and thus the computational challenge for large-scale applications still remains.

As evident from the above discussion, many existing estimation methods suffer from important limitations in practice. In Section \ref{s: mle}, we propose a scalable, easy-to-implement and efficient estimation method that can be applied to a wide range of models.

%%%%%%%%%%%%%%%%%%%%%%%%%%%%%%%%%%%%%%%%%%%
\section{Maximum Likelihood Estimation}
\label{s: mle}

In this section, we propose a general estimation procedure that can be applied to a wide range of ODE models. Here we use the ODE model in~(\ref{model: ltm + cox}) as an illustrative example, and the proposed estimation method can also be applied to other models such as those mentioned in Remark \ref{rm: other examples}.

We denote the event time as $T$, the censoring time as $C$. Let $Y = \min\{T, C\}$ and $\Delta=\mathbbm{1}(T\leq C)$, where $\mathbbm{1}(\cdot)$ denotes the indicator function. Our  data consist of $n$  independent and identically distributed observations $\{Y_i, \Delta_i, X_i, Z_i\},~ i = 1, \dots, n$.  Since $\alpha(\cdot)$ and $q(\cdot)$ in~(\ref{model: ltm + cox}) are positive, we set $\gamma(\cdot) = \log \alpha(\cdot)$ and $g(\cdot) = \log q(\cdot)$. Under the conditional independence between $T$ and $C$ given covariates~$(X, Z)$, the log-likelihood function of the parameters $(\beta, \gamma(\cdot), \boldsymbol{\eta}(\cdot), g(\cdot))$ is given by
\begin{align}
\label{obj}
	l_n(\beta, \gamma(\cdot), g(\cdot), \boldsymbol{\eta}(\cdot))=\frac{1}{n}\sum_{i=1}^n [ & \Delta_i\{ \gamma(Y_i)+ X_i^T\beta +Z_i^T\boldsymbol{\eta}(Y_i)+g(\Lambda_{i}(Y_i;\beta, \gamma, g, \boldsymbol{\eta}))\}\\
	& - \Lambda_{i}(Y_i;\beta, \gamma, \boldsymbol{\eta}, g) ], \nonumber 
\end{align}
where $\Lambda_{i}(t;\beta, \gamma, \boldsymbol{\eta}, g)$ denotes the solution of ODE (\ref{model: ltm + cox}) parameterized by $(\beta, \gamma, \boldsymbol{\eta}, g)$ given covariates $X=X_i$ and $Z=Z_i$. The log-likelihood function (\ref{obj}) includes both finite-dimensional parameter $\beta$ and infinite-dimensional parameters $\gamma, \boldsymbol{\eta}, g$.

We propose a sieve MLE that maximizes the log-likelihood over a sequence of finite-dimensional parameter spaces that are dense in the original parameter space as the sample size increases. The sieve space can be chosen as linear spans of many types of basis functions with desired properties \citep{Chen_2007}. In particular, we construct the sieve space using polynomial splines due to their capacity in approximating complex functions and the simplicity of their construction. Under suitable smoothness conditions, $\gamma_0(\cdot)$, $\boldsymbol{\eta}_0(\cdot)$, and $g_0(\cdot)$, \new{the true parameters associated with the data generating distribution}, can be well approximated by some functions in the space of polynomial splines as defined in \citet[page 108, Definition 4.1]{schumaker_2007}.  Further, there exists a group of spline bases such that functions in the space of polynomial splines can be written as linear combinations of the spline bases \citep[page 117, Corollary~4.10]{schumaker_2007}. Different groups of spline bases may be used for the estimation of different parameters $(\gamma, \boldsymbol{\eta})$ and $g$ because of their different domains. 

Specifically, we construct the proposed sieve estimator as follows. Let $\mc{B}\subset \reals^{d_1}$ be the parameter space of $\beta$.
Let $\{B_j^1, 1\leq j \leq q_n^1\}$ and $\{B_j^2, 1\leq j \leq q_n^2\}$ be two groups of spline bases that are used for the estimation of parameters $(\gamma, \boldsymbol{\eta})$ and $ g$ respectively. \new{Here the number of spline bases, $q^i_n$, should grow sublinearly in rate $O(n^{v_i})$ for some $v_i\in(0,0.5)$, $i=1,2$ for convergence guarantee (see Section~\ref{s: asymptotics} for rigorous definitions).} Overall, we wish to find $d_2+1$ members $(\gamma, \eta_1, \cdots, \eta_{d_2})$ from the space of polynomial splines associated with $\{B_j^1\}$, one member $g$ from that associated with $\{B_j^2\}$, along with $\beta \in \mc{B}$ to maximize the log-likelihood function~(\ref{obj}). Let $Z_{i0}= 1$, $Z_i =(Z_{i1},\cdots, Z_{id_2})^T$. Then the objective function can be written as
\begin{equation}
\label{eq: obj}
	l_n(\beta, a, b)=\frac{1}{n}\sum_{i=1}^n\left[\Delta_i\{X_i^T\beta + \sum_{l=0}^{d_2}\sum_{j=1}^{q_n^1}a_j^l B_j^1(Y_i)Z_{il}+ \sum_{j=1}^{q_n^2}b_j B_j^2(\Lambda_{i}(Y_i;\beta, a, b))\} - \Lambda_{i}(Y_i;\beta, a, b) \right], 
\end{equation}
where $a = \left(a_j^l \right)_{j=1,\cdots,q_n^1,l=0,\cdots,d_2}$ and $b = \left(b_j \right)_{j=1,\cdots,q_n^2}$ are the coefficients of the spline bases, and $\Lambda_{i}(t;\beta, a,b)$ is the solution of 
\begin{equation}
\label{ode: spline}
\left\{
\begin{array}{lr}
\Lambda'_{i}(t) = \exp(X_i^T\beta + \sum_{l=0}^{d_2}\sum_{j=1}^{q_n^1}a_j^l B_j^1(t)Z_{il}+\sum_{j=1}^{q_n^2}b_j B_j^2(\Lambda_{i}(t))), \\
\Lambda_{i}(0) =0.
\end{array}
\right.
\end{equation}
The proposed sieve estimators are given by $\hat{\beta}_n=\hat{\beta}$,   $\hat{\boldsymbol{\eta}}_{n}(\cdot)=\left(\sum_{j=1}^{q_n^1}\hat{a}_j^1 B_j^1(\cdot), \dots, \sum_{j=1}^{q_n^1}\hat{a}_j^{d_2} B_j^1(\cdot)\right)$, $\hat{\gamma}_n(\cdot)=\sum_{j=1}^{q_n^1}\hat{a}_j^0 B_j^1(\cdot)$, 
and $\hat{g}_n(\cdot)=\sum_{j=1}^{q_n^2}\hat{b}_j B_j^2(\cdot)$, where $(\hat{\beta}, \hat{a}, \hat{b})$ maximizes the objective function~(\ref{eq: obj}). 

Note that the objective function (\ref{eq: obj}) contains the solution of a parameterized ODE (i.e. (\ref{ode: spline})), and this is different from most traditional optimization problems. In particular, it is nontrivial to evaluate the objective function and its gradient with respect to parameters when there is no closed-form solution for the ODE. To address this optimization challenge, we develop a gradient-based optimization algorithm by taking advantage of local sensitivity analysis \citep{Dickinson_1976, petzold2006sensitivity} and well-implemented ODE solvers. Specifically, we evaluate the objective function and its gradient as follows:
\begin{enumerate}
	\item we numerically calculate $\Lambda_{i}(Y_i;\beta, a,b)$ by solving (\ref{ode: spline}) 
		given the current parameter estimates $\beta$, $a$, $b$ and covariates $X_i$, $Z_i$, the initial value at $t_0=0$, and the evaluating time $t=Y_i$;  
	\item we evaluate the derivative of $\Lambda_{i}(Y_i;\beta, a,b)$ with respect to the parameters $\beta$, $a$, and $b$ 
	through solving another ODE which is derived by local sensitivity analysis, and calculate the gradient of the objective function by the chain rule. 
\end{enumerate}

We summarize the results of the local sensitivity analysis in the following, and provide detailed derivations in the Supplemental Material.
The local sensitivity analysis is a technique that studies the rate of change in the solution of an ODE system with respect to the parameters. There are two ways to obtain the sensitivity: forward sensitivity analysis and adjoint sensitivity analysis. Both of them require solving another ODE with some fixed initial value. 
For example, we consider to compute the gradient of $\Lambda(y; \theta)$ with respect to its parameter $\theta$, where $\Lambda(t;\theta)$ is the solution of (\ref{ode: spline}) and $\theta$ consists of parameters $\beta$, $a$, and $b$ in our case. For presentation simplicity, we denote the right-hand side of (\ref{ode: spline}) by the function $f(t,\Lambda;\theta)$, i.e.
\[f(t, \Lambda; \theta) = \exp(X^T\beta + \sum_{l=0}^{d_2}\sum_{j=1}^{q_n^1}a_j^l B_j^1(t)Z_{j}+\sum_{j=1}^{q_n^2}b_j B_j^2(\Lambda)),\]
and its partial derivative with respect to $\theta$ and $\Lambda$ by $f'_{\theta}$ and $f'_{\Lambda}$ respectively. In forward sensitivity analysis, it can be shown that the partial derivative of $\Lambda(y;\theta)$ with respect to $\theta$ is given by the solution of (\ref{I}) at $t=y$, i.e. $\Lambda'_{\theta}(y;\theta)=F_1(y)$ with $F_1$ satisfying
\begin{equation}
\label{I}
\left\{
\begin{array}{lr}
F'_{1}(t) = f'_{\theta}(t, \Lambda; \theta) + f'_{\Lambda}(t, \Lambda;\theta) F_1,\\
F_1(0) =0.
\end{array}
\right.
\end{equation}
In the alternative adjoint sensitivity analysis, we can show that the partial derivative can also be obtained by evaluating the solution of (\ref{II}) at $t=0$, i.e. $\Lambda'_{\theta}(y;\theta) = F_2(0)$ with $F_2$ satisfying 
\begin{equation}
\label{II}
\left\{
\begin{array}{lr}
(\kappa(t); F'_{2}(t))  = (-\kappa\cdot f'_{\Lambda}(t, \Lambda;\theta); -\kappa\cdot f'_{\theta}(t, \Lambda;\theta) ), \\
(\kappa(t); F_{2}(t))|_{t=y} = (1;\mb{0}).
\end{array}
\right.
\end{equation}
Thus, after plugging the form of $f(t,\Lambda;\theta)$ into either (\ref{I}) or (\ref{II}), we can obtain the gradients through solving the corresponding ODE. 
In Remark~\ref{rmk: ode}, we compare the computational complexity of forward and adjoint sensitivity analyses and provide a general guidance on which sensitivity analysis to use when computing gradients under survival ODE models.

It is worth noting that the proposed estimation method can be easily implemented using existing computing packages. For example, the ``Optimization Toolbox'' in MATLAB contains ``fminunc'' for unconstrained optimization and ``fmincon'' for constrained optimization; both require initialization and the objective function. In our implementation, we also provide evaluation of the gradient for faster and more reliable computations. In particular, we compute both the objective function and the gradient by well-implemented ODE solvers in MATLAB. 
In addition, we construct the sieve space using B-splines for its numerical simplicity, whose implementation is available in the ``Curve Fitting Toolbox''.

\begin{remark}
\label{rmk: ode}
In general, forward sensitivity analysis is computationally more efficient when the dimension of the ODE system is relatively large and the number of parameters is small, while adjoint sensitivity analysis is best suited in the complementary scenario. See \citet{Dickinson_1976} and \citet{petzold2006sensitivity} for more details. 
\new{For a general ODE model such as (\ref{general ode}) where the size of the ODE system is 1 and the number of parameters increases as the sample size $n$ grows, 
we can use the adjoint sensitivity analysis along with parallel computing \new{for $n$ independent individuals}. Alternatively, if the memory permits, we can combine ODEs for $n$ individuals into a large ODE system with $n$ dimensions, which 
is larger than the number of parameters, and then the forward sensitivity analysis is preferred.}
\end{remark}

\begin{remark}
	\new{Moreover, we introduce a computational trick for the general class of ODE models in (\ref{model: ltm + cox}) that can significantly accelerate the evaluation of the objective and gradients, where we need to solve ODEs for $n$ independent individuals.  Specifically, the trick transforms the problem of solving $n$ different ODEs at their respective observed times into a problem of solving a single ODE at $n$ different time points. More generally, this trick can be applied to any ODE model where the right-hand side is separable in the way that $f(t,\Lambda_x;\theta, x) = f_1(t;\theta, x)f_2(\Lambda_x;\theta)$ with two functions $f_1$ and $f_2$. We refer to the Supplemental Material for more details about this computational trick.}
\end{remark}

\begin{remark}
The proposed sieve MLE can also be applied to many existing models. For example, for the time-varying Cox model where $q(\cdot)=1$, we can remove the function $g(\cdot)$ from the objective function~(\ref{obj}). For the semi-parametric AFT model where $Z$ is not considered and $\alpha(\cdot)=1$, we can just keep parameters $\beta$ and $g(\cdot)$ in (\ref{obj}). For the linear transformation model, if either $q(\cdot)$ or $\alpha(\cdot)$ is specified, we can replace the corresponding term in (\ref{obj}) with the specified finite-dimensional parametric form. Also note that in comparison to existing estimation methods in Section \ref{s:estimation}, the proposed estimation method allows parallel computing, which is especially important for large-scale applications. Specifically, since the log-likelihood of each individual only depends on its own observations, the evaluation for independent data points can be carried out simultaneously. Further, compared with the NPMLE where the number of optimization parameters is linear in $n$ \citep{murphy1997maximum, zeng2007maximum}, the number of optimization parameters used in sieve MLE increases more slowly with the sample size.
\end{remark}

\begin{remark}
The objective function~(\ref{eq: obj}) is convex with respect to $\beta$ and $a$ for the (time-varying) Cox model, where the parameter $b$ is not included, and the global optimum can be achieved quickly. 
For the general case, the objective function is nonconvex and the optimization algorithm may converge to a local optimum. Nevertheless, based on our extensive simulation studies, the algorithm generally performs well with appropriately chosen initialization, such as initializing the algorithm with the estimates from the Cox model. 
\end{remark}

\begin{remark}
\label{rmk: iden}
Note that different identifiability conditions are required for different survival models. Thus, we need to add corresponding constraints in the optimization algorithm. 
\begin{itemize}
    \item \new{For the general ODE model~(\ref{model: ltm + cox}) where both covariates $X$ (with time-independent effects) and $Z$ (with at least one non-zero time-varying effect) are considered, two groups of parameters $(\beta, \gamma, g, \boldsymbol{\eta})$ and $(\tilde{\beta}, \tilde{\gamma}, \tilde{g}, \tilde{\boldsymbol{\eta}})$ give the same survival distribution if and only if $\beta = \tilde{\beta}$, $\gamma=\tilde{\gamma}+c$, $g=\tilde{g}-c$, and $\boldsymbol{\eta}=\tilde{\boldsymbol{\eta}}$ for some constant $c$. To guarantee the identifiability, we can constrain either the value of $\gamma(\cdot)$ at a fixed time point $t^*$ or the norm of $\gamma(\cdot)$, in which the former leads to a linear constraint on the coefficients of spline bases. }
    \item \new{For the linear transformation model where the time-varying effects are not considered and at least one component of $X$ has a non-zero coefficient, parameters $(\beta, \gamma, g)$ are identifiable up to two scaling factors as shown in Proposition~\ref{identi}. To guarantee identifiability, we can put constraints on $\beta$ and $\gamma$. For~$\beta$, we can either constrain the first element of $\beta$ to be $1$ \citep{khan2007partial, Song_2006}, which can be naturally achieved by arranging covariates if we know which covariate has a non-zero effect, or set $\|\beta\|=1$. For $\gamma$, we can add a similar constraint as that for the general ODE model (\ref{model: ltm + cox}). Alternatively, we can put constraints on $\gamma$ and $g$ by setting $\int_0^{t^*} \exp(\gamma(s))\dev s = c_1  \text{ and } \int_0^{t^*} \exp(-g(s))\dev s = c_2,$
	with some positive constants $c_1\neq c_2>0$ and a fixed time point~$t^*$. In our implementation, we choose to use two linear constraints, i.e. set the first element of $\beta$ to 1 and $\gamma(t^*)=0$ for simplicity in optimization. }
\end{itemize}
\end{remark}

\section{Theoretical Properties}
\label{s: asymptotics}

In this section, we study the theoretical properties of the proposed sieve MLE. Although many works have investigated  asymptotic distributional theories for M-estimation with bundled parameters \citep{ai2003efficient, chen2003estimation, ding2011}, their results cannot be directly applied to our setting. In particular,  the nuisance parameters in existing works often take the form of an unknown function of only some finite-dimensional Euclidean parameters of interest. However, our work focuses on a more general scenario, where the nuisance parameter is an unknown function of not only the Euclidean parameters but also some other infinite-dimensional nuisance parameters. To deal with theoretical challenges due to the additional functional nuisance parameters, we develop a new sieve M-theorem for the asymptotic theory of a general family of semi-parametric M-estimators. 
Moreover, we apply the proposed general theorem to establish the asymptotic normality and semi-parametric efficiency of the proposed sieve MLE $\hat{\beta}_n$ when the convergence rate of the sieve estimator of the nuisance parameter can be slower than $\sqrt{n}$.  We present regularity conditions and main theorems in this section and give all the proofs in the Supplemental Material.

\new{For the simplicity of notation, we focus on model (\ref{model: ltm + cox}) without covariates $Z$, i.e. the linear transformation model (\ref{transformation ode}),  and the results can be similarly extended to the general case with additional regularity conditions on $Z$ (see Remark~\ref{rmk: full theory})}. Recall that we have set $\gamma(\cdot) = \log \alpha(\cdot)$ and $g(\cdot) = \log q(\cdot)$ to ensure the positivity of $\alpha(\cdot)$ and $q(\cdot)$ in (\ref{transformation ode}). Then we reformulate the ODE model as follows,
\begin{equation}
\label{ode: log double}
\left\{
\begin{array}{lr}
\Lambda'(t) = \exp(x^T\beta + \gamma(t)+g(\Lambda(t))) \\
\Lambda(0) =0
\end{array}
\right..
\end{equation}
Note that the parameter $\beta$ is identifiable when time-varying effects are considered, but in (\ref{ode: log double}) it is identifiable only up to a scaling factor when both $\gamma$ and $g$ are unknown as shown in Proposition~\ref{identi}. To guarantee the identifiability, we constrain the first element of $\beta$ to be $1$ and $\gamma(\new{t^*})=c$ with some constant $c$ for simplicity in optimization. Specifically, denote $X=(X_{(1)}, X_{(-1)})$, $\beta=(1, \bar{\beta}^T)^T$, $\bar{\gamma}(\cdot)=\gamma(\cdot)-\gamma(\new{t^*})$ with $\bar{\gamma}(\new{t^*})\equiv 0$, and $\bar{X}_{(1)}=X_{(1)}+\gamma(\new{t^*})$, then we have $X^T\beta + \gamma(t)=\bar{X}_{(1)}+ X_{(-1)}^T \bar{\beta}+ \bar{\gamma}(t)$. We substitute $\bar{\beta}$, $\bar{\gamma}$, and $\bar{X}_{(1)}$ by $\beta$, $\gamma$, and $X_{(1)}$ respectively for notational simplicity hereafter, and the ODE (\ref{ode: log double}) is then equivalent to  
\begin{equation}
\label{ode: log double iden}
\left\{
\begin{array}{lr}
\Lambda'(t) = \exp(x_{(1)}+ x_{(-1)}^T \beta + \gamma(t)+g(\Lambda(t))) \\
\Lambda(0) =0
\end{array}
\right.,
\end{equation}
with $\gamma(\new{t^*})\equiv 0$. Before stating the regularity conditions, we first introduce some notations. 
We denote the solution of (\ref{ode: log double iden}) by $\Lambda(t, x, \beta, \gamma, g)$ to explicitly indicate that the solution of (\ref{ode: log double iden}) depends on covariates $x$ and parameters $(\beta, \gamma, g)$. We denote the true parameters associated with the data generating distribution by $(\beta_0, \gamma_0, g_0)$ and simplify $\Lambda(t, x, \beta_0, \gamma_0, g_0)$ as $\Lambda_0(t,x)$. In addition, some commonly used notations in the empirical process literature will be used in this section as well. Let $Pf = \int f(x) Pr(\dev x)$, where $Pr$ is a probability measure, and denote the empirical probability measure as $\mathbb{P}_n$.  

Then we assume the following regularity conditions. 
\begin{enumerate}[label=(C\arabic*), ref=(C\arabic*)]
	\item\label{c: bounded b} The true parameter $\beta_0$ is an interior point of a compact set $\mc{B}\subset \reals^d$.
	\item\label{c: bounded x} The density of $X$ is bounded below by a constant $c>0$ over its domain $\mc{X}$, which is a compact subset of $\reals^{d+1}$, and $P (X_{(-1)}X_{(-1)}^T)$ is nonsingular.
	\item\label{c: bounded t} There exists a truncation time $\tau <\infty$ such that, for some positive constant $\delta_0$, $Pr(Y>\tau|X)\geq \delta_0$ almost surely with respect to the probability measure of $X$. Then there is a constant $\mu=\sup_{x\in\mc{X}}\Lambda_0(\tau, x) \leq - \log \delta_0 $ such that $\Lambda_0(\tau, X)= -\log Pr(T>\tau|X)\leq \mu$ almost surely with respect to the probability measure of $X$. 
	\item\label{c: smoothness} Let $S^p([a, b])$ be the collection of bounded functions $f$ on $[a, b]$ with bounded derivatives $f^{(j)}$, $j =1, \dots, k$, where the $k$th derivative $f^{(k)}$ satisfies the $m$-H\"older 
	continuity condition: 
	\[|f^{(k)}(s)- f^{(k)}(t)|\leq L|s-t|^m\text{\ \ \ \ \ \ for }s, t\in [a, b],\]
	where $k$ is a positive integer and $m\in (0,1]$ with $p=m+k$,
	and $L<\infty$ is a constant. The true function $\gamma_0(\cdot)$ belongs to $\Gamma^{p_1}= \{\gamma \in S^{p_1}([0, \tau]):\gamma(\new{t^*})=0\}$ with $p_1\geq 2$ and the true function $g_0(\cdot)$ belongs to $S^{p_2}([0, \mu+\delta_1]) = \mc{G}^{p_2}$ with some positive constant $\delta_1$ and $p_2 \geq 3$. 
	\item\label{c: lower bound1} Denote $R(t)=\int^t_0\exp(\gamma_0(s))\dev s$, $V=X_{(1)}+X_{(-1)}^T\beta_0$, and $U = e^{V}R(Y)$. There exists $\eta_1\in (0,1)$ such that for all $u\in \reals^d$ with $\|u\|=1$, 
	\[u^T Var(X_{(-1)}\mid U,V)u\geq\eta_1 u^T P(X_{(-1)}X_{(-1)}^T\mid U,V)u \ \ \text{ almost surely.}\]
	\item\label{c: lower bound2} Let $\psi(t, x, \beta, \gamma, g)= x_{(1)}+x_{(-1)}^T\beta + \gamma(t)+g(\Lambda(t, x, \beta, \gamma, g))$ and denote its functional derivatives with respect to $\gamma(\cdot)$ and $g(\cdot)$ along the direction $v(\cdot)$ and $w(\cdot)$ at the true parameter by $\psi'_{0\gamma}(t,x)[v]$ and $\psi'_{0g}(t,x)[w]$ respectively, whose rigorous definitions are given by (\ref{eq:grad psi gamma})-(\ref{eq:grad psi g}) in the Supplemental Material. For any $v(\cdot)\in \Gamma^{p_1}$ and $w(\cdot)\in \mc{G}^{p_2}$, there exists $ \eta_2 \in (0, 1)$ such that 
	\[(P\{\psi'_{0\gamma}(Y,X)[v]\psi'_{0g}(Y,X)[w]\,|\,\Delta=1\})^2\leq \eta_2 P\{(\psi'_{0\gamma}(Y,X)[v])^2\,|\,\Delta=1\} P\{(\psi'_{0g}(Y,X)[w])^2\,|\,\Delta=1\}\]
	almost surely.
\end{enumerate}

Conditions \ref{c: bounded b}-\ref{c: bounded t} are common regularity assumptions in survival analysis. 
Condition \ref{c: smoothness} requires $p_2\geq 3$ 
to control the error rates of the spline approximation for the true function 
$g_0$ and its first and second derivatives. Moreover, together with $p_1\geq 2$,   \ref{c: smoothness} will also be used to verify the assumptions (\ref{m:gc})-(\ref{m:smoothness}) for the general M-theorem (Theorem \ref{thm: m-theorem}) when we apply it to derive the asymptotic normality of the proposed sieve MLE (Theorem \ref{thm: asym normal}).  
A similar condition to  \ref{c: lower bound1} was  imposed by \citet{wellner2007} for the panel count data, by \citet{ding2011} for the linear transformation model with a known transformation, and by \citet{Zhao_2017} for the accelerated hazards model. When the transformation function is known, condition \ref{c: lower bound1} is equivalent to the assumption C7 in \citet{ding2011} and can be verified in many applications as shown in \citet{wellner2007}. For the general case where both the transformation function and the error distribution are unspecified, condition \ref{c: lower bound2} is assumed to avoid strong collinearity between $\psi'_{0\gamma}(Y,X)[v]$ and $\psi'_{0g}(Y,X)[w]$.

Note that the parameter $g(\cdot)$ takes $\Lambda(t, x, \beta, \gamma, g)$ as its argument in (\ref{ode: log double iden}), which involves the other parameters $\beta$ and $\gamma(\cdot)$. Thus, $\beta$, $\gamma(\cdot)$ and $g(\cdot)$ are bundled parameters. For any $g(\cdot) \in \mc{G}^{p_2}$, we directly consider the composite function $g(\Lambda(t, x, \beta, \gamma, g))$ as a function from $\mc{T}\times \mc{X}\times \mc{B} \times \Gamma^{p_1}$ to $\reals$. And we define the collection of functions
\begin{align*}
	\mc{H}^{p_2}=\{\zeta(\cdot, \beta, \gamma):  \zeta(t, x, \beta, \gamma)=& g(\Lambda(t, x, \beta, \gamma, g)), t\in [0, \tau], x\in \mc{X}, \beta\in \mc{B}, \gamma \in \Gamma^{p_1}, \\
	& g\in \mc{G}^{p_2} \text{ such that } \sup_{t\in[0,\tau],x\in\mc{X}}|\Lambda(t,x,\beta, \gamma, g)| \leq \mu+\delta_1\},
\end{align*}
with $\delta_1$ given in condition \ref{c: smoothness}. 
For any $\zeta(\cdot, \beta, \gamma) \in \mc{H}^{p_2}$, we define its norm as \[\|\zeta(\cdot, \beta, \gamma)\|_2 = \left[ \int_{\mc{X}} \int_{0}^{\tau} [\zeta(t, x, \beta, \gamma)]^2 \dev \Lambda_0(t, x) \dev F_X(x)\right]^{1/2},\] where $F_X(x)$ is the cumulative distribution function of $X$. 
Denote the parameter $\theta = (\beta, \gamma(\cdot), \zeta(\cdot, \beta, \gamma))$ and the true parameter $\theta_0 = (\beta_0, \gamma_0(\cdot), \zeta_0(\cdot, \beta_0, \gamma_0))$ with $\zeta_0(t,x, \beta_0, \gamma_0) =g_0(\Lambda(t,x, \beta_0, \gamma_0, g_0 ))$. 
Denote the parameter space by $\Theta=\mc{B} \times \Gamma^{p_1}\times \mc{H}^{p_2}$. For any $\theta_1$ and $\theta_2$ in $\Theta$, we define the distance \[d(\theta_1, \theta_2)= \left( \|\beta_1-\beta_2\|^2+\|\gamma_1-\gamma_2\|_{2}^2+\|\zeta_1(\cdot, \beta_1, \gamma_1) - \zeta_2(\cdot, \beta_2, \gamma_2)\|_2^2 \right)^{1/2},\] where $\|\cdot\|$ is the Euclidean norm and $\|\gamma\|_{2}=(\int^\tau_0(\gamma(t))^2\dev t)^{1/2}$ is the $L_2$ norm. 

Next, we construct the sieve space as follows. Let $0 = t_0 < t_1 < \cdots < t_{K_n^1}<t_{K_n^1+1}= \tau$ be a partition of $[0, \tau]$ with $K_n^1= O(n^{\nu_1})$ and $\max_{1\leq j\leq K_n^1+1} |t_j - t_{j-1}|=O(n^{-\nu_1})$ for some $\nu_1\in (0,0.5)$. Let $T_{K_n^1}=\{t_1, \cdots, t_{K_n^1}\}$ denote the set of partition points and $S_{n}(T_{K_n^1}, K_n^1, p_1)$ be the space of polynomial splines of order $p_1$ as defined in \citet[page 108, Definition 4.1]{schumaker_2007}. Similarly, let $T_{K_n^2}$ be a set of partition points of $[0,\mu]$ with $K_n^2= O(n^{\nu_2})$ and $\max_{1\leq j\leq K_n^2+1} |t_j - t_{j-1}|=O(n^{-\nu_2})$ for some $\nu_2\in (0,0.5)$, and $S_{n}(T_{K_n^2}, K_n^2, p_2)$ be the space of polynomial splines of order $p_2$. According to \citet[page~117, Corollary~4.10]{schumaker_2007}, there exist two sets of B-spline bases $\{B_j^1, 1\leq j \leq q_n^1\}$ with $q_n^1 = K_n^1+p_1$ and $\{B_j^2, 1\leq j \leq q_n^2\}$ with $q_n^2 = K_n^2+p_2$ such that for any $s_1 \in S_{n}(T_{K_n^1}, K_n^1, p_1)$ and $s_2 \in S_{n}(T_{K_n^2}, K_n^2, p_2)$, we can write $s_1(t) = \sum_{j=1}^{q_n^1}a_j B_j^1(t)$ and $s_2(t) = \sum_{j=1}^{q_n^2}b_j B_j^2(t)$.  Let $\Gamma^{p_1}_n = \{\gamma \in S_{n}(T_{K_{n}^1}, K_{n}^1, p_1): \gamma(0)=0\}$, $\mc{G}^{p_2}_n =S_{n}(T_{K_{n}^2}, K_{n}^2, p_2)$, and \[\mc{H}^{p_2}_n=\{\zeta(\cdot, \beta, \gamma): \zeta(t, x, \beta, \gamma)=g(\Lambda(t, x, \beta, \gamma, g)), g\in \mc{G}^{p_2}_n, t\in [0, \tau], x\in \mc{X}, \beta\in \mc{B}, \gamma \in \Gamma^{p_1}_n\}.\] Let $\Theta_n=\mc{B} \times \Gamma^{p_1}_n\times \mc{H}^{p_2}_n$ be the sieve space. It is not difficult to see that $\Theta_n \subset \Theta_{n+1}\subset \cdots \subset \Theta$. We consider the sieve estimator $\hat{\theta}_n = (\hat{\beta}_n, \hat{\gamma}_n(\cdot), \hat{\zeta}_n(\cdot, \hat{\beta}_n, \hat{\gamma}_n))$, where $ \hat{\zeta}_n(t, x, \hat{\beta}_n, \hat{\gamma}_n) = \hat{g}_n(\Lambda(t, x, \hat{\beta}_n, \hat{\gamma}_n, \hat{g}_n))$, that maximizes the log-likelihood (\ref{obj}) (without covariates $Z$ and parameter $\boldsymbol{\eta}$) over the sieve space $\Theta_n$. The consistency and convergence rate of the sieve MLE $\hat{\theta}_n$ are then established in the following theorem.

\begin{theorem}(Convergence rate of $\hat{\theta}_n$.)
	\label{thm: conv rate}
	Let $\nu_1$ and $\nu_2$ satisfy the restrictions $\max\{\frac{1}{2(2+p_1)},\frac{1}{2p_1}-\frac{\nu_2}{p_1} \}<\nu_1<\frac{1}{2p_1}$, $\max\{\frac{1}{2(1+p_2)},\frac{1}{2(p_2-1)}-\frac{2\nu_1}{p_2-1}\}<\nu_2<\frac{1}{2p_2}$,  and $2\min\{2\nu_1,\nu_2\} > \max\{\nu_1,\nu_2\}$. Suppose conditions \ref{c: bounded b}-\ref{c: lower bound2} hold, then we have
	\[d(\hat{\theta}_n, \theta_0)=O_p(n^{-\min\{p_1\nu_1, p_2\nu_2, \frac{1-\max\{\nu_1, \nu_2\}}{2}\}}).\] 
\end{theorem}

Theorem \ref{thm: conv rate} gives the convergence rate of the proposed estimator $\hat{\theta}_n$ to the true parameter $\theta_0$, and its proof is provided in the Supplemental Material  
by verifying the conditions in~\citet[Theorem 1]{shen1994}. Note the subscripts 1 and 2 correspond to the space of the spline approximation for two infinite-dimensional parameters $\gamma$ and $g$, respectively. 
The restrictions on $\nu_1$ and $\nu_2$ are feasible for $p_1$ and $p_2$ not far away from each other. For example, if $p_1=p_2=p$ and $\nu_1=\nu_2=\nu$, the restriction on~$\nu$ is equivalent to $\frac{1}{2(1+p)}< v<\frac{1}{2p}$, and the convergence rate becomes  $d(\hat{\theta}_n, \theta_0) = O_p(n^{-\min\{p\nu, \frac{1-\nu}{2}\}})$, which is the same as the case when there is only one infinite-dimensional parameter~\citep{ding2011, Zhao_2017}. Further, if $\nu=\frac{1}{1+2p}$, we have $d(\hat{\theta}_n, \theta_0) = O_p(n^{-\frac{p}{1+2p}})$, which achieves the optimal convergence rate in the nonparametric regression setting. 

Although the convergence rate for the nuisance parameter is slower than the typical rate $n^{1/2}$, we will show that the sieve MLE of the regression parameter, i.e. $\hat{\beta}_n$, is still asymptotically normal and achieves the semi-parametric efficiency~bound.  First, we introduce two additional regularity conditions which are stated below.

\begin{enumerate}[label=(C\arabic*), ref=(C\arabic*),resume]
	\item \label{c: pos info exist}
	There exist $\mb{v}^*=(v^*_1, \cdots, v^*_d)^T$ and $\mb{w}^*=(w^*_1, \cdots, w^*_d)^T$, where $v^*_j \in  \Gamma^{2}$ and $w^*_j \in \mc{G}^{2}$ for $j=1, \cdots, d$, such that $P\{\Delta \mb{A}^*(U, X)\psi'_{0\gamma}(Y,X)[v]\}=0$ and $P\{\Delta \mb{A}^*(U, X)\psi'_{0g}(Y,X)[w]\}=0$ 
	hold for any $v\in \Gamma^{p_1}$ and $w\in \mc{G}^{p_2}$. Here $U$ and $V$ are defined the same as in condition \ref{c: lower bound1} and
	\begin{align*}
		\mb{A}^*(t, X)=& -\left(g_0'(\tilde{\Lambda}_{0}(t))\exp(g_0(\tilde{\Lambda}_{0}(t)))t +1\right)X_{(-1)}\\
		& + g_0'(\tilde{\Lambda}_{0}(t))\exp(g_0(\tilde{\Lambda}_{0}(t))) \int^{t}_0 \mb{v}^*(R^{-1}(se^{-V}))\dev s + \mb{v}^*(R^{-1}(te^{-V}))\\
		& + g_0'(\tilde{\Lambda}_{0}(t))\exp(g_0(\tilde{\Lambda}_{0}(t)))\int^{\tilde{\Lambda}_{0}(t)}_0 \exp(-g_0(s))\mb{w}^*(s)\dev s + \mb{w}^*(\tilde{\Lambda}_{0}(t)),
	\end{align*}
	where $\tilde{\Lambda}_0(t)$ is the solution of $\tilde{\Lambda}_0'(t)=\exp(g_0(\tilde{\Lambda}_0))$ with $\tilde{\Lambda}_0(0)=0$.
	\item\label{c: pos fish} Let $\boldsymbol{l}^*(\beta_0, \gamma_0, \zeta_0;W) = \int \mb{A}^*(t, X)\dev M(t)$, where 
$M(t)=\Delta \mathbbm{1}(U \leq t) - \int_{0}^t \mathbbm{1}(U \geq s)\dev \tilde{\Lambda}_0(s)$ is the event counting process martingale. The information matrix $I(\beta_0) = P(\boldsymbol{l}^*(\beta_0, \gamma_0, \zeta_0;W)^{\otimes 2})$ is nonsingular. Here for a vector $a$, $a^{\otimes 2} = aa^T$.
\end{enumerate}

The additional condition \ref{c: pos info exist} essentially requires the existence of the least favorable direction that is used to establish the semi-parametric efficiency bound. The directions $\mb{v}^*$ and $\mb{w}^*$ may be found through the equations in \ref{c: pos info exist}. We illustrate how to construct $\mb{v}^*$ and $\mb{w}^*$ for the Cox model and   the linear transformation model with a known transformation respectively in Remark~\ref{rmk: least favorable direction}. Condition \ref{c: pos fish} is a natural assumption that requires the information matrix to be invertible. The following theorem establishes the asymptotic normality and semi-parametric efficiency of the sieve MLE $\hat{\beta}_n$ of the regression parameter for the general linear transformation model.

\begin{theorem}(Asymptotic normality of $\hat{\beta}_n$)
	\label{thm: asym normal}	 Suppose the conditions in Theorem \ref{thm: conv rate} and  \ref{c: pos info exist}-\ref{c: pos fish} hold, then we have
\begin{align*}
		\sqrt{n}(\hat{\beta}_n - \beta_0) & = \sqrt{n}I^{-1}(\beta_0)\mathbb{P}_n \boldsymbol{l}^*(\beta_0, \gamma_0, \zeta_0;W) + o_p(1) \rightarrow_d N(0, I^{-1}(\beta_0))
	\end{align*}
	with $I(\beta_0)$ given in condition \ref{c: pos fish} and $\rightarrow_d$ denoting convergence in distribution.
\end{theorem}

Theorem \ref{thm: asym normal} states that $\hat{\beta}_n$ is asymptotically normal with variance as the inverse of the information matrix. 
In practice, the information matrix can be approximated by the estimated information matrix of all parameters including the coefficients of spline bases. 

We note that the existing sieve M-theorem for bundled parameters \citep{ding2011,Zhao_2017} cannot be directly applied to prove Theorem \ref{thm: asym normal}, because it does not allow the infinite-dimensional nuance parameter  
to be a function of other infinite-dimensional nuance parameters. Therefore, to study the asymptotic distribution of $\hat{\beta}_n$, we first establish a new general M-theorem for bundled parameters where the infinite-dimensional nuisance parameter is a function of not only the Euclidean parameter of interest but also other infinite-dimensional nuisance parameters. The established M-theorem under such a general scenario then enables us to prove Theorem \ref{thm: asym normal} by verifying its assumptions for the linear transformation model.  The details are provided in the Supplemental Material.  Since the new M-theorem can be useful for developing the asymptotic normality of sieve estimators for other ODE models, we state it below for readers of interest.

We first introduce the general setting and notation for the proposed sieve M-theorem. 
Let $m(\theta; W)$ be an objective function of unknown parameters $\theta = (\beta, \boldsymbol\gamma(\cdot), \zeta(\cdot, \beta, \boldsymbol\gamma))$ given a single observation $W$, where $\beta$ is the finite-dimensional parameter of interest, \new{$\boldsymbol\gamma(\cdot) = (\gamma_1(\cdot), \dots, \gamma_{d_2}(\cdot))$ denotes infinite-dimensional nuisance parameters}, and $\zeta(\cdot,\beta,\boldsymbol\gamma)$ is another infinite-dimensional nuisance parameter that can be a function of $\beta$ and $\boldsymbol\gamma$. Here ``$\cdot$'' represents some components of $W$. Given i.i.d. observations $\{W_i\}^{n}_{i=1}$, the sieve estimator $\hat{\theta}_n = (\hat{\beta}_n, \hat{\boldsymbol\gamma}_n(\cdot), \hat{\zeta}_n(\cdot, \hat{\beta}_n, \hat{\boldsymbol\gamma}_n))$ maximizes the objective function, $\mathbb{P}_n m(\theta; W)$, over certain sieve space. For example, $\hat{\theta}_n$ becomes the sieve MLE if $m$ is the log-likelihood function. We denote the derivative of $m$ with respect to $\beta$ as $m'_{\beta}$, \new{the functional derivative of $m$ with respect to $\gamma_j$ along the direction $v(\cdot)$ as $m'_{\gamma_j}[v]$ for $1\le j\le d_2$}, and the functional derivative of $m$ with respect to $\zeta$ along the direction $h(\cdot)$ as $m'_{\zeta}[h]$, whose rigorous definitions are given in the Supplemental Material. The following theorem then establishes the asymptotic normality of the sieve estimator, $\hat{\beta}_n$, under the above general setting.

\begin{theorem}(A general M-theorem for bundled parameters.)
\label{thm: m-theorem}
Under assumptions (\ref{m:rate})-(\ref{m:smoothness}) in the Supplemental Material,  we have
	\begin{align*}
		\sqrt{n}(\hat{\beta}_n - \beta_0) & = A^{-1}\sqrt{n}\mathbb{P}_n \mb{m}^*(\beta_0, \boldsymbol\gamma_0(\cdot), \zeta_0(\cdot, \beta_0, \boldsymbol\gamma_0);W) + o_p(1) \\
		& \rightarrow_d N(0, A^{-1}B( A^{-1})^T),
	\end{align*}
	where 
	\begin{align*}
		\mb{m}^*(\beta_0, \boldsymbol\gamma_0(\cdot), \zeta_0(\cdot, \beta_0, \boldsymbol\gamma_0);W) & = m'_{\beta}(\beta_0, \boldsymbol\gamma_0(\cdot), \zeta_0(\cdot, \beta_0, \boldsymbol\gamma_0);W) -\new{\sum_{j=1}^{d_2} m'_{\gamma_j}(\beta_0, \boldsymbol\gamma_0(\cdot), \zeta_0(\cdot, \beta_0, \boldsymbol\gamma_0);W)[\mb{v}_j^*]} \\
		& \ \ \ - m'_{\zeta}(\beta_0, \boldsymbol\gamma_0(\cdot), \zeta_0(\cdot, \beta_0, \boldsymbol\gamma_0);W)[\mb{h}^*(\cdot, \beta_0, \boldsymbol\gamma_0)],\\
		B & = P\{\mb{m}^*(\beta_0, \boldsymbol\gamma_0(\cdot), \zeta_0(\cdot, \beta_0, \boldsymbol\gamma_0);W)\mb{m}^*(\beta_0, \boldsymbol\gamma_0(\cdot), \zeta_0(\cdot, \beta_0, \boldsymbol\gamma_0);W)^T\},
	\end{align*}
with $\mb{v}_j^*=(v^*_{j1}, \dots, v^*_{jd_1})^T$, $\mb{h}^*=(h^*_1,\dots, h^*_{d})^T$ and $A$ given in the assumption (\ref{m:positive info}). 
\end{theorem}

\begin{remark}
The assumptions needed in Theorem \ref{thm: m-theorem} are similar to those in \citet{ding2011} (see the Supplemental Material for details).  \new{However, our proposed theorem significantly differs from the main theorem in \citet{ding2011}, because the latter considers $\zeta(\cdot, \beta)$ to be a function of only the finite-dimensional parameter $\beta$, while we consider a more general scenario of bundled parameters, where the nuisance parameter $\zeta(\cdot, \beta,\boldsymbol\gamma)$ can be a function of both the finite-dimensional parameter $\beta$ and other infinite-dimensional nuisance parameters $\boldsymbol\gamma$.} The proposed theorem nontrivially extends the asymptotic distributional theories for M-estimation under this general scenario. 
\end{remark}

\begin{remark}
\label{rmk: least favorable direction}
We note that to find the least favorable directions $\mb{v}^*$ and $\mb{w}^*$ required in \ref{c: pos info exist}, we may solve the equations in \ref{c: pos info exist}, which can be simplified to equations (\ref{pos_info_1}) and (\ref{pos_info_22}) provided in the Supplemental Material. For illustration, we provide explicit constructions of the least favorable directions for the Cox model and for the linear transformation model with a known transformation respectively. Specifically, for the Cox model, we have $g_0\equiv 0$ and $\mb{v}^*$ can be derived as 
\[\mb{v}^*(t) = \frac{P\{\mathbbm{1}(Y\geq t)e^{X^T\beta_0}X\}}{P\{\mathbbm{1}(Y\geq t)e^{X^T\beta_0}\}};\]
for the linear transformation model where $\gamma_0$ is known, $\mb{w}^*$ can be obtained as  
\[\mb{w}^*(t) = \pmb{\phi}(t) - g_0'(t)\int^{t}_0\pmb{\phi}(s)\dev s,\]
where 
\[  \pmb{\phi}(t) = \left(g_0'(t)\exp(g_0(t))\tilde{\Lambda}_0^{-1}(t) +1\right) \frac{P\{\mathbbm{1}(\Lambda_0(Y,X)\geq t)X\}}{P\{\mathbbm{1}(\Lambda_0(Y,X)\geq t)\}}\]
with $\tilde{\Lambda}_0$ defined in \ref{c: pos info exist}.

\new{Given the above constructions of the least favorable directions, we can further simplify the non-singularity condition of the information matrix in \ref{c: pos fish}. For the Cox model, the information matrix can be derived as
$I(\beta_0) = \int_0^\infty P\left( \left[-X+ \boldsymbol\mu(t)\right]^{\otimes 2} \mathbbm{1}(U \geq t)\right) \dev t,$
where $\boldsymbol\mu(t) = P\{\mathbbm{1}(U\geq t)e^{X^T\beta_0}X\}/P\{\mathbbm{1}(U\geq t)e^{X^T\beta_0}\}$ with $U$ defined in \ref{c: lower bound1}. Respectively, for the linear transformation where $\gamma_0$ is known, the information matrix can be derived as
$I(\beta_0) = \int_0^\infty m^2(t) \cdot Var(X|U\geq t) \cdot P(U \geq t)\cdot \exp(g_0(\tilde{\Lambda}_{0}(t)))\dev t,$
where $m(t)=g_0'(\tilde{\Lambda}_{0}(t))\exp(g_0(\tilde{\Lambda}_{0}(t)))t +1$. The non-singularity condition requires the integral of a covariance matrix to be positive definite.}
\end{remark}

\begin{remark}
\label{rmk: full theory}
\new{
	Moreover, for the general class of ODE models that include covariates $Z$ with time-varying coefficients $\boldsymbol{\eta}(\cdot)$ in (\ref{model: ltm + cox}), we have further established the same convergence rate of the sieve estimator $\hat{\theta}_n$ in Theorem~\ref{thm:z conv rate} and the asymptotic normality of $\hat{\beta}_n$ in Theorem~\ref{thm:z asym normal} in the Supplemental Material. In particular, the conditions \ref{c: bounded b}-\ref{c: pos fish} have been revised to \ref{c:z bounded b}-\ref{c:z pos fish} with additional regularity conditions on covariates $Z$. We refer to the Supplemental Material for the full list of conditions, rigorous statements of theorems, and their proofs.}
\end{remark}

\section{Simulation Studies}
\label{s: simulation}
In this section, we use simulation studies to show the finite sample performance of the sieve MLE under the time-varying Cox model and the general linear transformation model.  

\subsection{Time-varying Cox model}
\label{subs: tinv cox}

We generate event times from the model \[\Lambda'_{x}(t) = \alpha(t) \exp(\beta_1 x_1 + \beta_2 x_2 + \beta_3 x_3 + \beta_4 x_4 + \eta(t) x_5 ),\]
where $(x_1, x_2, x_3, x_4, x_5)$ follows a multivariate normal distribution with mean $0$ and autoregressive covariance \new{truncated at $\pm 2$}, $\beta_1=\beta_4=1$, and $\beta_2=\beta_3=-1$. Let $\eta(t) = \sin(\frac{3}{4}\pi t)$ be a time-varying coefficient for $x_5$ and the coefficients of all other covariates be time-independent. The baseline hazard $\alpha(t)$ is set to 0.5. The censoring times are generated from an independent uniform distribution $U(0, 3)$, which leads to a censoring rate around 50\%. The sample size $N$ varies from $1,000$ to $8,000$. \new{We fit both the log-transformed baseline hazard function $\log\alpha(t)$ and time-varying coefficient $\eta(t)$ by cubic B-splines and set the number of knots $K_n=\lfloor N'^{\frac{1}{5}} \rfloor$, i.e., the largest integer smaller than $N'^{\frac{1}{5}}$, where $N'$ is the number of distinct observation time points. The interior knots are located at the $K_n$ quantiles of the $N'$ distinct observation time points.} We compare the estimation accuracy and the computing time of the proposed sieve MLE with those of the partial likelihood-based estimator implemented in the ``coxph'' function in R with the ``tt'' argument set as the same cubic B-spline transformation of time.

\begin{table}[!ht]
    \begin{center}
    \caption{Simulation results under time-varying Cox model.}
    \label{tab:time-varying cox1}
    \begin{threeparttable}
    \begin{tabular}{cc|cccc|cccc|cc}
    \toprule
        N & Method & \multicolumn{4}{c}{$\beta_1=1$} & \multicolumn{4}{c}{$\beta_2=-1$}& \multicolumn{2}{c}{IMSE($\eta(t)$)}\\
        & & Bias & SE & ESE & CP & Bias & SE & ESE & CP & Mean & SD \\
        \midrule
       \multirow{2}{*}{1000} &  ODE & .008& .070& .070 & .958 & -.012& .076& .078 & .955 & .053 & .041\\
        
        & Cox-MPLE & .006& .070& .068 & .952 & -.010& .075& .075 & .950 &  .109 & .094  \\ 
		\hline
       \multirow{2}{*}{2000} &  ODE & .004& .048& .048 & .958 & -.004& .053& .054 & .957 &  .029 & .021\\
        & Cox-MPLE & .002& .048& .048 & .956 & -.003& .053& .053 & .959 & .053 & .041  \\ 
		\hline
       \multirow{2}{*}{4000} &  ODE & .003& .033& .034 & .952 & -.003& .038& .038 & .938 &  .016 & .011\\
        & Cox-MPLE & .003& .033& .034 & .950 & -.002& .038& .037 & .936 & .026 & .020  \\ 
        \hline
        \multirow{2}{*}{8000} &  ODE & .000& .024& .024 & .962 & -.001& .026&  .026& .938 &  .009 & .006 \\
        & Cox-MPLE & .000& .023& .024 & .959 & -.001& .026& .026 & .936 & .013 & .009 \\
        \bottomrule
    \end{tabular}
    \begin{tablenotes}
  \item  \footnotesize{Bias is the difference between the mean of estimates and the true value, SE is the sample standard error of the estimates, Mean is the mean of IMSE, and SD is the standard deviation of IMSE. ESE is the mean of the standard error estimators by inverting the estimated information matrix of all parameters, including the coefficients of spline bases, and CP is the corresponding coverage proportion of 95\% confidence intervals.
  }
  \end{tablenotes}
  \end{threeparttable}
  \end{center}
\end{table}

\begin{figure}[!ht]
    \centering
    \includegraphics[width=\textwidth]{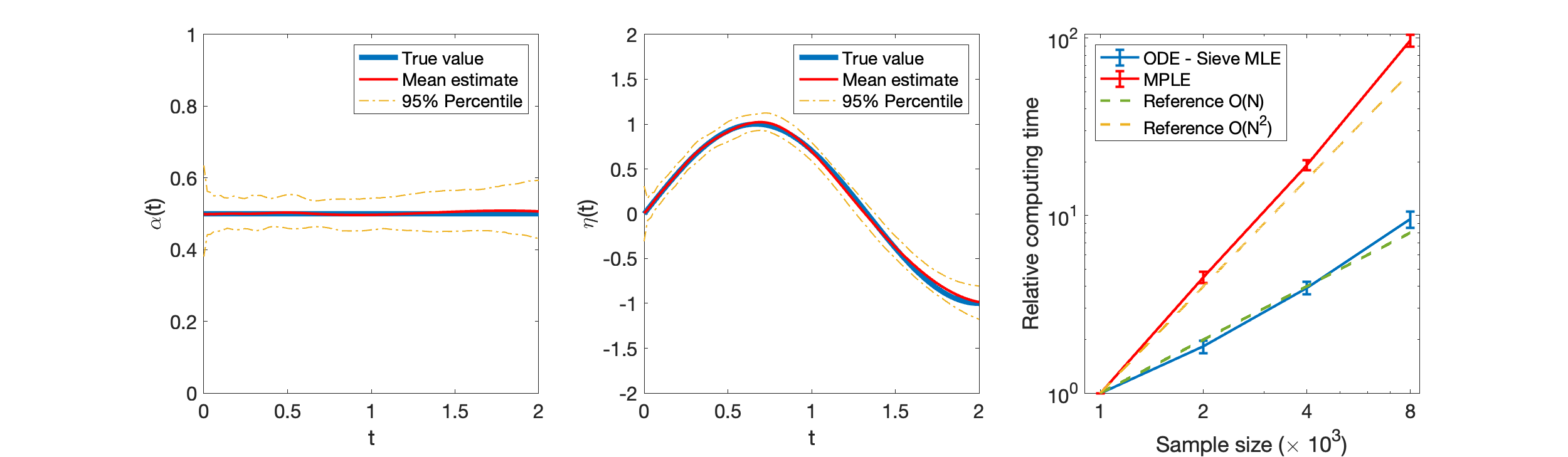}
    \caption{True $\alpha_0(t)$ and mean of $\hat{\alpha}(t)$ (left); true $\eta(t)$ and mean of $\hat{\eta}(t)$ (middle) with the sample size $N=8000$; \new{log-log plot  of mean relative computation time (right) with respect to the sample size \newnew{under the time-varying Cox model}.}} 
    \label{fig:cox_tv_N8000_ode}
\end{figure}

\new{Table~\ref{tab:time-varying cox1} summarizes the estimates of regression coefficients $\beta_1$ and $\beta_2$ based on $1000$ replications}. The estimates of the other two regression coefficients $\beta_3$ and $\beta_4$ perform similarly, and the results are included in the Supplemental Material. For the time-varying coefficient $\eta(t)$, we report the integrated mean square error (IMSE), which is the weighted sum of mean square error (MSE) of pointwise estimates over simulated time points from $0$ to $2$. As one can see, the mean and standard deviation of IMSE of the proposed sieve estimator decrease as the sample size increases. Remarkably, they are consistently smaller than those of the partial likelihood-based estimator. For time-independent coefficients, the proposed sieve estimator performs as well as the partial likelihood-based estimator. The mean of the standard error estimator, which is obtained by inverting the estimated information matrix of all parameters including the coefficients of spline bases, is approximate to the sample standard error, and the corresponding 95\% confidence interval achieves a proper coverage proportion.  From the left and middle panels of Figure~\ref{fig:cox_tv_N8000_ode}, we can see that the means of the estimated $\alpha(t)$ and $\eta(t)$ are close to the true functions, and the 95\% pointwise confidence bands cover the true functions well.

\new{It is also worth noting that, in comparison to the partial likelihood-based estimation method whose relative computing time with respect to that with the smallest sample size increases quickly as the sample size grows}, the proposed estimation method is computationally more efficient, especially when the sample size is large (see the right panel of Figure~\ref{fig:cox_tv_N8000_ode}). \new{When the number of knots increases with the sample size, the computation time of the proposed method grows at a rate slightly larger than the linear rate (but far below the quadratic rate).}

\subsection{Linear transformation model}
\label{subs: simu ltm}
We generate event times from the model $ \Lambda'_{x}(t) = q(\Lambda_x(t)) \exp(\beta_1x_1+\beta_2x_2+\beta_3x_3)\alpha(t)$. 
\new{The covariates are independent normal with mean $0$ and standard deviation $0.5$ truncated at $\pm 2$. We consider four different settings for $q(\cdot)$ and $\alpha(\cdot)$: 1) a constant $q(t)=1$ and a monotonic increasing $\alpha(t)=t^3$, in which case the Cox model is correctly specified; 2) a monotonic decreasing $q(t)=e^{-t}$ and a constant $\alpha(t)=2$; 3) a monotonic decreasing $q(t)=2/(1+t)$ and a constant $\alpha(t)=1$; 4) an increasing $q(t)=\log(1+t)+2$ and an increasing $\alpha(t)=log(1+t)$.} 
In each setting, we generate the censoring time from an independent uniform distribution $U(0, c)$, where $c$ is chosen to achieve approximately 25-30\% censoring rates. \new{The sample size $N$ varies from $1,000$ to $8,000$.} 

In setting 1), we compare the proposed sieve MLE for the ODE-Cox model, where the function $q(\cdot)$ is set to 1, with the partial-likelihood based estimator implemented using the R package {\it survival}. \new{We fit $\log\alpha(\cdot)$ by cubic B-splines with $\lfloor N'^{\frac{1}{5}} \rfloor$ interior knots that are located at the quantiles of the distinct observation time points.} 
\new{In setting 2), we compare the proposed sieve MLE for the ODE-LT model, where the function $q(\cdot)$ is set to $e^{-t}$, with the NPMLE for the equivalent logarithmic transformation model considered in \citet{zeng2007maximum}. We fit $\log\alpha(\cdot)$ by cubic B-splines with the same placement of interior knots.} In setting 3), we compare the proposed sieve MLE for the ODE-AFT model, where the function $\alpha$ is set to 1, with the rank-based estimation approach implemented using the R package {\it aftgee}. \new{We fit $\log q(t)$ by cubic B-splines with $\lfloor N^{\frac{1}{7}} \rfloor$ interior knots that are located at the quantiles of the estimated cumulative hazards under the Cox model. 
In setting 4) (as well as settings 1)-3)), we fit the general linear transformation model (ODE-Flex) where both $q(\cdot)$ and $\alpha(\cdot)$ are unspecified, and compare the sieve MLE with \new{the smoothed partial rank (SPR) method in~\citet{Song_2006}}. Both methods constrain $\beta_1=1$ for identifiability guarantee. For the sake of space, the results of the setting 4) are provided in the Supplemental Material.
}

\begin{table}[!t]
    \begin{center}
    \caption{Estimates of regression coefficients under correctly-specified ODE-Cox with $q(\cdot) \equiv 1$, \new{ODE-LT with $q(t) =e^{-t}$}, and ODE-AFT with $\alpha(\cdot)\equiv 1$. Bias, SE, ESE and CP contain the same meanings as those in Table \ref{tab:time-varying cox1}.}
    \label{tab:ltm}
    \begin{threeparttable}
    \begin{tabular}{cc|cccc|cccc|cccc}
    \toprule
         &  & \multicolumn{4}{c}{$\beta_1=1$} & \multicolumn{4}{c}{$\beta_2=1$} & \multicolumn{4}{c}{$\beta_3=1$}\\
        & Method & \small{Bias} & \small{SE} & \small{ESE} & \small{CP} & \small{Bias} & \small{SE} & \small{ESE} & \small{CP} &  \small{Bias} & \small{SE} & \small{ESE} & \small{CP} \\
        \midrule
        \multirow{2}{*}{1)} &  \small{MPLE} & \small{.002} & \small{.076} & \small{.075} & \small{.934} & \small{-.003} & \small{.075} & \small{.075} & \small{.941}  & \small{-.001} & \small{.074} & \small{.075} & \small{.954} \\
        & \small{ODE-Cox} & \small{.003} & \small{.076} & \small{.076} & \small{.936} & \small{-.002} & \small{.075} & \small{.076} & \small{.942} & \small{.000} & \small{.074} & \small{.076} & \small{.955} \\
        \hline
       \multirow{2}{*}{2)} &  \small{\new{NPMLE}} & \small{\new{.004}} & \small{\new{.117}} & \small{\new{.115}} & \small{\new{.949}} & \small{\new{-.001}} & \small{\new{.114}} & \small{\new{.115}} & \small{\new{.951}}  & \small{\new{.003}} & \small{\new{.113}} & \small{\new{.115}} & \small{\new{.960}} \\
        & \small{\new{ODE-LT}} & \small{\new{.005}} & \small{\new{.117}} & \small{\new{.115}} & \small{\new{.950}} & \small{\new{-.000}} & \small{\new{.114}} & \small{\new{.115}} & \small{\new{.951}} & \small{\new{.003}} & \small{\new{.113}} & \small{\new{.115}} & \small{\new{.961}}\\
        \hline
       \multirow{2}{*}{3)} &  \small{Rank-based} & \small{.004} & \small{.105} & \small{.102} & \small{.944} & \small{-.001} & \small{.102} & \small{.102} & \small{.950}  & \small{.002} & \small{.100} & \small{.103} & \small{.954} \\
        & \small{ODE-AFT} & \small{.000} & \small{.102} & \small{.097} & \small{.944} & \small{-.005} & \small{.100} & \small{.097} & \small{.944}  & \small{-.002} & \small{.097} & \small{.097} & \small{.950} \\	
        \bottomrule
    \end{tabular}
    \begin{tablenotes}
  \item  \footnotesize{Setting 1): the Cox model is correctly specified. \new{Setting 2): the logarithmic transformation model is correctly specified.} Setting 3): the AFT model is correctly specified.
  }
  \end{tablenotes}
  \end{threeparttable}
  \end{center}
\end{table}

\begin{figure}%[!h]
    \centering
    \includegraphics[width=\textwidth]{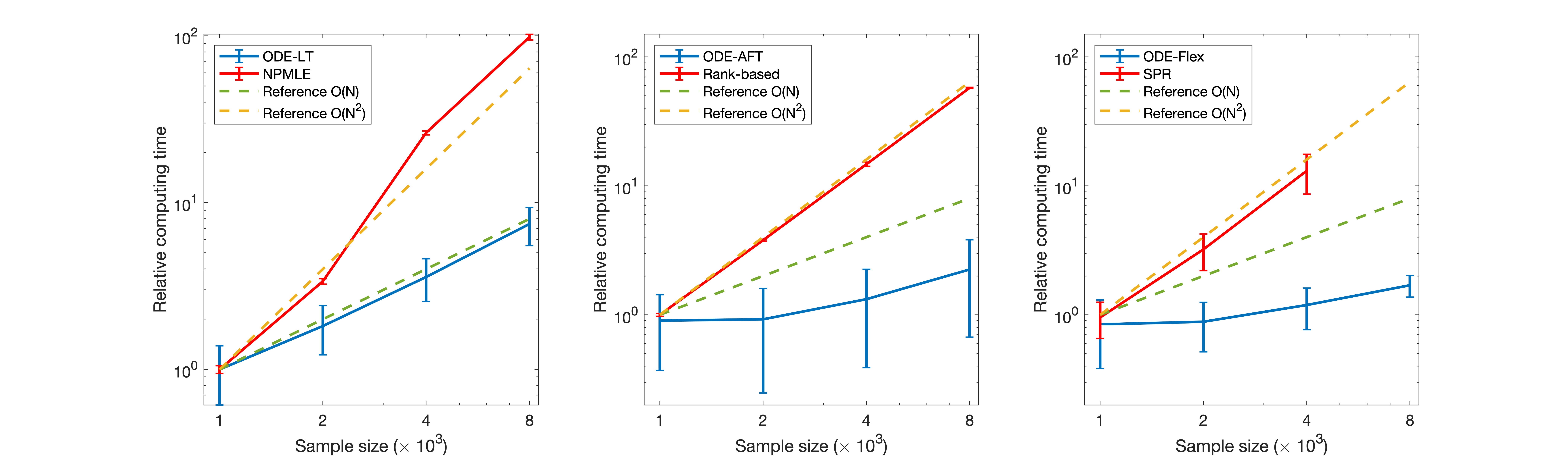}
    \caption{\new{The log-log plots of mean relative computing time with respect to the sample size under the ODE-LT, the ODE-AFT model, and the ODE-Flex model are provided from left to right respectively.}}
    \label{fig:npmle_rank_ode_computing_summary}
\end{figure}

\begin{table}[!h]
    \begin{center}
    \caption{Estimates of regression coefficients under the general linear transformation model ODE-Flex with both $q(\cdot)$ and $\alpha(\cdot)$ unspecified. Bias, SE, ESE and CP contain the same meanings as those in Table \ref{tab:time-varying cox1}.}
    \label{tab:ltm2}
    \begin{threeparttable}
    \begin{tabular}{c|cccc|cccc}
    \toprule
        & \multicolumn{4}{c}{$\beta_2=1$} & \multicolumn{4}{c}{$\beta_3=1$}\\
        Setting & Bias & SE & ESE & CP &  Bias & SE & ESE & CP \\
        \midrule
	\multirow{1}{*}{\small{1)}} &  \small{.008} & \small{.106} & \small{.107} & \small{.947}  & \small{.012} & \small{.104} & \small{.107} & \small{.959} \\ 
        \hline
	\multirow{1}{*}{\small{2)}} &  \small{-.019} & \small{.161} & \small{.151} & \small{.927} & \small{-.016} & \small{.159} & \small{.151} & \small{.938} \\
        \hline
	\multirow{1}{*}{\small{3)}} & \small{-.014} & \small{.134} & \small{.131} & \small{.941} & \small{-.012} & \small{.131} & \small{.132} & \small{.945} \\
       \hline
	\multirow{1}{*}{\small{4)}} & \small{.001} & \small{.092} & \small{.090} & \small{.939}  & \small{.005} & \small{.091} & \small{.090} & \small{.954} \\
        \bottomrule
    \end{tabular}
  \end{threeparttable}
  \end{center}
\end{table}

\begin{figure}[!h]
    \centering
    \includegraphics[width=\textwidth]{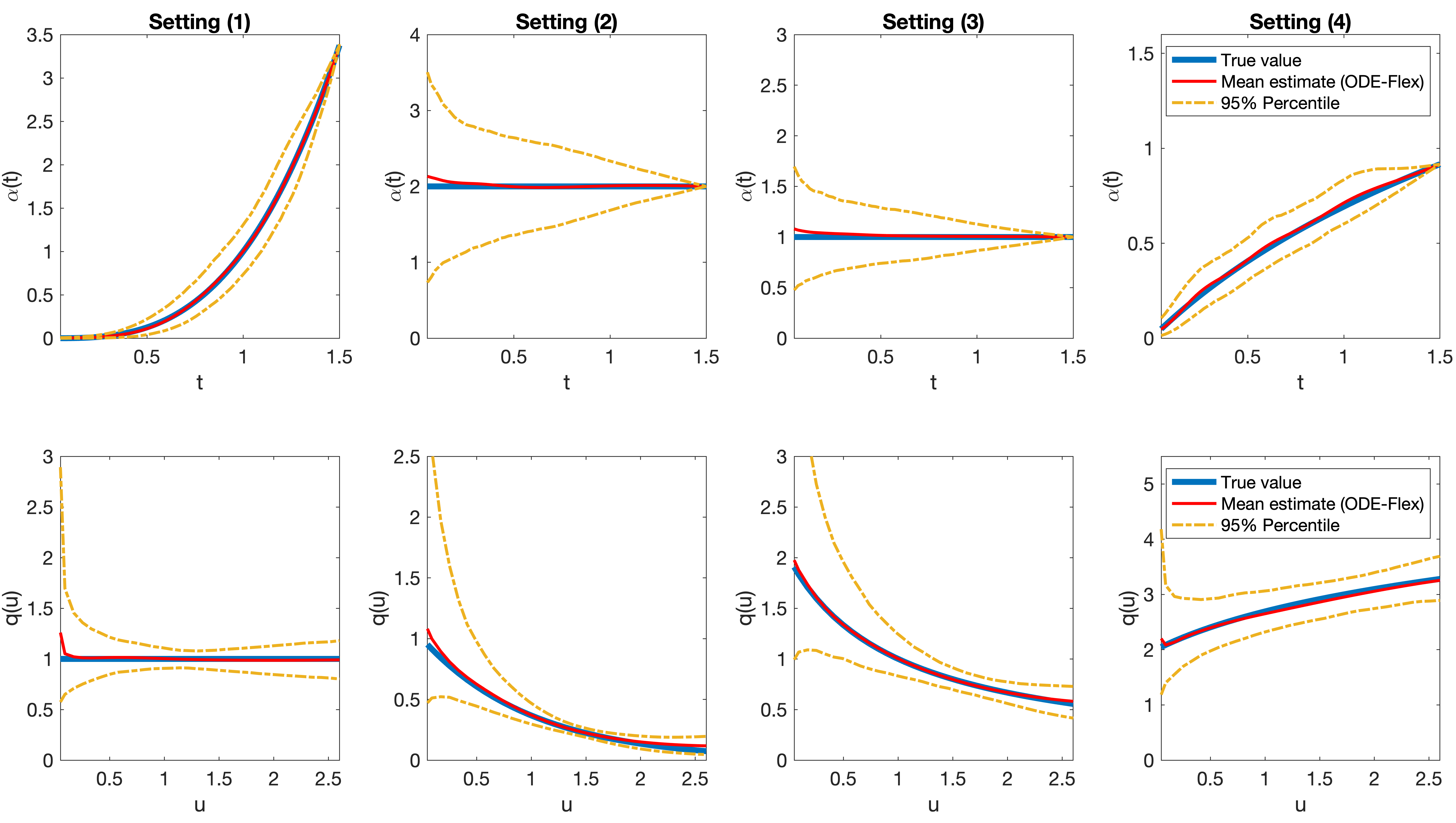}
    \caption{\new{The solid blue curves are the true $q(\cdot)$ (upper row) and $\alpha(\cdot)$ (lower row).  The solid red curves are the means of corresponding estimated $\hat{q}(\cdot)$ and $\hat{\alpha}(\cdot)$ under the general linear transformation model.  The dashed yellow curves represent 95\% pointwise confidence bands over $1,000$ replications.  From left to right, the four columns correspond to settings (1)-(4) respectively.}}
    \label{fig:ltm}
\end{figure}

Tables~\ref{tab:ltm} and \ref{tab:ltm2} summarize the estimates of regression coefficients with the sample size $N = 4,000$ based on \new{$1000$ replications}. \new{Full results for the other sample sizes are provided in the Supplemental Material.} 
\new{Table~\ref{tab:ltm} indicates that when any of the Cox model, the logarithmic transformation model, or the AFT model is correctly specified, the sieve estimator for the corresponding correctly specified ODE model achieves similar performance as the partial-likelihood based estimator for the Cox model, the NPMLE for the logarithmic transformation model, or the rank-based estimator for the AFT model. However, the relative computing time of the proposed ODE approach increases linearly as the sample size grows while that of the NPMLE for the logarithmic transformation model or the rank-based method for the AFT model increases in a quadratic rate as shown in Figure~\ref{fig:npmle_rank_ode_computing_summary}.} 

\new{For the general linear transformation model, we find that the proposed ODE-Flex method has advantages against the existing SPR method in terms of estimation accuracy, numerical stability, and computational efficiency. We refer to the Supplemental Material for detailed results and comparison with SPR.} 
From Table~\ref{tab:ltm2}, we can see that the bias of the ODE-Flex estimator is nearly negligible in all settings. The standard error estimators are close to the sample standard errors, and the corresponding 95\% confidence intervals achieve a reasonable coverage proportion. \new{When the Cox model, the logarithmic transformation model, or the AFT model is correctly specified, their estimators (in Table~\ref{tab:ltm}) achieve smaller standard errors than those for ODE-Flex (in Table~\ref{tab:ltm2}), which is expected because both $q(\cdot)$ and $\alpha(\cdot)$ are unspecified in ODE-Flex.} 
Figure~\ref{fig:ltm} shows the mean of $\hat{\alpha}(\cdot)$ and $\hat{q}(\cdot)$ respectively.  As one can see, 
the means of $\hat{\alpha}(\cdot)$ and $\hat{q}(\cdot)$ under the general linear transformation model are all close to the true functions. 
\new{Moreover, the relative computing time of ODE-Flex increases in a much smaller rate than that of SPR as the sample size grows as shown in the right panel of Figure~\ref{fig:npmle_rank_ode_computing_summary}.}

\new{Note we have also considered other alternative knots placements (see the Supplemental Material) and our numerical results suggest that knot selection does not appear critical for the proposed method.}

\section{Data Example}
\label{s: real data}

In this section, we apply the proposed method to a kidney post-transplantation mortality study. End-stage renal disease (ESRD) is one of the most deadly and costly diseases in the United States. From 2004 to 2016, ESRD incident cases increased from 345.6 to 373.4 per million people, with Medicare expenditures escalating from 18 to 35 billion dollars \citep[]{saran2019usrds}.
Kidney transplantation is the renal replacement therapy for  the majority of patients with ESRD. Successful kidney transplantation is associated with improved survival, improved quality of life, and health care cost savings when compared to dialysis.
However, despite aggressive efforts to increase the number of donor kidneys, the demand far exceeds the supply of donor kidneys for transplantation and hence, the donor waiting list is very long. 
Currently about 130,000 patients are waiting for lifesaving organ transplants in the U.S., among whom 100,000 await kidney transplants and fewer than 15\% of patients will receive transplants in their lifetime. To optimize the organ allocation, further research is essential to
determine the risk factor associated with post-transplant mortality.

To better understand this problem, we considered the data obtained from the Organ Procurement and Transplantation Network (OPTN).
There were 146,248 patients who received transplants between 1990 and 2008. 
Failure time (recorded in years) was defined as the time from transplantation to graft failure
or death, whichever occurred first, where graft failure was considered to occur when the transplanted kidney ceased to function. Patient survival was censored 6 year post-transplant or at the end of study (2008). 
The median follow-up time was around 6 years and the censoring rate was 62\%. Covariates included in this study were age at transplantation, race, gender, cold ischemic time, donation after cardiac death (DCD), BMI, expanded criteria donor (ECD), dialysis time, comorbidity conditions such as glomerulonephritis, polycystic kidney disease, diabetes, and hypertension.
Detecting and accounting for time-varying effects are particularly important in the context of kidney transplantation, as
non-proportional hazards have already been reported in the literature \citep[]{Wolfe1999,He2017}. Also, analyses with time-varying effects provide valuable clinical information that could be obscured otherwise.

\begin{figure}[!h]
     \centering
     \scalebox{0.55}
     {\includegraphics{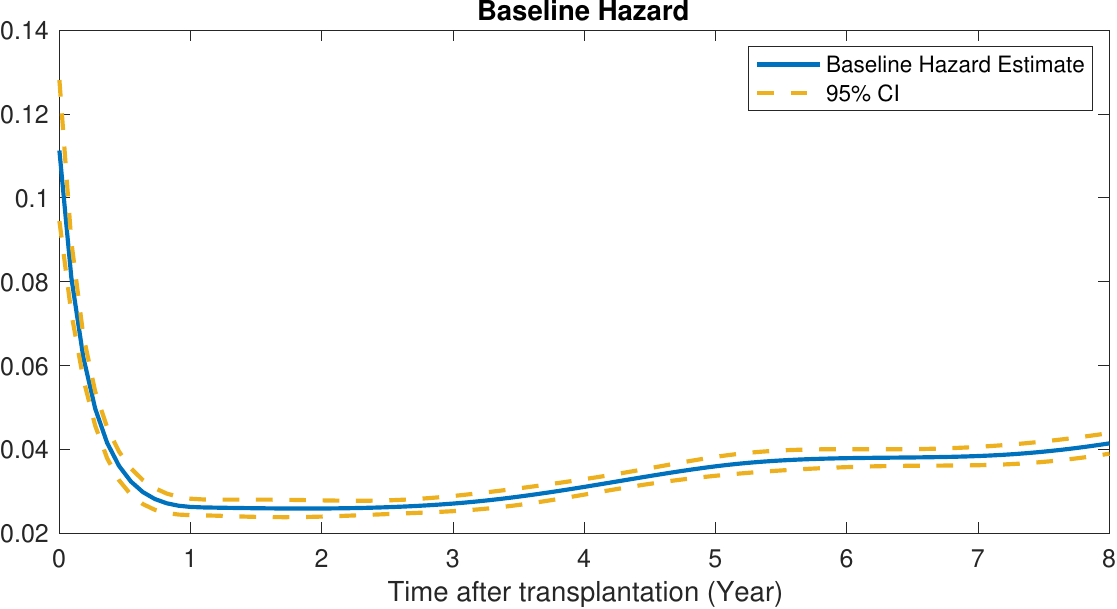}}
     \caption{Estimated baseline hazard $\hat{\alpha}(t)$ using the the proposed sieve MLE method for the kidney transplantation data.}
     \label{fig: baseline hazard}
 \end{figure}

\begin{figure}
    \centering
    \scalebox{0.65}
     {\includegraphics{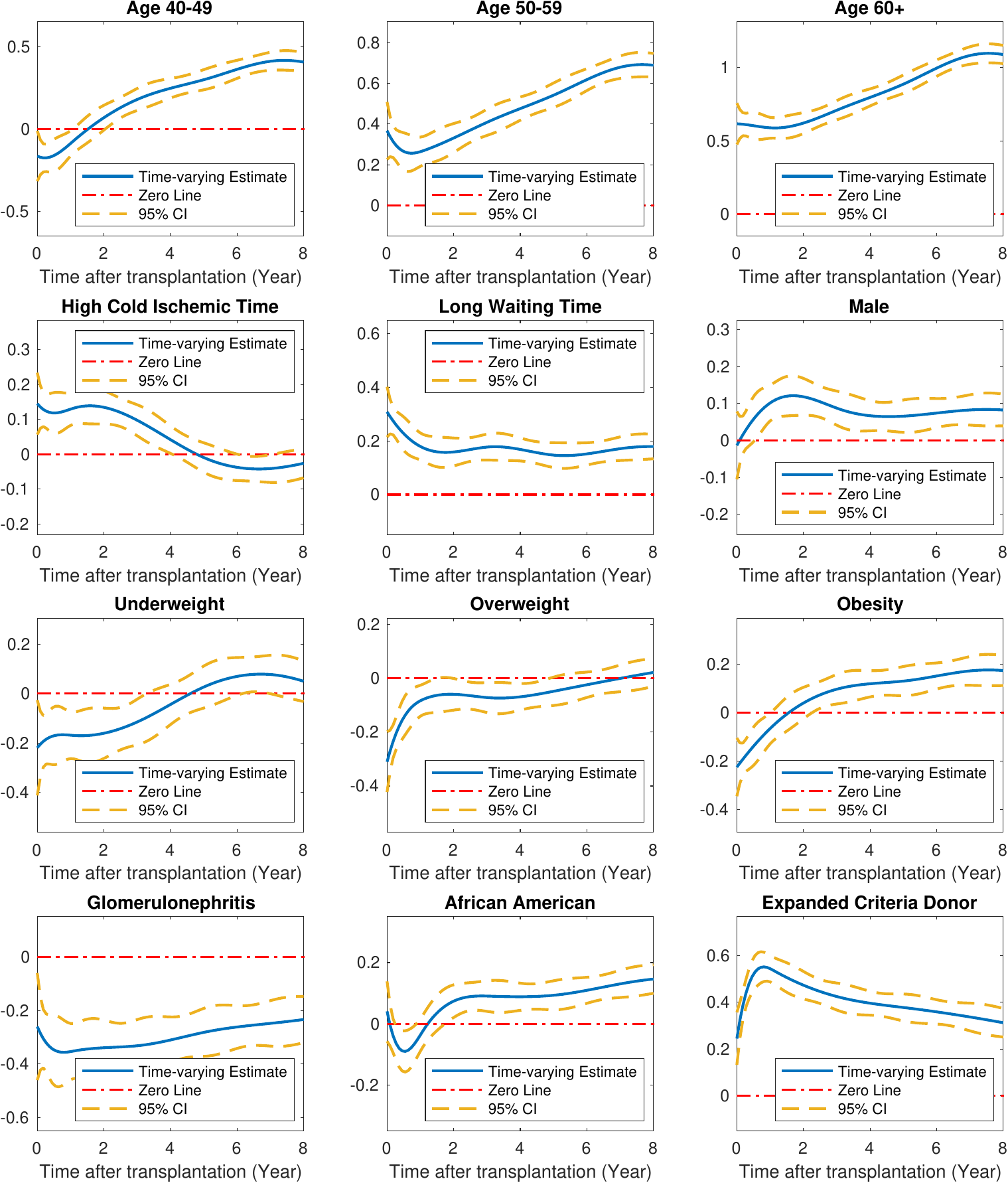}}
    \caption{Estimated time-varying effects using the proposed sieve MLE method for the kidney transplantation data.}
    \label{fig: time-varying effects}
\end{figure}

However, existing statistical softwares become computationally infeasible when fitting a time-varying effects model on a data set as large as what we have here. 
Thus, to estimate the potential time-varying effects, we fit the time-varying Cox model using the proposed sieve MLE, which is computationally scalable.
Specifically, based on previous studies, DCD, Polycystic, Diabetes and Hypertension are modeled with time-independent effects, and the remaining variables are estimated with time-varying effects. 
The time-varying effects are all implemented by cubic B-splines with 5 interior knots, \new{which is chosen based on the Bayesian information criterion.} 
Figure \ref{fig: baseline hazard} shows the estimated baseline hazard function. We can see that the post-transplant mortality is high in the short term after surgery, with a weakening association over time.  
Table \ref{tab:application} summarizes the estimated time-independent effects, and Figure \ref{fig: time-varying effects} shows examples of fitted time-varying effects with 95\% pointwise confidence intervals, where the standard error estimators were obtained by inverting the estimated information matrix of all parameters including time-independent coefficients and the coefficients of spline bases. As one can see, the effects of baseline age varied over time, resulting in an eventually strengthened association. Specifically, compared with the reference group (age at transplantation between 19-39), patients 40 to 49 years of age had a protective effect in the short term after transplantation.
We can also see that the high cold ischemic time is a risk factor for mortality in the short run, with a weakening association over time. Thus, special care should be dedicated to improve the short-term outcome.
As expected, longer waiting times on dialysis (greater than 5 years) negatively impact post-transplant survival, especially in the short run. Male gender was not significantly associated with mortality immediately after the renal transplantation but became a risk factor in the long run.  
As can be seen in Figure \ref{fig: time-varying effects}, underweight shows a protective effect in the short run, and then a slightly weakening association over time, which confirms the previous finding of \cite{Lafranca2015}.
The results regarding high BMI  should be interpreted with caution. Although higher levels of BMI in the general population are typically associated with high mortality, in chronic kidney diseases, such as patients with kidney dialysis and kidney transplantation, higher BMI has been associated with better survival, which has been labeled as reverse epidemiology \citep[]{Dekker2008, Kovesdy2010}.
Our results show that both overweight and obesity  improved survival in the short term after kidney transplantation, but obesity became a risk factor after long-term exposure.  One possible explanation is that BMI is a complex marker of visceral and nonvisceral adiposity and also of nutritional status including muscle mass \citep[]{Kovesdy2010}, and the improved short-term outcome associated with higher BMI may be related to differential benefits by one or more of these components.  Our findings indicate a need to critically reassess the role of BMI in the risk stratification of kidney transplantation. 
A further assessment (such as sub-group analysis) of high BMI that differentiates between visceral adiposity, nonvisceral adiposity and higher muscle mass may improve risk stratification in kidney transplant recipients. In addition, our
results show that graft survival for patients with Glomerulonephritis is better than patients with other primary diseases. 
Regarding racial disparities, the long-term survival outcomes for African Americans continue to lag behind non-African Americans.
Finally, as expected,  the effect
of expanded criteria donor (ECD) is not as good as optimal donor.  When a sub-optimal organ becomes available, patients and physicians must decide whether to accept the offer and special care  must be dedicated to improve the survival benefit.

\begin{table}[!h]
   \begin{center}
   \caption{Summary of estimates for time-independent effects in kidney post-transplantation mortality study} 
      \label{tab:application}
   \begin{threeparttable}
   \begin{tabular}{ccccc}
  \toprule
       Variables & DCD & Polycystic & Diabetes & Hypertension\\
      \midrule
        EST & $-0.081$ & $-0.511$ & $0.333$ & $-0.146$ \\
        ESE & $0.038$ & $0.021$ & $0.012$ & $0.014$  \\
        95\% CI & $[-0.156, -0.007]$ & $[-0.553, -0.469]$ & $[\ \ 0.310,\ \ 0.357]$ & $[-0.172, -0.119]$ \\
        p-value & $0.033$ & $<0.001$ & $<0.001$ & $<0.001$ \\
        \bottomrule
    \end{tabular}
    \begin{tablenotes}
  \item[*]  \footnotesize{EST is the estimated time-independent effect, ESE is the estimated standard error by inverting the estimated information matrix of all parameters including the coefficients of spline basis, and CI is the confidence interval.
  }
  \end{tablenotes}
  \end{threeparttable}
  \end{center}
\end{table}

\section{\new{Discussion}}
\label{s:discuss}
In this paper, we have proposed a novel ODE framework for survival analysis, which unifies the current literature, along with a general estimation procedure which is scalable and easy to implement. The ODE framework provides a new perspective for modeling censored data, which further allows us to utilize well-developed numerical solvers and local sensitivity analysis tools for ODEs in parameter estimation. Although we have only focused on one class of ODE models in this paper, the ODE framework and the estimation method offer new opportunities for investigating more flexible model structures.

\new{We note that a few recent works also use ODEs for survival analysis. Specifically,
	\citet{tang2020soden} model the cumulative hazard as in the ODE (1) with the function $f(\cdot)$ being a neural network to improve feature representation.
	The method proposed in \citet{tang2020soden} can be viewed as a neural-network-based extension of the general framework studied in this work, which demonstrates that the proposed ODE framework can be used to build flexible models. 
	\citet{groha2020neural} propose a neural-network-based ODE approach to model the Kolmogorov forward equation that  characterizes the transition probabilities for multi-state survival analysis. 
	Both the aforementioned works focus on developing flexible models with powerful representation learning via neural networks to improve prediction performance. In this work, instead, we focus on estimation and inference for a general class of semi-parametric ODE models, in which case the effects of certain covariates are often of interest. More importantly, we revisit the rich literature of survival analysis and provide a unified view of many existing survival models, 
	which is the key insight that differentiates this work and the aforementioned ones. This unification merit serves as the foundation of the proposed widely applicable estimation procedure. We also establish the consistency and semi-parametric efficiency of the proposed sieve estimator for a general class of semi-parametric ODE models, with a new general sieve M-theorem. }
	
The proposed general theory derives the asymptotic distribution of bundled parameters, where the nuisance parameter is a function of not only the regression parameters of interest but also other infinite-dimensional nuisance parameters. Though we have only illustrated the efficient estimation in the linear transformation model as an example to motivate such a theoretical development, the proposed general theory can be extended to other models.

\new{In addition, an interesting application of the unified ODE framework is to check the model specification. In particular, the estimation and inference for a general ODE model can help test whether a nested model is appropriate for a dataset. For example, Proposition~\ref{degeneration} implies that the function $q(\cdot)$ or $\alpha(\cdot)$ in the linear transformation model (\ref{transformation ode}) should be a power function when it coincides with the Cox or the AFT model. Though we have established the consistency of the functional parameters $q(\cdot)$ and $\alpha(\cdot)$ in the nonparametric linear transformation model,  it is worthwhile to further investigate their asymptotic distributional theory for model diagnostics as future work. As a preliminary study, we have explored a heuristic parametric approach for model diagnostics and provided its finite sample performance in the Supplemental Material.}

\new{Finally, we note that a few recent works have tried to address the computation burden of certain estimation methods for specific models on massive time-to-event data. 
In particular, 	\citet{10.1093/biostatistics/kxz036} proposed an efficient divide-and-conquer (DAC) algorithm for the sparse Cox model. 
	\citet{doi:10.1080/10618600.2020.1841650} developed an algorithm for reducing the computation cost of fitting the Fine-Gray~\citep{10.2307/2670170} proportional subdistributional hazards model by exploiting its special structure. 
	\citet{https://doi.org/10.1002/sim.8783} proposed a subsampling procedure to approximate the full-data estimator for the additive hazard model. 
	Note that most of these methods are tailored for a specific model while our method can be applied more broadly. Further, our estimation procedure and these methods are not competitors. In contrast, some of the techniques used in these methods, such as DAC, can be naturally integrated into the proposed estimation procedure, which is an interesting future direction to be explored.}

\bibliographystyle{chicago}
\bibliography{references}

\pagebreak
\begin{center}
\textbf{\Large Supplemental Material: Survival Analysis via Ordinary Differential Equations}
\end{center}
%%%%%%%%%% Merge with supplemental materials %%%%%%%%%%
%%%%%%%%%% Prefix a "S" to all equations, figures, tables and reset the counter %%%%%%%%%%
\setcounter{equation}{0}
\setcounter{figure}{0}
\setcounter{table}{0}
\setcounter{page}{1}
\setcounter{section}{0}
\makeatletter
\renewcommand{\theequation}{S\arabic{equation}}
\renewcommand{\thefigure}{S\arabic{figure}}
\renewcommand{\thetable}{S\arabic{table}}
\renewcommand{\bibnumfmt}[1]{[S#1]}
\renewcommand{\citenumfont}[1]{S#1}
%\renewcommand{\thesection}{S\arabic{section}}
%%%%%%%%%% Prefix a "S" to all equations, figures, tables and reset the counter %%%%%%%%%%

%\newpage
%\appendix
This supplementary material is structured as follows. We provide the detailed derivation of the local sensitivity analysis and optimization algorithm in Section \ref{appendx: algorithm}. We present the proposed general M-theorem for bundled parameters (Theorem \ref{thm: m-theorem}) and its proof in Section \ref{appendx: m-theorem}.
The proofs of Theorems \ref{thm: conv rate} and \ref{thm: asym normal} are given in Section \ref{appendx: proof}, those of Propositions~\ref{identi} and \ref{degeneration} are given in Section~\ref{s: iden}. 
\new{We further establish the convergence rate and the asymptotic normality of the proposed sieve estimator for the general class of ODE models in the presence of covariates $Z$ with time-varying coefficients in Section~\ref{s: extend thm}.}
Additional simulation studies are provided in Section \ref{appendx: simu}.

\section{Optimization Algorithm With Local Sensitivity Analysis}
\label{appendx: algorithm}

\new{In this section, we first derive two types of local sensitivity analysis that  can be used to compute the gradient of the log-likelihood function when it contains the solution of a general ODE. When the ODE is separable in the model formulation, we introduce a trick to further accelerate the evaluation of the objective for $n$ independent observations in subsection~\ref{appendix subs separable ODE}.}

We consider any parameterized survival model in the form of 
\begin{align}
\label{parametrized ode}
\left\{
\begin{array}{lr}
\dev \Lambda_{x}(t)/ \dev t = f(t, \Lambda_{x}(t);x, \theta) \\
\Lambda_{x}(t_0) = c(x, \theta)
\end{array}
\right.,
\end{align}
where $\theta$ denotes all the parameters. For example, for the general class of ODE models in  (\ref{ode: spline}), the function $f$ is given by the right hand side of (\ref{ode: spline}), the parameter $\theta$ consists of $\beta$, $a$, and $b$, the initial time point $t_0=0$, and the initial value $c(\cdot)$ equals to zero. Denote the solution of (\ref{parametrized ode}) by $\Lambda_x(t;\theta)$. Then under the non-informative censoring, the log-likelihood function is given by 
\[l_n(\theta)=\frac{1}{n}\sum_{i=1}^n \left[\Delta_i \log f\left(Y_i,\Lambda_{X_i}(Y_i;\theta);X_i, \theta \right) -\Lambda_{X_i}(Y_i;\theta)\right]. \] 
To obtain the maximum likelihood estimator, we propose a gradient-based optimization algorithm which utilizes the local sensitivity analysis to compute the gradient. By applying the chain rule, the gradient is given by
\new{\[\frac{\dev\  l_n(\theta)}{\dev \theta}=\frac{1}{n}\sum_{i=1}^n \left\{\left[\Delta_i \frac{ f'_2\left(Y_i,\Lambda_{X_i}(Y_i;\theta);X_i, \theta \right)}{f\left(Y_i,\Lambda_{X_i}(Y_i;\theta);X_i, \theta \right)} - 1 \right]\frac{\partial \Lambda_{x_i}(Y_i;\theta)}{\partial \theta} + \Delta_i \frac{ f'_4\left(Y_i,\Lambda_{X_i}(Y_i;\theta);X_i, \theta \right)}{f\left(Y_i,\Lambda_{X_i}(Y_i;\theta);X_i, \theta \right)}\right\},\]
where we use the subscript 2 and 4 in the derivatives to indicate that the derivatives are taken with respect to the first and the fourth argument of the function $f$ respectively. }
Then as long as we can derive the gradient of $\Lambda_x(y;\theta)$ with respect to $\theta$ for a given $y$, we can obtain the gradient of the likelihood function for faster gradient-based computations.

There are two commonly used types of local sensitivity analyses: forward sensitivity analysis and adjoint sensitivity analysis \citep{Dickinson_1976, petzold2006sensitivity}. We first derive the corresponding ODE for the forward sensitivity analysis. Denote the partial derivatives of $f(t, \Lambda;x, \theta)$ with respect to $\theta$ and $\Lambda$ by $f'_{\theta}$ and $f'_{\Lambda}$, respectively. Under certain smoothness condition of $f$, there is one unique solution $\Lambda_x(t;\theta)$ of (\ref{parametrized ode}) and it satisfies 
\[\Lambda_x(t;\theta)=\int^t_{t_0}f(s, \Lambda_x(s;\theta);x,\theta)\dev s + c(x,\theta). \]
By interchanging the integral and partial differential operators, it follows that  
\begin{align}
	\frac{\partial \Lambda_x(t;\theta)}{\partial \theta} &= \frac{\partial}{\partial \theta}\int^t_{t_0}f(s, \Lambda_x(s;\theta);x,\theta)\dev s + c'_{\theta}(x,\theta)\label{eq: two dev}\\
	& = \int^t_{t_0}\left(f'_{\theta}(s, \Lambda_x(s;\theta);x,\theta) +f'_{\Lambda}(s, \Lambda_x(s;\theta);x,\theta) \frac{\partial \Lambda_x(s;\theta)}{\partial \theta}\right) \dev s+ c'_{\theta}(x,\theta),\nonumber
\end{align}
where $c'_{\theta}(x,\theta)$ is the derivative of $c(x, \theta)$ with respect to $\theta$. Therefore, $\partial \Lambda_x(y;\theta)/\partial \theta=F_1(y)$ with $F_1$ satisfying 
\begin{align}
\label{forward sensitivity analysis}
\left\{
\begin{array}{lr}
\dev F_1(t)/ \dev t = f'_{\theta}(t, \Lambda_x(t;\theta);x,\theta) +f'_{\Lambda}(t, \Lambda_x(t;\theta);x,\theta)\cdot F_1\\
F_1(t_0) = c'_{\theta}(x, \theta)
\end{array}
\right..
\end{align}
After plugging $t_0=0$ and $c(\cdot)=0$, (\ref{forward sensitivity analysis}) becomes the initial value problem (\ref{I}) in Section \ref{s: mle}.

Next, we derive the corresponding ODE for the adjoint sensitivity analysis. Since $\Lambda_x(t;\theta)$ is solution of (\ref{parametrized ode}), for some appropriately chosen differentiable function $\kappa(t,\theta)$ to be specified later, we have
\begin{align*}
	\Lambda_x(t;\theta)&=\Lambda_x(t;\theta) - \int_{t_0}^t\kappa(s,\theta)\Big[\frac{\partial \Lambda_x(s;\theta)}{\partial s} - f(s, \Lambda_x(s,\theta); x,\theta)\Big]\dev s.
\end{align*}
By taking derivatives with respect to $\theta$ on both sides, it follows that
\begin{align*}
	\frac{\partial \Lambda_x(t;\theta)}{\partial \theta} & = \frac{\partial \Lambda_x(t;\theta)}{\partial \theta} - \frac{\partial}{\partial \theta}\int_{t_0}^t\kappa(s,\theta)\Big[\frac{\partial \Lambda_x(s;\theta)}{\partial s} - f(s, \Lambda_x(s,\theta); x,\theta)\Big]\dev s\\
	& = \frac{\partial \Lambda_x(t;\theta)}{\partial \theta} - \int_{t_0}^t\kappa(s,\theta)\frac{\partial}{\partial \theta}\Big[\frac{\partial \Lambda_x(s;\theta)}{\partial s} - f(s, \Lambda_x(s,\theta); x,\theta)\Big]\dev s\\
	& = \int^t_{t_0}(1+\kappa(s,\theta))\frac{\partial}{\partial \theta}f(s, \Lambda_x(s;\theta);x,\theta)\dev s +c'_{\theta}(x,\theta) - \int_{t_0}^t\kappa(s,\theta)\frac{\partial}{\partial s}\Big[\frac{\partial \Lambda_x(s;\theta)}{\partial \theta}\Big]\dev s,
\end{align*}
where the second equality holds because $$\int_{t_0}^t \frac{\partial \kappa(s,\theta)}{\partial \theta}\Big[\frac{\partial \Lambda_x(s;\theta)}{\partial s} - f(s, \Lambda_x(s,\theta); x,\theta)\Big]\dev s=0,$$ 
and the last equality holds by plugging (\ref{eq: two dev}) and exchanging the order of derivatives. Using integral by parts, we have
\[\int_{t_0}^t\kappa(s,\theta)\frac{\dev}{\dev s}\Big[\frac{\dev \Lambda_x(s;\theta)}{\dev \theta}\Big]\dev s + \int_{t_0}^t\frac{\dev\kappa(s,\theta)}{\dev s}\frac{\dev \Lambda_x(s;\theta)}{\dev \theta}\dev s = \Big(\kappa(s,\theta)\frac{\dev \Lambda_x(s;\theta)}{\dev \theta}\Big)\Big|^{t}_{t_0}.\]
Then it follows that
\begin{align*}
	\frac{\dev \Lambda_x(t;\theta)}{\dev \theta} & =\int^t_{t_0}(1+\kappa(s,\theta))\left(f'_{\theta}(s, \Lambda_x(s;\theta);x,\theta) +f'_{\Lambda}(s, \Lambda_x(s;\theta);x,\theta) \frac{\dev \Lambda_x(s;\theta)}{\dev \theta}\right)\dev s +c'_{\theta}(x,\theta)\\
	& \ \ \ \   +\int_{t_0}^t\frac{\dev\kappa(s,\theta)}{\dev s}\frac{\dev \Lambda_x(s;\theta)}{\dev \theta}\dev s - \Big(\kappa(s,\theta)\frac{\dev \Lambda_x(s;\theta)}{\dev \theta}\Big)\Big|^{t}_{t_0}\\
	& = \int^t_{t_0}(1+\kappa(s,\theta))f'_{\theta}(s, \Lambda_x(s;\theta);x,\theta) \dev s \\
	& \ \ \ \  + \int_{t_0}^t \frac{\dev \Lambda_x(s;\theta)}{\dev \theta} \left( \frac{\dev\kappa(s,\theta)}{\dev s} + (1+\kappa(s,\theta)) f'_{\Lambda}(s, \Lambda_x(s;\theta);x,\theta) \right)\dev s\\
	& \ \ \ \ + (1+\kappa(t_0, \theta))c'_{\theta}(x,\theta) - \kappa(t,\theta)\frac{\dev \Lambda_x(t;\theta)}{\dev \theta}.
\end{align*}
Denote  $\tilde{\kappa}(t,\theta)\triangleq \kappa(t,\theta)+1$ and choose proper $\tilde{\kappa}(t,\theta)$ that satisfies  
\begin{equation}
\label{adjoint}
\left\{
\begin{array}{lr}
\dev \tilde{\kappa}(t,\theta)/\dev t = -\tilde{\kappa}(t;\theta)f'_{\Lambda}(t, \Lambda_x(t,\theta);x,\theta)\\
\tilde{\kappa}(y;\theta) = 1
\end{array}
\right.,
\end{equation}
then the gradient of $\Lambda_x(t,\theta)$ with respect to $\theta$ is given by
\[\frac{\dev \Lambda_x(y;\theta)}{\dev \theta} = \int^y_{t_0}\tilde{\kappa}(s,\theta)f'_{\theta}(s, \Lambda_x(s;\theta);x,\theta) \dev s +\tilde{\kappa}(t_0, \theta)c'_{\theta}(x,\theta). \]
After plugging $t_0=0$ and $c(\cdot)=0$, the above equation becomes 
\[\frac{\dev \Lambda_x(y;\theta)}{\dev \theta}= \int^y_{0}\tilde{\kappa}(s,\theta) f'_{\theta}(s, \Lambda_x(s;\theta);x,\theta) \dev s.\]
Together with (\ref{adjoint}), it shows that the solution of (\ref{II}) at $t=0$ gives the gradient of $\Lambda_x(y;\theta)$ with respect to $\theta$. Note that  to solve (\ref{II}) at $t=0$, it requires evaluating the entire trajectory of $\Lambda_x(t,\theta)$ from $y$ to $0$. In our implementation, we combine ODEs (\ref{parametrized ode}) and (\ref{II}) into a larger ODE system, i.e.,
\begin{equation*}
\left\{
\begin{array}{lr}
(\Lambda'(t);\kappa'(t); F'_{2}(t))  = (f(t, \Lambda;\theta);-\kappa\cdot f'_{\Lambda}(t, \Lambda;\theta); -\kappa\cdot f'_{\theta}(t, \Lambda;\theta) ) \\
(\Lambda(t);\kappa(t); F_{2}(t))|_{t=y} = (\Lambda_x(y;\theta);1;\mb{0})
\end{array}
\right.,
\end{equation*}
and evaluate it at $t=0$, where $\Lambda_x(y;\theta)$ is available when computing the likelihood function.
As discussed in Section 3 of the main text, the proposed  estimation methods can be easily implemented using existing computing packages.

\subsection{\new{Acceleration trick for simultaneously solving separable ODEs for $n$ independent observations}}
\label{appendix subs separable ODE}

\new{Recall that evaluating the log-likelihood function requires solving ODEs for $n$ independent observations. For a general ODE model, as suggested in Remark~\ref{rmk: ode}, we can use either  the adjoint method along with parallel computing or the forward method by combining $n$ ODEs into a large ODE system with $n$ dimensions. The complexity of both methods scales linearly with the sample size. We further introduce a trick to reduce the absolute magnitude of computing time for separable ODEs, which cover the general class of ODE models in (\ref{model: ltm + cox}) as a special case.} 

\new{Specifically, we consider the separable ODE model in the form of 
\begin{align}
\label{parametrized separable ode}
\left\{
\begin{array}{lr}
\dev \Lambda_{x}(t)/ \dev t = f_1(t;x, \theta_1)\cdot f_2(\Lambda_x;\theta_2) \\
\Lambda_{x}(t_0) = c
\end{array}
\right.,
\end{align}
with two functions $f_1$ and $f_2$. In particular, for the general class of ODE models in (\ref{ode: spline}), $f_1(t;x,z,\theta_1)=\exp(x^T\beta + \sum_{l=0}^{d_2}\sum_{j=1}^{q_n^1}a_j^l B_j^1(t)z_{l})$ and $f_2(\Lambda_{x,z};\theta_2)= \exp(\sum_{j=1}^{q_n^2}b_j B_j^2(\Lambda_{x,z}(t)))$. For $n$ independent observations $\{\Delta_i, X_i, Y_i\}_{i=1}^n$, we need to evaluate the solution of $n$ different ODEs in~(\ref{parametrized separable ode}), each of which is associated with $X_i$, at their respective observed times $Y_i$. The acceleration trick is based on the key observation that solving (\ref{parametrized separable ode}) at $y$ is equivalent to solving the problem
\begin{align}
\label{repara separable ode}
\left\{
\begin{array}{lr}
\dev G(t)/ \dev t =  f_2(G;\theta_2) \\
G(t_0) = c
\end{array}
\right.
\end{align}
at $\int^{y}_{t_0} f_1(t;x, \theta_1) \dev t + t_0$, i.e., \[\Lambda_x(y;\theta_1, \theta_2) = G(\int^{y}_{t_0} f_1(t;x , \theta_1) \dev t + t_0 ;\theta_2).\] Therefore, we can instead solve a single ODE (\ref{repara separable ode}) at $n$ different points $\{\int^{Y_i}_{t_0} f_1(t;X_i, \theta_1) \dev t + t_0\}_{i=1}^n$ to compute $\Lambda_{X_i}(Y_i; \theta_1, \theta_2)$ for $1\le i \le n$. Moreover, given $\Lambda_{X_i}(Y_i; \theta_1, \theta_2)$, the gradient of $\Lambda_{X_i}(Y_i;\theta_1, \theta_2)$ with respect to $\theta_1$ can be computed by 
\[\frac{\partial \Lambda_{X_i}(Y_i;\theta_1, \theta_2) }{ \partial \theta_1} = f_2(\Lambda_{X_i}(Y_i;\theta_1, \theta_2);\theta_2) \int^{Y_i}_{t_0} \frac{\partial f_1(t;X_i, \theta_1)}{\partial \theta_1} \dev t.\]
And we can obtain the gradient of $\Lambda_{X_i}(Y_i;\theta_1, \theta_2)$ with respect to $\theta_2$ by solving another single ODE at $n$ different points: 
\[\frac{\partial \Lambda_{X_i}(Y_i;\theta_1, \theta_2) }{ \partial \theta_2} = G_2(\int^{y}_{t_0} f_1(t;x , \theta_1) \dev t + t_0; \theta_2),\]
where $\tilde{G}(\cdot; \theta_2)$ is the solution of
\begin{align*}
\left\{
\begin{array}{lr}
\dev \tilde{G}(t)/ \dev t = {f_{2}}'_{\theta_2}(G;\theta_2) +{f_2}'_{G}(G;\theta_2)\cdot \tilde{G}\\
\tilde{G}(t_0) = 0
\end{array}
\right..
\end{align*}
Based on our experiments, the proposed acceleration trick can significantly reduce the absolute computing time of simultaneously solving separable ODEs for $n$ independent observations.}

\section{The General Sieve M-theorem for Bundled Parameters (Theorem \ref{thm: m-theorem}) and Its Proof}
\label{appendx: m-theorem}
\new{In this section, we establish a new general sieve M-theorem for studying the asymptotic normality of M-estimators when the estimation criterion is parameterized with more general bundled parameters. Note that the proposed M-theorem significantly differs from Theorem 2.1 in \citet{ding2011} and Theorem 6.1 in \citet{wellner2007}. They consider either well-separated parameters \citep{wellner2007} or bundled parameters where the nuisance parameter can be a function of only the finite-dimensional parameters \citep{ding2011}; while we consider a more general scenario of bundled parameters where the nuisance parameter can be a function of both the finite-dimensional parameter $\beta$ and other infinite-dimensional parameters. Therefore, the proposed theorem nontrivially extends the asymptotic distributional theories for M-estimation under this general scenario and is crucial for studying the asymptotic normality of the sieve MLE for the general ODE model in (2).}

Specifically, given i.i.d. observations $W_1, \cdots, W_n \in \mathcal{W}$, we maximize an objective function \[\frac{1}{n}\sum_{1}^n m(\beta, \boldsymbol\gamma(\cdot), \zeta(\cdot, \beta, \boldsymbol\gamma);W_i)\] to estimate the unknown parameters $(\beta, \boldsymbol\gamma(\cdot), \zeta(\cdot, \beta, \boldsymbol\gamma))$. Here $\beta \in \mb{R}^{d_1}$ denotes the finite-dimensional parameter of interest, \new{$\boldsymbol\gamma(\cdot) = (\gamma_1(\cdot), \dots, \gamma_{d_2}(\cdot))$} denotes nuisance infinite-dimensional parameters and $\zeta(\cdot, \beta, \boldsymbol\gamma)$ denotes another nuisance infinite-dimensional parameter that is a function of $\beta$ and $\boldsymbol\gamma(\cdot)$. \new{To accommodate this different and challenging scenario bundled parameters, we develop a new general sieve M-theorem. We firstly introduce notation in Section~\ref{subs: m-theorem notation}, and establish the asymptotic normality of the sieve estimator that maximizes the objective function over some sieve parameter space in Section~\ref{subs: m-theorem}.}

\subsection{Notation}
\label{subs: m-theorem notation}
\new{Here we follow notation used in \citet{ding2011} and \citet{wellner2007}.} 
Let $\theta = (\beta, \boldsymbol\gamma(\cdot), \zeta(\cdot, \beta, \boldsymbol\gamma))$, $\beta \in \mathcal{B} \subset \mb{R}^{d_1}$, \new{$\boldsymbol\gamma \in \Gamma^{d_2}$}, and $\zeta \in \mathcal{H}$, where $\mathcal{B}$ is the parameter space of $\beta$, $\Gamma$ is a class of functions mapping from $\mathcal{W}$ to $\mb{R}$ and $\mathcal{H}$ is a class of functions mapping from $\mathcal{W} \times \mathcal{B} \times \Gamma^{d_2}$ to $\mb{R}$. Let $\Theta = \mathcal{B}\times \Gamma^{d_2} \times \mathcal{H}$ be the parameter space of $\theta$. The distance between $\theta_1$ and $\theta_2 \in \Theta$ is defined as \[d(\theta_1, \theta_2) = \{\|\beta_1-\beta_2\|^2+\sum_{j=1}^{d_2}\|\gamma_{1j}-\gamma_{2j}\|_{\Gamma}^2+ \|\zeta_1(\cdot, \beta_1, \boldsymbol\gamma_1) - \zeta_2(\cdot, \beta_2, \boldsymbol\gamma_2)\|_{\mc{H}}^2 \}^{1/2},\] where $\|\cdot\|$ is the Euclidean norm, $\|\cdot\|_{\Gamma}$ is some norm of $\Gamma$, and $\|\cdot \|_{\mc{H}}$ is some norm of $\mathcal{H}$. Let $\Theta_n$ be the sieve parameter space, where $\Theta_n\subset \Theta_{n+1}\subset \cdots \subset \Theta$ and the sequence becomes dense as $n\rightarrow \infty$. We obtain the sieve M-estimator $\hat{\theta}_n = (\hat{\beta}_n, \hat{\boldsymbol\gamma}_n, \hat{\zeta}_n(\cdot, \hat{\beta}_n, \hat{\boldsymbol\gamma}_n))\in \Theta_n$ by maximizing the objective function over the sieve parameter space. We study the asymptotic normality of the sieve M-estimator of the Euclidean parameter of interest, $\hat{\beta}_n$, as follows.

For any fixed $\gamma(\cdot)\in \Gamma$, let $\{\gamma_\eta(\cdot): \eta \text{ in a neighborhood of }0\in\mb{R}\}$ be a smooth curve in $\Gamma$ running through $\gamma(\cdot)$ at $\eta=0$, that is $\gamma_\eta(\cdot)|_{\eta=0}=\gamma(\cdot)$. Similarly, for any fixed $\zeta(\cdot, \beta, \boldsymbol\gamma) \in \mathcal{H}$, let $\{\zeta_{\eta}(\cdot, \beta, \boldsymbol\gamma): \eta$ in a neighborhood of $0\in \mb{R} \}$ be a smooth curve in $\mathcal{H}$ running through $\zeta(\cdot, \beta, \boldsymbol\gamma)$ at $\eta=0$, that is $\zeta_{\eta}(\cdot, \beta, \boldsymbol\gamma)|_{\eta=0} = \zeta(\cdot, \beta, \boldsymbol\gamma)$. Assume all $\zeta(\cdot, \beta, \boldsymbol\gamma) \in \mathcal{H}$ are twice Frechet differentiable with respect to $\beta$ and $\boldsymbol\gamma$, and denote 
\[\mathbb{V}=\{v: v(\cdot)=\frac{\partial \gamma_{\eta}(\cdot)}{\partial \eta}|_{\eta=0} , \gamma_{\eta}\in \Gamma \},\]
\[\mathbb{H}=\{h: h(\cdot, \beta, \boldsymbol\gamma)=\frac{\partial \zeta_{\eta}(\cdot, \beta, \boldsymbol\gamma)}{\partial \eta}|_{\eta=0} , \zeta_{\eta}\in \mc{H}, \beta\in \mc{B}, \boldsymbol\gamma \in \Gamma^{d_2}\}.\]
Assume the objective function $m$ is twice Frechet differentiable. For $1\le j\le d_2$, we use the subscript $1$, $2^{(j)}$ or $3$ in the derivatives to indicate that the derivatives are taken with respect to the first, the $j$-th component of the second or the third argument of the function, respectively. We use function $v$ or $h$ inside the square brackets to denote the direction of the functional derivative with respect to $\gamma_j$ or $\zeta$. Since for a small $\delta$, we have $\zeta(\cdot, \beta+\delta, \boldsymbol\gamma) - \zeta(\cdot, \beta, \boldsymbol\gamma)= \zeta'_{\beta}(\cdot, \beta, \boldsymbol\gamma)\delta+o(\delta)$, where $\zeta'_{\beta}(\cdot, \beta, \boldsymbol\gamma) =  {\partial \zeta(\cdot, \beta, \boldsymbol\gamma)}/{\partial \beta}$; then as shown in \citet{ding2011} on page 3036, it follows that 
\begin{align*}
	\lim_{\delta\rightarrow 0} ~& \frac{1}{\delta}\{ m(\beta, \boldsymbol\gamma(\cdot), \zeta(\cdot, \beta+\delta, \boldsymbol\gamma);W) - m(\beta, \boldsymbol\gamma(\cdot), \zeta(\cdot, \beta, \boldsymbol\gamma);W)\} \\
	& = m'_3(\beta, \boldsymbol\gamma(\cdot), \zeta(\cdot, \beta, \boldsymbol\gamma);W)[\zeta'_{\beta}(\cdot, \beta, \boldsymbol\gamma)],
\end{align*}
\begin{align*}
	\lim_{\delta\rightarrow 0}~ & \frac{1}{\delta}\{ m'_3(\beta, \boldsymbol\gamma(\cdot), \zeta(\cdot, \beta+\delta, \boldsymbol\gamma);W)[h(\cdot, \beta, \boldsymbol\gamma)] - m'_3(\beta, \boldsymbol\gamma(\cdot), \zeta(\cdot, \beta, \boldsymbol\gamma);W)[h(\cdot, \beta, \boldsymbol\gamma)]\} \\
	& = m''_{33}(\beta, \boldsymbol\gamma(\cdot), \zeta(\cdot, \beta, \boldsymbol\gamma);W)[h(\cdot, \beta, \boldsymbol\gamma),\zeta'_{\beta}(\cdot, \beta, \boldsymbol\gamma)],
\end{align*}
\begin{align*}
	\lim_{\delta\rightarrow 0}~ & \frac{1}{\delta}\{ m'_{2^{(j)}}(\beta, \boldsymbol\gamma(\cdot), \zeta(\cdot, \beta+\delta, \boldsymbol\gamma);W)[v] - m'_2(\beta, \boldsymbol\gamma(\cdot), \zeta(\cdot, \beta, \boldsymbol\gamma);W)[v]\} \\
	& = m''_{2^{(j)}3}(\beta, \boldsymbol\gamma(\cdot), \zeta(\cdot, \beta, \boldsymbol\gamma);W)[v,\zeta'_{\beta}(\cdot, \beta, \boldsymbol\gamma)],~\text{ for }1\le j\le d_2,
\end{align*}
and 
\begin{align*}
	\lim_{\delta\rightarrow 0}~ & \frac{1}{\delta}\{ m'_3(\beta, \boldsymbol\gamma(\cdot), \zeta(\cdot, \beta, \boldsymbol\gamma);W)[h(\cdot, \beta+\delta, \boldsymbol\gamma)] - m'_3(\beta, \boldsymbol\gamma(\cdot), \zeta(\cdot, \beta, \boldsymbol\gamma);W)[h(\cdot, \beta, \boldsymbol\gamma)]\} \\
	& = m'_3(\beta, \boldsymbol\gamma(\cdot), \zeta(\cdot, \beta, \boldsymbol\gamma);W)[h'_{\beta}(\cdot, \beta, \boldsymbol\gamma)].
\end{align*}

\new{Let $e_j = (0,\dots, 1, \dots, 0)\in \mb{R}^{d_2}$ with the $j$-th element being $1$.} For $1\le j\le d_2$, we have $\zeta(\cdot, \beta, \boldsymbol\gamma+v \cdot e_j)- \zeta(\cdot, \beta, \boldsymbol\gamma) = \zeta'_{\gamma_j}(\cdot, \beta, \boldsymbol\gamma)[v]+o(\|v\|_{\Gamma})$ for a small $v$; then by the definition of functional derivatives, it follows that, for $1\le j\le d_2,$
\begin{align*}
	 & m(\beta, \boldsymbol\gamma(\cdot), \zeta(\cdot, \beta, \boldsymbol\gamma+v\cdot e_j);W) - m(\beta, \boldsymbol\gamma(\cdot), \zeta(\cdot, \beta, \boldsymbol\gamma);W) \\
	 & = m(\beta, \boldsymbol\gamma(\cdot), \zeta(\cdot, \beta, \boldsymbol\gamma) + \zeta'_{\gamma_j}(\cdot, \beta, \boldsymbol\gamma)[v]+o(\|v\|_{\Gamma});W) - m(\beta, \boldsymbol\gamma(\cdot), \zeta(\cdot, \beta, \boldsymbol\gamma);W)\\
	 & = \{m(\beta, \boldsymbol\gamma(\cdot), \zeta(\cdot, \beta, \boldsymbol\gamma) + \zeta'_{\gamma_j}(\cdot, \beta, \boldsymbol\gamma)[v]+o(\|v\|_{\Gamma});W) - m(\beta, \boldsymbol\gamma(\cdot), \zeta(\cdot, \beta, \boldsymbol\gamma) + \zeta'_{\gamma_j}(\cdot, \beta, \boldsymbol\gamma)[v];W)\} \\
	 & \ \ \ \ + \{m(\beta, \boldsymbol\gamma(\cdot), \zeta(\cdot, \beta, \boldsymbol\gamma) + \zeta'_{\gamma_j}(\cdot, \beta, \boldsymbol\gamma)[v];W) - m(\beta, \boldsymbol\gamma(\cdot), \zeta(\cdot, \beta, \boldsymbol\gamma);W)\}\\
	 & = m'_3(\beta, \boldsymbol\gamma(\cdot), \zeta(\cdot, \beta, \boldsymbol\gamma) + \zeta'_{\gamma_j}(\cdot, \beta, \boldsymbol\gamma)[v];W)[o(\|v\|_{\Gamma})] + \\
	 & \ \ \ \ m'_3(\beta, \boldsymbol\gamma(\cdot), \zeta(\cdot, \beta, \boldsymbol\gamma);W)[\zeta'_{\gamma_j}(\cdot, \beta, \boldsymbol\gamma)[v]] + o(\|\zeta'_{\gamma_j}(\cdot, \beta, \boldsymbol\gamma)[v]\|_{\Gamma})\\
	 & = m'_3(\beta, \boldsymbol\gamma(\cdot), \zeta(\cdot, \beta, \boldsymbol\gamma);W)[\zeta'_{\gamma_j}(\cdot, \beta, \boldsymbol\gamma)[v]] + o(\|v\|_{\Gamma}),
\end{align*}
where the last equality holds because
\[\lim_{v\rightarrow 0}m(\beta, \boldsymbol\gamma(\cdot), \zeta(\cdot, \beta, \boldsymbol\gamma) + \zeta'_{\gamma_j}(\cdot, \beta, \boldsymbol\gamma)[v];W)\Big[\frac{o(\|v\|_{\Gamma})}{\|v\|_{\Gamma}}\Big] = 0,\]
and $o(\|\zeta'_{\gamma_j}(\cdot, \beta, \boldsymbol\gamma)[v]\|_{\Gamma}) = o(\|v\|_{\Gamma})$ for bounded functional derivatives.
Similarly we have for $1\le j,\ell\le d_2,$
\begin{align*}
	& m'_{2^{(j)}}(\beta, \boldsymbol\gamma(\cdot), \zeta(\cdot, \beta, \boldsymbol\gamma+v\cdot e_\ell);W)[v_1] - m'_{2^{(j)}}(\beta, \boldsymbol\gamma(\cdot), \zeta(\cdot, \beta, \boldsymbol\gamma);W)[v_1] \\
	& = m''_{2^{(j)}3}(\beta, \boldsymbol\gamma(\cdot), \zeta(\cdot, \beta, \boldsymbol\gamma);W)[v_1, \zeta'_{\gamma_\ell}(\cdot, \beta, \boldsymbol\gamma)[v]] + o(\|v\|_{\Gamma}),
\end{align*}
\begin{align*}
	& m'_3(\beta, \boldsymbol\gamma(\cdot), \zeta(\cdot, \beta, \boldsymbol\gamma+v\cdot e_j);W)[h(\cdot, \beta, \boldsymbol\gamma)] - m'_3(\beta, \boldsymbol\gamma(\cdot), \zeta(\cdot, \beta, \boldsymbol\gamma);W)[h(\cdot, \beta, \boldsymbol\gamma)]\\
	& = m''_{33}(\beta, \boldsymbol\gamma(\cdot), \zeta(\cdot, \beta, \boldsymbol\gamma);W)[h(\cdot, \beta, \boldsymbol\gamma), \zeta'_{\gamma_j}(\cdot, \beta, \boldsymbol\gamma)[v]] + o(\|v\|_{\Gamma}),
\end{align*}
\begin{align*}
	& m'_3(\beta, \boldsymbol\gamma(\cdot), \zeta(\cdot, \beta, \boldsymbol\gamma);W)[h(\cdot, \beta, \boldsymbol\gamma+v\cdot e_j)] - m'_3(\beta, \boldsymbol\gamma(\cdot), \zeta(\cdot, \beta, \boldsymbol\gamma);W)[h(\cdot, \beta, \boldsymbol\gamma)]\\
	& = m'_{3}(\beta, \boldsymbol\gamma(\cdot), \zeta(\cdot, \beta, \boldsymbol\gamma);W)[h'_{\gamma_j}(\cdot, \beta, \boldsymbol\gamma)[v]] + o(\|v\|_{\Gamma}).
\end{align*}
Based on the chain rule of the functional derivative, we have for $1\le j,\ell\le d_2,$
\begin{align*}
	m'_{\beta}(\beta, \boldsymbol\gamma(\cdot), \zeta(\cdot, \beta, \boldsymbol\gamma);W) & = \frac{\partial m(\beta, \boldsymbol\gamma(\cdot), \zeta(\cdot, \beta, \boldsymbol\gamma);W)}{\partial \beta}\\
	& = m'_1(\beta, \boldsymbol\gamma(\cdot), \zeta(\cdot, \beta, \boldsymbol\gamma);W) + m'_3(\beta, \boldsymbol\gamma(\cdot), \zeta(\cdot, \beta, \boldsymbol\gamma);W)[\zeta'_{\beta}(\cdot, \beta, \boldsymbol\gamma)],\\
m'_{\gamma_j}(\beta, \boldsymbol\gamma(\cdot), \zeta(\cdot, \beta, \boldsymbol\gamma);W)[v] &= m'_{2^{(j)}}(\beta, \boldsymbol\gamma(\cdot), \zeta(\cdot, \beta, \boldsymbol\gamma);W)[v]+m'_3(\beta, \boldsymbol\gamma(\cdot), \zeta(\cdot, \beta, \boldsymbol\gamma);W)[\zeta'_{\gamma_j}(\cdot, \beta, \boldsymbol\gamma)[v]],\\
	m'_{\zeta}(\beta, \boldsymbol\gamma(\cdot), \zeta(\cdot, \beta, \boldsymbol\gamma);W)[h] &= m'_3(\beta, \boldsymbol\gamma(\cdot), \zeta(\cdot, \beta, \boldsymbol\gamma);W)[h(\cdot, \beta, \boldsymbol\gamma)],
\end{align*}
\begin{align*}
	m''_{\beta\beta}(\beta, \boldsymbol\gamma(\cdot), \zeta(\cdot, \beta, \boldsymbol\gamma);W) & = \frac{\partial m'_{\beta}(\beta, \boldsymbol\gamma(\cdot), \zeta(\cdot, \beta, \boldsymbol\gamma);W)}{\partial \beta}\\
	& = m''_{11}(\beta, \boldsymbol\gamma(\cdot), \zeta(\cdot, \beta, \boldsymbol\gamma);W) + m''_{13}(\beta, \boldsymbol\gamma(\cdot), \zeta(\cdot, \beta, \boldsymbol\gamma);W)[\zeta'_{\beta}(\cdot, \beta, \boldsymbol\gamma)]\\
	& \ \ \ \ + m''_{31}(\beta, \boldsymbol\gamma(\cdot), \zeta(\cdot, \beta, \boldsymbol\gamma);W)[\zeta'_{\beta}(\cdot, \beta, \boldsymbol\gamma)] \\
	& \ \ \ \ + m''_{33}(\beta, \boldsymbol\gamma(\cdot), \zeta(\cdot, \beta, \boldsymbol\gamma);W)[\zeta'_{\beta}(\cdot, \beta, \boldsymbol\gamma), \zeta'_{\beta}(\cdot, \beta, \boldsymbol\gamma)]\\
	& \ \ \ \ + m'_{3}(\beta, \boldsymbol\gamma(\cdot), \zeta(\cdot, \beta, \boldsymbol\gamma);W)[\zeta''_{\beta\beta}(\cdot, \beta, \boldsymbol\gamma)],
\\
	m''_{\gamma_j \gamma_\ell}(\beta, \boldsymbol\gamma(\cdot), \zeta(\cdot, \beta, \boldsymbol\gamma);W)[v_1, v_2] & = m''_{2^{(j)}2^{(\ell)}}\beta, \boldsymbol\gamma(\cdot), \zeta(\cdot, \beta, \boldsymbol\gamma);W)[v_1, v_2] \\
	& \ \ \ \ + m''_{2^{(j)}3}(\beta, \boldsymbol\gamma(\cdot), \zeta(\cdot, \beta, \boldsymbol\gamma);W)[v_1, \zeta'_{\gamma_\ell}(\cdot, \beta, \boldsymbol\gamma)[v_2]]\\
	& \ \ \ \ + m''_{32^{(\ell)}}(\beta, \boldsymbol\gamma(\cdot), \zeta(\cdot, \beta, \boldsymbol\gamma);W)[\zeta'_{\gamma_j}(\cdot, \beta, \boldsymbol\gamma)[v_1], v_2] \\
	& \ \ \ \ + m''_{33}(\beta, \boldsymbol\gamma(\cdot), \zeta(\cdot, \beta, \boldsymbol\gamma);W)[\zeta'_{\gamma_j}(\cdot, \beta, \boldsymbol\gamma)[v_1], \zeta'_{\gamma_\ell}(\cdot, \beta, \boldsymbol\gamma)[v_2]]\\
	& \ \ \ \ + m'_{3}(\beta, \boldsymbol\gamma(\cdot), \zeta(\cdot, \beta, \boldsymbol\gamma);W)[\zeta''_{\gamma_j \gamma_\ell}(\cdot, \beta, \boldsymbol\gamma)[v_1, v_2]],
\\
m''_{\zeta\zeta}(\beta, \boldsymbol\gamma(\cdot), \zeta(\cdot, \beta, \boldsymbol\gamma);W)[h_1, h_2] &= m''_{33}(\beta, \boldsymbol\gamma(\cdot), \zeta(\cdot, \beta, \boldsymbol\gamma);W)[h_1(\cdot, \beta, \boldsymbol\gamma), h_2(\cdot, \beta, \boldsymbol\gamma)],
\end{align*}
\begin{align*}
	m''_{\gamma_j\beta}(\beta, \boldsymbol\gamma(\cdot), \zeta(\cdot, \beta, \boldsymbol\gamma);W)[v] & = \frac{\partial m'_{\gamma_j}(\beta, \boldsymbol\gamma(\cdot), \zeta(\cdot, \beta, \boldsymbol\gamma);W)[v]}{\partial \beta}\\
	& = m''_{2^{(j)}1}(\beta, \boldsymbol\gamma(\cdot), \zeta(\cdot, \beta, \boldsymbol\gamma);W)[v] \\
	& \ \ \ \ + m''_{2^{(j)}3}(\beta, \boldsymbol\gamma(\cdot), \zeta(\cdot, \beta, \boldsymbol\gamma);W)[v, \zeta'_{\beta}(\cdot, \beta, \boldsymbol\gamma)]\\
	& \ \ \ \ + m''_{31}(\beta, \boldsymbol\gamma(\cdot), \zeta(\cdot, \beta, \boldsymbol\gamma);W)[\zeta'_{\gamma_j}(\cdot, \beta, \boldsymbol\gamma)[v]]\\
	& \ \ \ \ + m''_{33}(\beta, \boldsymbol\gamma(\cdot), \zeta(\cdot, \beta, \boldsymbol\gamma);W)[\zeta'_{\gamma_j}(\cdot, \beta, \boldsymbol\gamma)[v], \zeta'_{\beta}(\cdot, \beta, \boldsymbol\gamma)]\\
	& \ \ \ \ + m'_{3}(\beta, \boldsymbol\gamma(\cdot), \zeta(\cdot, \beta, \boldsymbol\gamma);W)[\zeta''_{\gamma_j\beta}(\cdot, \beta, \boldsymbol\gamma)[v]]
\end{align*}
\begin{align*}
	m''_{\zeta\beta}(\beta, \boldsymbol\gamma(\cdot), \zeta(\cdot, \beta, \boldsymbol\gamma);W)[h] & = \frac{\partial m'_{\zeta}(\beta, \boldsymbol\gamma(\cdot), \zeta(\cdot, \beta, \boldsymbol\gamma);W)[h]}{\partial \beta}\\
	& = m''_{31}(\beta, \boldsymbol\gamma(\cdot), \zeta(\cdot, \beta, \boldsymbol\gamma);W)[h(\cdot, \beta, \boldsymbol\gamma)] \\
	& \ \ \ \ + m''_{33}(\beta, \boldsymbol\gamma(\cdot), \zeta(\cdot, \beta, \boldsymbol\gamma);W)[h(\cdot, \beta, \boldsymbol\gamma), \zeta'_{\beta}(\cdot, \beta, \boldsymbol\gamma)]\\
	& \ \ \ \ +  m'_3(\beta, \boldsymbol\gamma(\cdot), \zeta(\cdot, \beta, \boldsymbol\gamma);W)[h'_{\beta}(\cdot, \beta, \boldsymbol\gamma)],
\end{align*}
\begin{align*}
	m''_{\zeta\gamma_j}(\beta, \boldsymbol\gamma(\cdot), \zeta(\cdot, \beta, \boldsymbol\gamma);W)[h, v] & = m''_{32^{(j)}}(\beta, \boldsymbol\gamma(\cdot), \zeta(\cdot, \beta, \boldsymbol\gamma);W)[h(\cdot, \beta, \boldsymbol\gamma), v] \\
	& \ \ \ \ + m''_{33}(\beta, \boldsymbol\gamma(\cdot), \zeta(\cdot, \beta, \boldsymbol\gamma);W)[h(\cdot, \beta, \boldsymbol\gamma), \zeta'_{\gamma_j}(\cdot, \beta, \boldsymbol\gamma)[v]]\\
	& \ \ \ \ +  m'_3(\beta, \boldsymbol\gamma(\cdot), \zeta(\cdot, \beta, \boldsymbol\gamma);W)[h'_{\gamma_j}(\cdot, \beta, \boldsymbol\gamma)[v]],
\end{align*}
\begin{align*}
	m''_{\gamma_j\zeta}(\beta, \boldsymbol\gamma(\cdot), \zeta(\cdot, \beta, \boldsymbol\gamma);W)[v, h] & = m''_{2^{(j)}3}(\beta, \boldsymbol\gamma(\cdot), \zeta(\cdot, \beta, \boldsymbol\gamma);W)[v, h(\cdot, \beta, \boldsymbol\gamma)] \\
	& \ \ \ \ + m''_{33}(\beta, \boldsymbol\gamma(\cdot), \zeta(\cdot, \beta, \boldsymbol\gamma);W)[\zeta'_{\gamma_j}(\cdot, \beta, \boldsymbol\gamma)[v], h(\cdot, \beta, \boldsymbol\gamma)]\\
	& \ \ \ \ +  m'_3(\beta, \boldsymbol\gamma(\cdot), \zeta(\cdot, \beta, \boldsymbol\gamma);W)[h'_{\gamma_j}(\cdot, \beta, \boldsymbol\gamma)[v]].
\end{align*}
Following \citet{wellner2007}, we further define 
\[S'_{\beta}(\beta, \boldsymbol\gamma(\cdot), \zeta(\cdot, \beta, \boldsymbol\gamma))=Pm'_{\beta}(\beta, \boldsymbol\gamma(\cdot), \zeta(\cdot, \beta, \boldsymbol\gamma);W),\]
\[S'_{\gamma_j}(\beta, \boldsymbol\gamma(\cdot), \zeta(\cdot, \beta, \boldsymbol\gamma))[v] = Pm'_{\gamma_j}(\beta, \boldsymbol\gamma(\cdot), \zeta(\cdot, \beta, \boldsymbol\gamma);W)[v],\]
\[S'_{\zeta}(\beta, \boldsymbol\gamma(\cdot), \zeta(\cdot, \beta, \boldsymbol\gamma))[h] = Pm'_{\zeta}(\beta, \boldsymbol\gamma(\cdot), \zeta(\cdot, \beta, \boldsymbol\gamma);W)[h],\]
\[S'_{\beta,n}(\beta, \boldsymbol\gamma(\cdot), \zeta(\cdot, \beta, \boldsymbol\gamma))=\mathbb{P}_n m'_{\beta}(\beta, \boldsymbol\gamma(\cdot), \zeta(\cdot, \beta, \boldsymbol\gamma);W),\]
\[S'_{\gamma_j,n}(\beta, \boldsymbol\gamma(\cdot), \zeta(\cdot, \beta, \boldsymbol\gamma))[v] = \mathbb{P}_n m'_{\gamma_j}(\beta, \boldsymbol\gamma(\cdot), \zeta(\cdot, \beta, \boldsymbol\gamma);W)[v],\]
\[S'_{\zeta,n}(\beta, \boldsymbol\gamma(\cdot), \zeta(\cdot, \beta, \boldsymbol\gamma))[h] = \mathbb{P}_n m'_{\zeta}(\beta, \boldsymbol\gamma(\cdot), \zeta(\cdot, \beta, \boldsymbol\gamma);W)[h],\]
\[S''_{\beta\beta}(\beta, \boldsymbol\gamma(\cdot), \zeta(\cdot, \beta, \boldsymbol\gamma))=Pm''_{\beta\beta}(\beta, \boldsymbol\gamma(\cdot), \zeta(\cdot, \beta, \boldsymbol\gamma);W),\]
\[S''_{\gamma_j\gamma_\ell}(\beta, \boldsymbol\gamma(\cdot), \zeta(\cdot, \beta, \boldsymbol\gamma))[v_1, v_2] = Pm''_{\gamma_j\gamma_\ell}(\beta, \boldsymbol\gamma(\cdot), \zeta(\cdot, \beta, \boldsymbol\gamma);W)[v_1, v_2],\]
\[S''_{\zeta\zeta}(\beta, \boldsymbol\gamma(\cdot), \zeta(\cdot, \beta, \boldsymbol\gamma))[h_1, h_2] = Pm''_{\zeta\zeta}(\beta, \boldsymbol\gamma(\cdot), \zeta(\cdot, \beta, \boldsymbol\gamma);W)[h_1, h_2],\]
\[S''_{\gamma_j\beta}(\beta, \boldsymbol\gamma(\cdot), \zeta(\cdot, \beta, \boldsymbol\gamma))[v] = S''_{\beta\gamma_j}(\beta, \boldsymbol\gamma(\cdot), \zeta(\cdot, \beta, \boldsymbol\gamma))[v] = Pm''_{\gamma_j\beta}(\beta, \boldsymbol\gamma(\cdot), \zeta(\cdot, \beta, \boldsymbol\gamma);W)[v],\]
\[S''_{\zeta\beta}(\beta, \boldsymbol\gamma(\cdot), \zeta(\cdot, \beta, \boldsymbol\gamma))[h] = S''_{\beta\zeta}(\beta, \boldsymbol\gamma(\cdot), \zeta(\cdot, \beta, \boldsymbol\gamma))[h] = Pm''_{\zeta\beta}(\beta, \boldsymbol\gamma(\cdot), \zeta(\cdot, \beta, \boldsymbol\gamma);W)[h],\]
\[S''_{\zeta\gamma_j}(\beta, \boldsymbol\gamma(\cdot), \zeta(\cdot, \beta, \boldsymbol\gamma))[h, v] = Pm''_{\zeta\gamma_j}(\beta, \boldsymbol\gamma(\cdot), \zeta(\cdot, \beta, \boldsymbol\gamma);W)[h, v],\]
\[S''_{\gamma_j\zeta}(\beta, \boldsymbol\gamma(\cdot), \zeta(\cdot, \beta, \boldsymbol\gamma))[v, h] = Pm''_{\gamma_j\zeta}(\beta, \boldsymbol\gamma(\cdot), \zeta(\cdot, \beta, \boldsymbol\gamma);W)[v, h].\]
Furthermore, for $\mb{h}=(h_1, \cdots, h_{d_1})^T\in \mathbb{H}^{d_1}$ and  $\mb{v}=(v_1, \cdots, v_{d_1})^T\in \mathbb{V}^{d_1}$, denote that
\[m'_{\gamma_j}(\beta, \boldsymbol\gamma(\cdot), \zeta(\cdot, \beta, \boldsymbol\gamma);W)[\mb{v}] = (m'_{\gamma_j}(\beta, \boldsymbol\gamma(\cdot), \zeta(\cdot, \beta, \boldsymbol\gamma);W)[v_1], \cdots, m'_{\gamma_j}(\beta, \boldsymbol\gamma(\cdot), \zeta(\cdot, \beta, \boldsymbol\gamma);W)[v_{d_1}])^T,\]
\[m'_{\zeta}(\beta, \boldsymbol\gamma(\cdot), \zeta(\cdot, \beta, \boldsymbol\gamma);W)[\mb{h}] = (m'_{\zeta}(\beta, \boldsymbol\gamma(\cdot), \zeta(\cdot, \beta, \boldsymbol\gamma);W)[h_1], \cdots, m'_{\zeta}(\beta, \boldsymbol\gamma(\cdot), \zeta(\cdot, \beta, \boldsymbol\gamma);W)[h_{d_1}])^T,\]
\[m''_{\gamma_j\gamma_\ell}(\beta, \boldsymbol\gamma(\cdot), \zeta(\cdot, \beta, \boldsymbol\gamma);W)[\mb{v}, v] = (m''_{\gamma_j\gamma_\ell}(\beta, \boldsymbol\gamma(\cdot), \zeta(\cdot, \beta, \boldsymbol\gamma);W)[v_1, v], \cdots, m''_{\gamma_j\gamma_\ell}(\beta, \boldsymbol\gamma(\cdot), \zeta(\cdot, \beta, \boldsymbol\gamma);W)[v_{d_1}, v])^T,\]
\[m''_{\zeta\zeta}(\beta, \boldsymbol\gamma(\cdot), \zeta(\cdot, \beta, \boldsymbol\gamma);W)[\mb{h}, h] = (m''_{\zeta\zeta}(\beta, \boldsymbol\gamma(\cdot), \zeta(\cdot, \beta, \boldsymbol\gamma);W)[h_1, h], \cdots, m''_{\zeta\zeta}(\beta, \boldsymbol\gamma(\cdot), \zeta(\cdot, \beta, \boldsymbol\gamma);W)[h_{d_1}, h])^T,\]
\[m''_{\gamma_j\beta}(\beta, \boldsymbol\gamma(\cdot), \zeta(\cdot, \beta, \boldsymbol\gamma);W)[\mb{v}] = (m''_{\gamma_j\beta}(\beta, \boldsymbol\gamma(\cdot), \zeta(\cdot, \beta, \boldsymbol\gamma);W)[v_1], \cdots, m''_{\gamma_j\beta}(\beta, \boldsymbol\gamma(\cdot), \zeta(\cdot, \beta, \boldsymbol\gamma);W)[v_{d_1}])^T,\]
\[m''_{\zeta\beta}(\beta, \boldsymbol\gamma(\cdot), \zeta(\cdot, \beta, \boldsymbol\gamma);W)[\mb{h}] = (m''_{\zeta\beta}(\beta, \boldsymbol\gamma(\cdot), \zeta(\cdot, \beta, \boldsymbol\gamma);W)[h_1], \cdots, m''_{\zeta\beta}(\beta, \boldsymbol\gamma(\cdot), \zeta(\cdot, \beta, \boldsymbol\gamma);W)[h_{d_1}])^T,\]
\[m''_{\zeta\gamma_j}(\beta, \boldsymbol\gamma(\cdot), \zeta(\cdot, \beta, \boldsymbol\gamma);W)[\mb{h}, v] = (m''_{\zeta\gamma_j}(\beta, \boldsymbol\gamma(\cdot), \zeta(\cdot, \beta, \boldsymbol\gamma);W)[h_1, v], \cdots, m''_{\zeta\gamma_j}(\beta, \boldsymbol\gamma(\cdot), \zeta(\cdot, \beta, \boldsymbol\gamma);W)[h_{d_1}, v])^T,\]
\[m''_{\gamma_j\zeta}(\beta, \boldsymbol\gamma(\cdot), \zeta(\cdot, \beta, \boldsymbol\gamma);W)[\mb{v}, h] = (m''_{\gamma_j\zeta}(\beta, \boldsymbol\gamma(\cdot), \zeta(\cdot, \beta, \boldsymbol\gamma);W)[v_1, h], \cdots, m''_{\gamma_j\zeta}(\beta, \boldsymbol\gamma(\cdot), \zeta(\cdot, \beta, \boldsymbol\gamma);W)[v_{d_1}, h])^T.\]
We define correspondingly
\[S'_{\gamma_j}(\beta, \boldsymbol\gamma(\cdot), \zeta(\cdot, \beta, \boldsymbol\gamma))[\mb{v}]  = Pm'_{\gamma_j}(\beta, \boldsymbol\gamma(\cdot), \zeta(\cdot, \beta, \boldsymbol\gamma);W)[\mb{v}],\]
\[S'_{\zeta}(\beta, \boldsymbol\gamma(\cdot), \zeta(\cdot, \beta, \boldsymbol\gamma))[\mb{h}] = Pm'_{\zeta}(\beta, \boldsymbol\gamma(\cdot), \zeta(\cdot, \beta, \boldsymbol\gamma);W)[\mb{h}],\]
\[S'_{\gamma_j,n}(\beta, \boldsymbol\gamma(\cdot), \zeta(\cdot, \beta, \boldsymbol\gamma))[\mb{v}]  = \mathbb{P}_n m'_{\gamma_j}(\beta, \boldsymbol\gamma(\cdot), \zeta(\cdot, \beta, \boldsymbol\gamma);W)[\mb{v}],\]
\[S'_{\zeta,n}(\beta, \boldsymbol\gamma(\cdot), \zeta(\cdot, \beta, \boldsymbol\gamma))[\mb{h}] = \mathbb{P}_n m'_{\zeta}(\beta, \boldsymbol\gamma(\cdot), \zeta(\cdot, \beta, \boldsymbol\gamma);W)[\mb{h}],\]
\[S''_{\gamma_j\gamma_\ell}(\beta, \boldsymbol\gamma(\cdot), \zeta(\cdot, \beta, \boldsymbol\gamma))[\mb{v}, v] = P m''_{\gamma_j\gamma_\ell}(\beta, \boldsymbol\gamma(\cdot), \zeta(\cdot, \beta, \boldsymbol\gamma);W)[\mb{v}, v],\]
\[S''_{\zeta\zeta}(\beta, \boldsymbol\gamma(\cdot), \zeta(\cdot, \beta, \boldsymbol\gamma))[\mb{h}, h] = P m''_{\zeta\zeta}(\beta, \boldsymbol\gamma(\cdot), \zeta(\cdot, \beta, \boldsymbol\gamma);W)[\mb{h}, h],\]
\[S''_{\gamma_j\beta}(\beta, \boldsymbol\gamma(\cdot), \zeta(\cdot, \beta, \boldsymbol\gamma))[\mb{v}] = P m''_{\gamma_j\beta}(\beta, \boldsymbol\gamma(\cdot), \zeta(\cdot, \beta, \boldsymbol\gamma);W)[\mb{v}],\]
\[S''_{\zeta\beta}(\beta, \boldsymbol\gamma(\cdot), \zeta(\cdot, \beta, \boldsymbol\gamma))[\mb{h}] = Pm''_{\zeta\beta}(\beta, \boldsymbol\gamma(\cdot), \zeta(\cdot, \beta, \boldsymbol\gamma);W)[\mb{h}],\]
\[S''_{\zeta\gamma_j}(\beta, \boldsymbol\gamma(\cdot), \zeta(\cdot, \beta, \boldsymbol\gamma))[\mb{h}, v] = Pm''_{\zeta\gamma_j}(\beta, \boldsymbol\gamma(\cdot), \zeta(\cdot, \beta, \boldsymbol\gamma);W)[\mb{h}, v],\]
\[S''_{\gamma_j\zeta}(\beta, \boldsymbol\gamma(\cdot), \zeta(\cdot, \beta, \boldsymbol\gamma))[\mb{v}, h] = P m''_{\gamma_j\zeta}(\beta, \boldsymbol\gamma(\cdot), \zeta(\cdot, \beta, \boldsymbol\gamma);W)[\mb{v}, h].\]

\subsection{The general sieve M-theorem}
\label{subs: m-theorem}
\new{Recall that the sieve M-estimator $\hat{\theta}_n = (\hat{\beta}_n, \hat{\boldsymbol\gamma}_n, \hat{\zeta}_n(\cdot, \hat{\beta}_n, \hat{\boldsymbol\gamma}_n))\in \Theta_n$ maximizes the objective function over the sieve parameter space $\Theta_n$. Next, we establish the asymptotic normality of the sieve estimator $\hat{\beta}_n$. The key difference between the proposed new sieve M-theorem in this paper and Theorem 2.1 in \citet{ding2011} is that the nuisance parameter $\zeta(\cdot, \beta, \boldsymbol\gamma)$ can be a function of not only Euclidean parameter $\beta$ but also other nuisance parameters $\boldsymbol\gamma(\cdot)$.} 

To establish the asymptotic normality, we assume the following assumptions.

\begin{enumerate}[label=(A\arabic*), ref=A\arabic*]
\item \label{m:rate} (Rate of convergence) For an estimator   $\hat{\theta}_n = (\hat{\beta}_n, \hat{\boldsymbol\gamma}_n(\cdot), \hat{\zeta}_n(\cdot, \hat{\beta}_n, \hat{\boldsymbol\gamma}_n))\in \Theta_n$ and the true parameter  $\theta_0 = (\beta_0, \boldsymbol\gamma_0(\cdot), \zeta_0(\cdot, \beta_0, \boldsymbol\gamma_0))\in \Theta$, $d(\hat{\theta}_n, \theta_0)=O_p(n^{-\xi})$ for some positive $\xi$.
\item \label{m:dev at true} $S'_{\beta}(\beta_0, \boldsymbol\gamma_0(\cdot), \zeta_0(\cdot, \beta_0, \boldsymbol\gamma_0))=0$, $S'_{\gamma_j}(\beta_0, \boldsymbol\gamma_0(\cdot), \zeta_0(\cdot, \beta_0, \boldsymbol\gamma_0))[v]=0$ for all $v\in \Gamma^{p_1}$ and $1\le j\le d_2$, and 
	$S'_{\zeta} (\beta_0, \boldsymbol\gamma_0(\cdot), \zeta_0(\cdot, \beta_0, \boldsymbol\gamma_0))[h]=0$ for all $h\in \mathbb{H}$.
\item \label{m:positive info} (Positive information) There exists \new{$\mb{v}_j^*=(v^*_{j1}, \cdots, v^*_{jd_1})^T \in \mathbb{V}^{d_1}$, $1\le j \le d_2$}, and $\mb{h}^*=(h^*_1, \cdots, h^*_{d_1})^T \in \mathbb{H}^{d_1}$ such that for any $v\in  \mathbb{V}$ and $h\in  \mathbb{H}$, $1\le \ell \le d_2$
	\begin{align*}
		S''_{\beta\gamma_\ell}(\beta_0, \boldsymbol\gamma_0(\cdot), \zeta_0(\cdot, \beta_0, \boldsymbol\gamma_0))[v]& = \new{\sum_{j=1}^{d_2} S''_{\gamma_j\gamma_\ell}(\beta_0, \boldsymbol\gamma_0(\cdot), \zeta_0(\cdot, \beta_0, \boldsymbol\gamma_0))[\mb{v}_j^*,v]}\\
		& \ \ \ + S''_{\zeta\gamma_\ell}(\beta_0, \boldsymbol\gamma_0(\cdot), \zeta_0(\cdot, \beta_0, \boldsymbol\gamma_0))[\mb{h}^*,v],
\\
		S''_{\beta\zeta}(\beta_0, \boldsymbol\gamma_0(\cdot), \zeta_0(\cdot, \beta_0, \boldsymbol\gamma_0))[h] & = \new{\sum_{j=1}^{d_2} S''_{\gamma_j\zeta}(\beta_0, \boldsymbol\gamma_0(\cdot), \zeta_0(\cdot, \beta_0, \boldsymbol\gamma_0))[\mb{v}_j^*,h]}\\
		& \ \ \ + S''_{\zeta\zeta}(\beta_0, \boldsymbol\gamma_0(\cdot), \zeta_0(\cdot, \beta_0, \boldsymbol\gamma_0))[\mb{h}^*,h].
	\end{align*}
	Furthermore, the matrix 
	\begin{align*}
		A & = - S''_{\beta\beta}(\beta_0, \boldsymbol\gamma_0(\cdot), \zeta_0(\cdot, \beta_0, \boldsymbol\gamma_0)) + \new{\sum_{j=1}^{d_2} S''_{\gamma_j\beta}(\beta_0, \boldsymbol\gamma_0(\cdot), \zeta_0(\cdot, \beta_0, \boldsymbol\gamma_0))[\mb{v}_j^*]}\\
		&\ \ \ +S''_{\zeta\beta}(\beta_0, \boldsymbol\gamma_0(\cdot), \zeta_0(\cdot, \beta_0, \boldsymbol\gamma_0))[\mb{h}^*]\\
		& = - P\{m''_{\beta\beta}(\beta_0, \boldsymbol\gamma_0(\cdot), \zeta_0(\cdot, \beta_0, \boldsymbol\gamma_0);W) + \new{\sum_{j=1}^{d_2} m''_{\gamma_j\beta}(\beta_0, \boldsymbol\gamma_0(\cdot), \zeta_0(\cdot, \beta_0, \boldsymbol\gamma_0);W)[\mb{v}_j^*]}\\
		&\ \ \ \ \ \ \ \ +m''_{\zeta\beta}(\beta_0, \boldsymbol\gamma_0(\cdot), \zeta_0(\cdot, \beta_0, \boldsymbol\gamma_0);W)[\mb{h}^*]\}
	\end{align*}
	is nonsingular.
\item \label{m:gc} The estimator $\hat{\theta}_n = (\hat{\beta}_n, \hat{\boldsymbol\gamma}_n(\cdot), \hat{\zeta}_n(\cdot, \hat{\beta}_n, \hat{\boldsymbol\gamma}_n))$ satisfies $S'_{\beta,n}(\hat{\beta}_n, \hat{\boldsymbol\gamma}_n(\cdot), \hat{\zeta}_n(\cdot, \hat{\beta}_n, \hat{\boldsymbol\gamma}_n)) = o_p(n^{-1/2})$, \\ $S'_{\gamma_j,n}(\hat{\beta}_n, \hat{\boldsymbol\gamma}_n(\cdot), \hat{\zeta}_n(\cdot, \hat{\beta}_n, \hat{\boldsymbol\gamma}_n))[\mb{v}_j^*] = o_p(n^{-1/2})$ for $1\le j \le d_2$, and $S'_{\zeta,n}(\hat{\beta}_n, \hat{\boldsymbol\gamma}_n(\cdot), \hat{\zeta}_n(\cdot, \hat{\beta}_n, \hat{\boldsymbol\gamma}_n))[\mb{h}^*] = o_p(n^{-1/2})$.
\item \label{m:equicontinuity} (Stochastic equicontinuity) For some positive $C$, 
	\begin{align*}
		\sup_{d(\theta, \theta_0)\leq Cn^{-\xi}, \theta \in \Theta_n} & \| \sqrt{n}(S'_{\beta, n} - S'_{\beta})(\beta, \boldsymbol\gamma(\cdot), \zeta(\cdot, \beta, \boldsymbol\gamma)) \\
		& - \sqrt{n}(S'_{\beta, n} - S'_{\beta})(\beta_0, \boldsymbol\gamma_0(\cdot), \zeta_0(\cdot, \beta_0, \boldsymbol\gamma_0))\| = o_p(1),\\
		\sup_{d(\theta, \theta_0)\leq Cn^{-\xi}, \theta \in \Theta_n} & | \sqrt{n}(S'_{\gamma_j, n} - S'_{\gamma_j})(\beta, \boldsymbol\gamma(\cdot), \zeta(\cdot, \beta, \boldsymbol\gamma))[\mb{v}_j^*]\\
		& - \sqrt{n}(S'_{\gamma_j, n} - S'_{\gamma_j})(\beta_0, \boldsymbol\gamma_0(\cdot), \zeta_0(\cdot, \beta_0, \boldsymbol\gamma_0))[\mb{v}_j^*]| = o_p(1),~\text{ for }1\le j \le d_2,
	\end{align*}
	and 
	\begin{align*}
		\sup_{d(\theta, \theta_0)\leq Cn^{-\xi}, \theta \in \Theta_n} & | \sqrt{n}(S'_{\zeta, n} - S'_{\zeta})(\beta, \boldsymbol\gamma(\cdot), \zeta(\cdot, \beta, \boldsymbol\gamma))[\mb{h}^*(\cdot, \beta, \boldsymbol\gamma)] \\
		& - \sqrt{n}(S'_{\zeta, n} - S'_{\zeta})(\beta_0, \boldsymbol\gamma_0(\cdot), \zeta_0(\cdot, \beta_0, \boldsymbol\gamma_0))[\mb{h}^*(\cdot, \beta_0, \boldsymbol\gamma_0)]| = o_p(1).
	\end{align*}
\item \label{m:smoothness} (Smoothness of the model) For some $\alpha >1$ with $\alpha \xi >\frac{1}{2}$, and for $\theta \in \Theta_n$ satisfying $d(\theta, \theta_0)\leq Cn^{-\xi}$, 
	\begin{align*}
		& \|S'_{\beta}(\beta, \boldsymbol\gamma(\cdot), \zeta(\cdot, \beta, \boldsymbol\gamma)) - S'_{\beta}(\beta_0, \boldsymbol\gamma_0(\cdot), \zeta_0(\cdot, \beta_0, \boldsymbol\gamma_0)) - S''_{\beta\beta}(\beta_0, \boldsymbol\gamma_0(\cdot), \zeta_0(\cdot, \beta_0, \boldsymbol\gamma_0))(\beta-\beta_0)\\
		& \ \ \ \ \ \ \ \ - \new{\sum_{j=1}^{d_2}S''_{\beta\gamma_j}(\beta_0, \boldsymbol\gamma_0(\cdot), \zeta_0(\cdot, \beta_0, \boldsymbol\gamma_0))[e_j(\boldsymbol\gamma-\boldsymbol\gamma_0)^T]} \\
		& \ \ \ \ \ \ \ \ - S''_{\beta\zeta}(\beta_0, \boldsymbol\gamma_0(\cdot), \zeta_0(\cdot, \beta_0, \boldsymbol\gamma_0))[\zeta(\cdot, \beta, \boldsymbol\gamma) - \zeta_0(\cdot, \beta_0, \boldsymbol\gamma_0)]\|\\
		& \ \ \ \ \ \ \ \ \ \ \ \ \ \ \ \ = O(d^\alpha(\theta, \theta_0)),
\\
& |S'_{\gamma_j}(\beta, \boldsymbol\gamma(\cdot), \zeta(\cdot, \beta, \boldsymbol\gamma))[\mb{v}_j^*] - S'_{\gamma_j}(\beta_0, \boldsymbol\gamma_0(\cdot), \zeta_0(\cdot, \beta_0, \boldsymbol\gamma_0))[\mb{v}_j^*] - S''_{\gamma_j\beta}(\beta_0, \boldsymbol\gamma_0(\cdot), \zeta_0(\cdot, \beta_0, \boldsymbol\gamma_0))[\mb{v}_j^*](\beta-\beta_0)\\
		& \ \ \ \ \ \ \ \ - \new{\sum_{\ell=1}^{d_2}S''_{\gamma_j\gamma_\ell}(\beta_0, \boldsymbol\gamma_0(\cdot), \zeta_0(\cdot, \beta_0, \boldsymbol\gamma_0))[\mb{v}_j^*,e_\ell(\boldsymbol\gamma-\boldsymbol\gamma_0)^T]} \\
		& \ \ \ \ \ \ \ \ - S''_{\gamma_j\zeta}(\beta_0, \boldsymbol\gamma_0(\cdot), \zeta_0(\cdot, \beta_0, \boldsymbol\gamma_0))[\mb{v}_j^*,\zeta(\cdot, \beta, \boldsymbol\gamma) - \zeta_0(\cdot, \beta_0, \boldsymbol\gamma_0)]|\\
		& \ \ \ \ \ \ \ \ \ \ \ \ \ \ \ \ = O(d^\alpha(\theta, \theta_0)),~\text{ for }1\le j \le d_2,
	\end{align*}
	and
	\begin{align*}
		& |S'_{\zeta}(\beta, \boldsymbol\gamma(\cdot), \zeta(\cdot, \beta, \boldsymbol\gamma))[\mb{h}^*(\cdot, \beta, \boldsymbol\gamma)] - S'_{\zeta}(\beta_0, \boldsymbol\gamma_0(\cdot), \zeta_0(\cdot, \beta_0, \boldsymbol\gamma_0))[\mb{h}^*(\cdot, \beta_0, \boldsymbol\gamma_0)] \\
		& \ \ \ \ \ \ \ \ - S''_{\zeta\beta}(\beta_0, \boldsymbol\gamma_0(\cdot), \zeta_0(\cdot, \beta_0, \boldsymbol\gamma_0))[\mb{h}^*(\cdot, \beta_0, \boldsymbol\gamma_0)](\beta-\beta_0)\\
		& \ \ \ \ \ \ \ \ - \new{\sum_{j=1}^{d_2} S''_{\zeta\gamma_j}(\beta_0, \boldsymbol\gamma_0(\cdot), \zeta_0(\cdot, \beta_0, \boldsymbol\gamma_0))[\mb{h}^*(\cdot, \beta_0, \boldsymbol\gamma_0),e_j(\boldsymbol\gamma-\boldsymbol\gamma_0)^T]} \\
		& \ \ \ \ \ \ \ \ - S''_{\zeta\zeta}(\beta_0, \boldsymbol\gamma_0(\cdot), \zeta_0(\cdot, \beta_0, \boldsymbol\gamma_0))[\mb{h}^*(\cdot, \beta_0, \boldsymbol\gamma_0),\zeta(\cdot, \beta, \boldsymbol\gamma) - \zeta_0(\cdot, \beta_0, \boldsymbol\gamma_0)]|\\
		& \ \ \ \ \ \ \ \ \ \ \ \ \ \ \ \ = O(d^\alpha(\theta, \theta_0)).
	\end{align*}
\end{enumerate}

The convergence rate in (\ref{m:rate}) is a prerequisite for the asymptotic normality. Assumption (\ref{m:dev at true}) is a common regularity assumption when $m$ is the likelihood function, and it usually holds for the score functions. The direction $\mb{v}_j^*$ and $\mb{h}^*$ in (\ref{m:positive info}) are the least favorable directions for maximum likelihood estimation, which may be found through solving the equations in (\ref{m:positive info}). Assumptions (\ref{m:gc}) and (\ref{m:equicontinuity}) can be obtained by the maximal inequality in Lemma 3.4.2 of \citep[page 324]{billingsley2013convergence} and the Markov's inequality. Assumption (\ref{m:smoothness}) can be usually verified by the Taylor expansion. 
We repeat Theorem~\ref{thm: m-theorem} below for readers' convenience, which is a general sieve M-theorem for bundled parameters where the nuisance parameter $\zeta(\cdot, \beta, \boldsymbol\gamma)$ is a function of the Euclidean parameter $\beta$ and other nuisance parameters $\boldsymbol\gamma(\cdot)$.
\begin{theorem*}
%\label{thm: m-theorem}
	Suppose that assumptions (\ref{m:rate})-(\ref{m:smoothness}) hold, then
	\begin{align*}
		\sqrt{n}(\hat{\beta}_n - \beta_0) & = A^{-1}\sqrt{n}\mathbb{P}_n \mb{m}^*(\beta_0, \boldsymbol\gamma_0(\cdot), \zeta_0(\cdot, \beta_0, \boldsymbol\gamma_0);W) + o_p(1) \\
		& \rightarrow_d N(0, A^{-1}B( A^{-1})^T),
	\end{align*}
	where 
	\begin{align*}
		\mb{m}^*(\beta_0, \boldsymbol\gamma_0(\cdot), \zeta_0(\cdot, \beta_0, \boldsymbol\gamma_0);W) & = m'_{\beta}(\beta_0, \boldsymbol\gamma_0(\cdot), \zeta_0(\cdot, \beta_0, \boldsymbol\gamma_0);W) -\new{\sum_{j=1}^{d_2} m'_{\gamma_j}(\beta_0, \boldsymbol\gamma_0(\cdot), \zeta_0(\cdot, \beta_0, \boldsymbol\gamma_0);W)[\mb{v}_j^*]} \\
		& \ \ \ - m'_{\zeta}(\beta_0, \boldsymbol\gamma_0(\cdot), \zeta_0(\cdot, \beta_0, \boldsymbol\gamma_0);W)[\mb{h}^*(\cdot, \beta_0, \boldsymbol\gamma_0)],\\
		B & = P\{\mb{m}^*(\beta_0, \boldsymbol\gamma_0(\cdot), \zeta_0(\cdot, \beta_0, \boldsymbol\gamma_0);W)\mb{m}^*(\beta_0, \boldsymbol\gamma_0(\cdot), \zeta_0(\cdot, \beta_0, \boldsymbol\gamma_0);W)^T\},
	\end{align*}
and $A$ is given in the assumption (\ref{m:positive info}). 
\end{theorem*}

\begin{proof}[\textbf{Proof of Theorem \ref{thm: m-theorem}}]
	We prove the theorem by following the proof of Theorem 6.1 in~\citet{wellner2007} and Theorem 2.1 in~\citet{ding2011}. Assumptions (\ref{m:rate}) and (\ref{m:equicontinuity}) lead to 
	\[\sqrt{n}(S'_{\beta, n} - S'_{\beta})(\hat{\beta}_n, \hat{\boldsymbol\gamma}_n(\cdot), \hat{\zeta}_n(\cdot, \hat{\beta}_n, \hat{\boldsymbol\gamma}_n))
		- \sqrt{n}(S'_{\beta, n} - S'_{\beta})(\beta_0, \boldsymbol\gamma_0(\cdot), \zeta_0(\cdot, \beta_0, \boldsymbol\gamma_0)) = o_p(1).\]
	Note that $S'_{\beta}(\beta_0, \boldsymbol\gamma_0(\cdot), \zeta_0(\cdot, \beta_0, \boldsymbol\gamma_0))=0$ by (\ref{m:dev at true}), $S'_{\beta,n}(\hat{\beta}_n, \hat{\boldsymbol\gamma}_n(\cdot), \hat{\zeta}_n(\cdot, \hat{\beta}_n, \hat{\boldsymbol\gamma}_n)) = o_p(n^{-1/2})$ by (\ref{m:gc}), we have
	\begin{equation}
	\label{eq1: m-thoerem}
		\sqrt{n}S'_{\beta}(\hat{\beta}_n, \hat{\boldsymbol\gamma}_n(\cdot), \hat{\zeta}_n(\cdot, \hat{\beta}_n, \hat{\boldsymbol\gamma}_n))
		+ \sqrt{n}S'_{\beta, n}(\beta_0, \boldsymbol\gamma_0(\cdot), \zeta_0(\cdot, \beta_0, \boldsymbol\gamma_0)) = o_p(1).
	\end{equation}
	After combining the equation (\ref{eq1: m-thoerem}) and the equations in assumptions (\ref{m:dev at true}) and (\ref{m:smoothness}), we have
	\begin{align*}
		& S'_{\beta, n}(\beta_0, \boldsymbol\gamma_0(\cdot), \zeta_0(\cdot, \beta_0, \boldsymbol\gamma_0)) + S''_{\beta\beta}(\beta_0, \boldsymbol\gamma_0(\cdot), \zeta_0(\cdot, \beta_0, \boldsymbol\gamma_0))(\hat{\beta}_n-\beta_0)\\
		& \ \ \ \ \  + \sum_{j=1}^{d_2}S''_{\beta\gamma_j}(\beta_0, \boldsymbol\gamma_0(\cdot), \zeta_0(\cdot, \beta_0, \boldsymbol\gamma_0))[e_j(\hat{\boldsymbol\gamma}_n-\boldsymbol\gamma_0)^T] \\
		& \ \ \ \ \  + S''_{\beta\zeta}(\beta_0, \boldsymbol\gamma_0(\cdot), \zeta_0(\cdot, \beta_0, \boldsymbol\gamma_0))[\hat{\zeta}_n(\cdot, \hat{\beta}_n, \hat{\boldsymbol\gamma}_n) - \zeta_0(\cdot, \beta_0, \boldsymbol\gamma_0)]\\
		&= O(d^\alpha(\hat{\theta}_n, \theta_0)) + o_p(n^{-1/2}) = o_p(n^{-1/2}).
	\end{align*}
	The last equation holds because for $\alpha >1$ with $\alpha \xi >\frac{1}{2}$, assumption (\ref{m:rate}) implies that $$O(d^\alpha(\hat{\theta}_n, \theta_0)) = O_p(n^{-\alpha\xi})=o_p(n^{-1/2}).$$
	Similarly, we have for $1\le j \le d_2$
	\begin{align*}
		& S'_{\gamma_j, n}(\beta_0, \boldsymbol\gamma_0(\cdot), \zeta_0(\cdot, \beta_0, \boldsymbol\gamma_0))[\mb{v}_j^*] + S''_{\gamma_j\beta}(\beta_0, \boldsymbol\gamma_0(\cdot), \zeta_0(\cdot, \beta_0, \boldsymbol\gamma_0))[\mb{v}_j^*](\hat{\beta}_n-\beta_0)\\
		& \ \ \ \ \  + \sum_{\ell=1}^{d_2}S''_{\gamma_j\gamma_\ell}(\beta_0, \boldsymbol\gamma_0(\cdot), \zeta_0(\cdot, \beta_0, \boldsymbol\gamma_0))[\mb{v}_j^*,e_\ell(\hat{\boldsymbol\gamma}_n-\boldsymbol\gamma_0)^T] \\
		& \ \ \ \ \  + S''_{\gamma_j\zeta}(\beta_0, \boldsymbol\gamma_0(\cdot), \zeta_0(\cdot, \beta_0, \boldsymbol\gamma_0))[\mb{v}_j^*,\hat{\zeta}_n(\cdot, \hat{\beta}_n, \hat{\boldsymbol\gamma}_n) - \zeta_0(\cdot, \beta_0, \boldsymbol\gamma_0)]\\
		&= o_p(n^{-1/2})
	\end{align*}
	and
	\begin{align*}
		& S'_{\zeta, n}(\beta_0, \boldsymbol\gamma_0(\cdot), \zeta_0(\cdot, \beta_0, \boldsymbol\gamma_0))[\mb{h}^*(\cdot, \beta_0, \boldsymbol\gamma_0)] + S''_{\zeta\beta}(\beta_0, \boldsymbol\gamma_0(\cdot), \zeta_0(\cdot, \beta_0, \boldsymbol\gamma_0))[\mb{h}^*(\cdot, \beta_0, \boldsymbol\gamma_0)](\hat{\beta}_n-\beta_0)\\
		& \ \ \ \ \  + \sum_{j=1}^{d_2} S''_{\zeta\gamma_j}(\beta_0, \boldsymbol\gamma_0(\cdot), \zeta_0(\cdot, \beta_0, \boldsymbol\gamma_0))[\mb{h}^*(\cdot, \beta_0, \boldsymbol\gamma_0), e_j(\hat{\boldsymbol\gamma}_n-\boldsymbol\gamma_0)^T] \\
		& \ \ \ \ \  + S''_{\zeta\zeta}(\beta_0, \boldsymbol\gamma_0(\cdot), \zeta_0(\cdot, \beta_0, \boldsymbol\gamma_0))[\mb{h}^*(\cdot, \beta_0, \boldsymbol\gamma_0),\hat{\zeta}_n(\cdot, \hat{\beta}_n, \hat{\gamma}_n) - \zeta_0(\cdot, \beta_0, \boldsymbol\gamma_0)]\\
		&= o_p(n^{-1/2}).
	\end{align*}
	Combining these equations with assumption (\ref{m:positive info}) leads to
	\begin{align*}
		& \{S''_{\beta\beta}(\beta_0, \boldsymbol\gamma_0(\cdot), \zeta_0(\cdot, \beta_0, \boldsymbol\gamma_0)) - \sum_{j=1}^{d_2} S''_{\gamma_j\beta}(\beta_0, \boldsymbol\gamma_0(\cdot), \zeta_0(\cdot, \beta_0, \boldsymbol\gamma_0))[\mb{v}_j^*] \\
		& \ \ \ \ \ \ \ \ \ \ \ \ \ \ \ \ \ \ \ \ \ \ \  - S''_{\zeta\beta}(\beta_0, \boldsymbol\gamma_0(\cdot), \zeta_0(\cdot, \beta_0, \boldsymbol\gamma_0))[\mb{h}^*(\cdot, \beta_0, \boldsymbol\gamma_0)]\}(\hat{\beta}_n-\beta_0)\\
		& = - \{S'_{\beta, n}(\beta_0, \boldsymbol\gamma_0(\cdot), \zeta_0(\cdot, \beta_0, \boldsymbol\gamma_0)) -\sum_{j=1}^{d_2}S'_{\gamma_j, n}(\beta_0, \boldsymbol\gamma_0(\cdot), \zeta_0(\cdot, \beta_0, \boldsymbol\gamma_0))[\mb{v}_j^*] \\
		& \ \ \ \ \ \ \ \ \ \ \ \ \ \ \ \ \ \ \ \ \ \ \ - S'_{\zeta, n}(\beta_0, \boldsymbol\gamma_0(\cdot), \zeta_0(\cdot, \beta_0, \boldsymbol\gamma_0))[\mb{h}^*(\cdot, \beta_0, \boldsymbol\gamma_0)]\} \\
		& \ \ \ + o_p(n^{-1/2}),
	\end{align*}
	and equivalently, 
	\[-A(\hat{\beta}_n-\beta_0) = -\mathbb{P}_n \mb{m}^*(\beta_0, \boldsymbol\gamma_0(\cdot), \zeta_0(\cdot, \beta_0, \boldsymbol\gamma_0); W)+o_p(n^{-1/2}).\]
	Then under assumptions (\ref{m:gc}) and (\ref{m:equicontinuity}),
	\begin{align*}
		\sqrt{n}(\hat{\beta}_n - \beta_0) & = A^{-1}\sqrt{n}\mathbb{P}_n \mb{m}^*(\beta_0, \boldsymbol\gamma_0(\cdot), \zeta_0(\cdot, \beta_0, \boldsymbol\gamma_0);W) + o_p(1) \\
		& \rightarrow_d N(0, A^{-1}B( A^{-1})^T).
	\end{align*}
\end{proof}

\section{Proof of Theorems \ref{thm: conv rate} and \ref{thm: asym normal}}
\label{appendx: proof}

Without loss of generality, we prove Theorems \ref{thm: conv rate} and \ref{thm: asym normal} in the case that $X_{(1)}$ is not included in~(\ref{ode: log double iden}). The results in this section still hold if $X_{(1)}$ is included due to the boundedness of $X_{(1)}$. 
For notational simplicity, we further replace $X_{(-1)}$ by $X$ in (\ref{ode: log double iden}), which then becomes equivalent to  the ODE in (\ref{ode: log double}). 

We first introduce some common notations that will be used in the proof hereafter. For any fixed $\gamma(\cdot)\in \Gamma^{p_1}$, let $\{ \gamma_{\eta}(\cdot): \eta$ in a neighborhood of $0\in \mb{R} \}$ be a smooth curve in $\Gamma^{p_1}$ running through $\gamma(\cdot)$ at $\eta=0$, that is $\gamma_{\eta}(\cdot)|_{\eta=0} = \gamma(\cdot)$. Similarly, for any fixed $g(\cdot) \in \mathcal{G}^{p_2}$, let $\{g_{\eta}(\cdot): \eta$ in a neighborhood of $0\in \mb{R} \}$ be a smooth curve in $\mathcal{G}^{p_2}$ running through $g(\cdot)$ at $\eta=0$, that is $g_{\eta}(\cdot)|_{\eta=0} = g(\cdot)$. Denote 
\begin{equation*}
	\mathbb{V}=\{v: v(\cdot)=\frac{\partial \gamma_{\eta}(\cdot)}{\partial \eta}|_{\eta=0} , \gamma_{\eta}\in \Gamma^{p_1} \}
\end{equation*}
and 
\begin{equation*}
	\mathbb{W}=\{w: w(\cdot)=\frac{\partial g_{\eta}(\cdot)}{\partial \eta}|_{\eta=0} , g_{\eta}\in \mc{G}^{p_2} \}.
\end{equation*}
Recall that $\Lambda_0(t, x) = \Lambda(t, x, \beta_0, \gamma_0, g_0)$ and $R(t)=\int_0^t \exp(\gamma_0(s))\dev s$. Let $\tilde{\Lambda}_0(t)$ denote the solution of $\tilde{\Lambda}_0'(t)=\exp(g_0(\tilde{\Lambda}_0))$ with $\tilde{\Lambda}_0(0)=0$. It is straightforward to show that $\tilde{\Lambda}_0(\cdot)$ is the cumulative hazard function of $R(T)e^{X^T \beta_0}$ and $\Lambda_0(t, X) = \tilde{\Lambda}_0(R(t)e^{X^T \beta_0})$. We use symbol $\gtrsim$ to denote that the left side is bounded below by a constant times the right side. We also use symbol $\lesssim$ to denote that the left side is bounded above by a constant times the right side. If without further explanation, by default, the $L_2$ norm of a function $f(\cdot)$ of $t$ and $x$ is given by 
\[\|f(\cdot)\|_2 = \left[\int_{\mc{X}} \int_{0}^{\tau} (f(t, x))^2 \dev \Lambda_0(t, x) \dev F_X(x)\right]^{1/2},\]
and the supreme norm is given by $\|f(\cdot)\|_{\infty}=\sup_{t\in[0,\tau], x\in\mc{X}}|f(t,x)|$. For any $g \in \mc{G}^{p_2}$, the $L_2$ norm is given by $\|g\|_2 = (\int^\mu_0(g(t))^2\dev t)^{1/2}$ and the supreme norm is given by $\|g\|_{\infty} = \sup_{t\in[0,\mu]}|g(t)|$.

The rest of this section is structured as follows. Subsection \ref{appendix subs lemma} introduces several lemmas which will be used to prove Theorem \ref{thm: conv rate} and \ref{thm: asym normal}. Subsections \ref{appendix subs conv rate} and  \ref{appendix subs asym normal} provide  the proof of Theorem \ref{thm: conv rate} by checking the conditions C1-3 in \citet[Theorem 1]{shen1994}   and the proof of Theorem \ref{thm: asym normal} by verifying assumptions (\ref{m:rate})-(\ref{m:smoothness}) of the proposed general M-theorem, respectively. Furthermore, we derive  in subsection \ref{appendix subs condition exp} the equivalent but more feasible equations for finding the least favorable directions required in condition \ref{c: pos info exist} and provide explicit constructions for the Cox model and the linear transformation model with a known transformation as illustration. \new{Subsequently, we simplify the non-regularity assumption in  Condition \ref{c: pos fish} in subsection~\ref{appendix subs info condition exp}.}

\subsection{Lemmas}
\label{appendix subs lemma}
\begin{lemma}(Existence and uniqueness theorem.)
    \label{lemma: ode solutions} 
    For any $x\in \mc{X}, \beta\in \mc{B}, \gamma \in \Gamma^{p_1}, g\in \mc{G}^{p_2}$ under conditions \ref{c: bounded b}-\ref{c: smoothness}, the initial value problem (\ref{ode: log double}) has exactly one bounded and continuous solution $\Lambda(t, x, \beta, \gamma, g)$ on $[0, \tau]$. And its first and second derivatives with respect to $\beta\in \mc{B}, \gamma \in \Gamma^{p_1}$ and the first derivative with respect to $g\in \mc{G}^{p_2}$ are also bounded and continuous on $[0, \tau]$.
\end{lemma}
\begin{proof}[Proof of Lemma \ref{lemma: ode solutions}]
Let $f(t, \Lambda) = \exp(x^T \beta+ \gamma(t)+g(\Lambda))$, then by the mean value theorem
\[|f(t, \Lambda) - f(t, \tilde{\Lambda})| \leq \exp(x^T \beta+ \gamma(t)+g(c))|g'(c)|\cdot| \Lambda-\tilde{\Lambda}|\leq L | \Lambda-\tilde{\Lambda}|\] holds for any $(t, \Lambda)$ and $(t, \tilde{\Lambda})$ in $[0,\tau]\times [0, \mu]$, where $c \in [\Lambda, \tilde{\Lambda}]$ and $L<\infty$ under conditions \ref{c: bounded b}-\ref{c: smoothness}. This implies that $f(t, \Lambda)$ satisfies the Lipschitz condition with respect to $\Lambda$ in $[0,\tau]\times [0, \mu]$. By Theorem 10.VI in \citet[page 108]{Walter1998}, there is exactly one solution to the initial value problem (\ref{ode: log double}). The solution $\Lambda(t, x, \beta, \gamma, g)$ is bounded, continuous, and satisfies 
\begin{equation}
\label{ode int}
	\Lambda(t, x, \beta, \gamma, g)=\int_0^t \exp(x^T \beta+ \gamma(s)+g(\Lambda(s, x, \beta, \gamma, g)))\dev s.
\end{equation}
In the following, we write $\Lambda(t)=\Lambda(t, x, \beta, \gamma, g)$ for simplicity. Similarly to the above derivation, for any $\beta \in \mc{B},v\in \mathbb{V}, w\in \mathbb{W}$, we have unique, bounded, and continuous solutions of the following initial value problems: 
\begin{align}
	\frac{\dev \Lambda'_{\beta}(t)}{\dev t} & = \exp(x^T\beta + \gamma(t)+g(\Lambda(t)))\{x+g'(\Lambda(t)) \Lambda'_{\beta}(t)\}, \ \ \Lambda'_{\beta}(0) = 0,\label{eq: beta dev}\\
\frac{\dev \Lambda'_{\gamma}(t)[v]}{\dev t} &= \exp(x^T\beta + \gamma(t)+g(\Lambda(t)))\{v(t)+g'(\Lambda(t)) \Lambda'_{\gamma}(t)[v]\}, \ \ \Lambda'_{\gamma}(0)[v] = 0,\label{eq: gamma dev}\\
\frac{\dev \Lambda'_{g}(t)[w]}{\dev t} &= \exp(x^T\beta + \gamma(t)+g(\Lambda(t)))\{w(\Lambda(t))+g'(\Lambda(t)) \Lambda'_{g}(t)[w]\}, \ \ \Lambda'_{g}(0)[w] = 0,\label{eq: g dev}
\end{align}
\begin{align}
	\frac{\dev \Lambda''_{\beta\beta}(t)}{\dev t} &= \exp(x^T\beta + \gamma(t)+g(\Lambda(t))) \{[x+g'(\Lambda(t)) \Lambda'_{\beta}(t)][x+g'(\Lambda(t)) \Lambda'_{\beta}(t)]^T \nonumber \\
	& \ \ \ \ + g''(\Lambda(t))\Lambda'_{\beta}(t)\Lambda'_{\beta}(t)^T + g'(\Lambda(t))\Lambda''_{\beta\beta}(t)\}, \ \ \Lambda''_{\beta\beta}(0)=0, \label{eq: betabeta dev}\\
	\frac{\dev \Lambda''_{\gamma\gamma}(t)[v_1, v_2]}{\dev t} &= \exp(x^T\beta + \gamma(t)+g(\Lambda(t))) \{(v_1(t)+g'(\Lambda(t)) \Lambda'_{\gamma}(t)[v_1])(v_2(t)+g'(\Lambda(t)) \Lambda'_{\gamma}(t)[v_2]) \nonumber\\
	& \ \ \ \  + g''(\Lambda(t))\Lambda'_{\gamma}(t)[v_1]\Lambda'_{\gamma}(t)[v_2]\nonumber\\
	    & \ \ \ \ + g'(\Lambda(t))\Lambda''_{\gamma\gamma}(t)[v_1, v_2]\}, \ \ \Lambda''_{\gamma\gamma}(0)[v_1, v_2]=0, \label{eq: gammagamma dev}\\
	\frac{\dev \Lambda''_{\beta\gamma}(t)[v]}{\dev t} &= \exp(x^T\beta + \gamma(t)+g(\Lambda(t))) \{(v(t)+g'(\Lambda(t)) \Lambda'_{\gamma}(t)[v])(x+g'(\Lambda(t)) \Lambda'_{\beta}(t)) \nonumber\\
	& \ \ \ \  + g''(\Lambda(t))\Lambda'_{\beta}(t)\Lambda'_{\gamma}(t)[v]\nonumber\\
	& \ \ \ \  + g'(\Lambda(t))\Lambda''_{\beta\gamma}(t)[v]\}, \ \ \Lambda''_{\beta\gamma}(0)[v]=0,\label{eq: beta gamma dev}\\
	\frac{\dev \Lambda''_{g \beta}(t)[w]}{\dev t} &= \exp(x^T\beta + \gamma(t)+g(\Lambda(t))) \{(w(\Lambda(t))+g'(\Lambda(t)) \Lambda'_{g}(t)[w])(x+g'(\Lambda(t)) \Lambda'_{\beta}(t)) \nonumber\\
	& \ \ \ \  + w'(\Lambda(t))\Lambda'_{\beta}(t)+g''(\Lambda(t))\Lambda'_{\beta}(t)\Lambda'_{g}(t)[w]\nonumber\\
	& \ \ \ \  + g'(\Lambda(t))\Lambda''_{g\beta}(t)[w]\}, \ \ \Lambda''_{g \beta}(0)[w]=0,\label{eq: g beta dev}\\
	\frac{\dev \Lambda''_{g \gamma}(t)[w,v]}{\dev t} &= \exp(x^T\beta + \gamma(t)+g(\Lambda(t))) \{(w(\Lambda(t))+g'(\Lambda(t)) \Lambda'_{g}(t)[w])(v(t)+g'(\Lambda(t)) \Lambda'_{\gamma}(t)[v]) \nonumber\\
	& \ \ \ \  + w'(\Lambda(t))\Lambda'_{\gamma}(t)[v]+g''(\Lambda(t))\Lambda'_{\gamma}(t)[v]\Lambda'_{g}(t)[w]\nonumber\\
	& \ \ \ \  + g'(\Lambda(t))\Lambda''_{g\gamma}(t)[w,v]\}, \ \ \Lambda''_{g \gamma}(0)[w,v]=0.\label{eq: g gamma dev}
\end{align}
Next we verify that the derivative of $\Lambda(t, x, \beta, \gamma, g)$ with respect to $\beta$ follows the ODE (\ref{eq: beta dev}). By plugging in Equation (\ref{ode int}) and (\ref{eq: beta dev}), it follows that 
\begin{align*}
	& \limsup_{\delta\rightarrow 0}\frac{1}{|\delta|} |\Lambda(t, x, \beta+\delta, \gamma, g) - \Lambda(t) - \Lambda'_{\beta}(t)^T \delta|\\
	& = \limsup_{\delta\rightarrow 0}\frac{1}{|\delta|}\left|\right.\int_{0}^t \exp(x^T (\beta+\delta)+ \gamma(s)+g(\Lambda(s, x, \beta+\delta, \gamma, g))) \\
	& \ \ \ \ \ \ \ \ \ \ \ \ \ \ \  - \exp(x^T \beta+ \gamma(s)+g(\Lambda(s))) \\
	& \ \ \ \ \ \ \ \ \ \ \ \ \ \ \ \left. -  \exp(x^T \beta+ \gamma(s)+g(\Lambda(s)))(x^T\delta + g'(\Lambda(s))\Lambda'_{\beta}(s)^T \delta) \dev s\right|\\
	& \leq \limsup_{\delta\rightarrow 0}\frac{1}{|\delta|}\int_{0}^t \left|\right. \exp(x^T (\beta+\delta)+ \gamma(s)+g(\Lambda(s, x, \beta+\delta, \gamma, g))) \\
	& \ \ \ \ \ \ \ \ \ \ \ \ \ \ \  - \exp(x^T \beta+ \gamma(s)+g(\Lambda(s))) \\
	& \ \ \ \ \ \ \ \ \ \ \ \ \ \ \ \left. -  \exp(x^T \beta+ \gamma(s)+g(\Lambda(s)))(x^T\delta + g'(\Lambda(s))\Lambda'_{\beta}(s)^T \delta)\right| \dev s\\
	& \leq \int_{0}^t \limsup_{\delta\rightarrow 0}\frac{1}{|\delta|} \left|\right. \exp(x^T (\beta+\delta)+ \gamma(s)+g(\Lambda(s, x, \beta+\delta, \gamma, g))) \\
	& \ \ \ \ \ \ \ \ \ \ \ \ \ \ \  - \exp(x^T \beta+ \gamma(s)+g(\Lambda(s))) \\
	& \ \ \ \ \ \ \ \ \ \ \ \ \ \ \ \left. -  \exp(x^T \beta+ \gamma(s)+g(\Lambda(s)))(x^T\delta + g'(\Lambda(s))\Lambda'_{\beta}(s)^T \delta)\right| \dev s\\
	& =  \int_{0}^t \exp(x^T\beta + \gamma(s)+g(\Lambda(s)))\\
	&\ \ \ \ \ \ \ \ \cdot g'(\Lambda(s)) \{\limsup_{\delta\rightarrow 0}\frac{1}{|\delta|} |\Lambda(s, x, \beta+\delta, \gamma, g) - \Lambda(s) - \Lambda'_{\beta}(s)^T \delta| \}\dev s
\end{align*}
where the second inequality holds due to the reverse Fatou's lemma. Using the Gronwall's inequality, we have that
\[\limsup_{\delta\rightarrow 0}\frac{1}{|\delta|} |\Lambda(t, x, \beta+\delta, \gamma, g) - \Lambda(t) - \Lambda'_{\beta}(t)^T \delta| \leq 0,\]
which implies that the solution $\Lambda'_{\beta}(t)$ of (\ref{eq: beta dev}) is the derivative of $\Lambda(t, x, \beta, \gamma, g)$ with respect to $\beta$. The other first and second derivatives of of $\Lambda(t, x, \beta, \gamma, g)$ with respect to $\beta, \gamma, g$	can be similarly derived, and we omit the details here.
\end{proof}

\begin{lemma}
\label{lemma: bdd linear operators}	
Let $\psi(t, x, \beta, \gamma, g) = \log\lambda(t, x, \beta, \gamma, g)= x^T\beta + \gamma(t)+g(\Lambda(t, x, \beta, \gamma, g))$, and denote the first derivatives of $\psi(t, x, \beta, \gamma, g)$ with respect to $\gamma$ and $g$ at the true parameter $(\beta_0, \gamma_0, g_0)$ by $\psi_{0\gamma}'(t,x)[v]$ and $\psi_{0g}'(t,x)[w]$, respectively. For any $\psi_{0\gamma}'(\cdot)[v]\in \mc{E}_{\gamma} = \{\psi_{0\gamma}'(\cdot)[v]:\psi_{0\gamma}'(t,x)[v], t\in [0,\tau], x\in \mc{X}, v\in \Gamma^{p_1}\}$, the $L_2$ norm of $\psi_{0\gamma}'(\cdot)[v]$ is defined as
\[\|\psi_{0\gamma}'(\cdot)[v]\|_2 = \left[\int_{\mc{X}} \int_{0}^{\tau} (\psi_{0\gamma}'(t, x)[v])^2 \dev \Lambda_0(t, x) \dev F_X(x)\right]^{1/2}.\]
The $L_2$ norm of $\psi_{0g}'(\cdot)[w]\in \mc{E}_{g} = \{\psi_{0g}'(\cdot)[w]: \psi_{0g}'(t,x)[w], t\in[0, \tau], x\in \mc{X}, w\in \mc{G}^{p_2}\}$ is similarly defined. Under conditions \ref{c: bounded x}-\ref{c: smoothness}, $\psi_{0\gamma}'[\cdot]: v\rightarrow \psi_{0\gamma}'(\cdot)[v]$ and $\psi_{0g}'[\cdot]:w\rightarrow \psi_{0g}'(\cdot)[w]$ are bounded linear operators  (from $\Gamma^{p_1}$ to $\mc{E}_{\gamma}$ and from $\mc{G}^{p_2}$ to $\mc{E}_{g}$). In particular, the operators $\psi_{0\gamma}'[\cdot]$ and $\psi_{0g}'[\cdot]$ are bounded from below, i.e., %there exits constants $m_1, m_2>0$ such that 
\begin{equation}
\label{ineq: grad gamma pos}
	\|\psi_{0\gamma}'(\cdot)[v]\|_2 \gtrsim \|v\|_2, \text{ for any } v \in \Gamma^{p_1},
\end{equation}
and 
\begin{equation}
\label{ineq: grad g pos}
	\|\psi_{0g}'(\cdot)[w]\|_2 \gtrsim \|w\|_2, \text{ for any } w \in \mc{G}^{p_2}.
\end{equation}
\end{lemma}

\begin{proof}[Proof of Lemma \ref{lemma: bdd linear operators}]
By solving initial value problems in (\ref{eq: gamma dev})-(\ref{eq: g dev}), the first derivatives of $\psi(t, x, \beta, \gamma, g)$ with respect to $\gamma$ and $g$ at the true parameter $(\beta_0, \gamma_0, g_0)$ are given by
\begin{align}
	\psi_{0\gamma}'(t, x)[v] &= g'_0(\Lambda_0(t, x))\Lambda'_{0\gamma}(t, x)[v]+v(t) \nonumber\\
	& = g'_0(\Lambda_0(t, x))\exp(g_0(\Lambda_{0}(t,x)))e^{x^T\beta_0}\int^{t}_0 \exp(\gamma_0(s))v(s)\dev s + v(t),\label{eq:grad psi gamma}\\
	\psi_{0g}'(t, x)[w] &= g'_0(\Lambda_0(t, x))\Lambda'_{0g}(t, x)[w]+w(\Lambda_0(t, x)) \nonumber\\
	& = g'_0(\Lambda_0(t, x))\exp(g_0(\Lambda_{0}(t,x)))\int^{\Lambda_0(t, x)}_0 \exp(-g_0(s))w(s)\dev s + w(\Lambda_0(t, x)),\label{eq:grad psi g}
\end{align}

We first verify that $\psi_{0\gamma}'[\cdot]$ is a bounded linear operator. Using $(a+b)^2 \leq 2(a^2+b^2)$, the $L_2$ norm of $\psi_{0\gamma}'(\cdot)[v]$ is bounded by
\begin{align}
	\|\psi_{0\gamma}'(\cdot)[v]\|^2_2 \leq & 2\int_{\mc{X}} \int_{0}^{\tau} \left(g'_0(\Lambda_0(t, x))\exp(g_0(\Lambda_{0}(t,x)))e^{x^T\beta_0} \int^{t}_0 \exp(\gamma_0(s))v(s)\dev s\right)^2 \dev \Lambda_0(t, x) \dev F_X(x) \nonumber \\
	& + 2\int_{\mc{X}} \int_{0}^{\tau} v(t)^2 \dev \Lambda_0(t, x) \dev F_X(x) \nonumber \\
	= & 2\int_{\mc{X}} \int_{0}^{\tau} (g'_0(\Lambda_0(t, x)))^2\exp(2g_0(\Lambda_{0}(t,x)))e^{2x^T\beta_0} \left( \int^{t}_0 \exp(\gamma_0(s))v(s)\dev s\right)^2 \dev \Lambda_0(t, x) \dev F_X(x) \nonumber \\
	& + 2\int_{\mc{X}} \int_{0}^{\tau} v(t)^2 \dev \Lambda_0(t, x) \dev F_X(x) \label{ineq: upper bound of v}.
\end{align}
By the Cauchy-Schwarz inequality, we have for $t\in [0,\tau]$
\begin{align*}
	\left(\int^{t}_0 \exp(\gamma_0(s))v(s)\dev s\right)^2 & \leq \int^{t}_0 \left(v(s)\right)^2 \dev s \int^{t}_0 \exp(2\gamma_0(s)) \dev s\\
	& \leq \int^{\tau}_0 \left(v(s)\right)^2 \dev s \int^{\tau}_0 \exp(2\gamma_0(s))\dev s\\
	& \leq \|v\|_2^2 \tau e^{2c_1},
\end{align*}
where $c_1 = \max_{s\in[0,\tau]}\gamma_0(s) < \infty$ under \ref{c: smoothness}. 
It follows that the first term in (\ref{ineq: upper bound of v}) is bounded above by
\[2\|v\|_2^2 \tau e^{2c_1} \cdot  \int_{\mc{X}} \int_{0}^{\tau} (g'_0(\Lambda_0(t, x)))^2\exp(2g_0(\Lambda_{0}(t,x)))e^{2x^T\beta_0} \dev \Lambda_0(t, x) \dev F_X(x) \lesssim \|v\|_2^2,\]
because the integral is finite under \ref{c: bounded x}-\ref{c: smoothness}. The second term in (\ref{ineq: upper bound of v}) is also bounded by
\begin{align*}
	& 2\int_{\mc{X}} \int_{0}^{\tau} v(t)^2 \dev \Lambda_0(t, x) \dev F_X(x)\\
	= & 2\int_{\mc{X}} \int_{0}^{\tau} \exp(x^T\beta_0 + \gamma_0(t)+g_0(\Lambda_0(t,x)))v(t)^2 \dev t \dev F_X(x) \\
	\leq & 2\int_{\mc{X}} \int_{0}^{\tau} c_2 v(t)^2 \dev t \dev F_X(x) = 2c_2 \|v\|^2_2,
\end{align*}
where $c_2 = \max_{t\in[0,\tau],x\in\mc{X}}\exp(x^T\beta_0 + \gamma_0(t)+g_0(\Lambda_0(t,x)))< \infty$ under \ref{c: bounded x}-\ref{c: smoothness}. Therefore, $\|\psi_{0\gamma}'(\cdot)[v]\|_2 \lesssim \|v\|_2$ for any $v\in \Gamma^{p_1}$.

Similarly, we can show that $\psi_{0g}'[\cdot]$ is a bounded linear operator by
\begin{align*}
	\|\psi_{0g}'(\cdot)[w]\|^2_2 & =\int_{\mc{X}} \int_{0}^{\tau} (\psi_{0g}'(t, x)[w])^2 \dev \Lambda_0(t, x) \dev F_X(x)\\
	& = \int_{\mc{X}} \int_{0}^{\Lambda_0(\tau, x)}\left( g'_0(t) \exp(g_0(t))\int^{t}_0 \exp(-g_0(s))w(s)\dev s +w(t)\right)^2 \dev t \dev F_X(x)\\
	& \lesssim  \int_{\mc{X}} \int_{0}^{\Lambda_0(\tau, x)}\left( g'_0(t) \exp(g_0(t))\int^{t}_0 \exp(-g_0(s))w(s)\dev s\right)^2 \dev t \dev F_X(x)\\
	& \ \ \ \ +  \int_{\mc{X}} \int_{0}^{\Lambda_0(\tau, x)}\left(w(t)\right)^2 \dev t \dev F_X(x)\\
	& \lesssim \int_{\mc{X}} \int_{0}^{\Lambda_0(\tau, x)}(g'_0(t))^2 \exp(2g_0(t)) \left(\int^{t}_0 \exp(-2g_0(s))\dev s \right)  \left(\int^t_0 (w(s))^2\dev s\right) \dev t \dev F_X(x)\\
	& \ \ \ \ +  \int_{\mc{X}} \int_{0}^{\Lambda_0(\tau, x)}\left(w(t)\right)^2 \dev t \dev F_X(x)\\
	& \lesssim \int_{0}^{\mu}(g'_0(t))^2 \exp(2g_0(t)) \left(\int^{t}_0 \exp(-2g_0(s))\dev s \right)  \left(\int^t_0 (w(s))^2\dev s\right) \dev t \\
	& \ \ \ \ +  \int_{0}^{\mu}\left(w(t)\right)^2 \dev t ,
\end{align*}
where the second last inequality holds by the Cauchy-Schwarz inequality and $\mu=\max_{x\in\mc{X}}\Lambda_0(\tau, x)$ given in condition \ref{c: bounded t}. The first term is further bounded by 
\begin{align*}
	& \left(\int_{0}^{\mu} (w(t))^2\dev t\right)  \int_{0}^{\mu}(g'_0(t))^2 \exp(2g_0(t)) \left(\int^{t}_0 \exp(-2g_0(s))\dev s \right) \dev t \lesssim \int_{0}^{\mu} (w(t))^2\dev t = \|w\|_2^2,
\end{align*}
since the second integral is finite under conditions \ref{c: bounded x}-\ref{c: smoothness}. Thus, $\|\psi_{0g}'(\cdot)[w]\|_2 \lesssim  \|w\|_2$ for any $w\in \mc{G}^{p_2}$.

Next, we show that linear operators $\psi_{0\gamma}'[\cdot]$ and $\psi_{0g}'[\cdot]$ are bijective functions. 
Suppose that $\psi_{0\gamma}'(\cdot)[v_1]=\psi_{0\gamma}'(\cdot)[v_2]\in \mc{E}_\gamma$ holds almost surely with respect to the measure $\rho(t,x)=\Lambda_0(t,x)\times F_X(x)$. 
Using the ODE in (\ref{eq: gamma dev}), we have
\[v_i(t)= \psi_{0\gamma}'(t, x)[v_i]-g'_0(\Lambda_0(t, x))\int^t_0\psi_{0\gamma}'(s, x)[v_i]\dev \Lambda_0(s, x), \text{ for }i=1,2,\]
and then $v_1=v_2$ almost surely with respect to $\rho$, i.e., $\int_{\mc{X}} \int_{0}^{\tau} (v_1(t)-v_2(t))^2 \dev \rho(t, x)=0$. 
It follows that
\begin{align*}
	\int_{\mc{X}} \int_{0}^{\tau} (v_1(t)-v_2(t))^2 \dev \rho(t, x) & = \int_{\mc{X}} \int_{0}^{\tau}\exp(x^T\beta_0 + \gamma_0(t)+g_0(\Lambda_0(t,x))) (v_1(t)-v_2(t))^2 \dev t \dev F_X(x)\\
	& \geq \int_{\mc{X}} \int_{0}^{\tau} c_3 (v_1(t)-v_2(t))^2 \dev t \dev F_X(x) = c_3 \|v_1-v_2\|^2_2,
\end{align*}
where $c_3 = \min_{t\in[0,\tau],x\in\mc{X}}\exp(x^T\beta_0 + \gamma_0(t)+g_0(\Lambda_0(t,x)))< \infty$ under \ref{c: bounded x}-\ref{c: smoothness}, which implies that $\psi_{0\gamma}'[\cdot]$ is a bijective function from $\Gamma^{p_1}$ to $\mc{E}_{\gamma}$.

Similarly, suppose that $\psi_{0g}'(\cdot)[w_1]=\psi_{0g}'(\cdot)[w_2]\in \mc{E}_g$ holds almost surely with respect to the measure $\rho(t,x)$. Using the ODE in (\ref{eq: g dev}), we have
\[w_i(\Lambda_0(t, x))= \psi_{0g}'(t, x)[w_i]-g'_0(\Lambda_0(t, x))\int^t_0\psi_{0g}'(s, x)[w_i]\dev \Lambda_0(s, x), \text{ for }i=1,2,\]
and then $w_1(\Lambda_0(t, x))=w_2(\Lambda_0(t, x))$ almost surely with respect to $\rho$. It follows that
\begin{align*}
	0& = \int_{\mc{X}} \int_{0}^{\tau} (w_1(\Lambda_0(t, x))-w_2(\Lambda_0(t, x)))^2 \dev \rho(t, x)\\
	& = \int_{\mc{X}} \int_{0}^{\Lambda_0(\tau, x)} (w_1(t)-w_2(t))^2 \dev t \dev F_X(x)\\
	& \gtrsim \int_{0}^{\sup_{x\in\mc{X}}\Lambda_0(\tau, x)} (w_1(t)-w_2(t))^2 \dev t = \|w_1-w_2\|^2_2,
\end{align*}
where the last inequality holds under condition \ref{c: bounded x}. So $w_1=w_2\in \mc{G}^{p_2}$ and $\psi_{0g}'[\cdot]$ is a bijective function from $\mc{G}^{p_1}$ to $\mc{E}_{g}$.

By bounded inverse theorem, it follows that the bijective bounded linear operators $\psi_{0\gamma}'[\cdot]$ and $\psi_{0g}'[\cdot]$ have bounded inverse operator $(\psi_{0\gamma}')^{-1}[\cdot]$ and $(\psi_{0g}')^{-1}[\cdot]$. Then, there is a constant $0<L<\infty$ such that 
\[\|v\|_2 =\|(\psi_{0\gamma}')^{-1}\left[\psi_{0\gamma}'(\cdot)[v]\right] \|_2 \leq L \|\psi_{0\gamma}'(\cdot)[v]\|_2 ,\]
which implies that $\psi_{0\gamma}'[\cdot]$ is bounded from below since $\|\psi_{0\gamma}'(\cdot)[v]\|_2 \geq 1/L\|v\|_2 $. Analogously, $\psi_{0g}'[\cdot]$ is also bounded from below, which can be obtained using the same argument as above.
\end{proof}

\begin{lemma}
\label{lemma: score space}	Let $\zeta_{\eta}(\cdot, \beta, \gamma)$ be a smooth curve in $\mathcal{H}^{p_2}$ running through $\zeta(\cdot, \beta, \gamma)$ at $\eta=0$, that is $\zeta_{\eta}(\cdot, \beta, \gamma)|_{\eta=0} = \zeta(\cdot, \beta, \gamma)$.
For any score function $h(\cdot,\beta,  \gamma)$ with $\zeta(\cdot, \beta, \gamma) = g (\Lambda(\cdot, \beta, \gamma, g))$ in \[\mathbb{H}=\left\{ h:h(\cdot,\beta,  \gamma)=\frac{\partial \zeta_{\eta}(\cdot, \beta, \gamma)}{\partial \eta}|_{\eta=0}, \zeta_{\eta}\in \mc{H}^{p_2} \right\},\]
under conditions \ref{c: bounded b}-\ref{c: smoothness}, there exists $w\in \mathbb{W}$ such that   
\[h(\cdot,\beta,  \gamma)= w (\Lambda(\cdot, \beta, \gamma, g))+ g'(\Lambda(\cdot, \beta, \gamma, g))\Lambda'_{g}(\cdot, \beta, \gamma, g)[w].\]
%where \textcolor{red}{$w\in \mathbb{W}$}.
\end{lemma}

\begin{proof}[Proof of Lemma \ref{lemma: score space}]
Since $\zeta_{\eta}(\cdot, \beta, \gamma)$ is a smooth curve in $\mathcal{H}^{p_2}$ running through $\zeta(\cdot, \beta, \gamma)$ at $\eta=0$, we can rewrite it in the form of $\zeta_{\eta}(\cdot, \beta, \gamma) = g_{\eta} (\Lambda(\cdot, \beta, \gamma, g_{\eta}))$ where $g_{\eta}$ is a smooth curve in $\mc{G}^{p_2}$ running through $g$ at $\eta=0$. For a small $\eta$, we have $g_{\eta} = g + \eta w + o(\eta)$ with $w = \frac{\partial g_{\eta}}{\partial \eta}|_{\eta=0} \in \mathbb{W}$. It follows that
\[\lim_{\eta\rightarrow 0} \frac{g_{\eta} (\Lambda(\cdot, \beta, \gamma, g_{\eta})) - g(\Lambda(\cdot, \beta, \gamma, g_{\eta}))}{\eta} = w(\Lambda(\cdot, \beta, \gamma, g)).\]
Also, by the definition of functional derivatives, we have
\begin{align*}
	g (\Lambda(\cdot, \beta, \gamma, g_{\eta})) - g(\Lambda(\cdot, \beta, \gamma, g)) & = g (\Lambda(\cdot, \beta, \gamma, g + \eta w + o(\eta))) - g(\Lambda(\cdot, \beta, \gamma, g_{\eta})) \\
	 & =   g' (\Lambda(\cdot, \beta, \gamma, g))\Lambda'_g(\cdot, \beta, \gamma, g)[\eta w + o(\eta)] + o(\|\eta w + o(\eta)\|)\\
	 	 & = \eta g' (\Lambda(\cdot, \beta, \gamma, g))\Lambda'_g(\cdot, \beta, \gamma, g)[w]+o(\eta),
\end{align*}
where the last equality holds because
\[\lim_{\eta\rightarrow 0} \frac{g' (\Lambda(\cdot, \beta, \gamma, g))\Lambda'_g(\cdot, \beta, \gamma, g)[o(\eta)]}{\eta} = g' (\Lambda(\cdot, \beta, \gamma, g))\Lambda'_g(\cdot, \beta, \gamma, g)\Big[\lim_{\eta\rightarrow 0}\frac{o(\eta)}{\eta}\Big] = 0.\]
Combining these two equations together, we have,
\begin{align*}
	h(\cdot, \beta, \gamma) & = \lim_{\eta\rightarrow0}\frac{g_{\eta} (\Lambda(\cdot, \beta, \gamma, g_{\eta})) - g(\Lambda(\cdot, \beta, \gamma, g))}{\eta}\\
	& = \lim_{\eta\rightarrow0}\frac{g_{\eta} (\Lambda(\cdot, \beta, \gamma, g_{\eta})) - g(\Lambda(\cdot, \beta, \gamma, g_{\eta})) +g (\Lambda(\cdot, \beta, \gamma, g_{\eta})) - g(\Lambda(\cdot, \beta, \gamma, g))}{\eta} \\
	& = w (\Lambda(\cdot, \beta, \gamma, g))+ g'(\Lambda(\cdot, \beta, \gamma, g))\Lambda'_{g}(\cdot, \beta, \gamma, g)[w].
\end{align*}
\end{proof}

\begin{lemma}
	\label{lemma: devivatives}
	Denote 
	\begin{align*}
		l(\beta, \gamma, \zeta(\cdot, \beta, \gamma); W) &= \Delta \{X^T\beta + \gamma(Y)+g(\Lambda(Y, X, \beta, \gamma, g))\}-\Lambda(Y, X, \beta, \gamma, g)\\
		& = \Delta \{X^T\beta + \gamma(Y)+\zeta(Y, X, \beta, \gamma)\}-\int_0^Y \exp(X^T\beta + \gamma(t)+\zeta(t, X, \beta, \gamma)) \dev t.
	\end{align*}
Under conditions \ref{c: bounded b}-\ref{c: smoothness}
, $l(\beta, \gamma, \zeta(\cdot, \beta, \gamma); W)$ has bounded and continuous first and second derivatives with respect to $\beta \in \mc{B}$, $\gamma \in \Gamma^{p_2}$, and $\zeta(\cdot, \beta, \gamma) \in \mc{H}^{p_1}$.
\end{lemma}

\begin{proof}[Proof of Lemma \ref{lemma: devivatives}]
	\label{proof: devivatives}
	The derivatives with respect to the first, the second, and the third argument of the objective function are
	\[l'_1(\beta, \gamma, \zeta; W) = \Delta X - X\int_0^Y \exp(X^T\beta + \gamma(t)+\zeta(t, X, \beta, \gamma)) \dev t,\]
	\[l'_2(\beta, \gamma, \zeta; W)[v] = \Delta v(Y) - \int_0^Y \exp(X^T\beta + \gamma(t)+\zeta(t, X, \beta, \gamma))v(t) \dev t,\]
	\[l'_3(\beta, \gamma, \zeta; W)[h] = \Delta h(Y, X, \beta, \gamma) - \int_0^Y \exp(X^T\beta + \gamma(t)+\zeta(t, X, \beta, \gamma))h(t, X, \beta, \gamma) \dev t,\]
	\[l''_{11}(\beta, \gamma, \zeta; W) =- XX^T \int_0^Y \exp(X^T\beta + \gamma(t)+\zeta(t, X, \beta, \gamma)) \dev t,\]
	\[l''_{12}(\beta, \gamma, \zeta; W)[v] =- X\int_0^Y \exp(X^T\beta + \gamma(t)+\zeta(t, X, \beta, \gamma)) v(t)\dev t,\]
	\[l''_{13}(\beta, \gamma, \zeta; W)[h] =- X\int_0^Y \exp(X^T\beta + \gamma(t)+\zeta(t, X, \beta, \gamma)) h(t, X, \beta, \gamma)\dev t,\]
	\[l''_{23}(\beta, \gamma, \zeta; W)[v, h] =- \int_0^Y \exp(X^T\beta + \gamma(t)+\zeta(t, X, \beta, \gamma))v(t) h(t, X, \beta, \gamma)\dev t,\]
	\[l''_{22}(\beta, \gamma, \zeta; W)[v_1, v_2] =- \int_0^Y \exp(X^T\beta + \gamma(t)+\zeta(t, X, \beta, \gamma))v_1(t) v_2(t)\dev t,\]
	\[l''_{33}(\beta, \gamma, \zeta; W)[h_1, h_2] =- \int_0^Y \exp(X^T\beta + \gamma(t)+\zeta(t, X, \beta, \gamma))h_1(t, X, \beta, \gamma) h_2(t, X, \beta, \gamma)\dev t.\]
	The derivatives with respect to $\beta$ and $\gamma$ of $\zeta(\cdot, \beta, \gamma)$ are
	\begin{align*}
	  \zeta'_{\beta}(\cdot, \beta, \gamma)& = g'(\Lambda(\cdot, \beta, \gamma, g))\Lambda'_{\beta}(\cdot, \beta, \gamma, g),\\
	 \zeta'_{\gamma}(\cdot, \beta, \gamma)[v] &= g'(\Lambda(\cdot, \beta, \gamma, g))\Lambda'_{\gamma}(\cdot, \beta, \gamma, g)[v],\\
\zeta''_{\beta\beta}(\cdot, \beta, \gamma) &= g''(\Lambda(\cdot, \beta, \gamma, g))\Lambda'_{\beta}(\cdot, \beta, \gamma, g)\Lambda'_{\beta}(\cdot, \beta, \gamma, g)^T + g'(\Lambda(\cdot, \beta, \gamma, g))\Lambda''_{\beta\beta}(\cdot, \beta, \gamma, g),\\
	    \zeta''_{\gamma\gamma}(\cdot, \beta, \gamma)[v_1, v_2] & = g''(\Lambda(\cdot, \beta, \gamma, g))\Lambda'_{\gamma}(\cdot, \beta, \gamma, g)[v_1]\Lambda'_{\gamma}(\cdot, \beta, \gamma, g)[v_2]\\
	    & + g'(\Lambda(\cdot, \beta, \gamma, g))\Lambda''_{\gamma\gamma}(\cdot, \beta, \gamma, g)[v_1, v_2],\\
 	    \zeta''_{\beta\gamma}(\cdot, \beta, \gamma)[v] &= g''(\Lambda(\cdot, \beta, \gamma, g))\Lambda'_{\beta}(\cdot, \beta, \gamma, g)\Lambda'_{\gamma}(\cdot, \beta, \gamma, g)[v]\\
	    & + g'(\Lambda(\cdot, \beta, \gamma, g))\Lambda''_{\beta\gamma}(\cdot, \beta, \gamma, g)[v].
	\end{align*}
   After some calculations using the chain rule, we have
   \begin{align*}
     l'_{\beta}(\beta, \gamma, \zeta(\cdot, \beta, \gamma); W)  & = \Delta\{X+\zeta'_{\beta}(Y, X, \beta, \gamma)\} \\
     & \ \ \ \ - \int_0^Y \exp(X^T\beta + \gamma(t)+\zeta(t, X, \beta, \gamma))[\zeta'_{\beta}(t, X, \beta, \gamma)+X] \dev t\\
     & = \Delta\{X+g'(\Lambda(Y, X, \beta, \gamma, g))\Lambda'_{\beta}(Y, X, \beta, \gamma, g)\} - \Lambda'_{\beta}(Y, X, \beta, \gamma, g),
   \end{align*}
   \begin{align*}
     l'_{\gamma}(\beta, \gamma, \zeta(\cdot, \beta, \gamma); W)[v]  & = \Delta\{v(Y)+\zeta'_{\gamma}(Y, X, \beta, \gamma)[v]\} \\
     & \ \ \ \  - \int_0^Y \exp(X^T\beta + \gamma(t)+\zeta(t, X, \beta, \gamma))\{v(t)+ \zeta'_{\gamma}(t, X, \beta, \gamma)[v]\} \dev t\\
     & = \Delta\{v(Y)+g'(\Lambda(Y, X, \beta, \gamma, g))\Lambda'_{\gamma}(Y, X, \beta, \gamma, g)[v]\} - \Lambda'_{\gamma}(Y, X, \beta, \gamma, g)[v],
   \end{align*} 
\[ l'_{\zeta}(\beta, \gamma, \zeta(\cdot, \beta, \gamma); W)[h] = \Delta h(Y, X, \beta, \gamma) - \int_0^Y \exp(X^T\beta + \gamma(t)+\zeta(t, X, \beta, \gamma))h(t, X, \beta, \gamma) \dev t, \]
    \begin{align*}
     l''_{\beta\beta}(\beta, \gamma, \zeta(\cdot, \beta, \gamma); W)  & = \Delta \zeta''_{\beta\beta}(Y, X, \beta, \gamma) - \int_0^Y \exp(X^T\beta + \gamma(t)+\zeta(t, X, \beta, \gamma))\\
     & \ \ \ \ \cdot [(X+\zeta'_{\beta}(t, X, \beta, \gamma))(X+\zeta'_{\beta}(t, X, \beta, \gamma))^T +\zeta''_{\beta\beta}(t, X, \beta, \gamma) ] \dev t\\
     & = \Delta\{g''(\Lambda(Y, X, \beta, \gamma, g))\Lambda'_{\beta}(Y, X, \beta, \gamma, g)\Lambda'_{\beta}(Y, X, \beta, \gamma, g)^T \\
     & \ \ \ \ \ \ + g'(\Lambda(Y, X, \beta, \gamma, g))\Lambda''_{\beta\beta}(Y, X, \beta, \gamma, g)\} \\
     & \ \ \ \ - \Lambda''_{\beta\beta}(Y, X, \beta, \gamma, g),
  \end{align*}
  \begin{align*}
     l''_{\gamma\gamma}(\beta, \gamma, \zeta(\cdot, \beta, \gamma); W)[v_1, v_2]  & = \Delta \zeta''_{\gamma\gamma}(Y, X, \beta, \gamma)[v_1, v_2] - \int_0^Y \exp(X^T\beta + \gamma(t)+\zeta(t, X, \beta, \gamma))\\
     & \ \ \ \ \cdot \{(v_1(t)+\zeta'_{\gamma}(t, X, \beta, \gamma)[v_1])(v_2(t)+\zeta'_{\gamma}(t, X, \beta, \gamma)[v_2]) \\
     & \ \ \ \ \ \ +\zeta''_{\gamma\gamma}(t, X, \beta, \gamma)[v_1, v_2] \} \dev t\\
     & = \Delta\{g''(\Lambda(Y, X, \beta, \gamma, g))\Lambda'_{\gamma}(Y, X, \beta, \gamma, g)[v_1]\Lambda'_{\gamma}(Y, X, \beta, \gamma, g)[v_2]\\
     & \ \ \ \ \ \ + g'(\Lambda(Y, X, \beta, \gamma, g))\Lambda''_{\gamma\gamma}(Y, X, \beta, \gamma, g)[v_1, v_2]\} \\
     & \ \ \ \ - \Lambda''_{\gamma\gamma}(Y, X, \beta, \gamma, g)[v_1, v_2],
   \end{align*}
 \begin{align*}
     l''_{\gamma\beta}(\beta, \gamma, \zeta(\cdot, \beta, \gamma); W)[v]  & = \Delta \zeta''_{\gamma\beta}(Y, X, \beta, \gamma)[v] - \int_0^Y \exp(X^T\beta + \gamma(t)+\zeta(t, X, \beta, \gamma))\\
     &\ \ \ \  \cdot \{(v(t)+\zeta'_{\gamma}(t, X, \beta, \gamma)[v])(X+\zeta'_{\beta}(t, X, \beta, \gamma)) \\
     & \ \ \ \ \ \ +\zeta''_{\gamma\beta}(t, X, \beta, \gamma)[v] \} \dev t\\
     & = \Delta\{g''(\Lambda(Y, X, \beta, \gamma, g))\Lambda'_{\gamma}(Y, X, \beta, \gamma, g)[v]\Lambda'_{\beta}(Y, X, \beta, \gamma, g)\\
	    & \ \ \ \ \ \ + g'(\Lambda(Y, X, \beta, \gamma, g))\Lambda''_{\gamma\beta}(Y, X, \beta, \gamma, g)[v]\} \\
     & \ \ \ \ - \Lambda''_{\gamma\beta}(Y, X, \beta, \gamma, g)[v],
   \end{align*}
   \begin{align*}
     l''_{\zeta\beta}(\beta, \gamma, \zeta(\cdot, \beta, \gamma); W)[h]  & = \Delta h'_{\beta}(Y, X, \beta, \gamma) - \int_0^Y \exp(X^T\beta + \gamma(t)+\zeta(t, X, \beta, \gamma))\\
     &\ \ \ \  \cdot \{(h(t, X, \beta, \gamma))(X+\zeta'_{\beta}(t, X, \beta, \gamma))+h'_{\beta}(t, X, \beta, \gamma) \} \dev t\\
     & = \Delta\{w'(\Lambda(Y, X, \beta, \gamma, g)) \Lambda'_{\beta}(Y, X, \beta, \gamma, g)\\
     & \ \ \ \ \ \ + g''(\Lambda(Y, X, \beta, \gamma, g))\Lambda'_{g}(Y, X, \beta, \gamma, g)[w]
     \Lambda'_{\beta}(Y, X, \beta, \gamma, g)\\
	    & \ \ \ \ \ \ + g'(\Lambda(Y, X, \beta, \gamma, g))\Lambda''_{g\beta}(Y, X, \beta, \gamma, g)[w]\} \\
     & \ \ \ \ - \Lambda''_{g\beta}(Y, X, \beta, \gamma, g)[w],
   \end{align*}
   \begin{align*}
     l''_{\zeta\gamma}(\beta, \gamma, \zeta(\cdot, \beta, \gamma); W)[h, v]  & = \Delta h'_{\gamma}(Y, X, \beta, \gamma)[v] - \int_0^Y \exp(X^T\beta + \gamma(t)+\zeta(t, X, \beta, \gamma))\\
     &\ \ \ \  \cdot \{(h(t, X, \beta, \gamma))(v(t)+\zeta'_{\gamma}(t, X, \beta, \gamma)[v])+h'_{\gamma}(t, X, \beta, \gamma)[v] \} \dev t\\
     & = \Delta\{w'(\Lambda(Y, X, \beta, \gamma, g)) \Lambda'_{\gamma}(Y, X, \beta, \gamma, g)[v]\\
     & \ \ \ \ \ \ + g''(\Lambda(Y, X, \beta, \gamma, g))\Lambda'_{g}(Y, X, \beta, \gamma, g)[w]
     \Lambda'_{\gamma}(Y, X, \beta, \gamma, g)[v]\\
	    & \ \ \ \ \ \ + g'(\Lambda(Y, X, \beta, \gamma, g))\Lambda''_{g\gamma}(Y, X, \beta, \gamma, g)[w, v]\} \\
     & \ \ \ \ - \Lambda''_{g\gamma}(Y, X, \beta, \gamma, g)[w, v],
   \end{align*}
 \begin{align*}
 l''_{\zeta\zeta}(\beta, \gamma, \zeta(\cdot, \beta, \gamma); W)[h_1, h_2]  & = - \int_0^Y \exp(X^T\beta + \gamma(t)+\zeta(t, X, \beta, \gamma))h_1(t, X, \beta, \gamma)h_2(t, X, \beta, \gamma) \dev t,\\
     & = - \int_0^Y \exp(X^T\beta + \gamma(t)+\zeta(t, X, \beta, \gamma))\\
     & \ \ \ \ \cdot  \{w_1(\Lambda(t, X, \beta, \gamma, g)) + g'(\Lambda(t, X, \beta, \gamma, g))\Lambda'_{g}(t, X, \beta, \gamma, g)[w_1]\}\\
     & \ \ \ \  \cdot  \{w_2(\Lambda(t, X, \beta, \gamma, g)) + g'(\Lambda(t, X, \beta, \gamma, g))\Lambda'_{g}(t, X, \beta, \gamma, g)[w_2]\} \dev t,
 \end{align*}

All the above derivatives are bounded and continuous under conditions \ref{c: bounded b}-\ref{c: smoothness} by Lemma \ref{lemma: ode solutions}.
\end{proof}

\begin{lemma} (Spline approximation)
	\label{lemma: gnp and gp}
	For $\gamma_0 \in \Gamma^{p_1}$, there exists a function $\gamma_{0n}\in \Gamma^{p_1}_{n}$ such that \[\|\gamma_{0n}-\gamma_0\|_{\infty}=O(n^{-p_1\nu_1}).\]
	For $g_0 \in \mc{G}^{p_2}$, there exists a function $g_{0n}\in \mc{G}^{p_2}_{n}$ such that \[\|g_{0n}-g_0\|_{\infty}=O(n^{-p_2\nu_2}).\]
\end{lemma}
\begin{proof}[Proof of Lemma \ref{lemma: gnp and gp}]
Since $\gamma_0 \in \Gamma^{p_1}\subset S^{p_1}([0,\tau])$, by Corollary 6.21 in \citet{schumaker_2007}, there exists a function in the polynomial space with order $p_1$, i.e., $\tilde{\gamma}_{0n}\in S_{n}(T_{K_{n}^1}, K_{n}^1, p_1)$, such that $\|\tilde{\gamma}_{0n}-\gamma_0\|_{\infty}=O(n^{-p_1\nu_1})$. It follows that 
\begin{align*}
    \|(\tilde{\gamma}_{0n}(\cdot)-\tilde{\gamma}_{0n}(t^*)) - \gamma_0\|_{\infty} & \leq \|\tilde{\gamma}_{0n}-\gamma_0\|_{\infty} +  |\tilde{\gamma}_{0n}(t^*)| \\
    & = \|\tilde{\gamma}_{0n}-\gamma_0\|_{\infty} +  |\tilde{\gamma}_{0n}(t^*) - \gamma_0(t^*)|\\
    & \leq 2 \|\tilde{\gamma}_{0n}-\gamma_0\|_{\infty} = O(n^{-p_1\nu_1}),
\end{align*}
where the second equality holds because \new{$\gamma_0(t^*)=0$ for $\gamma_0 \in \Gamma^{p_1}$. 
Let $\gamma_{0n}(\cdot)=\tilde{\gamma}_{0n}(\cdot)-\tilde{\gamma}_{0n}(t^*)$, then $\gamma_{0n}(t^*)=0$} and thereby we find $\gamma_{0n}\in \Gamma^{p_1}_{n}$ such that $\|\gamma_{0n}-\gamma_0\|_{\infty}=O(n^{-p_1\nu_1})$. The second part is a direct result of Corollary 6.21 in \citet{schumaker_2007}.
\end{proof}

\begin{lemma} (Bracket number of $l(\theta;W)$)
	\label{lemma: bracket number of Fn}
	Let $\theta_{0n} = (\beta_0, \gamma_{0n}(\cdot), \zeta_{0n}(\cdot, \beta_0, \gamma_{0n}))$ with $$\zeta_{0n}(t, x, \beta_0, \gamma_{0n}) = g_{0n}(\Lambda(t, x, \beta_0, \gamma_{0n}, g_{0n})),$$ where $\gamma_{0n}$ and $g_{0n}$ are defined in Lemma \ref{lemma: gnp and gp}. Denote $\mc{F}_n=\{l(\theta;W)-l(\theta_{0n};W): \theta \in \Theta_n \}$. Under conditions \ref{c: bounded b}-\ref{c: smoothness}, the $\epsilon$-bracketing number associated with $\|\cdot\|_{\infty}$ for $\mc{F}_n$, denoted by $N_{[\ ]}(\epsilon, \mc{F}_n, \|\cdot\|_{\infty})$, has the following upper bound for some constants $c_1$ and $c_2$,
	%is bounded by %$O(\left(\frac{1}{\epsilon}\right)^{c_1 q_{n_1}+c_2q_{n_2}+d})$ for some constants $c_1$ and $c_2$, i.e.
	\[N_{[\ ]}(\epsilon, \mc{F}_n, \|\cdot\|_{\infty})\lesssim \left(\frac{1}{\epsilon}\right)^{c_1 q_{n_1}+c_2q_{n_2}+d}. \]
\end{lemma}
\begin{proof}[Proof of Lemma \ref{lemma: bracket number of Fn}]
	\label{proof: bracket number of Fn}
	Denote the ceiling of $x$ by $\ceil*{x}$. Following the calculation in \citet[Page 597]{shen1994}, we have that, for any $\epsilon>0$, there exists a set of $\epsilon$-brackets 
	\[\left\{[\gamma_l^L, \gamma_l^U]: \|\gamma_l^U-\gamma_l^L\|_{\infty}\leq \epsilon, l=1, \cdots, \ceil*{(\frac{1}{\epsilon})^{c_1q_{n_1}}}\right\}\]
	such that for any $\gamma \in \Gamma^{p_1}_{n}$, $\gamma_l^L(t)\leq \gamma(t) \leq \gamma_l^U(t)$ holds on $[0, \tau]$ for some $1\leq l\leq \ceil*{(\frac{1}{\epsilon})^{c_1q_{n_1}}}$. Similarly, there exists another set of $\epsilon$-brackets
	\[\left\{[g_i^L, g_i^U]: \|g_i^U-g_i^L\|_{\infty}\leq \epsilon, i=1, \cdots, \ceil*{(\frac{1}{\epsilon})^{c_2q_{n_2}}} \right\}\]
	such that for any $g \in \mc{G}^{p_1}_{n}$, $g_i^L(t)\leq g(t)\leq  g_i^U(t)$ holds on $[0, \mu]$ for some $1\leq i\leq \ceil*{(\frac{1}{\epsilon})^{c_2q_{n_2}}}$.
	Since $\mc{B} \subset \reals^d$ is compact, it can be covered by $\ceil*{c_3(\frac{1}{\epsilon})^d}$ balls with radius $\epsilon$, i.e. for any $\beta \in \mc{B}$, there exists $1\leq k \leq \ceil*{c_3(\frac{1}{\epsilon})^d}$ such that $\|\beta_k-\beta\|\leq \epsilon$. Hence, under condition \ref{c: bounded x}, $|X^T\beta-X^T\beta_k|\leq c_4 \epsilon$ for some constant $c_4>0$ and any $X\in \mc{X}$. By the mean value theorem, we have that
	\begin{align*}
	    & \ \ \ |\exp(g(\Lambda)+X^T\beta+\gamma(t))-\exp(g_i^L(\Lambda)+X^T\beta_k+\gamma_l^L(t))|\\
	    &= \exp(\tilde{\psi}(t, \Lambda))|g(\Lambda)+X^T\beta+\gamma(t)-g_i^L(\Lambda)+X^T\beta_k+\gamma_l^L(t)|\\
	    &\leq \exp(\tilde{\psi}(t, \Lambda)) (|g(\Lambda)-g_i^L(\Lambda)|+|X^T\beta-X^T\beta_k|+|\gamma(t)-\gamma_l^L(t)|)\\
	    & \leq \exp(\tilde{\psi}(t, \Lambda)) (\|g-g_i^L\|_{\infty}+|X^T\beta-X^T\beta_k|+\|\gamma-\gamma_l^L\|_{\infty}), 
	\end{align*}
	where $\tilde{\psi}(t, \Lambda)= g_i^L(\Lambda)+X^T\beta_k+\gamma_l^L(t) +\xi (g(\Lambda)-g_i^L(\Lambda)+X^T(\beta-\beta_k)+\gamma(t) -\gamma_l^L(t))$ for some $\xi\in(0,1)$ and is bounded under conditions \ref{c: bounded b}-\ref{c: smoothness}. Hence, $$|\exp(g(\Lambda)+X^T\beta+\gamma(t))-\exp(g_i^L(\Lambda)+ X^T\beta_k + \gamma_l^L(t))| \lesssim \epsilon$$ over $(t,\Lambda)\in [0, \tau]\times [0, b]$. Employing Theorem 12.V of continuous dependence in \citet[page 145]{Walter1998}, we have $|\Lambda(t, X, \beta, \gamma, g)-\Lambda(t, X, \beta_k, \gamma_l^L, g_i^L)|\leq c_5 \epsilon$ for some constant $c_5>0$ and any $t\in [0, \tau]$. Denote $\Lambda_{ilk}(t, x)=\Lambda(t, x, \beta_k, \gamma_l^L, g_i^L)$. Define 
	\begin{align*}
	    m(\theta;W)& =l(\theta;W)-l(\theta_{0n};W)\\
	    &=\Delta\{X^T \beta  +\gamma(Y)+g(\Lambda(Y, X, \beta, \gamma, g))\}-\Lambda(Y, X, \beta, \gamma, g)-l(\theta_{0n};W),
	\end{align*}
\[ m_{ilk}^L(W)=\Delta\{X^T \beta_k - c_4 \epsilon +\gamma_l^L(Y)+g_{i}^L (\xi_{ilk}^L)\}-\Lambda_{ilk}(Y, X)-c_5 \epsilon-l(\theta_{0n};W),\]
and
\[m_{ilk}^U(W)=\Delta\{X^T \beta_k + c_4 \epsilon +\gamma_l^U(Y)+g_{i}^U (\xi_{ilk}^U)\}-\Lambda_{ilk}(Y, X)+c_5 \epsilon-l(\theta_{0n};W),\]
where $\xi_{ilk}^L= \arg\min_{|s|\leq c_5 \epsilon} g_i^L(\Lambda_{ilk}(Y, X)+s)$ and $\xi_{ilk}^U= \arg\max_{|s|\leq c_5 \epsilon} g_i^U(\Lambda_{ilk}(Y, X)+s)$. 
	
Note that $[m_{ilk}^L(W), m_{ilk}^U(W)]$ is a $\epsilon$-bracket because
\begin{align*}
    |m_{ilk}^U(W)- m_{ilk}^L(W)| & = |\Delta\{2c_4\epsilon+ \gamma_l^U(Y) - \gamma_l^L(Y)+g_{i}^U (\xi_{ilk}^U) - g_{i}^L (\xi_{ilk}^L)\} + 2c_5\epsilon|\\
    & \leq 2c_4\epsilon + |\gamma_l^U(Y) - \gamma_l^L(Y)| + |g_{i}^U (\xi_{ilk}^U) - g_{i}^L (\xi_{ilk}^U)| + |g_{i}^L (\xi_{ilk}^U)- g_{i}^L (\xi_{ilk}^L)|+2c_5\epsilon\\
    & \leq 2c_4\epsilon+ \|\gamma_l^U-\gamma_l^L\|_{\infty} + \|g_i^U-g_i^L\|_{\infty} + c_7|\xi_{ilk}^U - \xi_{ilk}^L| +2c_5\epsilon\\
    & \leq 2c_4\epsilon+ \epsilon + \epsilon + 2c_7c_5\epsilon +2c_5\epsilon\lesssim \epsilon,
\end{align*}
where $c_7=\max_{t\in [0, b]}|(g_i^L)'(t)|$ in the second inequality. Hence $\| m_{ilk}^U- m_{ilk}^L\|_{\infty}\lesssim\epsilon$.
	
For any $\theta=(\beta,\gamma(\cdot), \zeta(\cdot, \beta, \gamma))$ with $\zeta(t, x, \beta, \gamma)=g(\Lambda(t, x, \beta, \gamma, g))$, there exits $1\leq i \leq \ceil*{(\frac{1}{\epsilon})^{c_2q_{n_2}}}, 1\leq l \leq \ceil*{(\frac{1}{\epsilon})^{c_1q_{n_1}}}, 1\leq k \leq \ceil*{c_3(\frac{1}{\epsilon})^d}$ such that $g_i^L(t)\leq g(t)\leq  g_i^U(t)$ on $t\in [0, \mu]$, $\gamma_l^L(t)\leq \gamma(t) \leq \gamma_l^U(t)$ on $t\in [0, \tau]$, and $|X^T\beta_k-X^T\beta|\leq c_4\epsilon$. It follows that
\begin{align*}
    m_{ilk}^U(W) & = \Delta\{(X^T \beta_k + c_4 \epsilon) +\gamma_l^U(Y)+g_{i}^U (\xi_{ilk}^U)\}+ (c_5\epsilon-\Lambda_{ilk}(Y, X))-l(\theta_{0n};W)\\
    & \geq \Delta\{X^T \beta +\gamma(Y)+g_{i}^U (\xi_{ilk}^U)\}+(c_5\epsilon-\Lambda_{ilk}(Y, X))-l(\theta_{0n};W)\\
    & \geq \Delta\{X^T \beta +\gamma(Y)+g_{i}^U (\Lambda(t, X, \beta, \gamma, g))\}- \Lambda(t, X, \beta, \gamma, g)-l(\theta_{0n};W)\\
    & \geq \Delta\{X^T \beta +\gamma(Y)+g (\Lambda(t, X, \beta, \gamma, g))\}- \Lambda(t, X, \beta, \gamma, g)-l(\theta_{0n};W)\\
    & = m(\theta;W),
\end{align*}
where the second inequality holds because $|\Lambda(Y, X, \beta, \gamma, g)-\Lambda_{ilk}(Y, X)|\leq c_5 \epsilon$. The other side can be verified similarly. Therefore, we have   
	\[N_{[\ ]}(\epsilon, \mc{F}_n, \|\cdot\|_{\infty}) \lesssim \left(\frac{1}{\epsilon}\right)^{c_1 q_{n_1}}\left(\frac{1}{\epsilon}\right)^{c_2 q_{n_2}}\left(\frac{1}{\epsilon}\right)^{d}= \left(\frac{1}{\epsilon}\right)^{c_1 q_{n_1}+c_2q_{n_2}+d}, \] which completes the proof. 
\end{proof}

\begin{lemma}
	\label{lemma: other bracket numbers for A4} For $1\leq j\leq d$, 
	denote $\mc{F}_{n,j}^{\gamma}(\eta)=\{l'_{\gamma}(\theta;W)[v_j^*-v_{j}]: \theta \in \Theta_n, v_j\in \Gamma^1_n, d(\theta, \theta_0)\leq \eta, \|v_j^*-v_{j}\|_{\infty}\leq \eta \}$ and $\mc{F}_{n,j}^{\zeta}(\eta)=\{l'_{\zeta}(\theta;W)[h^*_j-h_{j}]: \theta \in \Theta_n, h_j\in \mc{H}^2_n, d(\theta, \theta_0)\leq \eta, \|w_j^*-w_{j}\|_{\infty}\leq \eta \}$, where $v_j^*$ is defined in condition \ref{c: pos info exist} and $h_j^*(\cdot, \beta, \gamma)=w_j^* (\Lambda(\cdot, \beta, \gamma, g))+ g'(\Lambda(\cdot, \beta, \gamma, g))\Lambda'_{g}(\cdot, \beta, \gamma, g)[w_j^*]$ with $w_j^*$ given in condition \ref{c: pos info exist}. Then under conditions \ref{c: bounded b}-\ref{c: smoothness} and \ref{c: pos info exist}, we have 
	\[N_{[\ ]}(\epsilon, \mc{F}_{n,j}^{\gamma}(\eta), \|\cdot\|_{\infty})\lesssim \left(\frac{\eta}{\epsilon}\right)^{c_1 q_{n_1}+c_2q_{n_2}+d}\]
	and 
	\[N_{[\ ]}(\epsilon, \mc{F}_{n,j}^{\zeta}(\eta), \|\cdot\|_{\infty})\lesssim \left(\frac{\eta}{\epsilon}\right)^{c_3 q_{n_1}+c_4q_{n_2}+d}\]
	  for some constants $c_1, c_2, c_3,$ and $c_4$.
\end{lemma}

\begin{lemma}
	\label{lemma: other bracket numbers for A5} For $1\leq j\leq d$, denote
	 \[\mc{F}_{n,j}^{*\beta}(\eta)=\{l'_{\beta_j}(\theta;W)-l'_{\beta_j}(\theta_0;W): \theta \in \Theta_n, d(\theta, \theta_0)\leq \eta, \|g'(\Lambda(\cdot, \beta, \gamma, g))-g'_0(\Lambda_0(\cdot))\|_{2}\leq \eta \},\]
	 \[\mc{F}_{n,j}^{*\gamma}(\eta)=\{l'_{\gamma}(\theta;W)[v^*_j]-l'_{\gamma}(\theta_0;W)[v^*_j]: \theta \in \Theta_n, d(\theta, \theta_0)\leq \eta, \|g'(\Lambda(\cdot, \beta, \gamma, g))-g'_0(\Lambda_0(\cdot))\|_{2}\leq \eta \},\] 
	 and 
	 \[\mc{F}_{n,j}^{*\zeta}(\eta)=\{l'_{\zeta}(\theta;W)[h^*_j]-l'_{\zeta}(\theta_0;W)[h^*_j]: \theta \in \Theta_n, d(\theta, \theta_0)\leq \eta\},\] 
	 where $v_j^*$ is defined in condition \ref{c: pos info exist} and $h_j^*(\cdot, \beta, \gamma)=w_j^* (\Lambda(\cdot, \beta, \gamma, g))+ g'(\Lambda(\cdot, \beta, \gamma, g))\Lambda'_{g}(\cdot, \beta, \gamma, g)[w_j^*]$ with $w_j^*$ given in condition \ref{c: pos info exist}. Then under conditions \ref{c: bounded b}-\ref{c: smoothness} and \ref{c: pos info exist}, we have 
	\[N_{[\ ]}(\epsilon, \mc{F}_{n,j}^{*\beta}(\eta), \|\cdot\|_{\infty})\lesssim \left(\frac{\eta}{\epsilon}\right)^{c_1 q_{n_1}+c_2q_{n_2}+d},\]
	\[N_{[\ ]}(\epsilon, \mc{F}_{n,j}^{*\gamma}(\eta), \|\cdot\|_{\infty})\lesssim \left(\frac{\eta}{\epsilon}\right)^{c_3 q_{n_1}+c_4q_{n_2}+d},\]
	and 
	\[N_{[\ ]}(\epsilon, \mc{F}_{n,j}^{*\zeta}(\eta), \|\cdot\|_{\infty})\lesssim \left(\frac{\eta}{\epsilon}\right)^{c_5 q_{n_1}+c_6q_{n_2}+d}\]
	 for some constants $c_i$, $i=1,\dots,6$.
\end{lemma}

The proofs of Lemma \ref{lemma: other bracket numbers for A4} and \ref{lemma: other bracket numbers for A5} follow a  similar   calculation as in Lemma \ref{lemma: bracket number of Fn} and therefore are omitted here.  

\subsection{Proof of Theorem \ref{thm: conv rate}}
\label{appendix subs conv rate}

\begin{proof}[\textbf{Proof of Theorem \ref{thm: conv rate}}]
	We prove the theorem by checking the conditions C1-3 in \citet[Theorem 1]{shen1994}.
	Using the fact $P\{\int^Y_0 f(t, X)\dev \Lambda_0(t, X)\}=P\{\Delta f(Y, X)\}$, we have 
	\begin{align*}
	    Pl(\beta, \gamma, \zeta(\cdot, \beta, \gamma); W) = &  P\{\Delta [X^T \beta+ \gamma(Y)+g(\Lambda(Y, X, \beta, \gamma, g)) \\
	    & \ \ \  - \exp(X^T \beta+ \gamma(Y)+g(\Lambda(Y, X, \beta, \gamma, g))- X^T \beta_0- \gamma_0(Y)-g_0(\Lambda_0(Y, X)))]\}.
	\end{align*}
It follows that, by the Taylor expansion,
\begin{align}
    &\ \ \  Pl(\beta_0, \gamma_0, \zeta_0(\cdot, \beta_0, \gamma_0); W) - Pl(\beta, \gamma, \zeta(\cdot, \beta, \gamma); W) \nonumber \\
    & = P\{\Delta[\exp(X^T \beta+ \gamma(Y)+g(\Lambda(Y, X, \beta, \gamma, g))- X^T \beta_0- \gamma_0(Y)-g_0(\Lambda_0(Y, X)))\nonumber \\
    &  \ \ \ \ \ \ \ - 1 - (X^T \beta+ \gamma(Y)+g(\Lambda(Y, X, \beta, \gamma, g))- X^T \beta_0- \gamma_0(Y)-g_0(\Lambda_0(Y, X)))]\}\nonumber\\
    & =  \frac{1}{2} A + o(A),\label{eq: taylor exp}
\end{align}
where $A = P\{\Delta[X^T \beta+ \gamma(Y)+g(\Lambda(Y, X, \beta, \gamma, g))- X^T \beta_0- \gamma_0(Y)-g_0(\Lambda_0(Y, X))]^2\}$. 
After subtracting and adding the term $g(\Lambda_0(Y, X))$, we have
\begin{align*}
	A & =  P\{\Delta[X^T (\beta-\beta_0)+ \gamma(Y)-\gamma_0(Y)+g(\Lambda(Y, X, \beta, \gamma, g))- g(\Lambda_0(Y, X))\\
	& \ \ \ \ \ \ \ \ \ \ \ + g(\Lambda_0(Y, X))- g_0(\Lambda_0(Y, X))]^2\}\\
	& = P\{\Delta[(g'(\Lambda_0(Y, X))\Lambda'_{0\beta}(Y, X)+X)^T(\beta-\beta_0)\\
	& \ \ \ \ \ \ \ \ \ \ \ \ +g'(\Lambda_0(Y, X))\Lambda'_{0\gamma}(Y, X)[\gamma-\gamma_0]+ \gamma(Y) - \gamma_0(Y)\\
	&\ \ \ \ \ \ \ \ \ \ \ \  +g'(\Lambda_0(Y, X))\Lambda'_{0g}(Y, X)[g-g_0] + g(\Lambda_0(Y, X))- g_0(\Lambda_0(Y, X))\\
	&\ \ \ \ \ \ \ \ \ \ \ \  + o(\|\beta-\beta_0\|) + o(\|\gamma-\gamma_0\|_{2})+ o(\|g-g_0\|_{2})]^2\},
\end{align*}
where the second equality is obtained by using the Taylor expansion. Since $\Lambda'_{0\beta}(t, x)$ is bounded by Lemma \ref{lemma: ode solutions} and $\Lambda'_{0\gamma}(\cdot)[v]$ and $\Lambda'_{0g}(\cdot)[w]$ are bounded linear operators, which can be verified using the same arguments as in Lemma \ref{lemma: bdd linear operators}, we have 
\begin{align*}
	g'(\Lambda_0(Y, X))\Lambda'_{0\beta}(Y, X)^T(\beta-\beta_0) & = g_0'(\Lambda_0(Y, X))\Lambda'_{0\beta}(Y, X)^T(\beta-\beta_0)+o(\|\beta-\beta_0\|) +o(\|g-g_0\|_{2}),\\
	g'(\Lambda_0(Y, X))\Lambda'_{0\gamma}(Y, X)[\gamma-\gamma_0] & = g_0'(\Lambda_0(Y, X))\Lambda'_{0\gamma}(Y, X)[\gamma-\gamma_0] + o(\|\gamma-\gamma_0\|_{2})+ o(\|g-g_0\|_{2}),\\
	g'(\Lambda_0(Y, X))\Lambda'_{0g}(Y, X)[g-g_0] & = g_0'(\Lambda_0(Y, X))\Lambda'_{0g}(Y, X)[g-g_0] + o(\|g-g_0\|_{2}).
\end{align*}
Note that under conditions \ref{c: bounded b}-\ref{c: smoothness}, we have 
\begin{equation}
\label{norm ineq}
	d^2(\theta, \theta_0) \lesssim \|\beta-\beta_0\|^2 + \|\gamma-\gamma_0\|_{2}^2+ \|g-g_0\|_{2}^2 \lesssim d^2(\theta, \theta_0).
\end{equation}
Plugging these equations above into $A$, it follows that
	\begin{align}
	    A & \gtrsim P\{\Delta [(g_0'(\Lambda_0(Y, X))\Lambda'_{0\beta}(Y, X)+X)^T(\beta-\beta_0)\nonumber\\
	    & \ \ \ \ \ \ \ \ \ \ \ \ +g_0'(\Lambda_0(Y, X))\Lambda'_{0\gamma}(Y, X)[\gamma-\gamma_0]+ \gamma(Y) - \gamma_0(Y)\nonumber \\
	    &\ \ \ \ \ \ \ \ \ \ \ \  +g_0'(\Lambda_0(Y, X))\Lambda'_{0g}(Y, X)[g-g_0] + g(\Lambda_0(Y, X))- g_0(\Lambda_0(Y, X))]^2 \} \nonumber\\
	    & \ \ \ \ + o(d^2(\theta, \theta_0)).\label{first step}
	\end{align}
Then, by solving the initial value problem in (\ref{eq: beta dev}), we have
\begin{align}
g_0'(\Lambda_0(Y, X))\Lambda'_{0\beta}(Y, X)+X & = (g'_0(\Lambda_0(Y, X))\exp(g_0(\Lambda_{0}(Y,X)))R(Y)e^{X^T\beta_0} + 1)X \nonumber\\
& = (g_0'(\tilde{\Lambda}_{0}(U))\exp(g_0(\tilde{\Lambda}_{0}(U)))U +1)X\nonumber\\
& \triangleq \epsilon_1(U)X,\label{notation: epsilon1}
\end{align}
with $U$ given in condition \ref{c: lower bound1} and $\epsilon_1$ is a deterministic function. 

Note that using equations (\ref{eq:grad psi gamma}) and (\ref{eq:grad psi g}) in Lemma \ref{lemma: bdd linear operators}, we also have 
\begin{align}
	\psi_{0\gamma}'(Y, X)[\gamma-\gamma_0] & = g_0'(\Lambda_0(Y, X))\Lambda'_{0\gamma}(Y, X)[\gamma-\gamma_0]+ \gamma(Y) - \gamma_0(Y)  \nonumber\\
	& = g_0'(\tilde{\Lambda}_{0}(U)\exp(g_0(\tilde{\Lambda}_{0}(U)))\int^{U}_0 (\gamma-\gamma_0)(R^{-1}(se^{-V}))\dev s + (\gamma-\gamma_0)(R^{-1}(Ue^{-V}))\nonumber\\
	&\triangleq \epsilon_2(U, V)[\gamma-\gamma_0],\label{notation: epsilon2}
\end{align}
which is a deterministic function of $U$ and $V$ given in condition \ref{c: lower bound1}, and
\begin{align}
	\psi_{0g}'(Y, X)[g-g_0] &= g_0'(\Lambda_0(Y, X))\Lambda'_{0g}(Y, X)[g-g_0] + g(\Lambda_0(Y, X))- g_0(\Lambda_0(Y, X)) \nonumber\\
	& = g_0'(\tilde{\Lambda}_{0}(U)\exp(g_0(\tilde{\Lambda}_{0}(U)))\int^{\tilde{\Lambda}_{0}(U)}_0 \exp(-g_0(s))(g-g_0)(s)\dev s + (g-g_0)(\tilde{\Lambda}_{0}(U)) \nonumber\\
	& \triangleq \epsilon_3(U)[g-g_0],\label{notation: epsilon3}
\end{align}
which is a deterministic function of $U$. 

Then, it follows from (\ref{first step})
\begin{align}
    A & \gtrsim P\{\Delta [\epsilon_1(U) X^T(\beta-\beta_0)+\epsilon_2(U, V)[\gamma-\gamma_0]+\epsilon_3(U)[g-g_0]]^2 \}+ o(d^2(\theta, \theta_0))\nonumber\\
    & = P\{\Delta(\epsilon_1(U) X^T(\beta-\beta_0))^2\} +P\{\Delta (\epsilon_2(U,V)[\gamma-\gamma_0]+\epsilon_3(U)[g-g_0])^2\}\nonumber\\
    & \ \ \ +2P\{\Delta(\epsilon_1(U) X^T(\beta-\beta_0))(\epsilon_2(U,V)[\gamma-\gamma_0]+\epsilon_3(U)[g-g_0]) \} + o(d^2(\theta, \theta_0))\nonumber\\
    & \geq P\{\Delta(\epsilon_1(U) X^T(\beta-\beta_0))^2\} +P\{\Delta (\epsilon_2(U, V)[\gamma-\gamma_0]+\epsilon_3(U)[g-g_0])^2\}\nonumber\\
    & \ \ \ - 2|P\{\Delta(\epsilon_1(U) X^T(\beta-\beta_0))(\epsilon_2(U,V)[\gamma-\gamma_0]+\epsilon_3(U)[g-g_0]) \}| + o(d^2(\theta, \theta_0)).\label{second step}
\end{align}
By using the fact that $P\{\Delta f(U, X)\} =P\{\int^Y_0 f(R(t)e^{X^T\beta_0}, X)\dev \Lambda_0(t, X)\} =P\{\int^U_0 f(t, X)\dev \tilde{\Lambda}_0(t)\}$,
\begin{align*}
	& |P\{\Delta(\epsilon_1(U) X^T(\beta-\beta_0))(\epsilon_2(U,V)[\gamma-\gamma_0]+\epsilon_3(U)[g-g_0]) \}|^2\\ 
= & \left( P\left\{\int^U_0\epsilon_1(t) X^T(\beta-\beta_0)(\epsilon_2(t,V)[\gamma-\gamma_0]+\epsilon_3(t)[g-g_0])\dev \tilde{\Lambda}_0(t)\right\} \right)^2\\
= & \left( P\left\{\int^U_0\epsilon_1(t) P\{X^T(\beta-\beta_0)|U, V\}(\epsilon_2(t,V)[\gamma-\gamma_0]+\epsilon_3(t)[g-g_0])\dev \tilde{\Lambda}_0(t)\right\} \right)^2\\
\leq & P\left\{\int^U_0(\epsilon_1(t))^2 \left( P\{X^T(\beta-\beta_0)|U, V\}\right)^2\dev \tilde{\Lambda}_0(t)\right\} P\left\{\int^U_0(\epsilon_2(t,V)[\gamma-\gamma_0]+\epsilon_3(t)[g-g_0])^2\dev \tilde{\Lambda}_0(t)\right\},
\end{align*}
where the last step is obtained using the Cauchy-Schwartz inequality. Under condition \ref{c: lower bound1}, there exists $\eta_1\in(0,1)$ such that 
\[(1-\eta_1)(\beta-\beta_0)^T P\{XX^T|U, V\}(\beta-\beta_0) \geq (P\{X^T(\beta-\beta_0)|U, V\})^2,\]
since the first element of $\beta-\beta_0$ is zero with the identifiability constraint. Thus, we have 
\begin{align*}
	& |P\{\Delta(\epsilon_1(U) X^T(\beta-\beta_0))(\epsilon_2(U,V)[\gamma-\gamma_0]+\epsilon_3(U)[g-g_0]) \}|^2\\ 
	\leq & (1-\eta_1) P\left\{\int^U_0(\epsilon_1(t))^2 (\beta-\beta_0)^T P\{XX^T|U, V\}(\beta-\beta_0)\dev \tilde{\Lambda}_0(t)\right\} \\
	& \ \ \ \ \ \ \ \ \cdot P\left\{\int^U_0(\epsilon_2(t,V)[\gamma-\gamma_0]+\epsilon_3(t)[g-g_0])^2\dev \tilde{\Lambda}_0(t)\right\}\\
	= & (1-\eta_1)  P\left\{\int^U_0(\epsilon_1(t) X^T(\beta-\beta_0))^2\dev \tilde{\Lambda}_0(t)\right\}P\left\{\int^U_0(\epsilon_2(t,V)[\gamma-\gamma_0]+\epsilon_3(t)[g-g_0])^2\dev \tilde{\Lambda}_0(t)\right\}\\
	= & (1-\eta_1)P\{\Delta(\epsilon_1(U) X^T(\beta-\beta_0))^2\} P\{\Delta(\epsilon_2(U,V)[\gamma-\gamma_0]+\epsilon_3(U)[g-g_0])^2\},
\end{align*}
and it yields from (\ref{second step}) that
	\begin{align*}
	    A &  \geq P\{\Delta(\epsilon_1(U) X^T(\beta-\beta_0))^2\} +P\{\Delta (\epsilon_2(U,V)[\gamma-\gamma_0]+\epsilon_3(U)[g-g_0])^2\}\\
	    & \ \ \ - 2(1-\eta_1)^{1/2}(P\{\Delta(\epsilon_2(U,V)[\gamma-\gamma_0]+\epsilon_3(U)[g-g_0])^2\})^{1/2}(P\{\Delta(\epsilon_1(U) X^T(\beta-\beta_0))^2\})^{1/2}\\
	    & \geq (1- (1-\eta_1)^{1/2})\{P\{\Delta(\epsilon_1(U) X^T(\beta-\beta_0))^2\} +P\{\Delta (\epsilon_2(U,V)[\gamma-\gamma_0]+\epsilon_3(U)[g-g_0])^2\}\}\\
	    & \gtrsim P\{\Delta(\epsilon_1(U) X^T(\beta-\beta_0))^2\} +P\{\Delta (\epsilon_2(U,V)[\gamma-\gamma_0]+\epsilon_3(U)[g-g_0])^2\}\\
	    & =A_1 +A_2,
	\end{align*}
	where the second inequality is obtained by $2ab \leq a^2+b^2$.

For $A_1$, under condition \ref{c: bounded t}, we have for $t\in [0, \tau]$,
\[P\{\mathbbm{1}(Y > t)|X\}\geq P\{\mathbbm{1}((Y > \tau)|X\} \geq \delta_0.\]
Then it follows that,
\begin{align*}
	A_1 & =P\{\int^Y_0\exp(X^T\beta_0+\gamma_0(t)+g(\Lambda_0(t, X)))\left(\epsilon_1(R(t)e^{X^T\beta_0})X^T(\beta-\beta_0)\right)^2 \dev t\}\\
	& =P\{\int^\tau_0 P\{\mathbbm{1}(Y > t)|X\} \exp(X^T\beta_0+\gamma_0(t)+g(\Lambda_0(t, X)))\left(\epsilon_1(R(t)e^{X^T\beta_0})X^T(\beta-\beta_0)\right)^2 \dev t\}\\
	& \geq \delta_0 P\{\int^\tau_0 \exp(X^T\beta_0+\gamma_0(t)+g(\Lambda_0(t, X)))\left(\epsilon_1(R(t)e^{X^T\beta_0})X^T(\beta-\beta_0)\right)^2 \dev t\}\\
	& = \delta_0 P\{\int^{R(\tau)e^{X^T\beta_0}}_0  \exp(g(\tilde{\Lambda}_0(s)))\left(\epsilon_1(s)X^T(\beta-\beta_0)\right)^2 \dev s\}\\
	& \geq \delta_0 P\{\int^{cR(\tau)}_0  \exp(g(\tilde{\Lambda}_0(s)))\left(\epsilon_1(s)X^T(\beta-\beta_0)\right)^2 \dev s\}\\
	&= \delta_0 (\beta-\beta_0)^TP\{XX^T\}(\beta-\beta_0)  \int^{cR(\tau)}_0\exp(g(\tilde{\Lambda}_0(s)))(\epsilon_1(s))^2\dev s,
\end{align*}
where the fourth equality is derived by variable transformation $s=R(t)e^{X^T\beta_0}$ and $c=\min_{x\in \mc{X}}e^{x^T\beta_0}$, which is positive since $\mc{X}$ is bounded under condition \ref{c: bounded x}. As condition \ref{c: bounded x} implies that the smallest eigenvalue of $P\{XX^T\}$, denoted by $\lambda_1$, is positive as well, we have $(\beta-\beta_0)^TP\{XX^T\}(\beta-\beta_0) \geq \lambda_1 \|\beta-\beta_0\|^2$. Also, by definition $\epsilon_1(s)$ satisfies the equation $g_0'(\tilde{\Lambda}_{0}(t))\int_{0}^t \exp(g_0(\tilde{\Lambda}_0(s)))\epsilon_1(s)\dev s +1 = \epsilon_1(t)$, thus it can not be a constant zero and $\int^{cR(\tau)}_0\exp(g(\tilde{\Lambda}_0(s)))(\epsilon_1(s))^2\dev s$ is bounded away from $0$ below. Hence, $A_1 \gtrsim \|\beta-\beta_0\|^2$.

For $A_2$, it is bounded below by
\begin{align*}
    & \ \ \ P\{\Delta (\epsilon_2(U, V)[\gamma-\gamma_0]+\epsilon_3(U)[g-g_0])^2\}\\
    & \geq P\{\Delta (\epsilon_2(U, V)[\gamma-\gamma_0])^2\} +P\{\Delta(\epsilon_3(U)[g-g_0])^2\} - 2|P\{\Delta (\epsilon_2(U, V)[\gamma-\gamma_0])(\epsilon_3(U)[g-g_0])\}|\\
    & \geq P\{\Delta (\epsilon_2(U, V)[\gamma-\gamma_0])^2\} +P\{\Delta(\epsilon_3(U)[g-g_0])^2\}\\
    & \ \ \ - 2\eta_2^{1/2}P\{\Delta\}(P\{\Delta (\epsilon_2(U, V)[\gamma-\gamma_0])^2\})^{1/2}(P\{\Delta(\epsilon_3(U)[g-g_0])^2\})^{1/2}\\
    & \geq (1- \eta_2^{1/2}P\{\Delta\}) \{P\{\Delta (\epsilon_2(U, V)[\gamma-\gamma_0])^2\} +P\{\Delta(\epsilon_3(U)[g-g_0])^2\}\}\\
    & \gtrsim P\{\Delta (\epsilon_2(U, V)[\gamma-\gamma_0])^2\} +P\{\Delta(\epsilon_3(U)[g-g_0])^2\},
\end{align*}
where the second inequality holds under condition \ref{c: lower bound2} because there exists some $\eta_2 \in (0, 1)$ such that \[(P\{\epsilon_2(U,Y)[\gamma-\gamma_0]\epsilon_3(U)[g-g_0]|\Delta=1\})^2\leq \eta_2 P\{(\epsilon_2(U,Y)[\gamma-\gamma_0])^2|\Delta=1\} P\{(\epsilon_3(U)[g-g_0])^2|\Delta=1\}.\]
Furthermore, the first term is bounded under condition \ref{c: bounded t}
\begin{align*}
	P\{\Delta (\epsilon_2(U, V)[\gamma-\gamma_0])^2\}& = P\{\Delta (\psi_{0\gamma}'(Y, X)[\gamma-\gamma_0])^2\}\\
	& = P\{\int^\tau_0 P\{\mathbbm{1}(Y>t)|X\} (\psi_{0\gamma}'(t, X)[\gamma-\gamma_0])^2 \dev \Lambda_0(t, X)\}\\
	& \geq \delta_0 P\{\int^\tau_0 (\psi_{0\gamma}'(t, X)[\gamma-\gamma_0])^2 \dev \Lambda_0(t, X)\}\\
	& \gtrsim \|\gamma-\gamma_0\|_2^2,
\end{align*}
where the second inequality is obtained by Lemma \ref{lemma: bdd linear operators} because $\gamma-\gamma_0\in\Gamma^{p_1}$. Using the same argument, we have $P\{\Delta(\epsilon_3(U)[g-g_0])^2\} \gtrsim \|g-g_0\|_2^2$. 
Therefore,
\begin{align*}
    Pl(\beta_0, \gamma_0, \zeta_0(\cdot, \beta_0, \gamma_0); W) - Pl(\beta, \gamma, \zeta(\cdot, \beta, \gamma); W) & =  \frac{1}{2} A + o(A)\\
    & \gtrsim \|\beta-\beta_0\|^2+\|\gamma-\gamma_0\|_{2}^2 + \|g-g_0\|_{2}^2\\
    & \gtrsim d^2(\theta, \theta_0),
\end{align*}
which implies that 
\[\inf_{d(\theta, \theta_0)\geq \epsilon, \theta\in \Theta_n} Pl(\beta_0, \gamma_0, \zeta_0(\cdot, \beta_0, \gamma_0); W) - Pl(\beta, \gamma, \zeta(\cdot, \beta, \gamma); W) \gtrsim  \epsilon^2.\]
Hence the condition C1 in \citet[Theorem 1]{shen1994} holds with $\alpha=1$ in their notation.

Next, we verify the condition C2 in \citet[Theorem 1]{shen1994}. It follows that
\begin{align}
\label{thm1: upper bound}
    & \ \ \ (l(\beta, \gamma, \zeta(\cdot, \beta, \gamma); W) - l(\beta_0, \gamma_0, \zeta_0(\cdot, \beta_0, \gamma_0); W))^2 \nonumber \\
    &= \{ \Delta X^T (\beta-\beta_0)+ \Delta[\gamma(Y)-\gamma_0(Y)]+\Delta[g(\Lambda(Y, X, \beta, \gamma, g))-g_0(\Lambda_0(Y, X))] \nonumber \\
    & \ \ \  - \int_{0}^Y [\exp(X^T \beta+ \gamma(t)+g(\Lambda(t, X, \beta, \gamma, g)))- \exp(X^T \beta_0+ \gamma_0(t)+g_0(\Lambda_0(t, X)))]\dev t \}^2\nonumber \\
    & \lesssim (X^T (\beta-\beta_0))^2+ \Delta(\gamma(Y)-\gamma_0(Y))^2+\Delta[g(\Lambda(Y, X, \beta, \gamma, g))-g_0(\Lambda_0(Y, X))]^2\nonumber \\
    & \ \ \  + \{\int_{0}^Y [\exp(X^T \beta+ \gamma(t)+g(\Lambda(t, X, \beta, \gamma, g)))- \exp(X^T \beta_0+ \gamma_0(t)+g_0(\Lambda_0(t, X)))]\dev t \}^2\nonumber \\
    & \lesssim \|\beta-\beta_0\|^2 +\Delta(\gamma(Y)-\gamma_0(Y))^2+\Delta[g(\Lambda(Y, X, \beta, \gamma, g))-g_0(\Lambda_0(Y, X))]^2 \nonumber \\
    & \ \ \  + \int_{0}^\tau [\exp(X^T \beta+ \gamma(t)+g(\Lambda(t, X, \beta, \gamma, g)))- \exp(X^T \beta_0+ \gamma_0(t)+g_0(\Lambda_0(t, X)))]^2\dev t,
\end{align}
where the second inequality is obtained by the condition \ref{c: bounded x} and the Cauchy-Schwartz inequality
\begin{align*}
    & \ \ \ \{\int_{0}^Y [\exp(X^T \beta+ \gamma(t)+g(\Lambda(t, X, \beta, \gamma, g)))- \exp(X^T \beta_0+ \gamma_0(t)+g_0(\Lambda_0(t, X)))]\dev t \}^2\\
    & = \{\int_{0}^\tau [1(Y\geq t)\exp(X^T \beta+ \gamma(t)+g(\Lambda(t, X, \beta, \gamma, g)))- \exp(X^T \beta_0+ \gamma_0(t)+g_0(\Lambda_0(t, X)))]\dev t \}^2\\
    & \leq \int_{0}^\tau 1(Y\geq t)\dev t \int_{0}^\tau [\exp(X^T \beta+ \gamma(t)+g(\Lambda(t, X, \beta, \gamma, g)))- \exp(X^T \beta_0+ \gamma_0(t)+g_0(\Lambda_0(t, X)))]^2\dev t\\
    & \leq \tau \int_{0}^\tau [\exp(X^T \beta+ \gamma(t)+g(\Lambda(t, X, \beta, \gamma, g)))- \exp(X^T \beta_0+ \gamma_0(t)+g_0(\Lambda_0(t, X)))]^2\dev t.
\end{align*}
For the second term in (\ref{thm1: upper bound}), we have
\begin{align}
	& \ \ \ \  P\{\Delta(\gamma(Y)-\gamma_0(Y))^2\}\nonumber \\
	& = P\int_0^\tau 1(Y\geq t) \exp(X^T \beta_0+ \gamma_0(t)+g_0(\Lambda_0(t, X))) (\gamma(t)-\gamma_0(t))^2\dev t\nonumber \\
	& \leq \int_0^\tau P\{\exp(X^T \beta_0+ \gamma_0(t)+g_0(\Lambda_0(t, X)))\} (\gamma(t)-\gamma_0(t))^2\dev t\nonumber\\
	& \lesssim \|\gamma - \gamma_0\|_2^2,\label{ineq: gamma}
\end{align}
where the last inequality holds because $\exp(X^T \beta_0+ \gamma_0(t)+g_0(\Lambda_0(t, X)))$ is bounded under conditions \ref{c: bounded b}-\ref{c: smoothness}. For the third term in (\ref{thm1: upper bound}), we have
\begin{align}
    & \ \ \ \ P\{\Delta[g(\Lambda(Y, X, \beta, \gamma, g))-g_0(\Lambda_0(Y, X))]^2\}\nonumber\\
    & = P\int_0^Y [g(\Lambda(t, X, \beta, \gamma, g))-g_0(\Lambda_0(t, X))]^2 \dev \Lambda_0(t, X)\nonumber\\
    & = P\int_0^\tau 1(Y\geq t)[g(\Lambda(t, X, \beta, \gamma, g))-g_0(\Lambda_0(t, X))]^2 \dev \Lambda_0(t, X)\nonumber\\
    & \leq P\int_0^\tau [g(\Lambda(t, X, \beta, \gamma, g))-g_0(\Lambda_0(t, X))]^2 \dev \Lambda_0(t, X)\nonumber\\
    & = \|\zeta(\cdot, \beta, \gamma)-\zeta_0(\cdot, \beta_0, \gamma_0)\|_2^2,\label{ineq: g}
\end{align}
For the fourth term in (\ref{thm1: upper bound}), using the mean value theorem, it follows that
\begin{align*}
    & \ \ \ \ P\{\int_{0}^\tau [\exp(X^T \beta+ \gamma(t)+g(\Lambda(t, X, \beta, \gamma, g)))- \exp(X^T \beta_0+ \gamma_0(t)+g_0(\Lambda_0(t, X)))]^2\dev t \}\\
    & = P\{\int_{0}^\tau \exp(2\tilde{\psi}(t, X))[ X^T (\beta-\beta_0)+ (\gamma(t)-\gamma_0(t))+g(\Lambda(t, X, \beta, \gamma, g))-g_0(\Lambda_0(t, X))]^2\dev t \}\\
    & \lesssim P\{\int_{0}^\tau \exp(2\tilde{\psi}(t, X))\{[ X^T (\beta-\beta_0)]^2+ [\gamma(t)-\gamma_0(t)]^2+[g(\Lambda(t, X, \beta, \gamma, g))-g_0(\Lambda_0(t, X))]^2\}\dev t \}\\
    & = I_1+I_2+I_3,
\end{align*}
where $\tilde{\psi}(t, X)= X^T \beta_0+ \gamma_0(t)+g_0(\Lambda_0(t, X)) + \xi (X^T (\beta-\beta_0)+ \gamma(t)-\gamma_0(t)+g(\Lambda(t, X, \beta, \gamma, g))-g_0(\Lambda_0(t, X)))$ for some $\xi\in (0, 1)$ and is bounded under conditions \ref{c: bounded b}-\ref{c: smoothness}. Hence, 
\[I_1\lesssim (\beta-\beta_0)^T P(XX^T)(\beta-\beta_0) \leq \lambda_d \|\beta-\beta_0\|^2,\] 
where $\lambda_d$ is the largest eigenvalue of $P(XX^T)$,
\[I_2 \lesssim \|\gamma-\gamma_0\|^2_{2},\]
and
\begin{align*}
    I_3 & = P\{\int_{0}^\tau \exp(2\tilde{\psi}(t, X)- X^T \beta_0- \gamma_0(t)-g_0(\Lambda_0(t, X)))\\
    & \ \ \ \ \ \ \ \ \ \ \ \ \ \ \  \cdot[g(\Lambda(t, X, \beta, \gamma, g))-g_0(\Lambda_0(t, X))]^2\dev \Lambda_0(t, X) \}\\
    & \lesssim P\{\int_{0}^\tau [g(\Lambda(t, X, \beta, \gamma, g))-g_0(\Lambda_0(t, X))]^2\dev \Lambda_0(t, X) \}\\
    & = \|\zeta(\cdot, \beta, \gamma)-\zeta_0(\cdot, \beta_0, \gamma_0)\|_2^2.
\end{align*}
Therefore, we have 
\begin{align*}
    P(l(\beta, \gamma, \zeta(\cdot, \beta, \gamma); W) - l(\beta_0, \gamma_0, \zeta_0(\cdot, \beta_0, \gamma_0); W))^2 & \lesssim \|\beta-\beta_0\|^2+\|\gamma-\gamma_0\|^2_{2} + \|\zeta(\cdot, \beta, \gamma)-\zeta_0(\cdot, \beta_0, \gamma_0)\|_2^2\\
    & \lesssim  d^2(\theta, \theta_0),
\end{align*} 
which implies that 
\begin{align*}
    &\ \ \ \sup_{d(\theta, \theta_0)\leq \epsilon, \theta\in \Theta_n} Var\{l(\beta, \gamma, \zeta(\cdot, \beta, \gamma); W) - l(\beta_0, \gamma_0, \zeta_0(\cdot, \beta_0, \gamma_0); W)\} \\
    & \leq \sup_{d(\theta, \theta_0)\leq \epsilon, \theta\in \Theta_n} P\{l(\beta, \gamma, \zeta(\cdot, \beta, \gamma); W) - l(\beta_0, \gamma_0, \zeta_0(\cdot, \beta_0, \gamma_0); W)\}^2  \lesssim \epsilon^2.
\end{align*}
Thus the condition C2 in \citet[Theorem 1]{shen1994} holds with $\beta =1$ in their notation.

Next we verify the condition C3 in \citet[Theorem 1]{shen1994}. By Lemma \ref{lemma: bracket number of Fn}, we have
\[H(\epsilon, \mc{F}_n, \|\cdot\|_{\infty}) = \log(N(\epsilon, \mc{F}_n, \|\cdot\|_{\infty}))\lesssim (c_1 q_{n_1}+c_2q_{n_2}+d)\log(1/\epsilon)\lesssim n^{\max\{\nu_1, \nu_2\}}\log (1/\epsilon).\]
So the C3 holds with constants $2r_0=\max\{\nu_1, \nu_2\}$ and $r=0^{+}$ in their notations, which leads to $\tau=\frac{1-\max\{\nu_1, \nu_2\}}{2}-\frac{\log\log n}{2\log n}$ in their main result. We can select slightly large $\tilde{\nu}_1$ and $\tilde{\nu}_2$ such that $\frac{1-\max\{\tilde{\nu}_1, \tilde{\nu}_2\}}{2}\leq \frac{1-\max\{\nu_1, \nu_2\}}{2}-\frac{\log\log n}{2\log n}$ for sufficiently large n and still denote $\tilde{\nu}_i$ by $\nu_i$ for $i=1,2$. Then, $\tau=\frac{1-\max\{\nu_1, \nu_2\}}{2}$. Also, since the sieve estimator $\hat{\theta}_n$ maximizes the empirical log-likelihood over the sieve space $\Theta_n$, the inequality (1.1) in \citet{shen1994} holds with $\eta_n = 0$. Therefore, by Theorem 1 in \citet{shen1994}, we have 
\begin{equation*}
	\label{shen's rate}
	d(\hat{\theta}_{n}, \theta_0) = O_p(\max\{n^{-\frac{1-\max\{\nu_1, \nu_2\}}{2}}, d(\theta_{0n}, \theta_0), K^{1/2}(\theta_{0n}, \theta_0)\}),
\end{equation*}
where $ K(\theta_{0n}, \theta_0) = P\{l(\theta_0;W)-l(\theta_{0n};W)\}$. Further, using the Taylor expansion for $P\{l(\theta_0;W)-l(\theta_{0n};W)\}$ in (\ref{eq: taylor exp}), we have 
\begin{align*}
	K(\theta_{0n}, \theta_0)& =\frac{1}{2}P\{\Delta[\gamma_0(Y)+g_0(\Lambda(Y, X, \beta_0, \gamma_0, g_0)\\
	&\ \ \ \ \ \ \ \ \ \  -\gamma_{0n}(Y)-g_{0n}(\Lambda(Y, X, \beta_0, \gamma_{0n}, g_{0n}))]^2\} + o(d^2(\theta_{0n}, \theta_0))\\
	& \leq P\{\Delta[g_0(\Lambda(Y, X, \beta_0, \gamma_0, g_0))-g_{0n}(\Lambda(Y, X, \beta_0, \gamma_{0n}, g_{0n}))]^2\}\\
	& \ \ \ \   + P\{\Delta (\gamma_0(Y)-\gamma_{0n}(Y))^2\} + o(d^2(\theta_{0n}, \theta_0))\\
	& \lesssim \|\zeta_0(\cdot, \beta_0, \gamma_0)-\zeta_{0n}(\cdot, \beta_{0n}, \gamma_{0n})\|_2^2 + \|\gamma_0-\gamma_{0n}\|^2_{2} + o(d^2(\theta_{0n}, \theta_0))\\
	& = O(d^2(\theta_{0n}, \theta_0)),
\end{align*}
where the first inequality is obtained by the fact $(a+b)^2 \leq 2(a^2+b^2)$ and the second inequality holds by using the same argument as in (\ref{ineq: gamma}) and (\ref{ineq: g}). Moreover, $d^2(\theta_{0n}, \theta_0) \lesssim \|\gamma_0-\gamma_{0n}\|^2_{2} + \|g_0-g_{0n}\|^2_{2}\lesssim \|\gamma_0-\gamma_{0n}\|^2_{\infty} + \|g_0-g_{0n}\|^2_{\infty} = O(n^{-2\min\{p_1\nu_1, p_2\nu_2\}})$
due to inequality (\ref{norm ineq}) and Lemma \ref{lemma: gnp and gp}. 
Thus, we have \[d(\hat{\theta}_{n}, \theta_0) = O_p(\max\{n^{-\frac{1-\max\{\nu_1, \nu_2\}}{2}}, n^{-\min\{p_1\nu_1, p_2\nu_2\}}\}) = O_p(n^{-\min\{p_1\nu_1, p_2\nu_2, \frac{1-\max\{\nu_1, \nu_2\}}{2}\}}),\]
which completes the proof.
\end{proof}

\subsection{Proof of Theorem \ref{thm: asym normal}}
\label{appendix subs asym normal}
\begin{proof}[\textbf{Proof of Theorem \ref{thm: asym normal}}]

We prove the theorem by verifying assumptions (\ref{m:rate})-(\ref{m:smoothness}) in Appendix \ref{appendx: m-theorem}. By Theorem \ref{thm: conv rate} we know that assumption (\ref{m:rate}) holds with $\xi=\min\{p_1\nu_1, p_2\nu_2, \frac{1-\max\{\nu_1, \nu_2\}}{2}\}$. It is straightforward to show that assumption (\ref{m:dev at true}) holds based on the fact that score functions have zero mean. To verify assumption (\ref{m:positive info}), first, we will find $\mb{v}^*=(v^*_1, \cdots, v^*_d)'$ and $\mb{h}^*=(h^*_1, \cdots, h^*_d)'$ with $\mb{h}^*(\cdot) = \mb{w}^*(\Lambda_0(\cdot))+g_0'(\Lambda_0(\cdot))\Lambda'_{0g}(\cdot)[\mb{w}^*]$ such that for any $v\in  \mathbb{V}$ and $h\in  \mathbb{H}$ with $h(\cdot) = w(\Lambda_0(\cdot))+g_0'(\Lambda_0(\cdot))\Lambda'_{0g}(\cdot)[w]$,
\begin{align}
		S''_{\beta\gamma}(\beta_0, \gamma_0(\cdot), \zeta_0(\cdot, \beta_0, \gamma_0))[v]& = S''_{\gamma\gamma}(\beta_0, \gamma_0(\cdot), \zeta_0(\cdot, \beta_0, \gamma_0))[\mb{v}^*,v]\nonumber\\
		& \ \ \ + S''_{\zeta\gamma}(\beta_0, \gamma_0(\cdot), \zeta_0(\cdot, \beta_0, \gamma_0))[\mb{h}^*,v],\label{eq: pos info 1}\\
		S''_{\beta\zeta}(\beta_0, \gamma_0(\cdot), \zeta_0(\cdot, \beta_0, \gamma_0))[h] & = S''_{\gamma\zeta}(\beta_0, \gamma_0(\cdot), \zeta_0(\cdot, \beta_0, \gamma_0))[\mb{v}^*,h]\nonumber\\
		& \ \ \ + S''_{\zeta\zeta}(\beta_0, \gamma_0(\cdot), \zeta_0(\cdot, \beta_0, \gamma_0))[\mb{h}^*,h].\label{eq: pos info 2}
\end{align}
By Lemma \ref{lemma: devivatives} and the property $P\{\int^Y_0 f(t, X)\dev \Lambda_0(t, X)\}=P\{\Delta f(Y, X)\}$, for any $\mb{v}\in\mathbb{V}^d, v\in\mathbb{V}$ and $\mb{h}\in \mathbb{H}^d$ with $\mb{h}(\cdot) = \mb{w}(\Lambda_0(\cdot))+g_0'(\Lambda_0(\cdot))\Lambda'_{0g}(\cdot)[\mb{w}]$, we have
\begin{align}
	& S''_{\beta\gamma}(\beta_0, \gamma_0, \zeta_0)[v] - S''_{\gamma\gamma}(\beta_0, \gamma_0, \zeta_0)[\mb{v},v] - S''_{\zeta\gamma}(\beta_0, \gamma_0, \zeta_0)[\mb{h},v]\nonumber\\
	=&  P\{l''_{\beta\gamma}(\beta_0, \gamma_0, \zeta_0;W)[v]-l''_{\gamma\gamma}(\beta_0, \gamma_0, \zeta_0;W)[\mb{v}, v]- l''_{\zeta\gamma}(\beta_0, \gamma_0, \zeta_0;W)[\mb{h}, v ]\}\nonumber\\
	=& P\{\Delta \left[ g_0'(\Lambda_0(Y, X))\Lambda'_{0\beta}(Y, X)+X - g_0'(\Lambda_0(Y, X))\Lambda'_{0\gamma}(Y, X)[\mb{v}]- \mb{v}(Y) \right.\nonumber\\
	& \ \ \ \ \  \left. - g_0'(\Lambda_0(Y, X))\Lambda'_{0g}(Y, X)^T[\mb{w}] - \mb{w}(\Lambda_0(Y, X)) \right] (g_0'(\Lambda_0(Y, X))\Lambda'_{0\gamma}(Y, X)[v]+ v(Y)) \}\nonumber\\
	=& P\{\Delta \left(\epsilon_1(U)X -\epsilon_2(U, V)[\mb{v}]-\epsilon_3(U)[\mb{w}]\right)\psi'_{0\gamma}(Y,X)[v]\},\label{eq: pos info 11}
\end{align} 
where the last equality holds with $\epsilon_1, \epsilon_2, \epsilon_3,\psi'_{0\gamma}$ given in (\ref{notation: epsilon1})-(\ref{notation: epsilon3}) and $U$ given in the condition \ref{c: lower bound1}. Similarly, for any $\mb{v}\in\mathbb{V}^d$, $\mb{h}\in \mathbb{H}^d$ and $h\in \mathbb{H}$ with $h(\cdot) = w(\Lambda_0(\cdot))+g_0'(\Lambda_0(\cdot))\Lambda'_{0g}(\cdot)[w]$, we have
\begin{align}
	 & S''_{\beta\zeta}(\beta_0, \gamma_0, \zeta_0)[h] - S''_{\gamma\zeta}(\beta_0, \gamma_0, \zeta_0)[\mb{v},h] - S''_{\zeta\zeta}(\beta_0, \gamma_0, \zeta_0)[\mb{h},h]\nonumber\\
	=&P\{\Delta \left(\epsilon_1(U)X -\epsilon_2(U, V)[\mb{v}]-\epsilon_3(U)[\mb{w}]\right)\psi'_{0g}(Y,X)[w]\}.\label{eq: pos info 22}
\end{align}
Note that under condition \ref{c: pos info exist}, there exists $\mb{v}^*=(v^*_1, \cdots, v^*_d)^T$ and $\mb{w}^*=(w^*_1, \cdots, w^*_d)^T$, where $v^*_j \in  \Gamma^{2}$ and $w^*_j \in \mc{G}^{2}$ for $j=1, \cdots, d$, such that $P\{\Delta \mb{A}^*(U, X)\psi'_{0\gamma}(Y,X)[v]\}=0$ and $P\{\Delta \mb{A}^*(U, X)\psi'_{0g}(Y,X)[w]\}=0$ 
	hold for any $v\in \Gamma^{p_1}$ and $w\in \mc{G}^{p_2}$. Since $\mb{A}^*(U,X)=\epsilon_1(U)X -\epsilon_2(U, V)[\mb{v}^*]-\epsilon_3(U)[\mb{w}^*]$, plugging $\mb{v}=\mb{v}^*$ in (\ref{eq: pos info 11}) and $\mb{w}=\mb{w}^*$ in (\ref{eq: pos info 22}) we have equations (\ref{eq: pos info 1}) and (\ref{eq: pos info 2}) hold with $\mb{v}^*$ and $\mb{w}^*$ given in condition \ref{c: pos info exist}. Then it follows that
\begin{align*}
	&l'_{\beta}(\beta_0, \gamma_0, \zeta_0;W) -l'_{\gamma}(\beta_0, \gamma_0, \zeta_0;W)[\mb{v}^*] - l'_{\zeta}(\beta_0, \gamma_0, \zeta_0;W)[\mb{h}^*(\cdot, \beta_0, \gamma_0)]\\
	=	& \Delta \mb{A}^*(U, X) - \int^Y_0 \mb{A}^*(R(t)e^{X^T\beta_0}, X)\dev \Lambda_0(t, X)\\
	=	& \Delta \mb{A}^*(U, X) - \int^{R(Y)e^{X^T\beta_0}}_0 \mb{A}^*(t, X)\dev \tilde{\Lambda}_0(t)\\
	=	& \int \mb{A}^*(t, X)\dev M(t) = \boldsymbol{l}^*(\beta_0, \gamma_0, \zeta_0;W),
\end{align*} 
with $M(t)$ and $\boldsymbol{l}^*$ given in condition \ref{c: pos fish}. 
Based on the zero-mean property of score function together with the facts in (\ref{eq: pos info 1}) and (\ref{eq: pos info 2}), the matrix A in assumption (\ref{m:positive info}) is given by
\begin{align*}
	A & =  - S''_{\beta\beta}(\beta_0, \gamma_0, \zeta_0) + S''_{\gamma\beta}(\beta_0, \gamma_0, \zeta_0)[\mb{v}^*]+ S''_{\zeta\beta}(\beta_0, \gamma_0, \zeta_0)[\mb{h}^*]\\
	& \ \ \  - S''_{\gamma\gamma}(\beta_0, \gamma_0, \zeta_0)[\mb{v}^*,\mb{v}^*] + S''_{\beta\gamma}(\beta_0, \gamma_0, \zeta_0)[\mb{v}^*]- S''_{\zeta\gamma}(\beta_0, \gamma_0, \zeta_0)[\mb{h}^*, \mb{v}^*]\\
	& \ \ \  - S''_{\zeta\zeta}(\beta_0, \gamma_0, \zeta_0)[\mb{h}^*,\mb{h}^*] + S''_{\beta\zeta}(\beta_0, \gamma_0, \zeta_0)[\mb{h}^*]- S''_{\gamma\zeta}(\beta_0, \gamma_0, \zeta_0)[\mb{v}^*, \mb{h}^*]\\
	& = P\{(l'_{\beta}(\beta_0, \gamma_0, \zeta_0;W) -l'_{\gamma}(\beta_0, \gamma_0, \zeta_0;W)[\mb{v}^*] - l'_{\zeta}(\beta_0, \gamma_0, \zeta_0;W)[\mb{h}^*])^{\otimes 2}\}\\
	& = P \{\boldsymbol{l}^*(\beta_0, \gamma_0, \zeta_0;W)^{\otimes 2}\},
\end{align*}
which is the information matrix for $\beta_0$ and is nonsingular under condition \ref{c: pos fish}. Thus, assumption (\ref{m:positive info}) holds.

To verify assumption (\ref{m:gc}), we first note that the first part holds because $\hat{\beta}_n$ satisfies $S'_{\beta,n}(\hat{\theta}_n)=0$ where $\hat{\theta}_n=(\hat{\beta}_n, \hat{\gamma}_n(\cdot), \hat{\zeta}_n(\cdot, \hat{\beta}_n, \hat{\gamma}_n))$. Next we need to show that $S'_{\gamma,n}(\hat{\theta}_n)[v_j^*] = o_p(n^{-1/2})$. Since $v_j^*\in \Gamma^2$, by Lemma \ref{lemma: gnp and gp} there exists $v_{j,n}^* \in \Gamma^2_n$ such that $\|v_{j,n}^* - v^*_{j}\|_{\infty}=O(n^{-2\nu_1})$. Based on the fact that $v_{j,n}^*$ can be written as the linear combination of basis functions $B_k^1$ for $k=1, \dots, q_n^1$, we have $S'_{\gamma,n}(\hat{\theta}_n)[v_{j,n}^*] = 0$.\footnotemark[1]\footnotetext[1]{Note that we constrain the parameter $\gamma(t^*)=0$ for identifiability guarantee. For any $\gamma \in \Gamma_n^{p_1}$ in the sieve space, the constraint can be achieved by fixing the coefficient of one specific B-spline basis (suppose it is indexed as the first basis and let $a_1 \equiv 0$) and leaving coefficients of other bases as free optimization parameters.  Since $\hat{\theta}_n$ maximizes $l_n(\theta)$ in the sieve space and $v_{j,n}^* \in \Gamma^2_n$ can be written as the linear combination of bases with the first coefficient $a_1$ fixed as $0$, we have the gradient of $l_n(\theta)$ with respect to $\gamma$ along the direction $v_{j,n}^*$ at $\hat{\theta}_n$ equal to zero, i.e.,  $S'_{\gamma,n}(\hat{\theta}_n)[v_{j,n}^*] = 0$.} Since $S'_{\gamma}(\beta_0, \gamma_0(\cdot), \zeta_0(\cdot, \beta_0, \gamma_0))[v_j^*-v_{j,n}^*] = 0$, it suffices to show that for each $1\leq j\leq d$,
		\begin{align*}
			& \ \ \ \ P\{l'_{\gamma}(\hat{\theta}_n;W)[v_j^*-v_{j,n}^*]-l'_{\gamma}(\theta_0;W)[v_j^*-v_{j,n}^*]\}  + (\mathbb{P}_n-P)\{l'_{\gamma}(\hat{\theta}_n;W)[v_j^*-v_{j,n}^*]\}\\
			& = I_{1n} + I_{2n} = o_p(n^{-1/2}).
		\end{align*}
We will first show that $I_{1n}$ is $o_p(n^{-1/2})$. Using the Taylor expansion for $l'_{\gamma}(\hat{\theta}_n)[v_j^*-v_{j,n}^*]$ at $\theta_0$, we have 
		\begin{align*}
			I_{1n} =& P\{(\hat{\beta}_n-\beta_0)^T l''_{\beta\gamma}(\tilde{\beta}_n, \tilde{\gamma}_n(\cdot), \tilde{\zeta}_n(\cdot, \tilde{\beta}_n, \tilde{\gamma}_n);W)[v_j^*-v_{j,n}^*] \\
			& \ \ \ \ \ + l''_{\gamma\gamma}(\tilde{\beta}_n, \tilde{\gamma}_n(\cdot), \tilde{\zeta}_n(\cdot, \tilde{\beta}_n, \tilde{\gamma}_n);W)[v_j^*-v_{j,n}^*, \hat{\gamma}_n-\gamma_0]\\
			& \ \ \ \ \ + l''_{\gamma\zeta}(\tilde{\beta}_n, \tilde{\gamma}_n(\cdot), \tilde{\zeta}_n(\cdot, \tilde{\beta}_n, \tilde{\gamma}_n);W)[v_j^*-v_{j,n}^*, \hat{\zeta}_n-\zeta_0]\} ,
		\end{align*}
		where $(\tilde{\beta}_n, \tilde{\gamma}_n(\cdot), \tilde{\zeta}_n(\cdot, \tilde{\beta}_n, \tilde{\gamma}_n))$ is some point between $\theta_0$ and $\hat{\theta}_n$. Let $\tilde{\Lambda}(t, x) = \Lambda(t, x, \tilde{\beta}_n, \tilde{\gamma}_n, \tilde{g}_n)$. 
Note that by solving initial value problems in Lemma \ref{lemma: ode solutions}, we have $\tilde{\Lambda}'_{\beta}(t, x)$ and $\tilde{\Lambda}''_{\beta\beta}(t, x)$ are bounded on $t\in[0,\tau]$ and $x\in\mc{X}$ based on the boundedness of $\tilde{\gamma}_n$, $\tilde{g}_n$, $\tilde{g}'_n$ and $\tilde{g}''_n$. Also, we have $\|\tilde{\Lambda}'_{\gamma}(\cdot)[v]\|_{\infty} \lesssim\|v\|_{\infty}$ and $\sup_{t\in[0,\tau], x\in\mc{X}}\|\tilde{\Lambda}''_{\beta\gamma}(t, x)[v]\|\lesssim \|v\|_{\infty}$. It follows that 
\begin{align*}
	& \sup_{t\in[0,\tau], x\in\mc{X}}\|\tilde{\zeta}''_{\beta\gamma}(t, x, \tilde{\beta}_n, \tilde{\gamma}_n)[v_j^*-v_{j,n}^*]\| \\
	=& \sup_{t\in[0,\tau], x\in\mc{X}}\|\tilde{g}''_n(\tilde{\Lambda}(t, x))\tilde{\Lambda}'_{\beta}(t, x)\tilde{\Lambda}'_{\gamma}(t, x)[v_j^*-v_{j,n}^*]+ \tilde{g}'_n(\tilde{\Lambda}(t, X))\tilde{\Lambda}''_{\beta\gamma}(t, x) [v_j^*-v_{j,n}^*]\|\\
	\lesssim & \|v_j^*-v_{j,n}^*\|_{\infty},
\end{align*}
and
\begin{align*}
	& P\{\|l''_{\beta\gamma}(\tilde{\beta}_n, \tilde{\gamma}_n(\cdot), \tilde{\zeta}_n(\cdot, \tilde{\beta}_n, \tilde{\gamma}_n);W)[v_j^*-v_{j,n}^*]\|\}\\
	=& P\{\Big\|\int^\tau_0\tilde{\zeta}''_{\beta\gamma}(t, X, \tilde{\beta}_n, \tilde{\gamma}_n)[v_j^*-v_{j,n}^*]\mathbbm{1}(Y\geq t)\dev \Lambda_0(t, X) -  \int_0^\tau\exp(X^T\tilde{\beta}_n + \tilde{\gamma}_n(t)+\tilde{g}_n(\tilde{\Lambda}(t, X))) \\
     &\ \ \ \ \ \ \  \cdot \{(v_j^*(t)-v_{j,n}^*(t)+\tilde{g}'_n(\tilde{\Lambda}(t, X))\tilde{\Lambda}'_{\gamma}(t, X)[v_j^*-v_{j,n}^*])(X+\tilde{g}'_n(\tilde{\Lambda}(t, X))\tilde{\Lambda}'_{\beta}(t, X)) \\
     & \ \ \ \ \ \ \ \ \ \ \ +\tilde{\zeta}''_{\beta\gamma}(t, X, \tilde{\beta}_n, \tilde{\gamma}_n)[v_j^*-v_{j,n}^*] \} \mathbbm{1}(Y\geq t) \dev t \Big\|\}\\
     \lesssim &  \sup_{t\in[0,\tau], x\in\mc{X}}\|\tilde{\zeta}''_{\beta\gamma}(t, x, \tilde{\beta}_n, \tilde{\gamma}_n)[v_j^*-v_{j,n}^*]\| + \|v_j^*-v_{j,n}^*\|_{\infty}\lesssim \|v_j^*-v_{j,n}^*\|_{\infty}.
\end{align*}
Therefore, the first term in $I_{1n}$ is dominated by 
\begin{align*}
	&P\{|(\hat{\beta}_n-\beta_0)^T l''_{\beta\gamma}(\tilde{\beta}_n, \tilde{\gamma}_n(\cdot), \tilde{\zeta}_n(\cdot, \tilde{\beta}_n, \tilde{\gamma}_n);W)[v_j^*-v_{j,n}^*]|\}\\
	\leq & \|\hat{\beta}_n-\beta_0\|P\{\|l''_{\beta\gamma}(\tilde{\beta}_n, \tilde{\gamma}_n(\cdot), \tilde{\zeta}_n(\cdot, \tilde{\beta}_n, \tilde{\gamma}_n);W)[v_j^*-v_{j,n}^*]\|\}\\
	\lesssim &\|\hat{\beta}_n-\beta_0\|\|v_j^*-v_{j,n}^*\|_{\infty} \leq d(\hat{\theta}_n, \theta_0)\|v_j^*-v_{j,n}^*\|_{\infty} \\
	=& O_p(n^{-\min\{p_1\nu_1, p_2\nu_2, \frac{1-\max\{\nu_1, \nu_2\}}{2}\}})\cdot O(n^{-2\nu_1}) \\
	=& O_p(n^{-\min\{(p_1+2)\nu_1, p_2\nu_2+2\nu_1, \frac{1-\max\{\nu_1, \nu_2\}}{2}+2\nu_1\}}).
\end{align*}
By solving initial value problems in (\ref{eq: gamma dev}) and (\ref{eq: gammagamma dev}) and the Cauchy-Schwarz inequality (similar arguments are used in Lemma \ref{lemma: bdd linear operators} to prove that linear operators are bounded above), we have $\|\tilde{\Lambda}'_{\gamma}(\cdot)[v]\|_2\lesssim \|v\|_2$ and $\|\tilde{\Lambda}''_{\gamma\gamma}(\cdot)[\nu_1,\nu_2]\|_2 \lesssim \|\nu_1\|_{\infty}\|\nu_2\|_2$. It follows that 
\begin{align*}
	& \|\tilde{\zeta}''_{\gamma\gamma}(\cdot, \tilde{\beta}_n, \tilde{\gamma}_n)[v_j^*-v_{j,n}^*,\hat{\gamma}_n-\gamma_0]\|_2 \\
	=&\|\tilde{g}''_n(\tilde{\Lambda}(\cdot))\tilde{\Lambda}'_{\gamma}(\cdot)[\hat{\gamma}_n-\gamma_0]\tilde{\Lambda}'_{\gamma}(\cdot)[v_j^*-v_{j,n}^*]+ \tilde{g}'_n(\tilde{\Lambda}(\cdot))\tilde{\Lambda}''_{\gamma\gamma}(\cdot) [v_j^*-v_{j,n}^*,\hat{\gamma}_n-\gamma_0]\|_2\\
	\lesssim & \|\tilde{\Lambda}'_{\gamma}(\cdot)[v_j^*-v_{j,n}^*]\|_{\infty}\|\tilde{\Lambda}'_{\gamma}(\cdot)[\hat{\gamma}_n-\gamma_0]\|_2 + \|\tilde{\Lambda}''_{\gamma\gamma}(\cdot) [v_j^*-v_{j,n}^*,\hat{\gamma}_n-\gamma_0]\|_2\\
	\lesssim & \|v_j^*-v_{j,n}^*\|_{\infty}\cdot \|\hat{\gamma}_n-\gamma_0\|_2,
\end{align*}
and by the Cauchy-Schwarz inequality the second term in $I_{1n}$ is bounded by
\begin{align*}
	&(P\{\big|l''_{\gamma\gamma}(\tilde{\beta}_n, \tilde{\gamma}_n(\cdot), \tilde{\zeta}_n(\cdot, \tilde{\beta}_n, \tilde{\gamma}_n);W)[v_j^*-v_{j,n}^*, \hat{\gamma}_n-\gamma_0]\big|\})^2\\
	\leq & P\{\big|l''_{\gamma\gamma}(\tilde{\beta}_n, \tilde{\gamma}_n(\cdot), \tilde{\zeta}_n(\cdot, \tilde{\beta}_n, \tilde{\gamma}_n);W)[v_j^*-v_{j,n}^*, \hat{\gamma}_n-\gamma_0]\big|^2\}\\
	=& P\Big\{\Big|\Delta\tilde{\zeta}''_{\gamma\gamma}(Y, X, \tilde{\beta}_n, \tilde{\gamma}_n)[v_j^*-v_{j,n}^*, \hat{\gamma}_n-\gamma_0] \\
	& - \int_0^\tau\mathbbm{1}(Y\geq t) \exp(X^T\tilde{\beta}_n + \tilde{\gamma}_n(t)+\tilde{g}_n(\tilde{\Lambda}(t, X)))\cdot\{ \tilde{\zeta}''_{\gamma\gamma}(t, X, \tilde{\beta}_n, \tilde{\gamma}_n)[v_j^*-v_{j,n}^*, \hat{\gamma}_n-\gamma_0]\\
     &\ \ \ \ \ + ((v_j^*-v_{j,n}^*)(t)+\tilde{g}'_n(\tilde{\Lambda}(t, X))\tilde{\Lambda}'_{\gamma}(t, X)[v_j^*-v_{j,n}^*])\\
     & \ \ \ \ \ \ \cdot ((\hat{\gamma}_n-\gamma_0)(t)+\tilde{g}'_n(\tilde{\Lambda}(t, X))\tilde{\Lambda}'_{\gamma}(t, X)[\hat{\gamma}_n-\gamma_0]) \}\dev t \Big|^2\Big\}\\    
     \lesssim & \|\tilde{\zeta}''_{\gamma\gamma}(\cdot, \tilde{\beta}_n, \tilde{\gamma}_n)[v_j^*-v_{j,n}^*,\hat{\gamma}_n-\gamma_0]\|_2^2 + \|v_j^*-v_{j,n}^*\|_{\infty}^2 \cdot(\|\hat{\gamma}_n-\gamma_0\|_2^2+ \|\tilde{\Lambda}'_{\gamma}(\cdot)[\hat{\gamma}_n-\gamma_0]\|_2^2)\\
     \lesssim & \|v_j^*-v_{j,n}^*\|^2_{\infty}\cdot \|\hat{\gamma}_n-\gamma_0\|_2^2\leq \|v_j^*-v_{j,n}^*\|^2_{\infty}\cdot d^2(\hat{\theta}_n, \theta_0).
\end{align*}
So $P\{\big|l''_{\gamma\gamma}(\tilde{\beta}_n, \tilde{\gamma}_n(\cdot), \tilde{\zeta}_n(\cdot, \tilde{\beta}_n, \tilde{\gamma}_n);W)[v_j^*-v_{j,n}^*, \hat{\gamma}_n-\gamma_0]\big|\}= O_p(n^{-\min\{(p_1+2)\nu_1, p_2\nu_2+2\nu_1, \frac{1-\max\{\nu_1, \nu_2\}}{2}+2\nu_1\}})$. 
Also, by subtracting and adding some terms and using $\|a+b\|_2\leq \|a\|_2+\|b\|_2$, we have
\begin{align*}
	&\|\hat{\zeta}'_{n,\gamma}(\cdot, \hat{\beta}_n, \hat{\gamma}_n)[v_j^*-v_{j,n}^*] -\zeta'_{0\gamma}(\cdot, \beta_0, \gamma_0)[v_j^*-v_{j,n}^*]  \|_2\\
	= & \|\hat{g}'_n(\Lambda(\cdot, \hat{\beta}_n, \hat{\gamma}_n, \hat{g}_n))\Lambda'_{\gamma}(\cdot, \hat{\beta}_n, \hat{\gamma}_n, \hat{g}_n)[v_j^*-v_{j,n}^*] - g'_0(\Lambda_0(\cdot))\Lambda'_{0\gamma}(\cdot)[v_j^*-v_{j,n}^*]\|_2\\
	\leq & \|\hat{g}'_n(\Lambda(\cdot, \hat{\beta}_n, \hat{\gamma}_n, \hat{g}_n))\Lambda'_{\gamma}(\cdot, \hat{\beta}_n, \hat{\gamma}_n, \hat{g}_n)[v_j^*-v_{j,n}^*] - g'_0(\Lambda_0(\cdot))\Lambda'_{\gamma}(\cdot, \hat{\beta}_n, \hat{\gamma}_n, \hat{g}_n)[v_j^*-v_{j,n}^*]\|_2\\
	& + \|g'_0(\Lambda_0(\cdot))\Lambda'_{\gamma}(\cdot, \hat{\beta}_n, \hat{\gamma}_n, \hat{g}_n)[v_j^*-v_{j,n}^*] - g'_0(\Lambda_0(\cdot))\Lambda'_{\gamma}(\cdot, \beta_0, \hat{\gamma}_n, \hat{g}_n)[v_j^*-v_{j,n}^*]\|_2\\
	& + \|g'_0(\Lambda_0(\cdot))\Lambda'_{\gamma}(\cdot, \beta_0, \hat{\gamma}_n, \hat{g}_n)[v_j^*-v_{j,n}^*] - g'_0(\Lambda_0(\cdot))\Lambda'_{\gamma}(\cdot, \beta_0, \gamma_0, \hat{g}_n)[v_j^*-v_{j,n}^*]\|_2\\
	& + \|g'_0(\Lambda_0(\cdot))\Lambda'_{\gamma}(\cdot, \beta_0, \gamma_0, \hat{g}_n)[v_j^*-v_{j,n}^*] - g'_0(\Lambda_0(\cdot))\Lambda'_{0\gamma}(\cdot)[v_j^*-v_{j,n}^*]\|_2\\
	= & J_{1}+J_2+J_3+J_4.
\end{align*} 
For $J_1$, since $\hat{\gamma}_n$, $\hat{g}_n$ and $\hat{g}'_n$ are bounded, we have $\|\Lambda'_{\gamma}(\cdot, \hat{\beta}_n, \hat{\gamma}_n, \hat{g}_n)[v_j^*-v_{j,n}^*]\|_{\infty}\lesssim \|v_j^*-v_{j,n}^*\|_{\infty}$ and it follows that
\begin{align*}
	J_1& \leq \|\hat{g}'_n(\Lambda(\cdot, \hat{\beta}_n, \hat{\gamma}_n, \hat{g}_n)) - g'_0(\Lambda_0(\cdot))\|_2 \cdot \|\Lambda'_{\gamma}(\cdot, \hat{\beta}_n, \hat{\gamma}_n, \hat{g}_n)[v_j^*-v_{j,n}^*]\|_{\infty}\\
	& \lesssim \|\hat{g}'_n(\Lambda(\cdot, \hat{\beta}_n, \hat{\gamma}_n, \hat{g}_n)) - g'_0(\Lambda_0(\cdot))\|_2 \cdot \|v_j^*-v_{j,n}^*\|_{\infty}\\
	& =O_p(n^{-\min\{p_1\nu_1, (p_2-1)\nu_2, \frac{1-\max\{\nu_1, \nu_2\}}{2}\}})\cdot O(n^{-2\nu_1}) \\
	& =O_p(n^{-\min\{(p_1+2)\nu_1, (p_2-1)\nu_2+2\nu_1, \frac{1-\max\{\nu_1, \nu_2\}}{2}+2\nu_1\}}),
\end{align*}
where the third equality holds based on the same argument of \cite{ding2011} on their page 3058.
For~$J_2$, by using the mean value theorem, it follows that 
\begin{align*}
	J_2& =\|g'_0(\Lambda_0(\cdot))(\Lambda''_{\gamma\beta}(\cdot, \tilde{\beta}_n, \hat{\gamma}_n, \hat{g}_n)[v_j^*-v_{j,n}^*])^T(\hat{\beta}_n-\beta_0)\|_2\\
	& \lesssim \|\Lambda''_{\gamma\beta}(\cdot, \tilde{\beta}_n, \hat{\gamma}_n, \hat{g}_n)[v_j^*-v_{j,n}^*]\|_2 \|\hat{\beta}_n-\beta_0\|\\
	& \lesssim \|v_j^*-v_{j,n}^*\|_{\infty} \cdot \|\hat{\beta}_n-\beta_0\|\\
	& = O_p(n^{-\min\{(p_1+2)\nu_1, p_2\nu_2+2\nu_1, \frac{1-\max\{\nu_1, \nu_2\}}{2}+2\nu_1\}}),
\end{align*}
where $\tilde{\beta}_n$ is a point between $\hat{\beta}_n$ and $\beta_0$, the second inequality is based on the boundedness of $g'_0$, and the third inequality is obtained by solving the initial value problem in (\ref{eq: beta gamma dev}) along with the boundedness of $\hat{\gamma}_n$, $\hat{g}_n$, $\hat{g}'_n$, $\hat{g}''_n$ and $\Lambda'_{\beta}(\cdot, \tilde{\beta}_n, \hat{\gamma}_n, \hat{g}_n)$. By a similar argument that we used for the second term in $I_{1n}$, we have for $J_3$,
\begin{align*}
	J_3 &= \|g'_0(\Lambda_0(\cdot))\Lambda''_{\gamma\gamma}(\cdot, \beta_0, \tilde{\gamma}_n, \hat{g}_n)[v_j^*-v_{j,n}^*, \hat{\gamma}_n-\gamma_0]\|_2\\
	& \lesssim \|v_j^*-v_{j,n}^*\|_{\infty}\cdot \|\hat{\gamma}_n-\gamma_0\|_2  = O_p(n^{-\min\{(p_1+2)\nu_1, p_2\nu_2+2\nu_1, \frac{1-\max\{\nu_1, \nu_2\}}{2}+2\nu_1\}}),
\end{align*}
and for $J_4$
\begin{align*}
	J_4& =\|g'_0(\Lambda_0(\cdot))\Lambda''_{\gamma g}(\cdot, \beta_0, \gamma_0, \tilde{g}_n)[v_j^*-v_{j,n}^*,\hat{g}_n-g_0]\|_2\\
	& \lesssim \|\Lambda''_{\gamma g}(\cdot, \beta_0, \gamma_0, \tilde{g}_n)[v_j^*-v_{j,n}^*,\hat{g}_n-g_0]\|_2\\
	& \lesssim (\|\hat{g}_n(\Lambda(\cdot, \hat{\beta}_n, \hat{\gamma}_n, \hat{g}_n)) - g_0(\Lambda_0(\cdot))\|_2  + \|\hat{g}'_n(\Lambda(\cdot, \hat{\beta}_n, \hat{\gamma}_n, \hat{g}_n)) - g'_0(\Lambda_0(\cdot))\|_2 )\cdot \|v_j^*-v_{j,n}^*\|_{\infty}\\
	& = O_p(n^{-\min\{p_1\nu_1, (p_2-1)\nu_2, \frac{1-\max\{\nu_1, \nu_2\}}{2}\}})\cdot O(n^{-2\nu_1}),
\end{align*}
where $\tilde{\gamma}_n$ is a point between $\hat{\gamma}_n$ and $\gamma_0$ and $\tilde{g}_n$ is a point between $\hat{g}_n$ and $g_0$. 
Thus, we have 
\begin{align*}
	&\|\hat{\zeta}'_{n,\gamma}(\cdot, \hat{\beta}_n, \hat{\gamma}_n)[v_j^*-v_{j,n}^*] -\zeta'_{0\gamma}(\cdot, \beta_0, \gamma_0)[v_j^*-v_{j,n}^*]  \|_2 \\
	\lesssim &  O_p(n^{-\min\{(p_1+2)\nu_1, (p_2-1)\nu_2+2\nu_1, \frac{1-\max\{\nu_1, \nu_2\}}{2}+2\nu_1\}}),
\end{align*}
and it follows that for the third term in $I_{1n}$ is bounded by
\begin{align*}
	& (P\{\big|l''_{\gamma\zeta}(\tilde{\beta}_n, \tilde{\gamma}_n(\cdot), \tilde{\zeta}_n(\cdot, \tilde{\beta}_n, \tilde{\gamma}_n);W)[v_j^*-v_{j,n}^*, \hat{\zeta}_n-\zeta_0]\big|\})^2\\
	\leq & P\{\big|l''_{\gamma\zeta}(\tilde{\beta}_n, \tilde{\gamma}_n(\cdot), \tilde{\zeta}_n(\cdot, \tilde{\beta}_n, \tilde{\gamma}_n);W)[v_j^*-v_{j,n}^*, \hat{\zeta}_n-\zeta_0]\big|^2\}\\
	=& P\Big\{\Big|\Delta (\hat{\zeta}'_{n,\gamma}(Y,X, \hat{\beta}_n, \hat{\gamma}_n)[v_j^*-v_{j,n}^*] -\zeta'_{0\gamma}(Y,X, \beta_0, \gamma_0)[v_j^*-v_{j,n}^*]) \\
	& - \int_0^\tau \mathbbm{1}(Y\geq t) \exp(X^T\tilde{\beta}_n + \tilde{\gamma}_n(t)+\tilde{g}_n(\tilde{\Lambda}(t, X)))\cdot\{  (\hat{\zeta}_{n}(t,X, \hat{\beta}_n, \hat{\gamma}_n) -\zeta_{0}(t,X, \beta_0, \gamma_0)) \\
     &\ \ \ \ \ \cdot ((v_j^*-v_{j,n}^*)(t)+\tilde{g}'_n(\tilde{\Lambda}(t, X))\tilde{\Lambda}'_{\gamma}(t, X)[v_j^*-v_{j,n}^*])\\
     & \ \ \ \ \ + \hat{\zeta}'_{n,\gamma}(t,X, \hat{\beta}_n, \hat{\gamma}_n)[v_j^*-v_{j,n}^*] -\zeta'_{0\gamma}(t,X, \beta_0, \gamma_0)[v_j^*-v_{j,n}^*] \}\dev t \Big|^2\Big\}\\    
     \lesssim & \|\hat{\zeta}'_{n,\gamma}(\cdot, \hat{\beta}_n, \hat{\gamma}_n)[v_j^*-v_{j,n}^*] -\zeta'_{0\gamma}(\cdot, \beta_0, \gamma_0)[v_j^*-v_{j,n}^*]\|_2^2 + \|\hat{\zeta}_{n}(\cdot, \hat{\beta}_n, \hat{\gamma}_n) -\zeta_{0}(\cdot, \beta_0, \gamma_0)\|^2_2 \cdot \|v_j^*-v_{j,n}^*\|_{\infty}^2 \\
     = & O_p(n^{-2\min\{(p_1+2)\nu_1, (p_2-1)\nu_2+2\nu_1, \frac{1-\max\{\nu_1, \nu_2\}}{2}+2\nu_1\}}).
\end{align*}
Thus, we have $I_{1n}=O_p(n^{-\min\{(p_1+2)\nu_1, (p_2-1)\nu_2+2\nu_1, \frac{1-\max\{\nu_1, \nu_2\}}{2}+2\nu_1\}}) = o_p(n^{-1/2})$, because $(p_1+2)\nu_1>1/2$, $(p_2-1)\nu_2+2\nu_1>1/2$, and $4\nu_1>\max\{\nu_1, \nu_2\}$ under the restrictions listed in Theorem \ref{thm: conv rate}.

Next we will use the maximal inequality in Lemma 3.4.2 of \cite{van1996weak} (on  page 324) and the Markov's inequality to show that $I_{2n}=o_p(n^{-1/2})$. By Lemma \ref{lemma: other bracket numbers for A4}, the $\epsilon$-bracketing number associated with $\|\cdot\|_{\infty}$ norm for the class $\mc{F}_{n,j}^{\gamma}(\eta)$ is bounded by $\left(\eta/\epsilon \right)^{c_1 q_{n_1}+c_2q_{n_2}+d}$, which implies that 
\[\log N_{[\ ]}(\epsilon, \mc{F}_{n,j}^{\gamma}(\eta), L_2(P))\leq \log N_{[\ ]}(\epsilon, \mc{F}_{n,j}^{\gamma}(\eta), \|\cdot\|_{\infty}) \lesssim (c_1 q_{n_1}+c_2q_{n_2})\log(\eta/\epsilon). \]
It follows that the bracketing integral satisfies 
\[J_{[\ ]}(\epsilon, \mc{F}_{n,j}^{\gamma}(\eta), L_2(P)) = \int_0^{\eta}\sqrt{1+\log N_{[\ ]}(\epsilon, \mc{F}_{n,j}^{\gamma}(\eta), L_2(P))}\dev \epsilon \lesssim (c_1 q_{n_1}+c_2q_{n_2})^{1/2}\eta.\]
Here we choose $\eta_n = O(n^{-\min\{2\nu_1, p_2\nu_2, \frac{1-\max\{\nu_1, \nu_2\}}{2}\}})$ such that $\|v_j^*-v_{j,n}^*\|_{\infty}= O(n^{-2\nu_1})\leq \eta_n$ and $d(\hat{\theta}_n, \theta_0) = O_p(n^{-\min\{p_1\nu_1, p_2\nu_2, \frac{1-\max\{\nu_1, \nu_2\}}{2}\}})\leq \eta_n$ for $p_1\geq 2$, then $l'_{\gamma}(\hat{\theta}_n);W)[v_j^*-v_{j,n}^*] \in \mc{F}_{n,j}^{\gamma}(\eta_n)$. For any $l'_{\gamma}(\theta;W)[v_j^*-v_{j}] \in \mc{F}_{n,j}^{\gamma}(\eta_n)$, we have
\begin{align*}
	& P\{l'_{\gamma}(\theta;W)[v_j^*-v_{j}]\}^2\\
	=& P\{\Delta((v_j^*-v_{j})(Y)+g'(\Lambda(Y,X, \beta,\gamma,g))\Lambda'_{\gamma}(Y, X, \beta, \gamma,g)[v_j^*-v_{j}]) \\
     & \ \ \  - \int_0^Y \exp(X^T\beta + \gamma(t)+g(\Lambda(t,X, \beta,\gamma,g)))\{(v_j^*-v_{j})(t)+ \zeta'_{\gamma}(t, X, \beta, \gamma)[v_j^*-v_j]\} \dev t\}^2\\
	\lesssim & \|v_j^*-v_j\|^2_{\infty} + \|\Lambda'_{\gamma}(\cdot, \beta, \gamma,g)[v_j^*-v_{j}]\|^2_{\infty}\\
	\lesssim & \|v_j^*-v_j\|^2_{\infty}.
\end{align*}
Also, $\sup_{\theta: d(\theta,\theta_0)\leq \eta_n; v_j:\|v_j^*-v_j\|_{\infty}\leq \eta_n}|l'_{\gamma}(\theta;W)[v_j^*-v_{j}]|$ is bounded by some constant $0<M<\infty$ (or slowly growing with $n$ and it can be treated as bounded by the same argument used in \citet[page 591]{shen1994}). By the maximal inequality, it follows that 
\begin{align*}
				E_{P}\|\mathbb{G}_n\|_{\mc{F}_{n,j}^{\gamma}(\eta_n)}& \lesssim J_{[\ ]}(\epsilon, \mc{F}_{n,j}^{\gamma}(\eta_n), L_2(P))\left(1+\frac{J_{[\ ]}(\epsilon, \mc{F}_{n,j}^{\gamma}(\eta_n), L_2(P))}{\eta_n^2\sqrt{n}}M\right)\\
				& \lesssim (c_1 q_{n_1}+c_2q_{n_2})^{1/2}\eta_n + (c_1 q_{n_1}+c_2q_{n_2})n^{-1/2}\\
				& = O(n^{\frac{\max\{\nu_1, \nu_2\}}{2}})\cdot O(n^{-\min\{2\nu_1, p_2\nu_2, \frac{1-\max\{\nu_1, \nu_2\}}{2}\}}) + O(n^{\max\{\nu_1,\nu_2\}-1/2})\\
				& = O(n^{-\min\{2\nu_1-\frac{\max\{\nu_1, \nu_2\}}{2}, p_2\nu_2-\frac{\max\{\nu_1, \nu_2\}}{2}, 1/2-\max\{\nu_1, \nu_2\}\}}) + O(n^{\max\{\nu_1,\nu_2\}-1/2})\\
				& =o(1),
\end{align*}
where $\mathbb{G}_n = \sqrt{n}(\mathbb{P}_n-P)$ and the last equality holds because $0<\nu_1,\nu_2<1/2$,  $4\nu_1>\max\{\nu_1,\nu_2\}$, and $p_2\nu_2 >2\nu_2 >\max\{\nu_1,\nu_2\}$. Then by the Markov's inequality, we have \[I_{2n} = n^{-1/2}\mathbb{G}_n l'_{\gamma}(\hat{\theta}_n;W)[v_j^*-v_{j,n}^*] = o_p(n^{-1/2}).\]
By combining $I_{1n}=o_p(n^{-1/2})$ and $I_{2n}=o_p(n^{-1/2})$, we have  $S'_{\gamma,n}(\hat{\theta}_n)[v_j^*] = o_p(n^{-1/2})$.

Next, to verify the last part of (\ref{m:gc}), we need to show that $S'_{\zeta,n}(\hat{\theta}_n)[h_j^*] = o_p(n^{-1/2})$ with $h^*_j (\cdot, \hat{\beta}_n, \hat{\gamma}_n) = w^*_j(\hat{\Lambda}(\cdot))+\hat{g}'_n(\hat{\Lambda}(\cdot))\hat{\Lambda}'_{g}(\cdot)[w^*_j]$, where we write $\hat{\Lambda}(\cdot) = \Lambda(\cdot,\hat{\beta}_n, \hat{\gamma}_n, \hat{g}_n)$ for notational simplicity. Since $w_j^* \in \mc{G}^2$, by Lemma \ref{lemma: gnp and gp} there exists $w_{j,n}^* \in \mc{G}^2_n$ such that $\|w_{j,n}^* - w^*_{j}\|_{\infty}=O(n^{-2\nu_2})$. It follows that %TODO: \wj{add details} 
$S'_{\zeta,n}(\hat{\beta}_n, \hat{\gamma}_n(\cdot), \hat{\zeta}_n(\cdot, \hat{\beta}_n, \hat{\gamma}_n))[h_{j,n}^*] = 0$ with $h^*_{j,n} (\cdot, \hat{\beta}_n, \hat{\gamma}_n) = w^*_{j,n}(\hat{\Lambda}(\cdot))+\hat{g}_n'(\hat{\Lambda}(\cdot))\hat{\Lambda}'_{g}(\cdot)[w^*_{j,n}]$. Then it suffices to show that for each $1\leq j\leq d$,
\begin{align*}
	S'_{\zeta,n}(\hat{\theta}_n)[h_j^*] &= S'_{\zeta,n}(\hat{\theta}_n)[h_j^*-h^*_{j,n}]\\
	& = P\{l'_{\zeta}(\hat{\theta}_n;W)[h_j^*-h_{j,n}^*]-l'_{\zeta}(\theta_0;W)[h_j^*-h_{j,n}^*]\} +  (\mathbb{P}_n-P)\{l'_{\zeta}(\hat{\zeta}_n;W)[h_j^*-h_{j,n}^*]\} \\
	& = I_{3n} + I_{4n} = o_p(n^{-1/2}),
\end{align*}
since $S'_{\zeta}(\theta_0)[h_j^*-h_{j,n}^*] = 0$. We will take the similar arguments used in the proof of $S'_{\gamma,n}(\hat{\theta}_n)[v_j^*] = o_p(n^{-1/2})$ to show that both $I_{3n}$ and $I_{4n}$ equal to $o_p(n^{-1/2})$.

For $I_{3n}$, using the Taylor expansion for $l'_{\zeta}(\hat{\theta}_n)[h_j^*-h_{j,n}^*]$ at $\theta_0$,  we have 
		\begin{align*}
			I_{3n} =& P\{(\hat{\beta}_n-\beta_0)^T l''_{\beta\zeta}(\tilde{\beta}_n, \tilde{\gamma}_n(\cdot), \tilde{\zeta}_n(\cdot, \tilde{\beta}_n, \tilde{\gamma}_n);W)[h_j^*-h_{j,n}^*] \\
			& \ \ \ \ \ + l''_{\zeta\gamma}(\tilde{\beta}_n, \tilde{\gamma}_n(\cdot), \tilde{\zeta}_n(\cdot, \tilde{\beta}_n, \tilde{\gamma}_n);W)[h_j^*-h_{j,n}^*, \hat{\gamma}_n-\gamma_0]\\
			& \ \ \ \ \ + l''_{\zeta\zeta}(\tilde{\beta}_n, \tilde{\gamma}_n(\cdot), \tilde{\zeta}_n(\cdot, \tilde{\beta}_n, \tilde{\gamma}_n);W)[h_j^*-h_{j,n}^*, \hat{\zeta}_n-\zeta_0]\}
		\end{align*}
		where $(\tilde{\beta}_n, \tilde{\gamma}_n(\cdot), \tilde{\zeta}_n(\cdot, \tilde{\beta}_n, \tilde{\gamma}_n))$ is some point between $\theta_0$ and $\hat{\theta}_n$. Let $\tilde{\Lambda}(t, x) = \Lambda(t, x, \tilde{\beta}_n, \tilde{\gamma}_n, \tilde{g}_n)$. 
Note that by solving initial value problems in Lemma \ref{lemma: ode solutions}, we have $\tilde{\Lambda}'_{\beta}(t, x)$ is bounded on $t\in[0,\tau]$ and $x\in\mc{X}$ based on the boundedness of $\tilde{\gamma}_n$, $\tilde{g}_n$, and $\tilde{g}'_n$. Also, we have $\|\tilde{\Lambda}'_{g}(\cdot)[w]\|_{\infty} \lesssim\|w\|_{\infty}$, $\|\tilde{\Lambda}'_{\gamma}(\cdot)[v]\|_{2} \lesssim\|v\|_{2}$, and furthermore, $\sup_{t\in[0,\tau], x\in\mc{X}}\|\tilde{\Lambda}''_{g\beta}(t, x)[w]\|\lesssim \|w\|_{\infty}+\|w'\|_{\infty}$ and $\|\tilde{\Lambda}''_{g\gamma}(\cdot)[w,v]\|_2\lesssim (\|w\|_{\infty}+\|w'\|_{\infty})\|v\|_{2}$. Using the triangle inequality, it follows that 
\begin{align*}
	 \|(h^*_{j} - h^*_{j,n})(\cdot, \tilde{\beta}_n, \tilde{\gamma}_n)\|_{\infty} &= \|(w^*_j-w^*_{j,n})(\tilde{\Lambda}(\cdot))+\tilde{g}'_n(\tilde{\Lambda}(\cdot))\tilde{\Lambda}'_{g}(\cdot)[w^*_j-w^*_{j,n}]\|_{\infty}\\
	& \lesssim \|w^*_j-w^*_{j,n}\|_{\infty}, \\
	 \sup_{t\in[0,\tau], x\in\mc{X}}\|(h^*_{j} - h^*_{j,n})'_{\beta}(t, x, \tilde{\beta}_n, \tilde{\gamma}_n)\| & =\sup_{t\in[0,\tau], x\in\mc{X}} \|(w^{*}_j-w^*_{j,n})'(\tilde{\Lambda}(t, x))\tilde{\Lambda}'_{\beta}(t, x) \\
	 & \ \ \ \ \ \ \ \ + \tilde{g}'_n(\tilde{\Lambda}(t, x))\tilde{\Lambda}''_{g\beta}(t, x)[w^*_j-w^*_{j,n}]\\
	 & \ \ \ \ \ \ \ \ + \tilde{g}''_n(\tilde{\Lambda}(t, x))\tilde{\Lambda}'_{g}(t, x)[w^*_j-w^*_{j,n}]\tilde{\Lambda}'_{\beta}(t, x)\|\\
	 & \lesssim \|w^*_j-w^*_{j,n}\|_{\infty} + \|(w^*_j-w^*_{j,n})'\|_{\infty},
\end{align*}
and 
\begin{align*}
	\|(h^*_{j} - h^*_{j,n})'_{\gamma}(\cdot, \tilde{\beta}_n, \tilde{\gamma}_n)[\hat{\gamma}_n-\gamma_0]\|_2 & = \|(w^{*}_j-w^*_{j,n})'(\tilde{\Lambda}(\cdot))\tilde{\Lambda}'_{\gamma}(\cdot)[\hat{\gamma}_n-\gamma_0] \\
	 & \ \ \ \ + \tilde{g}'_n(\tilde{\Lambda}(\cdot))\tilde{\Lambda}''_{g\gamma}(\cdot)[w^*_j-w^*_{j,n}, \hat{\gamma}_n-\gamma_0]\\
	 & \ \ \ \  + \tilde{g}''_n(\tilde{\Lambda}(\cdot))\tilde{\Lambda}'_{g}(\cdot)[w^*_j-w^*_{j,n}]\tilde{\Lambda}'_{\gamma}(\cdot)[\hat{\gamma}_n-\gamma_0]\|_2\\
	 & \lesssim (\|w^*_j-w^*_{j,n}\|_{\infty} + \|(w^*_j-w^*_{j,n})'\|_{\infty})\|\hat{\gamma}_n-\gamma_0\|_2.
\end{align*}
Therefore, by plugging the derivatives in Lemma \ref{lemma: devivatives} and using the triangle inequality and the Cauchy-Schwarz inequality, $I_{3n}$ is dominated by 
\begin{align*}
I_{3n}  \lesssim &  \sup_{t\in[0,\tau], x\in\mc{X}}|(\hat{\beta}_n-\beta_0)^T(h^*_{j} - h^*_{j,n})'_{\beta}(t, x, \tilde{\beta}_n, \tilde{\gamma}_n)|  + \|(h^*_{j} - h^*_{j,n})'_{\gamma}(\cdot, \tilde{\beta}_n, \tilde{\gamma}_n)[\hat{\gamma}_n-\gamma_0]\|_2\\
& + \|(h^*_{j} - h^*_{j,n})(\cdot, \tilde{\beta}_n, \tilde{\gamma}_n)\|_{\infty} \cdot P\Big\{ \int^{\tau}_0  \exp(X^T\tilde{\beta}_n + \tilde{\gamma}_n(t)+\tilde{g}_n(\tilde{\Lambda}(t, X))) \\
& \ \ \ \ \ \ \cdot \big( (X+\tilde{g}'_n(\tilde{\Lambda}(t, X))^T(\hat{\beta}_n-\beta_0)+\hat{\gamma}_n(t)-\gamma_0(t)+\tilde{g}'_n(\tilde{\Lambda}(t, X))\tilde{\Lambda}'_{\gamma}(t, X)[\hat{\gamma}_n-\gamma_0]\\
& \ \ \ \ \ \ \ \ \ + \hat{\zeta}_n(t, X)-\zeta_0(t, X)\big)^2 \dev t \Big\}^{1/2} \\
\lesssim &~ \|\hat{\beta}_n-\beta_0\|\cdot \sup_{t\in[0,\tau], x\in\mc{X}}\|(h^*_{j} - h^*_{j,n})'_{\beta}(t, x, \tilde{\beta}_n, \tilde{\gamma}_n)\|+\|(h^*_{j} - h^*_{j,n})'_{\gamma}(\cdot, \tilde{\beta}_n, \tilde{\gamma}_n)[\hat{\gamma}_n-\gamma_0]\|_2\\
& + \|(h^*_{j} - h^*_{j,n})(\cdot, \tilde{\beta}_n, \tilde{\gamma}_n)\|_{\infty} \cdot \left(\|\hat{\beta}_n-\beta_0\|^2+ \|\hat{\gamma}_n-\gamma_0\|_2^2 + \|\hat{\zeta}_n-\zeta_0\|^2_2 \right)^{1/2}\\
\lesssim & ~ (\|w^*_j-w^*_{j,n}\|_{\infty} + \|(w^*_j-w^*_{j,n})'\|_{\infty}) d(\hat{\theta}_n, \theta_0).
\end{align*}
Based on the Corollary 6.21 in \citet{schumaker_2007}, we have $\|(w^*_j-w^*_{j,n})'\|_{\infty}=O(n^{-\nu_2})$ and 
\begin{align*}
	I_{3n} & = O(n^{-\nu_2})\cdot O_p(n^{-\min\{p_1\nu_1, p_2\nu_2, \frac{1-\max\{\nu_1, \nu_2\}}{2}\}})\\
	&= O_p(n^{-\min\{p_1\nu_1+\nu_2, (p_2+1)\nu_2, \frac{1-\max\{\nu_1, \nu_2\}}{2}+\nu_2\}})\\
	& = o_p(n^{-1/2}),
\end{align*}
where the last equality holds because $p_1\nu_1+\nu_2 >1/2$, $(p_2+1)\nu_2>1/2$, and $2\nu_2>\max\{\nu_1,\nu_2\}$. 

Next, we use the maximal inequality and the Markov's inequality to show that $I_{4n}=o_p(n^{-1/2})$. By Lemma \ref{lemma: other bracket numbers for A4}, the $\epsilon$-bracketing number associated with $\|\cdot\|_{\infty}$ norm for the class $\mc{F}_{n,j}^{\zeta}(\eta)$ is bounded by $\left(\eta/\epsilon \right)^{c_3 q_{n_1}+c_4q_{n_2}+d}$, which implies that 
\[\log N_{[\ ]}(\epsilon, \mc{F}_{n,j}^{\zeta}(\eta), L_2(P))\leq \log N_{[\ ]}(\epsilon, \mc{F}_{n,j}^{\zeta}(\eta), \|\cdot\|_{\infty}) \lesssim (c_3 q_{n_1}+c_4q_{n_2})\log(\eta/\epsilon). \]
It follows that the bracketing integral satisfies 
\[J_{[\ ]}(\epsilon, \mc{F}_{n,j}^{\zeta}(\eta), L_2(P)) = \int_0^{\eta}\sqrt{1+\log N_{[\ ]}(\epsilon, \mc{F}_{n,j}^{\zeta}(\eta), L_2(P))}\dev \epsilon \lesssim (c_3 q_{n_1}+c_4q_{n_2})^{1/2}\eta.\]
Here we choose $\eta_n = O(n^{-\min\{p_1\nu_1, 2\nu_2, \frac{1-\max\{\nu_1, \nu_2\}}{2}\}})$ such that $\|w_j^*-w_{j,n}^*\|_{\infty}= O(n^{-2\nu_2})\leq \eta_n$ and $d(\hat{\theta}_n, \theta_0) = O_p(n^{-\min\{p_1\nu_1, p_2\nu_2, \frac{1-\max\{\nu_1, \nu_2\}}{2}\}})\leq \eta_n$ for $p_2\geq 3$, then $l'_{\zeta}(\hat{\theta}_n;W)[h_j^*-h_{j,n}^*] \in \mc{F}_{n,j}^{\zeta}(\eta_n)$. For any $l'_{\zeta}(\theta;W)[h_j^*-h_{j}] \in \mc{F}_{n,j}^{\zeta}(\eta_n)$, we have
\begin{align*}
	& P\{l'_{\zeta}(\theta;W)[h_j^*-h_{j}]\}^2\\
	=& P\{\Delta((w_j^*-w_{j})(Y)+g'(\Lambda(Y,X, \beta,\gamma,g))\Lambda'_{g}(Y, X, \beta, \gamma,g)[w_j^*-w_{j}]) \\
     & \ \ \  - \int_0^Y \exp(X^T\beta + \gamma(t)+g(\Lambda(t,X, \beta,\gamma,g)))\{(w_j^*-w_{j})(t)+ \zeta'_{g}(t, X, \beta, \gamma)[w_j^*-w_j]\} \dev t\}^2\\
	\lesssim & \|w_j^*-w_j\|^2_{\infty} + \|\Lambda'_{g}(\cdot, \beta, \gamma,g)[w_j^*-w_{j}]\|^2_{\infty}\\
	\lesssim & \|w_j^*-w_j\|^2_{\infty} \leq \eta_n.
\end{align*}
Also, $\sup_{\theta: d(\theta,\theta_0)\leq \eta_n; w_j:\|w_j^*-w_j\|_{\infty}\leq \eta_n}|l'_{\zeta}(\theta;W)[h_j^*-h_{j}]|$ is bounded by some constant $0<M<\infty$. By the maximal inequality, it follows that 
\begin{align*}
				E_{P}\|\mathbb{G}_n\|_{\mc{F}_{n,j}^{\zeta}(\eta_n)}& \lesssim J_{[\ ]}(\epsilon, \mc{F}_{n,j}^{\zeta}(\eta_n), L_2(P))\left(1+\frac{J_{[\ ]}(\epsilon, \mc{F}_{n,j}^{\zeta}(\eta_n), L_2(P))}{\eta_n^2\sqrt{n}}M\right)\\
				& \lesssim (c_3 q_{n_1}+c_4q_{n_2})^{1/2}\eta_n + (c_3 q_{n_1}+c_4q_{n_2})n^{-1/2}\\
				& = O(n^{\frac{\max\{\nu_1, \nu_2\}}{2}})\cdot O(n^{-\min\{p_1\nu_1, 2\nu_2, \frac{1-\max\{\nu_1, \nu_2\}}{2}\}}) + O(n^{\max\{\nu_1,\nu_2\}-1/2})\\
				& = O(n^{-\min\{p_1\nu_1-\frac{\max\{\nu_1, \nu_2\}}{2}, 2\nu_2-\frac{\max\{\nu_1, \nu_2\}}{2}, 1/2-\max\{\nu_1, \nu_2\}\}}) + O(n^{\max\{\nu_1,\nu_2\}-1/2})\\
				& =o(1),
\end{align*}
where the last equality holds because $0<\nu_1,\nu_2<1/2$,  $2\nu_2>\max\{\nu_1,\nu_2\} > \max\{\nu_1,\nu_2\}/2$, and $p_1\nu_1 \geq 2\nu_1 >\max\{\nu_1,\nu_2\}/2$ for $p_1 \geq 2$. Then by the Markov's inequality, we have \[I_{4n} = n^{-1/2}\mathbb{G}_n l'_{\zeta}(\hat{\theta}_n;W)[h_j^*-h_{j,n}^*] = o_p(n^{-1/2}).\]
By combining $I_{3n}=o_p(n^{-1/2})$ and $I_{4n}=o_p(n^{-1/2})$, we verify that $S'_{\zeta,n}(\hat{\theta}_n)[h_j^*] = o_p(n^{-1/2})$. This completes the verification of the assumption (\ref{m:gc}).

Now we verify assumption (\ref{m:equicontinuity}). Since the proofs of three stochastic equicontinuity equations are essentially based on the identical arguments, we only present the proof of the first equation as follows. First, by Lemma \ref{lemma: other bracket numbers for A5}, the $\epsilon$-bracketing number associated with $\|\cdot\|_{\infty}$ norm for the class $\mc{F}_{n,j}^{*\beta}(\eta)$ is bounded by $\left(\eta/\epsilon \right)^{c_1 q_{n_1}+c_2q_{n_2}+d}$, which implies that the bracketing integral is bounded by $(c_1 q_{n_1}+c_2q_{n_2})^{1/2}\eta$, i.e. 
\[J_{[\ ]}(\epsilon, \mc{F}_{n,j}^{*\beta}(\eta), L_2(P)) \lesssim (c_1 q_{n_1}+c_2q_{n_2})^{1/2}\eta.\]
For any $l'_{\beta_j}(\theta;W) -l'_{\beta_j}(\theta_0;W)\in \mc{F}_{n,j}^{*\beta}(\eta_n)$, by taking the Taylor expansion at $\theta_0$, it follows that
\begin{align*}
			l'_{\beta_j}(\theta;W) -l'_{\beta_j}(\theta_0;W) =& (\beta-\beta_0)^T l''_{\beta_j \beta}(\tilde{\beta}, \tilde{\gamma}(\cdot), \tilde{\zeta}(\cdot, \tilde{\beta}, \tilde{\gamma});W) \\
			& + l''_{\beta_j \gamma}(\tilde{\beta}, \tilde{\gamma}(\cdot), \tilde{\zeta}(\cdot, \tilde{\beta}, \tilde{\gamma});W)[\gamma-\gamma_0]\\
			& + l''_{\beta_j \zeta}(\tilde{\beta}, \tilde{\gamma}(\cdot), \tilde{\zeta}(\cdot, \tilde{\beta}, \tilde{\gamma});W)[\zeta-\zeta_0] 
\end{align*}
where $(\tilde{\beta}, \tilde{\gamma}(\cdot), \tilde{\zeta}(\cdot, \tilde{\beta}, \tilde{\gamma}))$ is some point between $\theta_0$ and $\theta$. By applying the triangle inequality and the Cauchy-Schwarz inequality, we have 
\begin{align*}
	P\{l'_{\beta_j}(\theta;W) -l'_{\beta_j}(\theta_0;W)\}^2 \leq & \|\beta-\beta_0\|^2 P\{\|l''_{\beta_j \beta}(\tilde{\beta}, \tilde{\gamma}(\cdot), \tilde{\zeta}(\cdot, \tilde{\beta}, \tilde{\gamma});W) \|^2\} \\
	& +P\{ l''_{\beta_j \gamma}(\tilde{\beta}, \tilde{\gamma}(\cdot), \tilde{\zeta}(\cdot, \tilde{\beta}, \tilde{\gamma});W)[\gamma-\gamma_0]\}^2\\
	& + P\{l''_{\beta_j \zeta}(\tilde{\beta}, \tilde{\gamma}(\cdot), \tilde{\zeta}(\cdot, \tilde{\beta}, \tilde{\gamma});W)[\zeta-\zeta_0] \}^2\\
	= & B_1+B_2+B_3.
\end{align*}
For $B_1$, by Lemma \ref{lemma: devivatives}, $l''_{\beta_j \beta}(\tilde{\theta};W)$ is bounded and it follows that $B_1 \lesssim \|\beta-\beta_0\|^2$.
For $B_2$, since $\tilde{g}$, $\tilde{g}'$, $\tilde{g}''$, $\tilde{\Lambda}'_{\beta_j}(t, x)$ are bounded and $\|\tilde{\Lambda}''_{\beta_j \gamma}(\cdot)[v]\|_2 \lesssim \|v\|_2$, by applying the Cauchy-Schwarz inequality and the same arguments that are used in Lemma~\ref{lemma: bdd linear operators} to prove that linear operators are bounded above, it follows that 
\begin{align*}
	B_2 =&  P\Big\{\Delta \tilde{\zeta}''_{\beta_j \gamma}(Y, X)[\gamma-\gamma_0] - \int^Y_0 \Big(\tilde{\zeta}''_{\beta_j \gamma}(t, X)[\gamma-\gamma_0] + (X_j+ \tilde{g}'(\tilde{\Lambda}(t, X))\tilde{\Lambda}'_{\beta_j}(t, X))\\
	& \ \ \ \ \ \ \ \ \ \ \ \ \ \ \ \ \ \ \ \ \ \ \ \cdot (\gamma(t)-\gamma_0(t)+\tilde{g}'(\tilde{\Lambda}(t, X))\tilde{\Lambda}'_{\gamma}(t, X)[\gamma-\gamma_0])  \Big) \dev \tilde{\Lambda}(t, X) \Big\}^2\\
	\lesssim & P\{\int^Y_0 (\tilde{\zeta}''_{\beta_j \gamma}(t, X)[\gamma-\gamma_0])^2\dev \Lambda_0(t, X)\} \\
	& + P\{ \int^Y_0(\gamma(t)-\gamma_0(t)+\tilde{g}'(\tilde{\Lambda}(t, X))\tilde{\Lambda}'_{\gamma}(t, X)[\gamma-\gamma_0])^2  \dev \tilde{\Lambda}(t, X)\}\\
	\lesssim & P\{\int^Y_0(\tilde{g}'(\tilde{\Lambda}(t, X))\tilde{\Lambda}''_{\beta_j \gamma}(t, X)[\gamma-\gamma_0])^2\dev \Lambda_0(t, X)\}\\
	& + P\{\int^Y_0(\tilde{g}''(\tilde{\Lambda}(t, X))\tilde{\Lambda}'_{\gamma}(t, X)[\gamma-\gamma_0]\tilde{\Lambda}'_{\beta_j}(t, X))^2\dev \Lambda_0(t, X)\}\\
	& + P\{ \int^Y_0(\gamma(t)-\gamma_0(t)+\tilde{g}'(\tilde{\Lambda}(t, X))\tilde{\Lambda}'_{\gamma}(t, X)[\gamma-\gamma_0])^2  \dev \tilde{\Lambda}(t, X)\}\\
	\lesssim & \|\gamma-\gamma_0\|_2^2 \leq \eta^2.
\end{align*}
For $B_2$, similarly, we can show that
\begin{align*}
	B_3  = & P\Big\{- \int^Y_0 \Big((\zeta'_{\beta_j}-\zeta'_{0\beta_j})(t, X) + (X_j+ \tilde{g}'(\tilde{\Lambda}(t, X))\tilde{\Lambda}'_{\beta_j}(t, X))(\zeta-\zeta_{0})(t, X)\dev \tilde{\Lambda}(t, X) \\
	& + \Delta (\zeta'_{\beta_j}-\zeta'_{0\beta_j})(Y, X) \Big\}^2 \\
	\lesssim & \|\zeta-\zeta_0\|_2^2 + \|\zeta'_{\beta_j}-\zeta'_{0\beta_j}\|_2^2 \leq \eta^2+ \|\zeta'_{\beta_j}-\zeta'_{0\beta_j}\|_2^2.
\end{align*}
Furthermore, by using the triangle inequality together with the boundedness of $\Lambda'_{\beta_j}$ and $g_0'$, it follows that 
\begin{align*}
	\|\zeta'_{\beta_j}-\zeta'_{0\beta_j}\|_2^2 &= \|g'(\Lambda(\cdot, \beta, \gamma, g))\Lambda'_{\beta_j}(\cdot, \beta, \gamma, g) - g'_0(\Lambda_0(\cdot))\Lambda'_{0\beta_j}(\cdot)\|_2^2\\
	& \leq \|g'(\Lambda(\cdot, \beta, \gamma, g))\Lambda'_{\beta_j}(\cdot, \beta, \gamma, g) - g'_0(\Lambda_0(\cdot))\Lambda'_{\beta_j}(\cdot, \beta, \gamma, g)\|_2^2 \\
	& \ \ +  \|g'_0(\Lambda_0(\cdot))\Lambda'_{\beta_j}(\cdot, \beta, \gamma, g) - g'_0(\Lambda_0(\cdot))\Lambda'_{0\beta_j}(\cdot)\|_2^2\\
	& \lesssim \|g'(\Lambda(\cdot, \beta, \gamma, g)) - g'_0(\Lambda_0(\cdot))\|_2^2 + \|\Lambda'_{\beta_j}(\cdot, \beta, \gamma, g) - \Lambda'_{0\beta_j}(\cdot)\|^2_2\\
	& \lesssim \|g'(\Lambda(\cdot, \beta, \gamma, g)) - g'_0(\Lambda_0(\cdot))\|_2^2 + d^2(\theta, \theta_0) \leq \eta^2.
\end{align*}
Therefore, we have $P\{l'_{\beta_j}(\theta;W) -l'_{\beta_j}(\theta_0;W)\}^2 \lesssim \eta^2$. By Lemma \ref{lemma: devivatives}, we also have $\|l'_{\beta_j}(\theta;W) -l'_{\beta_j}(\theta_0;W)\|_{\infty}$ is bounded. We choose $\eta_n = O(n^{-\min\{p_1\nu_1, (p_2-1)\nu_2, \frac{1-\max\{\nu_1, \nu_2\}}{2}\}})$. Then by the maximal inequality, it follows that 
\begin{align*}
		E_{P}\|\mathbb{G}_n\|_{\mc{F}_{n,j}^{*\beta}(\eta_n)}&  \lesssim (c_1 q_{n_1}+c_2q_{n_2})^{1/2}\eta_n + (c_1 q_{n_1}+c_2q_{n_2})n^{-1/2}\\
		& = O(n^{\frac{\max\{\nu_1, \nu_2\}}{2}})\cdot O(n^{-\min\{p_1\nu_1, (p_2-1)\nu_2, \frac{1-\max\{\nu_1, \nu_2\}}{2}\}}) + O(n^{\max\{\nu_1,\nu_2\}-1/2})\\
		& = O(n^{-\min\{p_1\nu_1-\frac{\max\{\nu_1, \nu_2\}}{2}, (p_2-1)\nu_2-\frac{\max\{\nu_1, \nu_2\}}{2}, 1/2-\max\{\nu_1, \nu_2\}\}}) + O(n^{\max\{\nu_1,\nu_2\}-1/2})\\
		& =o(1),
\end{align*}
where the last equality holds because $p_1\nu_1 \geq \nu_1 > \max\{\nu_1, \nu_2\} / 2$, $(p_2-1)\nu_2 \geq 2\nu_2 > \max\{\nu_1, \nu_2\} / 2$ for $p_2 \geq 3$, and $0<\nu_1,\nu_2<1/2$. Thus, for $\xi=\min\{p_1\nu_1, p_2\nu_2, \frac{1-\max\{\nu_1, \nu_2\}}{2}\}$ and $Cn^{-\xi} = O(n^{-\min\{p_1\nu_1, p_2\nu_2, \frac{1-\max\{\nu_1, \nu_2\}}{2}})$, by Markov's inequality, we have 
\[\sup_{d(\theta, \theta_0)\leq Cn^{-\xi}, \theta\in\Theta_n}|\mathbb{G}_n \{l'_{\beta_j}(\theta;W) -l'_{\beta_j}(\theta_0;W)\}|=o_p(1),\]
which completes the verification of the first equation in the assumption (\ref{m:equicontinuity}). The other two stochastic equicontinuity equations in (\ref{m:equicontinuity}) can be verified using the same arguments. 

Finally, we verify   assumption (\ref{m:smoothness}) using the Taylor expansion. Similarly, we just prove the first equation, since the proofs of the other two equations are based on the same arguments. By taking the Taylor expansion of $l'_{\beta}(\theta;W)$ at $\theta_0$, it follows that 
\begin{align*}
			l'_{\beta}(\theta;W) -l'_{\beta}(\theta_0;W) =&  l''_{\beta \beta}(\tilde{\theta};W)(\beta-\beta_0)+ l''_{\beta \gamma}(\tilde{\theta};W)[\gamma-\gamma_0] + l''_{\beta \zeta}(\tilde{\theta};W)[\zeta-\zeta_0] 
\end{align*}
where $\tilde{\theta} = (\tilde{\beta}, \tilde{\gamma}(\cdot), \tilde{\zeta}(\cdot, \tilde{\beta}, \tilde{\gamma}))$ is a point between $\theta$ and $\theta_0$. Thus,
\begin{align*}
	P\{& l'_{\beta}(\theta;W) -l'_{\beta}(\theta_0;W)-  l''_{\beta \beta}(\theta_0;W)(\beta-\beta_0)- l''_{\beta \gamma}(\theta_0;W)[\gamma-\gamma_0] - l''_{\beta \zeta}(\theta_0;W)[\zeta-\zeta_0]\}\\
	& = P\Big\{(l''_{\beta \beta}(\tilde{\theta};W)-l''_{\beta \beta}(\theta_0;W))(\beta-\beta_0)\Big\}+ P\Big\{ l''_{\beta \gamma}(\tilde{\theta};W)[\gamma-\gamma_0] -l''_{\beta \gamma}(\theta_0;W)[\gamma-\gamma_0] \Big\}\\
	& \ \ \ + P\Big\{ l''_{\beta \zeta}(\tilde{\theta};W)[\zeta-\zeta_0] - l''_{\beta \zeta}(\theta_0;W)[\zeta-\zeta_0]\Big\}
\end{align*}
After some direct calculation, we have
\begin{align*}
	& \Big|P\Big\{l''_{\beta \beta}(\tilde{\theta};W) -l''_{\beta \beta}(\theta_0;W)\Big\}\Big|  \\
	\leq &P\Big\{  \int^Y_0\Big| \left( \exp(X^T\beta_0 +\gamma_0(t)+\zeta_0(t, X))- \exp(X^T\tilde{\beta} +\tilde{\gamma}(t)+\tilde{\zeta}(t, X))\right) \tilde{\zeta}''_{\beta\beta}(t, X)\Big|\dev t\Big\}\\
	& + P\Big\{\Big|\int^Y_0  (X+\zeta'_{0\beta}(t, X))(X+\zeta'_{0\beta}(t, X))^T -(X+\tilde{\zeta}'_{\beta}(t, X))(X+\tilde{\zeta}'_{\beta}(t, X))^T \dev \Lambda_0(t, X)\Big|\Big\}\\
	& + P\Big\{  \int^Y_0  \Big|\left( \exp(X^T\beta_0 +\gamma_0(t)+\zeta_0(t, X))- \exp(X^T\tilde{\beta} +\tilde{\gamma}(t)+\tilde{\zeta}(t, X))\right) \\
	&\ \ \ \ \ \ \ \ \ \ \ \ \ \cdot (X+\tilde{\zeta}'_{\beta}(t, X))(X+\tilde{\zeta}'_{\beta}(t, X))^T\Big|\dev t\Big\}\\
	= & K_1+K_2+K_3.
\end{align*}
For $K_1$, by the mean value theorem and the Cauchy-Schwarz inequality, it follows that 
\begin{align*}
	K_1 & = P\Big\{  \int^Y_0\Big| \exp(\tilde{\psi}(t, X)) \left( X^T(\beta_0-\tilde{\beta}) +(\gamma_0-\tilde{\gamma})(t)+\zeta_0(t, X)-\tilde{\zeta}(t, X)\right) \tilde{\zeta}''_{\beta\beta}(t, X)\Big|\dev t\Big\}\\
	& \lesssim \|\beta_0-\tilde{\beta}\|+\|\gamma_0-\tilde{\gamma}\|_2+\|\zeta_0-\tilde{\zeta}\|_2 \leq d(\theta_0, \theta)\\
	& = O(n^{-\min\{p_1\nu_1, p_2\nu_2, \frac{1-\max\{\nu_1, \nu_2\}}{2}\}}),
\end{align*}
where $\tilde{\psi}(t, X)= X^T \beta_0+ \gamma_0(t)+\zeta_0(t, X) + \xi (X^T (\tilde{\beta}-\beta_0)+ \tilde{\gamma}(t)-\gamma_0(t)+\tilde{\zeta}(t, X)-\zeta_0(t, X))$ for some $\xi\in (0, 1)$ and is bounded. 
For $K_2$, by the Cauchy-Schwarz inequality and the same arguments that are used to verify assumption (\ref{m:gc}), we have 
\begin{align*}
	K_2 & \lesssim P\Big\{\int^\tau_0 \left|(\zeta'_{0\beta}(t, X)-\tilde{\zeta}'_{\beta}(t, X)) (X+\zeta'_{0\beta}(t, X)+\tilde{\zeta}'_{\beta}(t, X))^T \right|^2  \dev \Lambda_0(t, X)\Big\}^{1/2}\\
	& \lesssim \|\zeta'_{0\beta}(\cdot)-\tilde{\zeta}'_{\beta}(\cdot)\|_2\\
	& \lesssim d(\theta_0, \theta) + \|g'_0(\Lambda_0(\cdot))-g(\Lambda(\cdot, \beta, \gamma, g))\|_2\\
	& = O(n^{-\min\{p_1\nu_1, (p_2-1)\nu_2, \frac{1-\max\{\nu_1, \nu_2\}}{2}\}}).
\end{align*}
For $K_3$, by applying the same arguments for $K_1$, we can show that 
\begin{align*}
	K_3 \lesssim \|\beta_0-\tilde{\beta}\|+\|\gamma_0-\tilde{\gamma}\|_2+\|\zeta_0-\tilde{\zeta}\|_2 = O(n^{-\min\{p_1\nu_1, p_2\nu_2, \frac{1-\max\{\nu_1, \nu_2\}}{2}\}}).
\end{align*}
Therefore, 
\begin{align*}
	P\Big\{& \Big|(l''_{\beta \beta}(\tilde{\theta};W)-l''_{\beta \beta}(\theta_0;W))(\beta-\beta_0)\Big|\Big\}\\
	& = O(n^{-\min\{p_1\nu_1, (p_2-1)\nu_2, \frac{1-\max\{\nu_1, \nu_2\}}{2}\}})\cdot O(n^{-\min\{p_1\nu_1, p_2\nu_2, \frac{1-\max\{\nu_1, \nu_2\}}{2}\}})\\
	& = O(n^{-\min\{2p_1\nu_1, p_1\nu_1+(p_2-1)\nu_2, (2p_2-1)\nu_2,\frac{1}{2} + p_1\nu_1-\frac{\max\{\nu_1, \nu_2\}}{2}, \frac{1}{2}+(p_2-1)\nu_2-\frac{\max\{\nu_1, \nu_2\}}{2}, 1-\max\{\nu_1, \nu_2\} \}})\\
	& = o(n^{-1/2}),
\end{align*}
where the last equality holds because $p_1\geq 2$ and $p_2\geq 3$, thus $2p_1\nu_1 > p_1/(p_1+2)\geq 1/2$, 
$p_1\nu_1+(p_2-1)\nu_2 >\frac{p_1}{2(p_1+2)}+\frac{p_2-1}{2(p_2+1)} \geq \frac{1}{2\cdot 2}+\frac{1}{2\cdot 2} = \frac{1}{2}$, 
$(2p_2-1)\nu_2 > \frac{2p_2-1}{2(p_2+1)}>\frac{1}{2}$, 
$ p_1\nu_1\geq 2\nu_1>\frac{\max\{\nu_1, \nu_2\}}{2}$, 
$(p_2-1)\nu_2 >\nu_2>\frac{\max\{\nu_1, \nu_2\}}{2}$, and $\max\{\nu_1, \nu_2\}<1/2$. 
Similarly, we can show that 
\begin{align*}
	P\Big\{& \Big|l''_{\beta \gamma}(\tilde{\theta};W)[\gamma-\gamma_0] -l''_{\beta \gamma}(\theta_0;W)[\gamma-\gamma_0]\Big|\Big\}\\
	& = O(n^{-\min\{2p_1\nu_1, p_1\nu_1+(p_2-1)\nu_2, (2p_2-1)\nu_2,\frac{1}{2} + p_1\nu_1-\frac{\max\{\nu_1, \nu_2\}}{2}, \frac{1}{2}+(p_2-1)\nu_2-\frac{\max\{\nu_1, \nu_2\}}{2}, 1-\max\{\nu_1, \nu_2\} \}})\\
	& = o(n^{-1/2})
\end{align*}
and 
\begin{align*}
	P\Big\{& \Big|l''_{\beta \zeta}(\tilde{\theta};W)[\zeta-\zeta_0] - l''_{\beta \zeta}(\theta_0;W)[\zeta-\zeta_0]\Big|\Big\}\\
	& = O(n^{-\min\{2p_1\nu_1, p_1\nu_1+(p_2-1)\nu_2, (2p_2-1)\nu_2,\frac{1}{2} + p_1\nu_1-\frac{\max\{\nu_1, \nu_2\}}{2}, \frac{1}{2}+(p_2-1)\nu_2-\frac{\max\{\nu_1, \nu_2\}}{2}, 1-\max\{\nu_1, \nu_2\} \}})\\
	& = o(n^{-1/2}).
\end{align*}
Thus, it follows that
\begin{align*}
	&P\{ l'_{\beta}(\theta;W) -l'_{\beta}(\theta_0;W)-  l''_{\beta \beta}(\theta_0;W)(\beta-\beta_0)- l''_{\beta \gamma}(\theta_0;W)[\gamma-\gamma_0] - l''_{\beta \zeta}(\theta_0;W)[\zeta-\zeta_0]\}\\
	& = O(n^{-\min\{2p_1\nu_1, p_1\nu_1+(p_2-1)\nu_2, (2p_2-1)\nu_2,\frac{1}{2} + p_1\nu_1-\frac{\max\{\nu_1, \nu_2\}}{2}, \frac{1}{2}+(p_2-1)\nu_2-\frac{\max\{\nu_1, \nu_2\}}{2}, 1-\max\{\nu_1, \nu_2\} \}})= O(n^{-\alpha \xi})
\end{align*}
where $\alpha= \min\{2p_1\nu_1, p_1\nu_1+(p_2-1)\nu_2, (2p_2-1)\nu_2,\frac{1}{2} + p_1\nu_1-\frac{\max\{\nu_1, \nu_2\}}{2}, \frac{1}{2}+(p_2-1)\nu_2-\frac{\max\{\nu_1, \nu_2\}}{2}, 1-\max\{\nu_1, \nu_2\} \} / \min\{p_1\nu_1, p_2\nu_2, \frac{1-\max\{\nu_1, \nu_2\}}{2}\}>1 $ and $\alpha\xi>1/2$. This completes the verification of (\ref{m:smoothness}).

Therefore, we have verified (\ref{m:rate})-(\ref{m:smoothness}) and by Theorem \ref{thm: m-theorem}, we have 
\[\sqrt{n}(\hat{\beta}_n - \beta_0)  = A^{-1}\sqrt{n}\mathbb{P}_n \boldsymbol{l}^*(\beta_0, \gamma_0, \zeta_0;W) + o_p(1)\rightarrow_d N(0, A^{-1}B( A^{-1})^T),\]
where $\boldsymbol{l}^*(\beta_0, \gamma_0, \zeta_0;W) = l'_{\beta}(\beta_0, \gamma_0, \zeta_0;W) -l'_{\gamma}(\beta_0, \gamma_0, \zeta_0;W)[\mb{v}^*] - l'_{\zeta}(\beta_0, \gamma_0, \zeta_0;W)[\mb{h}^*(\cdot, \beta_0, \gamma_0)]$ and $A$ is given by $P \{\boldsymbol{l}^*(\beta_0, \gamma_0, \zeta_0;W)^{\otimes 2}\}=I(\beta_0)$, as shown in the above verification of   (\ref{m:positive info}). Thus, $A=B=I(\beta_0)$ and $A^{-1}B( A^{-1})^T = I^{-1}(\beta_0)$. Therefore, we have
\[\sqrt{n}(\hat{\beta}_n - \beta_0)  = \sqrt{n}I^{-1}(\beta_0)\mathbb{P}_n \boldsymbol{l}^*(\beta_0, \gamma_0, \zeta_0;W) + o_p(1)\rightarrow_d N(0, I^{-1}(\beta_0)),\]
which completes the proof.
\end{proof}

\subsection{Explanation of Condition \ref{c: pos info exist}}
\label{appendix subs condition exp}

Condition \ref{c: pos info exist} assumes the existence of the least favorable directions which are essential for semi-parametric efficiency. We may find $\mb{v}^*$ and $\mb{w}^*$ through equations in \ref{c: pos info exist}. Specifically, $\mb{v}^*$ and $\mb{w}^*$ need to satisfy   $P\{\Delta \mb{A}^*(U, X)\psi'_{0\gamma}(Y,X)[v]\}=0$ and $P\{\Delta \mb{A}^*(U, X)\psi'_{0g}(Y,X)[w]\}=0$ 
	  for any $v\in \Gamma^{p_1}$ and $w\in \mc{G}^{p_2}$.

For the first equation, using the fact of $P\{\int^Y_0 f(t, X)\dev \Lambda_0(t, X)\}=P\{\Delta f(Y, X)\}$ and the equations in (\ref{eq:grad psi gamma}), we have for any $v\in \Gamma^{p_1}$
\begin{align}
	 & P\{\Delta \mb{A}^*(U, X)\psi'_{0\gamma}(Y,X)[v]\}\nonumber\\
	 = & P\{\int^{Y}_0 \mb{A}^*(R(t)e^{X^T\beta_0}, X)\nonumber\\
	& \ \ \ \ \ \ \ \ \ \ \cdot \left(g_0'(\Lambda_0(t, X))\exp(g_0(\Lambda_0(t, X)))e^{X^T \beta_0} \int^t_0 \exp(\gamma_0(s)) v(s)\dev s +v(t) \right)\dev \Lambda_0(t, X)\}\nonumber\\
	=&  P\{\int^{R(Y)e^{X^T\beta_0}}_0\exp(g_0(\tilde{\Lambda}_0(t)))\mb{A}^*(t, X)\nonumber\\
	& \ \ \ \ \ \ \ \ \ \ \cdot \left(g_0'(\tilde{\Lambda}_0(t))\exp(g_0(\tilde{\Lambda}_0(t)))\int^t_0v(R^{-1}(e^{-X^T\beta_0} s))\dev s +v(R^{-1}(e^{-X^T\beta_0} t) \right)\dev t\}\nonumber\\
	=&  P\{\int^{R(Y)e^{X^T\beta_0}}_0 v(R^{-1}(e^{-X^T\beta_0} s))\int^{R(Y)e^{X^T\beta_0}}_s g_0'(\tilde{\Lambda}_0(t))\exp(2g_0(\tilde{\Lambda}_0(t)))\mb{A}^*(t, X)\dev t\dev s \nonumber\\
	& \ \ \ \ \ + \int^{R(Y)e^{X^T\beta_0}}_0 v(R^{-1}(e^{-X^T\beta} s)) \exp(g_0(\tilde{\Lambda}_0(s)))\mb{A}^*(s, X)\dev s \}\nonumber\\
	=&  P\{\int^{\infty}_0 \mathbbm{1}(R(Y) \geq s) \cdot v(R^{-1}(s))  \cdot e^{X^T\beta_0}\nonumber\\
	& \ \ \ \ \ \cdot \left( \int^{R(Y)e^{X^T\beta_0}}_{s e^{X^T\beta_0}} g_0'(\tilde{\Lambda}_0(t))\exp(2g_0(\tilde{\Lambda}_0(t)))\mb{A}^*(t, X)\dev t + \exp(g_0(\tilde{\Lambda}_0(s e^{X^T\beta_0})))\mb{A}^* (s e^{X^T\beta_0}, X) \right)\dev s\}\nonumber\\
	=&  \int^{\infty}_0   v(R^{-1}(s)) \cdot  P\Big\{ \mathbbm{1}(R(Y) \geq s) \cdot e^{X^T\beta_0}\nonumber\\
	& \ \ \ \ \ \cdot \left( \int^{R(Y)e^{X^T\beta_0}}_{s e^{X^T\beta_0}} g_0'(\tilde{\Lambda}_0(t))\exp(2g_0(\tilde{\Lambda}_0(t)))\mb{A}^*(t, X)\dev t + \exp(g_0(\tilde{\Lambda}_0(s e^{X^T\beta_0})))\mb{A}^* (s e^{X^T\beta_0}, X) \right)\Big\}\dev s,\label{eq: gamma req}
\end{align}
where the second equality is obtained by the variable transformation $\tilde{t}=R(t)x^{X^T\beta_0}$ and further replacing the notation  $\tilde{t}$ with $t$ in the integral, and the third equality holds by switching the order of integration. To make the equation (\ref{eq: gamma req})   equal to zero for any $v\in \Gamma^{p_1}$, we can take $\mb{v}^*$ and $\mb{w}^*$ satisfying
\begin{align}
\label{pos_info_1}
	& P\{\int^{R(Y)e^{X^T\beta_0}}_{s e^{X^T\beta_0}} g_0'(\tilde{\Lambda}_0(t))\exp(2g_0(\tilde{\Lambda}_0(t)))\mb{A}^*(t, X)e^{X^T\beta_0}\dev t\}\nonumber\\
	& \ \ \ \ \ \ \ \ =- P\{\mathbbm{1}(R(Y)\geq s) \exp(g_0(\tilde{\Lambda}_0(s e^{X^T\beta_0})))\mb{A}^* (s e^{X^T\beta_0}, X)e^{X^T\beta_0} \}.
\end{align}

For the second equation in \ref{c: pos info exist}, similarly, we have 
\begin{align*}
	& P\{\Delta \mb{A}^*(U, X)\psi'_{0g}(Y,X)[w]\}\nonumber\\
	 = & P\{\int^{Y}_0 \mb{A}^*(R(t)e^{X^T\beta_0}, X)\nonumber\\
	& \ \ \ \ \ \ \ \ \ \ \cdot \left(g_0'(\Lambda_0(t, X))\exp(g_0(\Lambda_0(t, X))) \int^{\Lambda_0(t, X)}_0 \exp(-g_0(s)) w(s)\dev s +w(\Lambda_0(t, X)) \right)\dev \Lambda_0(t, X)\}\nonumber\\
	=& P\{\int^{R(Y)e^{X^T\beta_0}}_0\exp(g_0(\tilde{\Lambda}_0(t)))\mb{A}^*(t, X)\nonumber\\
	& \ \ \ \ \ \ \ \ \ \ \cdot \left(g_0'(\tilde{\Lambda}_0(t))\exp(g_0(\tilde{\Lambda}_0(t)))\int^t_0 w(\tilde{\Lambda}_0(s))\dev s +w(\tilde{\Lambda}_0(t)) \right)\dev t\}\nonumber\\
	=& P\{\int^{U}_0 \left( \int^{U}_s g_0'(\tilde{\Lambda}_0(t))\exp(2g_0(\tilde{\Lambda}_0(t)))\mb{A}^*(t, X)\dev t+ \exp(g_0(\tilde{\Lambda}_0(\eta)))\mb{A}^*(t, X)\right) \cdot w(\tilde{\Lambda}_0(s))\dev s \}\nonumber\\
	=& \int^{\infty}_0 w(\tilde{\Lambda}_0(s)) \cdot P\{\int^{U}_s g_0'(\tilde{\Lambda}_0(t))\exp(2g_0(\tilde{\Lambda}_0(t)))\mb{A}^*(t, X)\dev t + \mathbbm{1}(U\geq s)\exp(g_0(\tilde{\Lambda}_0(s)))\mb{A}^*(s, X) \} \dev s.
\end{align*}
To make it equal to zero for any $w\in \mc{G}^{p_2}$, we can take $\mb{v}^*$ and $\mb{w}^*$ such that, for any $\eta$, $\mb{A}^*(t, x)$ satisfies
\begin{align}
\label{pos_info_2}
	& \int^{\infty}_s P\{\mathbbm{1}(U\geq t)\mb{A}^*(t, X)\}g_0'(\tilde{\Lambda}_0(t))\exp(2g_0(\tilde{\Lambda}_0(t)))\dev t \nonumber\\
	&\ \ \ \ \ \ \ \ = - \exp(g_0(\tilde{\Lambda}_0(s)))P\{\mathbbm{1}(U\geq s)\mb{A}^*(s, X)\}.
\end{align}
By taking derivatives with respect to $s$ on both sides, we have 
\[\exp(g_0(\tilde{\Lambda}_0(s))) \frac{\dev P\{\mathbbm{1}(U\geq s)\mb{A}^*(s, X)\}}{\dev s}=0,\]
which implies that $P\{\mathbbm{1}(U\geq s)\mb{A}^*(s, X)\}$ is a constant. Then equation (\ref{pos_info_2}) holds only if 
\begin{equation}
\label{pos_info_22}
	P\{\mathbbm{1}(U\geq s)\mb{A}^*(s, X)\}=0.
\end{equation}
Therefore, we can take $\mb{v}^*$ and $\mb{w}^*$ such that $\mb{A}^*(t, x)$ satisfies equations (\ref{pos_info_1}) and (\ref{pos_info_22}).

Next, we provide solutions for the Cox model and the linear transformation model with a known transformation function as illustration. 

For the Cox model where $g_0\equiv 0$, it suffices to find $\mb{v}^*$ such that the equation in (\ref{pos_info_1}) holds with $\mb{A}^*(t, x) = - x+\mb{v}^*(R^{-1}(te^{-x^T\beta_0}))$, which implies that $P\{\mathbbm{1}(R(Y)\geq t)e^{X^T\beta_0}(\mb{v}^*(R^{-1}(t))-X)\}=0$. We can take \[\mb{v}^*(t) = \frac{P\{\mathbbm{1}(Y\geq t)e^{X^T\beta_0}X\}}{P\{\mathbbm{1}(Y\geq t)e^{X^T\beta_0}\}}.\]

For the linear transformation model where $\gamma_0$ is known, it suffices to find $\mb{w}^*$ such that the equation in (\ref{pos_info_22}) holds with 
\begin{align*}
		\mb{A}^*(t, x)=& -(g_0'(\tilde{\Lambda}_{0}(t))\exp(g_0(\tilde{\Lambda}_{0}(t)))t +1)x\\
		& + g_0'(\tilde{\Lambda}_{0}(t)\exp(g_0(\tilde{\Lambda}_{0}(t)))\int^{\tilde{\Lambda}_{0}(t)}_0 \exp(-g_0(s))\mb{w}^*(s)\dev s + \mb{w}^*(\tilde{\Lambda}_{0}(t)).
\end{align*}
It follows that $\mb{w}^*$ satisfies 
\begin{align*}
		&g_0'(\tilde{\Lambda}_{0}(t)\exp(g_0(\tilde{\Lambda}_{0}(t)))\int^{\tilde{\Lambda}_{0}(t)}_0 \exp(-g_0(s))\mb{w}^*(s)\dev s + \mb{w}^*(\tilde{\Lambda}_{0}(t))\\
		& \ \ \ \ \ =(g_0'(\tilde{\Lambda}_{0}(t))\exp(g_0(\tilde{\Lambda}_{0}(t)))t +1)\frac{P\{\mathbbm{1}(U\geq t)X\}}{P\{\mathbbm{1}(U\geq t)\}}.
\end{align*}
By taking the variable transformation $\tilde{t}=\tilde{\Lambda}_0(t)$ and further replacing $\tilde{t}$ with $t$, it is sufficient to take $\mb{w}^*$ such that $g'_0(t)\exp(g_0(t))\int^t_0 \exp(-g_0(s))\mb{w}^*(s)\dev s +\mb{w}^*(t)=\pmb{\phi}(t)$ where $\pmb{\phi}(t)$ is given by
\[  \pmb{\phi}(t) = \left(g_0'(t)\exp(g_0(t))\tilde{\Lambda}_0^{-1}(t) +1\right) \frac{P\{\mathbbm{1}(\Lambda_0(Y,X)\geq t)X\}}{P\{\mathbbm{1}(\Lambda_0(Y,X)\geq t)\}}.\]
It is straightforward to verify that $\mb{w}^*$ can be taken as $\mb{w}^*(t) = \pmb{\phi}(t) - g_0'(t)\int^{t}_0\pmb{\phi}(s)\dev s$.

\subsection{\new{Simplification of Condition \ref{c: pos fish}}}
\label{appendix subs info condition exp}

\new{Condition \ref{c: pos fish} assumes non-singularity assumption of the information matrix. We may simplify it to some sufficient conditions if we can find the least favorable directions required in the condition (C7). Recall that we have provided explicit constructions of the least favorable directions for the Cox model and for the linear transformation model with a known transformation respectively in Section~\ref{appendix subs condition exp}. We further reduce the non-singularity assumption for the above two cases as follows.
}

\new{For the Cox model, we have $g_0\equiv 0$, $\tilde{\Lambda}_0(t)\equiv t$, and the least favorable function $\mb{v}^*$ can be derived as 
			\[\mb{v}^*(t) = \frac{P\{\mathbbm{1}(Y\geq t)e^{X^T\beta_0}X\}}{P\{\mathbbm{1}(Y\geq t)e^{X^T\beta_0}\}}.\]
			It follows that the efficient score for $\beta$ is 
			\[\boldsymbol{l}^*(\beta_0, \gamma_0;W) = \int_0^\infty \mb{A}^*(t, x)\dev M(t)=\int_0^\infty [-X+ \frac{P\{\mathbbm{1}(U\geq t)e^{X^T\beta_0}X\}}{P\{\mathbbm{1}(U\geq t)e^{X^T\beta_0}\}}]\dev M(t),\]
			where $U=e^{X^T\beta_0}\int^Y_0\exp(\gamma_0(s))\dev s $ as defined in (C5) and $M(t)=\Delta \mathbbm{1}(U \leq t) - \int_{0}^t \mathbbm{1}(U \geq s)\dev s$ is the event counting process martingale. Let $\boldsymbol\mu(t) = \frac{P\{\mathbbm{1}(U\geq t)e^{X^T\beta_0}X\}}{P\{\mathbbm{1}(U\geq t)e^{X^T\beta_0}\}}$. Then by the property of martingale, the information matrix is given by
			\begin{align*}
				I(\beta_0) &= P(\boldsymbol{l}^*(\beta_0, \gamma_0;W)^{\otimes 2})= P\left( \int_0^\infty \left[-X+ \boldsymbol\mu(t)\right]^{\otimes 2} \mathbbm{1}(U \geq t)\dev t\right)\\
				& = \int_0^\infty P\left( \left[-X+ \boldsymbol\mu(t)\right]^{\otimes 2} \mathbbm{1}(U \geq t)\right) \dev t,
			\end{align*}
			which reduces to the same information matrix of the MPLE for the Cox model. The above information matrix is similarly assumed to be positive definite in \citet[page 175]{kalbfleisch2011statistical}.
			The non-singularity condition in (C8) can be satisfied if $P\left( \left[-X+ \boldsymbol\mu(t)\right]^{\otimes 2} \mathbbm{1}(U \geq t)\right)$ is positive definite over a set of $t$ with non-zero measure.
}

\new{For the linear transformation model with a known transformation, i.e. $\gamma_0$ is known, given the least favorable direction $\mb{w}^*$ in Remark 8, the efficient score for $\beta$ is 
			\[\boldsymbol{l}^*(\beta_0, \zeta_0(\cdot, \beta_0);W) = \int_0^\infty m(t)\left[P(X|U\geq t)-X\right]\dev M(t),\]
			with $m(t)=g_0'(\tilde{\Lambda}_{0}(t))\exp(g_0(\tilde{\Lambda}_{0}(t)))t +1$, and the information matrix is 
			\begin{align*}
				I(\beta_0) & = P\left(\int_0^\infty m(t)^2\left[P(X|U\geq t)-X\right]^{\otimes 2}\mathbbm{1}(U \geq t) \ \dev \tilde{\Lambda}_{0}(t)\right)\\
				& = \int_0^\infty m^2(t) \cdot P\left(\left[P(X|U\geq t)-X\right]^{\otimes 2} \mathbbm{1}(U \geq t)\right)\cdot \exp(g_0(\tilde{\Lambda}_{0}(t)))\dev t\\
				& = \int_0^\infty m^2(t) \cdot Var(X|U\geq t) \cdot P(U \geq t)\cdot \exp(g_0(\tilde{\Lambda}_{0}(t)))\dev t.
			\end{align*}
			The information matrix takes a similar form as that in \citet{ding2011}, where it is assumed to be positive definite. Here we further investigate some sufficient conditions for its non-singularity.  
			The condition (C8) can be satisfied if $m^2(t) \cdot Var(X|U\geq t) \cdot P(U \geq t)$ is positive definite over a set of $t$ with non-zero measure. In particular, when the event time follows the AFT model with a Weibull error, i.e., $\gamma_0 \equiv 0$ and $\tilde{\Lambda}_0(t)=kt^v$, the information matrix becomes
			\[I(\beta_0) = \int_0^\infty v^2 \cdot Var(X|Ce^{X^T\beta_0}\geq t) \cdot P(Ce^{X^T\beta_0} \geq t) \ \dev F_0(t),\]
			where $F_0(t)=1-\exp(-kt^v)$ and $C$ is the censoring time. This information matrix is nonsingular if the conditional variance $Var(X|Ce^{X^T\beta_0}\geq t)$ is positive definite for~$t$ over certain interval.
}
\section{Proof of Propositions~\ref{identi} and \ref{degeneration}}
\label{s: iden}
The proof of Proposition~\ref{identi} is based on the existing identifiability conditions for the linear transformation model \citep{horowitz1996semiparametric} when both the transformation function and the error distribution are unknown. 

\begin{proof}[Proof of Proposition~\ref{identi}]
Suppose two groups of parameters $(q_i(\cdot), \beta_i, \alpha_i(\cdot))$ for $i=1, 2$ give the same survival distribution. Let $H_i(u) = \int_0^{-\ln u} {q_i^{-1}(v)}\dev v$, $ G_i(u) = H_i^{-1}(u)$, and $\varphi_i(t) = \log \int_0^t \alpha_i(s)\dev s$ for $i=1, 2$. In the equivalent linear regression representation, we have that $\varphi_i(T)=-x^T\beta_i +\epsilon_i$ specifies the same distribution of event time $T$ for $i=1, 2$, where the survival function of $\exp(\epsilon_i)$ is given by~$G_i$. 
\newnew{Note that, for the linear transformation model $\varphi(T)=-x^\top \beta +\epsilon$ with both $\varphi$ and the distribution of $\epsilon$ unspecified, \citet[page 105]{horowitz1996semiparametric} stated that the model parameters are identifiable up to a scale and a location normalization when at least one of the covariates $x$ has a non-zero $\beta$ coefficient and the conditional probability distribution of this covariate given the remaining covariates is absolutely continuous with respect to Lebesgue measure. Since we assume that there is at least one of the covariates in $x$ is continuous and this covariate has a non-zero coefficient, following the identifiability conditions stated in \citet{horowitz1996semiparametric}, there exist constants $c_1>0$ and $c_2$ such that $\beta_1 = c_1 \beta_2$, $\varphi_1(t)=c_1 \varphi_2(t)+c_2$ for any $t>0$, and $\epsilon_1$ has the same distribution as $c_1\epsilon_2 +c_2$,~i.e., 
\[G_1(t)=Pr(\exp(\epsilon_1)>t)=Pr(\exp(c_1 \epsilon_2 +c_2)>t)=Pr(\exp(\epsilon_2) > (t e^{-c_2})^{1/c_1}))=G_2((t e^{-c_2})^{1/c_1}).\]}
After plugging the definitions of $\varphi_i$ along with some calculations, we have for any $t>0$
\[\int_0^t \alpha_1(s)\dev s = e^{c_2} \left(\int_0^t \alpha_2(s)\dev s\right)^{c_1}.\]
Let $\exp(-s)= G_1(t)=G_2((t e^{-c_2})^{1/c_1})$. Then by the definitions of $G_i$ we have
\[t=H_1(\exp(-s))= \int_0^{s} {q_1^{-1}(v)}\dev v \ \text{ and } \ (t e^{-c_2})^{1/c_1} = H_2(\exp(-s))= \int_0^{s} {q_2^{-1}(v)}\dev v.\]
It follows that $\int_0^{s} {q_1^{-1}(v)}\dev v = e^{c_2} \left(\int_0^{s} {q_2^{-1}(v)}\dev v \right)^{c_1}$ for any $s>0$, which completes the proof.
\end{proof}

As a direct result of Proposition~\ref{identi}, Proposition \ref{degeneration} provides the necessary and sufficient degeneration condition for AFT and Cox models. 

\begin{proof}[Proof of Proposition~\ref{degeneration}]
The linear transformation model in~(\ref{transformation ode}) coincides with the Cox model if and only if there exists some positive function $\tilde{\alpha}$ such that parameters $(1, \tilde{\beta}, \tilde{\alpha}(\cdot))$ and $(q(\cdot), \beta, \alpha(\cdot))$ give the same survival distribution. By Proposition~\ref{identi}, there exists positive constants $c_1$ and $c_2$ such that \[\int_0^{t}q^{-1}(s)\dev s =c_2 t^{c_1}, \beta=c_1 \tilde{\beta}, \text{ and } \int_0^t \alpha(s)\dev s = c_2 \left(\int_0^t \tilde{\alpha}(s)\dev s\right)^{c_1}.\] It implies that the function $q$ satisfies $q(t)=\frac{1}{c_1c_2}t^{1-c_1}$. Similarly, when the linear transformation model coincides with the AFT model, there exists some positive function $\tilde{q}$ such that parameters $(\tilde{q}(\cdot), \tilde{\beta}, 1)$ and $(q(\cdot), \beta, \alpha(\cdot))$ give the same survival distribution. 
By Proposition~\ref{identi}, there exists positive constants $c_1$ and $c_2$ such that \[\int_0^{t}q^{-1}(s)\dev s =c_2 \left(\int_0^t \tilde{q}(s)\dev s\right)^{c_1}, \beta=c_1 \tilde{\beta}, \text{ and } \int_0^t \alpha(s)\dev s = c_2 t^{c_1}.\]
It follows that the function $\alpha$ takes the form $\alpha(t)=c_1c_2 t^{c_1-1}$, which completes the proof.
\end{proof}

\section{\new{Theoretical Properties for the General Class of ODE Models and Their Proofs}}
\label{s: extend thm}
\new{In this section, we further establish the convergence rate and the asymptotic normality of the proposed sieve estimator for the general class of ODE models in the presence of covariates $Z$ with time-varying coefficients. We reformulate the model to ensure the positivity of $\alpha(\cdot)$ and $q(\cdot)$ in (\ref{model: ltm + cox}) below,
\begin{equation}
\label{ode: log general model}
\left\{
\begin{array}{lr}
\Lambda'(t) = \exp(x^T \beta + \gamma(t) + z^T \boldsymbol{\eta}(t) +g(\Lambda(t))) \\
\Lambda(0) =0
\end{array}
\right.,
\end{equation}
where $\gamma(\cdot) = \log \alpha(\cdot)$ and $g(\cdot) = \log q(\cdot)$. Recall that, when there is at least one non-zero time-varying effect, i.e.,  $\boldsymbol{\eta}(t)\neq 0$, two groups of parameters $(\beta, \gamma, g, \boldsymbol{\eta})$ and $(\tilde{\beta}, \tilde{\gamma}, \tilde{g}, \tilde{\boldsymbol{\eta}})$ give the same survival distribution if only if $\beta = \tilde{\beta}$, $\gamma=\tilde{\gamma}+c$, $g=\tilde{g}-c$, and $\boldsymbol{\eta}=\tilde{\boldsymbol{\eta}}$ for some constant $c$. To guarantee the identifiability, we constrain $\gamma(t^*)=0$ with some fixed time point $t^*$.
}

\new{
Before stating the regularity conditions and main theorems, we firstly update the notation to make them consistent with the model in~(\ref{ode: log general model}). Let $Z \in \mb{R}^{d_2+1}$ substitute $(1, Z^T)^T$ and $\boldsymbol{\gamma}(\cdot)$ substitute $ (\gamma(\cdot), \eta_1(\cdot), \dots, \eta_{d_2}(\cdot))^T$ for notational simplicity, then the general class of ODE models is equivalent to 
\begin{equation}
\label{ode: log general model + replace}
\left\{
\begin{array}{lr}
\Lambda'(t) = \exp(x^T \beta + z^T \boldsymbol{\gamma}(t) +g(\Lambda(t))) \\
\Lambda(0) =0
\end{array}
\right.,
\end{equation}
with the first component of $\boldsymbol{\gamma}$ fixed at the time point $t^*$, i.e., $\gamma_1(t^*)=c$.
We denote the solution of (\ref{ode: log general model + replace}) by $\Lambda(t, x, z, \beta, \boldsymbol\gamma, g)$ and the true parameters associated with the data generating distribution by $(\beta_0, \gamma_0, g_0)$ and simplify $\Lambda(t, x, z,\beta_0, \boldsymbol\gamma_0, g_0)$ as $\Lambda_0(t,x,z)$.}

\new{To accommodate covariates $Z$ with time-varying coefficients, we update the conditions \ref{c: bounded b}-\ref{c: pos fish} to \ref{c:z bounded b}-\ref{c:z pos fish} with additional regularity conditions on covariates $Z$ and provide the theorem statements and the sketch of proof in the following subsections.}

\subsection{\new{Regularity conditions and main theorems}}
\new{We assume additional regularity conditions on $Z$ and list the updated conditions below.
\begin{enumerate}[label=(C\arabic*$'$), ref=(C\arabic*$'$)]
	\item\label{c:z bounded b} The true parameter $\beta_0$ is an interior point of a compact set $\mc{B}\subset \reals^{d_1}$.
	\item\label{c:z bounded x} The joint density of $X$ and $Z$ is bounded below by a constant $c>0$ over the compact domain $\mc{X}\times \mc{Z} \subset \reals^{d_1+d_2+1}$. $P (XX^T)$ and $P (ZZ^T)$ are nonsingular.
	\item\label{c:z bounded t} There exists a truncation time $\tau <\infty$ such that, for some positive constant $\delta_0$, $Pr(Y>\tau|X,Z)\geq \delta_0$ almost surely with respect to the joint probability measure of $X$ and $Z$. Then there is a constant $\mu=\sup_{x\in\mc{X}, z\in \mc{Z}}\Lambda_0(\tau, x, z) \leq - \log \delta_0 $ such that $\Lambda_0(\tau, X, Z)= -\log Pr(T>\tau|X, Z)\leq \mu$ almost surely with respect to the joint probability measure of $X$ and $Z$. 
	\item\label{c:z smoothness} Let $S^p([a, b])$ denote the collection of bounded functions $f$ on $[a, b]$ defined in \ref{c: smoothness}. The true function $\boldsymbol\gamma_0(\cdot)$ belongs to $\Gamma^{p_1}_{t^*}\times \underbrace{\Gamma^{p_1} \times \cdots \times \Gamma^{p_1}}_{d_2}$, where $\Gamma^{p_1}:= S^{p_1}([0, \tau])$ and $\Gamma^{p_1}_{t^*}:= \{\gamma \in S^{p_1}([0, \tau]): \gamma(t^*)=0\}$ with $p_1\geq 2$, and the true function $g_0(\cdot)$ belongs to $\mc{G}^{p_2}:=S^{p_2}([0, \mu+\delta_1])$ with some positive constant $\delta_1$ and $p_2 \geq 3$. 
	\item\label{c:z lower bound1} Denote $R_z(t)=\int^t_0\exp(z^T \boldsymbol\gamma_0(s))\dev s$, $V=X^T\beta_0$, and $U = e^{V}R_Z(Y)$. There exists $\eta_1\in (0,1)$ such that for all $u\in \reals^{d_1}$ with $\|u\|=1$, 
	\[u^T Var(X\mid U,V, Z)u\geq\eta_1 u^T P(XX^T\mid U,V,Z)u \ \ \text{ almost surely.}\]
	\item\label{c:z lower bound2} Let $\psi(t, x, z, \beta, \boldsymbol\gamma, g)= x^T\beta + z^T \boldsymbol\gamma(t)+g(\Lambda(t, x, z, \beta, \boldsymbol\gamma, g))$ and denote its functional derivatives with respect to the entirety $\bar{\gamma}(\cdot)=z^T \boldsymbol\gamma(\cdot)$ and $g(\cdot)$ along the direction $v(\cdot)$ and $w(\cdot)$ at the true parameter by $\psi'_{0\bar{\gamma}}(t,x,z)[v]$ and $\psi'_{0g}(t,x, z)[w]$ respectively, whose rigorous definitions are given by (\ref{eq:z grad psi gamma})-(\ref{eq:z grad psi g}). For any $\boldsymbol{v}(\cdot)=(v_1,\dots, v_{d_2+1})^T$ with $v_j\in \Gamma^{p_1}$, $1\le j \le d_2+1$, and $w(\cdot)\in \mc{G}^{p_2}$, there exists $ \eta_2 \in (0, 1)$ such that 
	\begin{align*}
		&(P\{\psi'_{0\bar{\gamma}}(Y,X,Z)[Z^T \boldsymbol{v}]\psi'_{0g}(Y,X,Z)[w]\,|\,\Delta=1\})^2\\
		&\ \ \ \ \ \ \leq  \eta_2 P\{(\psi'_{0\bar{\gamma}}(Y,X,Z)[Z^T \boldsymbol{v}])^2\,|\,\Delta=1\} P\{(\psi'_{0g}(Y,X,Z)[w])^2\,|\,\Delta=1\}
	\end{align*}
	almost surely.
	\item \label{c:z pos info exist}
	There exist $\mb{v}_j^*=(v^*_{j1}, \cdots, v^*_{jd_1})^T$ and $\mb{w}^*=(w^*_1, \cdots, w^*_{d_1})^T$, where $v^*_{jk} \in  \Gamma^{2}$ and $w^*_k \in \mc{G}^{2}$ for $1\le j \le d_2+1, 1\le k \le d_1$, such that \[P\{\Delta \mb{A}^*(U, X, Z)\psi'_{0\gamma_\ell}(Y,X,Z)[v]\}=0 \text{ and } P\{\Delta \mb{A}^*(U, X,Z)\psi'_{0g}(Y,X,Z)[w]\}=0\] 
	hold for any $v\in \Gamma^{p_1}$, $1\le \ell \le d_2+1$, and $w\in \mc{G}^{p_2}$. Here $\psi'_{0\gamma_\ell}(t,x,z)[v]$ denotes the functional derivative with respect to the $\ell$-th component of $\boldsymbol{\gamma}$ along the direction $v(\cdot)$ at the true parameter, $U$ and $V$ are defined in condition \ref{c:z lower bound1}, and
	\begin{align*}
		\mb{A}^*(t, X,Z)=& -\left(g_0'(\tilde{\Lambda}_{0}(t))\exp(g_0(\tilde{\Lambda}_{0}(t)))t +1\right)X\\
		& + \sum_{j=1}^{d_2+1} \left[g_0'(\tilde{\Lambda}_{0}(t))\exp(g_0(\tilde{\Lambda}_{0}(t))) \int^{t}_0 Z_j\mb{v}_j^*(R_Z^{-1}(se^{-V}))\dev s + Z_j\mb{v}_j^*(R_Z^{-1}(te^{-V}))\right]\\
		& + g_0'(\tilde{\Lambda}_{0}(t))\exp(g_0(\tilde{\Lambda}_{0}(t)))\int^{\tilde{\Lambda}_{0}(t)}_0 \exp(-g_0(s))\mb{w}^*(s)\dev s + \mb{w}^*(\tilde{\Lambda}_{0}(t)),
	\end{align*}
	where $\tilde{\Lambda}_0(t)$ is the solution of $\tilde{\Lambda}_0'(t)=\exp(g_0(\tilde{\Lambda}_0))$ with $\tilde{\Lambda}_0(0)=0$.
	\item\label{c:z pos fish} Let $\boldsymbol{l}^*(\beta_0, \boldsymbol\gamma_0, \zeta_0;W) = \int \mb{A}^*(t, X,Z)\dev M(t)$, where 
$M(t)=\Delta \mathbbm{1}(U \leq t) - \int_{0}^t \mathbbm{1}(U \geq s)\dev \tilde{\Lambda}_0(s)$ is the event counting process martingale. The information matrix $I(\beta_0) = P(\boldsymbol{l}^*(\beta_0, \boldsymbol\gamma_0, \zeta_0;W)^{\otimes 2})$ is nonsingular. Here for a vector $a$, $a^{\otimes 2} = aa^T$.
\end{enumerate}
}

\new{
In the presence of covariates $Z$ with time-varying coefficients, conditions \ref{c:z bounded x}-\ref{c:z bounded t} contain additional common regularity assumptions for $Z$ in survival analysis. Condition \ref{c:z smoothness} controls the error rates of the spline approximation for the true time-varying coefficients. The expectation in condition \ref{c:z lower bound1} is further conditioned on covariates $Z$. Condition \ref{c:z lower bound2} is similarly assumed to avoid strong collinearity between $\psi'_{0\bar{\gamma}}(Y,X,Z)[v]$ and $\psi'_{0g}(Y,X,Z)[w]$ while $\bar{\gamma}$ denotes the linear combination $z^T \boldsymbol{\gamma}$. Condition~\ref{c:z pos info exist} additionally requires the existence of the least favorable directions for time-varying coefficients and the information matrix in \ref{c:z pos fish} also depends on the additional least favorable directions. In particular, conditions \ref{c:z bounded b}-\ref{c:z pos fish} are equivalent to conditions \ref{c: bounded b}-\ref{c: pos fish} respectively when $Z$ only contains the intercept.}

\new{Given the above regularity conditions, for the general class of ODE models in (\ref{model: ltm + cox}), we can establish the same convergence rate of the sieve estimator as that in Theorem~\ref{thm: conv rate} and the asymptotic normality as in Theorem~\ref{thm: asym normal}. Since the theory is investigated with the fixed number of covariates $d_1$ and $d_2$ as the sample size $n$ grows, including additional covariates $Z$ with time-varying coefficients does not change the nature of the proof. For presentation integrity, we provide rigorous definitions of the corresponding parameter space, the sieve space, theorem statements, and a sketch of proof that summarizes the main steps in the following subsection.}

\new{First, we define the parameter space and the associated distance when including covariates $Z$ with time-varying coefficients.
We similarly define the collection of functions 
\begin{align*}
	\mc{H}^{p_2}=\{\zeta(\cdot, \beta, &\boldsymbol\gamma):  \zeta(t, x, z, \beta, \boldsymbol\gamma)= g(\Lambda(t, x, z, \beta, \boldsymbol\gamma, g)), t\in [0, \tau], x\in \mc{X}, z\in \mc{Z}, \beta\in \mc{B}, \\
	& \boldsymbol\gamma \in \Gamma^{p_1}_{t^*}\times \underbrace{\Gamma^{p_1} \times \cdots \times \Gamma^{p_1}}_{d_2},  g\in \mc{G}^{p_2} \text{ such that } \sup_{t\in[0,\tau],x\in\mc{X},z\in\mc{Z}}|\Lambda(t,x,z,\beta, \boldsymbol\gamma, g)| \leq \mu+\delta_1\},
\end{align*}
with $\delta_1$ given in condition \ref{c:z smoothness}. 
For any $\zeta(\cdot, \beta, \boldsymbol\gamma) \in \mc{H}^{p_2}$, we define its norm as \[\|\zeta(\cdot, \beta, \boldsymbol\gamma)\|_2 = \left[ \int_{\mc{X}\times \mc{Z}} \int_{0}^{\tau} [\zeta(t, x, z, \beta, \boldsymbol\gamma)]^2 \dev \Lambda_0(t, x,z) \dev F_{X,Z}(x,z)\right]^{1/2},\] where $F_{X,Z}(x,z)$ is the cumulative distribution function of $(X,Z)$. 
Denote the parameter $\theta = (\beta, \boldsymbol\gamma(\cdot), \zeta(\cdot, \beta, \boldsymbol\gamma))$ and the true parameter $\theta_0 = (\beta_0, \boldsymbol\gamma_0(\cdot), \zeta_0(\cdot, \beta_0, \boldsymbol\gamma_0))$ with \[\zeta_0(t,x, z,\beta_0, \boldsymbol\gamma_0) =g_0(\Lambda(t,x, z,\beta_0, \boldsymbol\gamma_0, g_0 )).\] 
Denote the parameter space by $\Theta=\mc{B} \times \Gamma^{p_1}_{t^*}\times \underbrace{\Gamma^{p_1} \times \cdots \times \Gamma^{p_1}}_{d_2}\times \mc{H}^{p_2}$. For any $\theta_1$ and $\theta_2$ in $\Theta$, we define the distance \[d(\theta_1, \theta_2)= \left( \|\beta_1-\beta_2\|^2+\|\boldsymbol\gamma_1-\boldsymbol\gamma_2\|_{2}^2+\|\zeta_1(\cdot, \beta_1, \boldsymbol\gamma_1) - \zeta_2(\cdot, \beta_2, \boldsymbol\gamma_2)\|_2^2 \right)^{1/2},\] where $\|\cdot\|$ is the Euclidean norm and $\|\boldsymbol\gamma\|_{2}=(\sum_{j=1}^{d_2+1}\int^\tau_0(\gamma_j(t))^2\dev t)^{1/2}$.} 

\new{Next, we construct the sieve space by using the space of polynomial splines in a similar way.   
Let $\Gamma^{p_1}_n= S_{n}(T_{K_{n}^1}, K_{n}^1, p_1)$, $\Gamma^{p_1}_{t^*,n}= \{\gamma \in S_{n}(T_{K_{n}^1}, K_{n}^1, p_1): \gamma(t^*)=0\}$, $\mc{G}^{p_2}_n =S_{n}(T_{K_{n}^2}, K_{n}^2, p_2)$, and 
\begin{align*}
	\mc{H}^{p_2}_n=\{\zeta(\cdot, \beta, \boldsymbol\gamma):  \zeta(t, x, z, \beta, \boldsymbol\gamma)= g(\Lambda(t, x, z, \beta, \boldsymbol\gamma, g)), & ~t\in [0, \tau], x\in \mc{X}, z\in \mc{Z}, \beta\in \mc{B}, \\
	& \boldsymbol\gamma \in \Gamma^{p_1}_{t^*,n}\times \underbrace{\Gamma^{p_1}_n \times \cdots \times \Gamma^{p_1}_n}_{d_2},  g\in \mc{G}^{p_2}_n \}.
\end{align*}
Let $\Theta_n=\mc{B} \times \Gamma^{p_1}_{t^*,n}\times \underbrace{\Gamma^{p_1}_n \times \cdots \times \Gamma^{p_1}_n}_{d_2} \times \mc{H}^{p_2}_n$ be the sieve space. The sieve estimator $\hat{\theta}_n = (\hat{\beta}_n, \hat{\boldsymbol\gamma}_n(\cdot), \hat{\zeta}_n(\cdot, \hat{\beta}_n, \hat{\boldsymbol\gamma}_n))$ maximizes the log-likelihood (\ref{obj}) over the sieve space $\Theta_n$. The convergence rate of the sieve MLE $\hat{\theta}_n$ and the asymptotic normality of the sieve MLE $\hat{\beta}_n$ of the regression parameter are then established in Theorem~\ref{thm:z conv rate} and Theorem~\ref{thm:z asym normal} respectively.}

\new{
\begin{theorem}(Convergence rate of $\hat{\theta}_n$.)
	\label{thm:z conv rate}
	Let $\nu_1$ and $\nu_2$ satisfy the restrictions $\max\{\frac{1}{2(2+p_1)},\frac{1}{2p_1}-\frac{\nu_2}{p_1} \}<\nu_1<\frac{1}{2p_1}$, $\max\{\frac{1}{2(1+p_2)},\frac{1}{2(p_2-1)}-\frac{2\nu_1}{p_2-1}\}<\nu_2<\frac{1}{2p_2}$,  and $2\min\{2\nu_1,\nu_2\} > \max\{\nu_1,\nu_2\}$. Suppose conditions \ref{c:z bounded b}-\ref{c:z lower bound2} hold, then we have
	\[d(\hat{\theta}_n, \theta_0)=O_p(n^{-\min\{p_1\nu_1, p_2\nu_2, \frac{1-\max\{\nu_1, \nu_2\}}{2}\}}).\] 
\end{theorem}
\begin{theorem}(Asymptotic normality of $\hat{\beta}_n$)
	\label{thm:z asym normal}	 Suppose the conditions in Theorem \ref{thm:z conv rate} and  \ref{c:z pos info exist}-\ref{c:z pos fish} hold, then we have
\begin{align*}
		\sqrt{n}(\hat{\beta}_n - \beta_0) & = \sqrt{n}I^{-1}(\beta_0)\mathbb{P}_n \boldsymbol{l}^*(\beta_0, \gamma_0, \zeta_0;W) + o_p(1) \rightarrow_d N(0, I^{-1}(\beta_0))
	\end{align*}
	with $I(\beta_0)$ given in condition \ref{c:z pos fish} and $\rightarrow_d$ denoting convergence in distribution.
\end{theorem}
}

\subsection{\new{Sketch of proof}}
\new{Given the updated conditions \ref{c:z bounded b}-\ref{c:z pos fish}, the proof of Theorems~\ref{thm:z conv rate} and~\ref{thm:z asym normal} is based on the similar techniques and arguments as that of Theorems~\ref{thm: conv rate} and~\ref{thm: asym normal}. We provide the sketch of proof and highlight their main differences below.}

\paragraph{Lemmas.} \new{The corresponding Lemmas~\ref{lemma: ode solutions}-\ref{lemma: other bracket numbers for A5} in the presence of covariates $Z$ still hold under new conditions \ref{c:z bounded b}-\ref{c:z pos info exist}, which are used to prove  Theorems~\ref{thm:z conv rate} and~\ref{thm:z asym normal}. Specifically,
\begin{itemize}
	\item The existence and uniqueness of the solution $\Lambda(t, x, z, \beta, \boldsymbol\gamma, g)$ of the initial value problem in (\ref{ode: log general model}) along with its derivatives in Lemma~\ref{lemma: ode solutions}, and the boundedness and continuity of derivatives of $l(\beta, \boldsymbol{\gamma}, \zeta;W)$ in Lemma~\ref{lemma: devivatives} both hold due to the boundedness of $Z$ and the smoothness of $\boldsymbol{\eta}$ under conditions \ref{c:z bounded b}-\ref{c:z smoothness}. In particular, the derivatives are characterized by the corresponding updated initial value problems with covariates $Z$. For example, initial value problems (\ref{eq: beta dev})-(\ref{eq: g dev}) become (\ref{eq:z beta dev})-(\ref{eq:z g dev}) respectively as follows
		\begin{align}
			\frac{\dev \Lambda'_{\beta}(t)}{\dev t} & = \exp(x^T\beta + z^T\boldsymbol\gamma(t)+g(\Lambda(t)))\{x+g'(\Lambda(t)) \Lambda'_{\beta}(t)\}, \ \ \Lambda'_{\beta}(0) = 0,\label{eq:z beta dev}\\
		\frac{\dev \Lambda'_{\gamma_j}(t)[v]}{\dev t} &= \exp(x^T\beta + z^T\boldsymbol\gamma(t)+g(\Lambda(t)))\{z_jv(t)+g'(\Lambda(t)) \Lambda'_{\gamma_j}(t)[v]\}, \ \ \Lambda'_{\gamma_j}(0)[v] = 0,\label{eq:z gamma dev}\\
		\frac{\dev \Lambda'_{g}(t)[w]}{\dev t} &= \exp(x^T\beta + z^T\boldsymbol\gamma(t)+g(\Lambda(t)))\{w(\Lambda(t))+g'(\Lambda(t)) \Lambda'_{g}(t)[w]\}, \ \ \Lambda'_{g}(0)[w] = 0.\label{eq:z g dev}
		\end{align}
	\item In Lemma~\ref{lemma: bdd linear operators}, we show that the operators $\psi_{0\bar{\gamma}}'[\cdot]$ and $\psi_{0g}'[\cdot]$ are bounded from below by the continuous dependence of the IVP solution on parameters in \citet[page 145]{Walter1998}, where $\psi_{0\bar{\gamma}}'[\cdot]$ denotes the functional derivatives with respect to the entirety $\bar{\gamma}(\cdot)=z^T \boldsymbol\gamma(\cdot)$. By solving initial value problem in (\ref{eq:z gamma dev}), the first derivatives of $\psi(t, x, \beta, \gamma, g)$ with respect to $\bar{\gamma}$ and $g$ at the true parameter $(\beta_0, \gamma_0, g_0)$ are updated  as
		\begin{align}
			\psi_{0\bar{\gamma}}'(t, x, z)[v] &= g'_0(\Lambda_0(t, x, z))\Lambda'_{0\bar{\gamma}}(t, x)[v]+v(t) \nonumber\\
			& = g'_0(\Lambda_0(t, x, z))\exp(g_0(\Lambda_{0}(t,x,z)))e^{x^T\beta_0}\int^{t}_0 \exp(z^T\boldsymbol\gamma_0(s))v(s)\dev s + v(t),\label{eq:z grad psi gamma}\\
			\psi_{0g}'(t, x, z)[w] &= g'_0(\Lambda_0(t, x, z))\Lambda'_{0g}(t, x, z)[w]+w(\Lambda_0(t, x, z)) \nonumber\\
			& = g'_0(\Lambda_0(t, x, z))\exp(g_0(\Lambda_{0}(t,x,z)))\int^{\Lambda_0(t, x,z)}_0 \exp(-g_0(s))w(s)\dev s + w(\Lambda_0(t, x,z)).\label{eq:z grad psi g}
		\end{align}
	\item The upper bounds of the $\epsilon$-bracketing numbers associated with $\mc{F}_n$, $\mc{F}_{n,j}^{\gamma_\ell}(\eta)$, $\mc{F}_{n,j}^\zeta(\eta)$, $\mc{F}_{n,j}^{*\beta}(\eta)$, $\mc{F}_{n,j}^{*\gamma_\ell}(\eta)$, $\mc{F}_{n,j}^{*\zeta}(\eta)$ for $1\le  \ell \le d_2+1, 1\le j \le d_1$  in Lemmas~\ref{lemma: bracket number of Fn}-\ref{lemma: other bracket numbers for A5} are updated as $(\frac{1}{\epsilon})^{c_1 q_{n_1} (d_2+1) + c_2 q_{n_2} + d_1}$ and $(\frac{\eta}{\epsilon})^{c_1 q_{n_1} (d_2+1) + c_2 q_{n_2} + d_1}$, where $d_1$ and $d_2$ are dimensions of covariates $X$ and $Z$ respectively. Since we consider the number of covariates $d_i$ fixed as the sample size increases, the updated upper bounds in the presence of $Z$ would not change the convergence rate of the sieve estimator and the nature of the proof.  
\end{itemize}
}

\paragraph{Proof of Theorem~\ref{thm:z conv rate}.} \new{To establish the overall convergence rate of the sieve MLE $\hat{\theta}_n$ in Theorem~\ref{thm:z conv rate}, we verify three conditions C1-C3 required in the main theorem in \citet{shen1994}. Specifically,
\begin{itemize}
	\item The condition C1 in \citet{shen1994} specifies the increasing rate of the expected log-likelihood ratio as the parameter $\theta$ moves away from the true value $\theta_0$. We will prove that
	\[\inf_{d(\theta, \theta_0)\geq \epsilon, \theta\in \Theta_n} Pl(\beta_0, \gamma_0, \zeta_0(\cdot, \beta_0, \gamma_0); W) - Pl(\beta, \gamma, \zeta(\cdot, \beta, \gamma); W) \gtrsim  \epsilon^2.\]
	In the presence of covariate $Z$, we update 
	\begin{align*}
	    Pl(\beta, & \boldsymbol\gamma, \zeta(\cdot, \beta, \boldsymbol\gamma) ; W) =   P\{\Delta [X^T \beta+ Z^T \boldsymbol\gamma(Y)+g(\Lambda(Y, X, Z, \beta, \boldsymbol\gamma, g)) \\
	    & \   - \exp(X^T \beta+ Z^T \boldsymbol\gamma(Y)+g(\Lambda(Y, X,Z, \beta, \boldsymbol\gamma, g))- X^T \beta_0- Z^T \boldsymbol\gamma_0(Y)-g_0(\Lambda_0(Y, X,Z)))]\}.
	\end{align*}
	Using the Taylor expansion along with the same arguments, we have 
	\begin{align*}
	    &\ \ \  Pl(\beta_0, \boldsymbol\gamma_0, \zeta_0(\cdot, \beta_0, \boldsymbol\gamma_0); W) - Pl(\beta, \boldsymbol\gamma, \zeta(\cdot, \beta, \boldsymbol\gamma); W) \\
	    & \gtrsim P\{\Delta [(g_0'(\Lambda_0(Y, X,Z))\Lambda'_{0\beta}(Y, X,Z)+X)^T(\beta-\beta_0)\\
	    & \ \ \ \ \ \ \ \ \ \ \ \ +\sum_{j=1}^{d_2+1}g_0'(\Lambda_0(Y, X, Z))\Lambda'_{0\gamma_j}(Y, X, Z)[(\boldsymbol\gamma-\boldsymbol\gamma_0)^T e_j]+ Z^T (\boldsymbol\gamma(Y) - \boldsymbol\gamma_0(Y)) \\
	    &\ \ \ \ \ \ \ \ \ \ \ \  +g_0'(\Lambda_0(Y, X, Z))\Lambda'_{0g}(Y, X, Z)[g-g_0] + g(\Lambda_0(Y, X, Z))- g_0(\Lambda_0(Y, X, Z))]^2 \} \\
	    & \ \ \ \ + o(d^2(\theta, \theta_0)) \\
	    & = P\{\Delta [\epsilon_1(U) X^T(\beta-\beta_0)+\epsilon_2(U, V, Z)[(\boldsymbol\gamma(Y) - \boldsymbol\gamma_0(Y))^T Z]+\epsilon_3(U)[g-g_0]]^2 \}+ o(d^2(\theta, \theta_0)),
	\end{align*}
	where $\epsilon_1$, $\epsilon_2$, and $\epsilon_3$ are deterministic functions of $Z$, $U$, $V$ given in condition \ref{c:z lower bound1}. Under the updated conditions \ref{c:z lower bound1}-\ref{c:z lower bound2}, we can similarly derive that  
	\begin{align*}
	    &\ \ \  Pl(\beta_0, \boldsymbol\gamma_0, \zeta_0(\cdot, \beta_0, \boldsymbol\gamma_0); W) - Pl(\beta, \boldsymbol\gamma, \zeta(\cdot, \beta, \boldsymbol\gamma); W) \\
	    & \gtrsim P\{\Delta(\epsilon_1(U) X^T(\beta-\beta_0))^2\} +P\{\Delta (\epsilon_2(U, V, Z)[(\boldsymbol\gamma(Y) - \boldsymbol\gamma_0(Y))^T Z])^2\}\\
	    & \ \  \ \ \ \ +P\{\Delta(\epsilon_3(U)[g-g_0])^2\}+ o(d^2(\theta, \theta_0)).
	\end{align*}
	Given the boundedness of $Z$, the first and third terms are similarly bounded below by $\|\beta-\beta_0\|^2$ and $\|g-g_0\|_2^2$ respectively. The second term is bounded below by
	\begin{align*}
	   P\{\Delta (\epsilon_2(U, V, Z)[(\boldsymbol\gamma(Y) - \boldsymbol\gamma_0(Y))^T Z])^2\} & \gtrsim \|(\boldsymbol\gamma(Y) - \boldsymbol\gamma_0(Y))^T Z\|_2^2 \\
	    & = \int_0^\tau (\boldsymbol\gamma - \boldsymbol\gamma_0)(t)^T P\{ZZ^T\} (\boldsymbol\gamma - \boldsymbol\gamma_0)(t) \dev t+\\
	    &\ge \int_0^\tau \lambda_1^{(Z)}(\boldsymbol\gamma - \boldsymbol\gamma_0)(t)^T (\boldsymbol\gamma - \boldsymbol\gamma_0)(t) \dev t \\
	    &=\lambda_1^{(Z)} \|\boldsymbol\gamma - \boldsymbol\gamma_0\|_2^2,
	\end{align*}
	where $\lambda_1^{(Z)}$ is the smallest eigenvalue of $P\{ZZ^T\}$, which is positive due to the nonsingularity in the updated condition~\ref{c:z bounded x}. Therefore, we have
	\[Pl(\beta_0, \boldsymbol\gamma_0, \zeta_0(\cdot, \beta_0, \boldsymbol\gamma_0); W) - Pl(\beta, \boldsymbol\gamma, \zeta(\cdot, \beta, \boldsymbol\gamma); W) \gtrsim \|\beta-\beta_0\|^2 + \|\boldsymbol\gamma - \boldsymbol\gamma_0\|_2^2 + \|g-g_0\|_2^2\gtrsim d^2(\theta, \theta_0).\]
	\item The condition C2 in \citet{shen1994} controls the decreasing rate of the variance of the log-likelihood ratio as the parameter $\theta$ approaches the true value $\theta_0$. We use the same arguments to show that
	\[\sup_{d(\theta, \theta_0)\leq \epsilon, \theta\in \Theta_n} Var\{l(\beta, \boldsymbol\gamma, \zeta(\cdot, \beta, \boldsymbol\gamma); W) - l(\beta_0, \boldsymbol\gamma_0, \zeta_0(\cdot, \beta_0, \boldsymbol\gamma_0); W)\} \lesssim \epsilon^2.\] Note that the second term in~(\ref{thm1: upper bound}) is replaced and upper bounded by
	\begin{align*}
		& \ \ \ \  P\{\Delta\left(Z^T(\boldsymbol\gamma(Y)-\boldsymbol\gamma_0(Y))\right)^2\} \\
		& = P\int_0^\tau 1(Y\geq t) \exp(X^T \beta_0+ Z^T \boldsymbol\gamma_0(t)+g_0(\Lambda_0(t, X,Z))) \left(Z^T (\boldsymbol\gamma(t)-\boldsymbol\gamma_0(t))\right)^2\dev t \\
		& \leq \int_0^\tau \sup_{x\in \mc{X},z\in \mc{Z}, t\in [0,\tau]}\{\exp(x^T \beta_0+ Z^T \boldsymbol\gamma_0(t)+g_0(\Lambda_0(t, x, z)))\} P \left(Z^T (\boldsymbol\gamma(t)-\boldsymbol\gamma_0(t))\right)^2\dev t\\
		& \leq \int_0^\tau \sup_{x\in \mc{X},z\in \mc{Z}, t\in [0,\tau]}\{\exp(x^T \beta_0+ Z^T \boldsymbol\gamma_0(t)+g_0(\Lambda_0(t, x, z)))\} \lambda^{(Z)}_{d_2+1} \left\| \boldsymbol\gamma(t)-\boldsymbol\gamma_0(t)\right\|^2\dev t\\
		& \lesssim \|\boldsymbol\gamma - \boldsymbol\gamma_0\|_2^2,
	\end{align*}
	where $\lambda^{(Z)}_{d_2+1}$ is the largest eigenvalue of $P(ZZ^T)$.	
	\item The condition C3 in \citet{shen1994} bounds the size of the space of log-likelihood ratio induced by $\theta$, i.e.,  $\mc{F}_n=\{l(\theta;W)-l(\theta_{0n};W): \theta \in \Theta_n \}$. By Lemma 6, we have the $L_{\infty}$-metric entropy of the space $\mc{F}_n$ bounded by
	\[H(\epsilon, \mc{F}_n, \|\cdot\|_{\infty}) = \log(N(\epsilon, \mc{F}_n, \|\cdot\|_{\infty}))\lesssim c_1 q_{n_1} (d_2+1) + c_2 q_{n_2} + d_1 \lesssim n^{\max\{\nu_1, \nu_2\}}\log (1/\epsilon),\]
	as the number of covariates $d_i$ is considered as fixed. 
\end{itemize}
After verifying the conditions C1-C3, by Theorem 1 in \citet{shen1994}, we have for the sieve MLE $\hat{\theta}_n$ 
	\begin{equation*}
	\label{shen's rate}
	d(\hat{\theta}_{n}, \theta_0) = O_p(\max\{n^{-\frac{1-\max\{\nu_1, \nu_2\}}{2}}, d(\theta_{0n}, \theta_0), K^{1/2}(\theta_{0n}, \theta_0)\}),
\end{equation*}
where $ K(\theta_{0n}, \theta_0) = P\{l(\theta_0;W)-l(\theta_{0n};W)\}$. We can similarly show that $K(\theta_{0n}, \theta_0)\lesssim O(d^2(\theta_{0n}, \theta_0))$ by the Taylor expansion, so the convergence rate of $\hat{\theta}_n$ depends on the sieve approximation error $d(\theta_{0n}, \theta_0)$. 
Here $\theta_{0n} =(\beta_0, \boldsymbol\gamma_{0n}(\cdot), \zeta_{0n}(\cdot, \beta_0, \boldsymbol\gamma_{0n}))\in \Theta_n$ with $\zeta_{0n}(t, x, z, \beta_0, \boldsymbol\gamma_{0n}) = g_{0n}(\Lambda(t, x, z, \beta_0, \boldsymbol\gamma_{0n}, g_{0n}))$. Note that $\gamma_{0n,j}\in \Gamma^{p_1}_{n}$ and $g_{0n}\in \mc{G}^{p_2}$ are defined in Lemma 5 such that $\|\gamma_{0n,j}-\gamma_{0,j}\|_{\infty}=O(n^{-p_1\nu_1})$ and $\|g_{0n}-g_0\|_{\infty}=O(n^{-p_2\nu_2})$, which is based on the existing spline approximation error in Corollary 6.21 in \citet{schumaker_2007}. Since $d^2(\theta_{0n}, \theta_0) \lesssim \|\beta_0-\beta_0\|^2 + \|\boldsymbol\gamma_{0n}-\boldsymbol\gamma_0\|_{2}^2+ \|g_{0n}-g_0\|_{2}^2 \lesssim \|\boldsymbol\gamma_0-\boldsymbol\gamma_{0n}\|^2_{\infty} + \|g_0-g_{0n}\|^2_{\infty} = O(n^{-2\min\{p_1\nu_1, p_2\nu_2\}})$, it follows that \[d(\hat{\theta}_{n}, \theta_0) =  O_p(n^{-\min\{p_1\nu_1, p_2\nu_2, \frac{1-\max\{\nu_1, \nu_2\}}{2}\}}).\]}

\paragraph{Proof of Theorem~\ref{thm:z asym normal}.} \new{To establish the asymptotic normality in Theorem~\ref{thm:z asym normal}, we similarly verify the assumptions (\ref{m:rate})-(\ref{m:smoothness}) for the proposed general M-theorem in Theorem~\ref{thm: m-theorem} under the updated conditions \ref{c:z bounded b}-\ref{c:z pos fish}. For example, to verify assumption (\ref{m:positive info}), first, we need to find $\mb{v}_j^*=(v^*_{j1}, \cdots, v^*_{jd_1})'$, $1\le j \le d_2+1$, and $\mb{h}^*=(h^*_1, \cdots, h^*_{d_1})'$ with $\mb{h}^*(\cdot) = \mb{w}^*(\Lambda_0(\cdot))+g_0'(\Lambda_0(\cdot))\Lambda'_{0g}(\cdot)[\mb{w}^*]$ such that for any $v\in  \mathbb{V}$ and $h\in  \mathbb{H}$ with $h(\cdot) = w(\Lambda_0(\cdot))+g_0'(\Lambda_0(\cdot))\Lambda'_{0g}(\cdot)[w]$,
\begin{align}
		S''_{\beta\gamma_\ell}(\beta_0, \boldsymbol\gamma_0(\cdot), \zeta_0(\cdot, \beta_0, \boldsymbol\gamma_0))[v]& = \sum_{j=1}^{d_2+1} S''_{\gamma_j \gamma_\ell}(\beta_0, \boldsymbol\gamma_0(\cdot), \zeta_0(\cdot, \beta_0, \boldsymbol\gamma_0))[\mb{v}_j^*,v]\nonumber\\
		& \ \ \ + S''_{\zeta\gamma_l}(\beta_0, \boldsymbol\gamma_0(\cdot), \zeta_0(\cdot, \beta_0, \boldsymbol\gamma_0))[\mb{h}^*,v],\label{eq:z pos info 1}\\
		S''_{\beta\zeta}(\beta_0, \boldsymbol\gamma_0(\cdot), \zeta_0(\cdot, \beta_0, \boldsymbol\gamma_0))[h] & = \sum_{j=1}^{d_2+1} S''_{\gamma_j \zeta}(\beta_0, \boldsymbol\gamma_0(\cdot), \zeta_0(\cdot, \beta_0, \boldsymbol\gamma_0))[\mb{v}_j^*,h]\nonumber\\
		& \ \ \ + S''_{\zeta\zeta}(\beta_0, \boldsymbol\gamma_0(\cdot), \zeta_0(\cdot, \beta_0, \boldsymbol\gamma_0))[\mb{h}^*,h].\label{eq:z pos info 2}
\end{align}
By Lemma \ref{lemma: devivatives} and the property $P\{\int^Y_0 f(t, X,Z)\dev \Lambda_0(t, X,Z)\}=P\{\Delta f(Y, X,Z)\}$, for any $\mb{v}_j\in\mathbb{V}^{d_1}, v\in\mathbb{V}$ and $\mb{h}\in \mathbb{H}^{d_1}$ with $\mb{h}(\cdot) = \mb{w}(\Lambda_0(\cdot))+g_0'(\Lambda_0(\cdot))\Lambda'_{0g}(\cdot)[\mb{w}]$, we have for $1\le \ell \le d_2+1$
\begin{align}
	& S''_{\beta\gamma_\ell}(\beta_0, \boldsymbol\gamma_0(\cdot), \zeta_0)[v] - \sum_{j=1}^{d_2+1} S''_{\gamma_j \gamma_\ell}(\beta_0, \boldsymbol\gamma_0(\cdot), \zeta_0)[\mb{v}_j,v] - S''_{\zeta\gamma_l}(\beta_0, \boldsymbol\gamma_0(\cdot), \zeta_0)[\mb{h},v]\nonumber\\
	=&  P\{l''_{\beta\gamma_\ell}(\beta_0, \boldsymbol\gamma_0, \zeta_0;W)[v]-\sum_{j=1}^{d_2+1}l''_{\gamma_j \gamma_\ell}(\beta_0, \boldsymbol\gamma_0, \zeta_0;W)[\mb{v}_j, v]- l''_{\zeta\gamma_\ell}(\beta_0, \boldsymbol\gamma_0, \zeta_0;W)[\mb{h}, v ]\}\nonumber\\
	=& P\left\{\Delta \left[ g_0'(\Lambda_0(Y, X, Z))\Lambda'_{0\beta}(Y, X, Z)+X - \sum_{j=1}^{d_2+1}\left(g_0'(\Lambda_0(Y, X, Z))\Lambda'_{0\gamma_j}(Y, X, Z)[\mb{v}_j]+ \mb{v}_j(Y)Z_j \right) \right.\right.\nonumber\\
	& \ \ \ \ \   - g_0'(\Lambda_0(Y, X, Z))\Lambda'_{0g}(Y, X,Z)^T[\mb{w}] - \mb{w}(\Lambda_0(Y, X,Z)) \Bigg] \left(g_0'(\Lambda_0(Y, X, Z))\Lambda'_{0\gamma_\ell}(Y, X, Z)[v]+ v(Y)Z_\ell\right) \Bigg\}\nonumber.
\end{align}
Under the updated condition~\ref{c:z pos info exist}, there exist $\mb{v}_j^*=(v^*_{j1}, \cdots, v^*_{jd_1})^T$ and $\mb{w}^*=(w^*_1, \cdots, w^*_{d_1})^T$, where $v^*_{jk} \in  \Gamma^{2}$ and $w^*_k \in \mc{G}^{2}$ for $1\le j \le d_2+1, 1\le k \le d_1$, such that $P\{\Delta \mb{A}^*(U, X, Z)\psi'_{0\gamma_\ell}(Y,X,Z)[v]\}=0$ hold for any $v\in \Gamma^{p_1}$, $1\le \ell \le d_2+1$. Therefore, we have that the equation (\ref{eq:z pos info 1}) holds with $\mb{v}^*_j$ and $\mb{w}^*$ given in condition \ref{c:z pos info exist}. Similarly, we can show that the equation (\ref{eq:z pos info 2}) holds as well.}

\section{Additional Simulation Studies}
\label{appendx: simu}
	In this section, we provide full results of simulation studies with various sample sizes and \new{investigate 1) how the numerical performance of the proposed method depends on the knot selection by comparing multiple natural knot selections; 2) a heuristic parametric approach that applies the unified ODE framework along with the proposed estimation and inference procedure for model diagnostics.}

\subsection{Time-varying Cox model}
	Table \ref{tab:time-varying cox2} summarizes the estimates of regression coefficients $\beta_3$ and $\beta_4$ in the time-varying Cox model that is considered in subsection \ref{subs: tinv cox}. The proposed sieve estimators for $\beta_3$ and $\beta_4$ perform similarly to those for  $\beta_1$ and $\beta_2$ as shown in Table \ref{tab:time-varying cox1}. The bias of the estimators for $\beta_3$ and $\beta_4$ decreases and becomes negligible as the sample size increases. The estimated standard error by inverting the estimated information matrix for all parameters including the coefficients of spline basis are close to the sample standard error and the corresponding 95\% confidence intervals obtain reasonable coverage proportion. 
	
	\begin{table}[!ht]
	    \begin{center}
	    \caption{Simulation results under time-varying Cox model.}
	    \label{tab:time-varying cox2}
	    \begin{threeparttable}
	    \begin{tabular}{cccccccccc}
	    \toprule
	        N & Method & \multicolumn{4}{c}{$\beta_3=-1$} & \multicolumn{4}{c}{$\beta_4=1$}\\
	        & & Bias & SE & ESE & CP & Bias & SE & ESE & CP \\
	        \midrule
	       \multirow{2}{*}{1000} &  ODE & -.009& .076& .078 & .942 & .009& .068& .070 & .948 \\
	        & Cox-MPLE & -.007 & .076 & .075 & .938 & .007 & .068 & .068 & .943  \\ 
			\hline
	       \multirow{2}{*}{2000} &  ODE & -.004 & .052& .054 & .965 & .005 & .047& .048 & .955 \\
	        & Cox-MPLE & -.003 & .052 & .053 & .966 & .004 & .047 & .048 & .952  \\ 
			\hline
	       \multirow{2}{*}{4000} &  ODE & -.003& .037 & .038 & .951 & .004 & .034& .034 & .951 \\
	        & Cox-MPLE & -.003& .037 & .037 & .950 & .003& .034 & .034 & .950  \\ 
	        \hline
	        \multirow{2}{*}{8000} &  ODE & .000 & .026 & .026 & .959 & -.001 & .024&  .024& .947  \\
	        & Cox-MPLE & .000 & .026& .026 & .952 & -.001 & .024 & .024 & .949  \\
	        \bottomrule
	    \end{tabular}
	    \begin{tablenotes}
	  \item  \footnotesize{Bias is the difference between mean of estimates and the true value; SE is the sample standard error of the estimates; ESE is the mean of the standard error estimators by inverting the estimated information matrix of all parameters including the coefficients of spline basis, and CP is the corresponding coverage proportion of 95\% confidence intervals.
	  }
	  \end{tablenotes}
	  \end{threeparttable}
	  \end{center}
	\end{table}

\subsection{\new{Comparison with the method in \citet{https://doi.org/10.1002/sim.1203} under the Cox model}}
	\new{In setting 1), we compare the proposed sieve MLE under the Cox model with the parametric method in \citet{https://doi.org/10.1002/sim.1203}, where the log-transformed baseline cumulative hazard is modeled as a natural cubic spline function of the log-transformed time. We implement it using the ``flexsurvspline'' function in the R package \textit{flexsurv} with the same number of interior knots, i.e., $\lfloor N'^{\frac{1}{5}} \rfloor$. The sample size $N$ varies from $1000$ to $8000$.}
		
\begin{figure}[!h]
    \centering
    \includegraphics[width=0.8\textwidth]{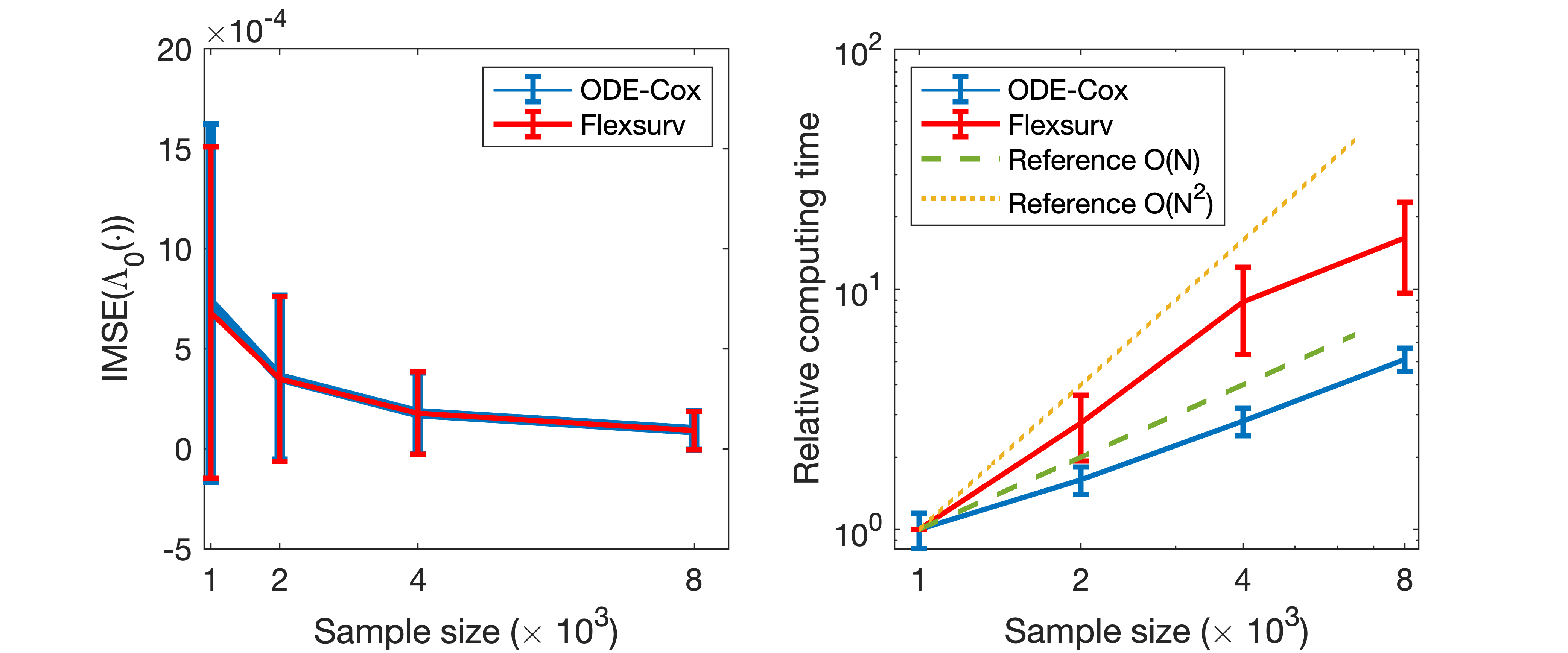}
    \caption{Integrated mean square error (IMSE) of estimated baseline cumulative hazard functions and the log-log plot of mean relative computing time with respect to the sample size under the Cox model are provided from left to right.}
    \label{fig:cox_ode_summary}
	\end{figure}

	\new{Table~\ref{tab: cox} summarizes the estimates of regression coefficients based on $1000$ replicates. We can see that both the proposed estimation method (ODE-Cox) and the method in \citet{https://doi.org/10.1002/sim.1203} (flexsurv) perform similarly to maximum partial likelihood estimation (MPLE) in terms of estimation accuracy. As shown in Figure~\ref{fig:cox_ode_summary}, the proposed method ODE-Cox achieves comparable integrated mean square errors (IMSE) of the estimated cumulative hazard function to those of ``flexsurv''. In addition, the relative computing time (the computing time with respect to that with the smallest sample size $1000$) of proposed method ODE-Cox increases slowly than  that of ``flexsurv'' as the sample size grows. We note that the increasing rate of the relative computing time of the ODE-Cox is even slower than the linear rate, which may be benefited from efficient implementation of existing numerical ODE solvers.}
	
	\begin{table*}[!ht]
	    \begin{center}
	    \caption{Simulation results under the Cox model.}
	    \label{tab: cox}
	    \begin{threeparttable}
	    \begin{tabular}{cc|cccc|cccc|cccc}
	    \toprule
	        N & Method & \multicolumn{4}{c}{$\beta_1=1$} & \multicolumn{4}{c}{$\beta_2=1$} & \multicolumn{4}{c}{$\beta_3=1$}\\
	        & & \small{Bias} & \small{SE} & \small{ESE} & \small{CP} & \small{Bias} & \small{SE} & \small{ESE} & \small{CP} &  \small{Bias} & \small{SE} & \small{ESE} & \small{CP} \\
	        \midrule
	        \multirow{3}{*}{1000} & \small{MPLE} & \small{.006} & \small{.153} & \small{.152} & \small{.948} & \small{.010} & \small{.157} & \small{.152} & \small{.944}  & \small{.004} & \small{.152} & \small{.152} & \small{.950} \\
	         & \small{ODE-Cox} & \small{.009} & \small{.153} & \small{.157} & \small{.952} & \small{.013} & \small{.157} & \small{.157} & \small{.952} & \small{.007} & \small{.152} & \small{.158} & \small{.961} \\ 
	         & \small{Flexsurv} & \small{.007} & \small{.153} & \small{.152} & \small{.948} & \small{.011} & \small{.156} & \small{.151} & \small{.943} & \small{.005} & \small{.151} & \small{.152} & \small{.952} \\
	         \midrule
	        \multirow{3}{*}{2000} & \small{MPLE} & \small{.005} & \small{.106} & \small{.107} & \small{.954} & \small{-.002} & \small{.107} & \small{.107} & \small{.949}  & \small{.006} & \small{.105} & \small{.107} & \small{.958} \\
	         & \small{ODE-Cox} & \small{.007} & \small{.106} & \small{.109} & \small{.956} & \small{-.001} & \small{.107} & \small{.109} & \small{.955} & \small{.007} & \small{.105} & \small{.109} & \small{.961} \\ 
	         & \small{Flexsurv} & \small{.006} & \small{.105} & \small{.107} & \small{.956} & \small{-.001} & \small{.107} & \small{.107} & \small{.950} & \small{.007} & \small{.105} & \small{.107} & \small{.955} \\
	         \midrule
	        \multirow{3}{*}{4000} & \small{MPLE} & \small{.002} & \small{.076} & \small{.075} & \small{.934} & \small{-.003} & \small{.075} & \small{.075} & \small{.941}  & \small{-.001} & \small{.074} & \small{.075} & \small{.954} \\
	         & \small{ODE-Cox} & \small{.003} & \small{.076} & \small{.076} & \small{.936} & \small{-.002} & \small{.075} & \small{.076} & \small{.942} & \small{.000} & \small{.074} & \small{.076} & \small{.955} \\ 
	         & \small{Flexsurv} & \small{.002} & \small{.076} & \small{.075} & \small{.934} & \small{-.002} & \small{.075} & \small{.075} & \small{.942} & \small{-.001} & \small{.074} & \small{.075} & \small{.953} \\
	         \midrule
	        \multirow{3}{*}{8000} & \small{MPLE} & \small{-.002} & \small{.053} & \small{.053} & \small{.953} & \small{.000} & \small{.052} & \small{.053} & \small{.954}  & \small{-.001} & \small{.053} & \small{.053} & \small{.944} \\
	         & \small{ODE-Cox} & \small{-.002} & \small{.053} & \small{.054} & \small{.953} & \small{-.000} & \small{.052} & \small{.054} & \small{.957} & \small{-.002} & \small{.054} & \small{.054} & \small{.947} \\ 
	         & \small{Flexsurv} & \small{-.001} & \small{.053} & \small{.053} & \small{.954} & \small{.000} & \small{.052} & \small{.053} & \small{.952} & \small{-.001} & \small{.053} & \small{.053} & \small{.944} \\
	        \bottomrule
	    \end{tabular}
	    \begin{tablenotes}
	 	 \item  \footnotesize{Bias is the difference between the mean of estimates and the true value, and SE is the sample standard error of the estimates. ESE is the mean of the standard error estimators, and CP is the corresponding coverage proportion of 95\% confidence intervals.
	  	}
	  		\end{tablenotes}
	  		\end{threeparttable}
	 	 \end{center}
	\end{table*}

\subsection{\new{Comparison with the NPMLE~\citep{zeng2007maximum} under the linear transformation model}}
	\new{We have compared the proposed ODE approach and the NPMLE for the logarithmic transformation model in~\citep{zeng2007maximum}. Specifically, in the simulation setting (2), we generate event times from the ODE 
		\[\Lambda'_{x}(t) = q(\Lambda_x(t)) \exp(\beta_1x_1+\beta_2x_2+\beta_3x_3)\alpha(t),\]
		where functions $q(t)=\exp(-t)$ and $\alpha(t)=2$. It is equivalent to generate event times  with the cumulative hazard function 
		\[\Lambda_x(t)=G\{\exp(\beta_1x_1+\beta_2x_2+\beta_3x_3)\Lambda_0(t)\},\]
		where $G(u)=\log(1+u)$ and $\Lambda_0(t)=\int^t_0\alpha(s)\dev s=2t$. For the NPMLE in \citet{zeng2007maximum}, note that the function $G(\cdot)$ is known and the baseline cumulative hazard $\Lambda_0(\cdot)$ is unknown. An EM algorithm was implemented in Matlab to compute the NPMLE. To make fair comparison,  we set the function $q(\cdot)$ known, i.e., $q(t)=\exp(-t)$, and the function $\alpha(\cdot)$ unknown for the ODE-LT. We fit $\log\alpha(\cdot)$ by cubic B-splines and set the number of knots $K_n$ as the largest integer below $N'^{\frac{1}{5}}$, where $N'$ is the number of distinct observation time points. The sample size $N$ varies from $1,000$ to $8,000$. }
		
\begin{figure}[!h]
    \centering
    \includegraphics[width=0.8\textwidth]{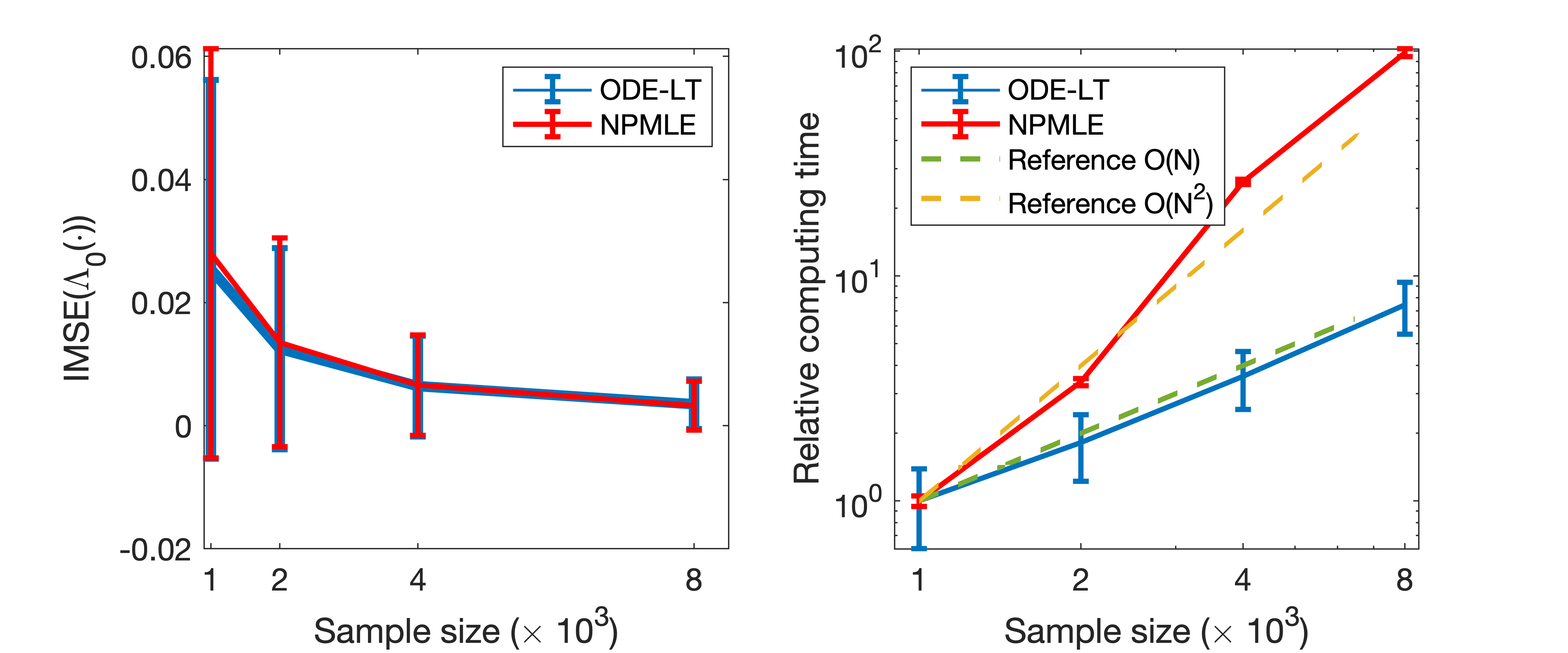}
    \caption{Integrated mean square error (IMSE) of estimated baseline cumulative hazard functions and the log-log plot of mean relative computing time with respect to the sample size under the linear transformation model are provided from left to right.}
    \label{fig:lt_ode_summary}
	\end{figure}

	\new{Table~\ref{tab: lt} summarizes the estimates of regression coefficients $\beta$ based on $1000$ replicates. The proposed estimation method (ODE-LT) achieves similar estimation accuracy of both $\beta$ and the cumulative hazard (shown in the left panel of Figure~\ref{fig:lt_ode_summary}) as the NPMLE. However, the relative computing time of the proposed method ODE-LT increase linearly as the sample size grows while that of the NPMLE increases in a quadratic rate as shown in the right panel of Figure~\ref{fig:lt_ode_summary}.}
	
	\begin{table*}[!h]
	    \begin{center}
	    \caption{Simulation results under the linear transformation model.}
	    \label{tab: lt}
	    \begin{threeparttable}
	    \begin{tabular}{cc|cccc|cccc|cccc}
	    \toprule
	        N & Method & \multicolumn{4}{c}{$\beta_1=1$} & \multicolumn{4}{c}{$\beta_2=1$} & \multicolumn{4}{c}{$\beta_3=1$} \\
	        & & \small{Bias} & \small{SE} & \small{ESE} & \small{CP} & \small{Bias} & \small{SE} & \small{ESE} & \small{CP} &  \small{Bias} & \small{SE} & \small{ESE} & \small{CP}\\
	        \midrule
	        \multirow{2}{*}{1000} & \small{NPMLE} & \small{.003} & \small{.227} & \small{.230} & \small{.954} & \small{.003} & \small{.236} & \small{.230} & \small{.949} & \small{.003} & \small{.229} & \small{.230} & \small{.954}  \\
	       & \small{ODE-LT} & \small{.005} & \small{.227} & \small{.231} & \small{.956} & \small{.005} & \small{.237} & \small{.231} & \small{.949} & \small{.004} & \small{.229} & \small{.231} & \small{.955}  \\
	        \midrule
	        \multirow{2}{*}{2000} & \small{NPMLE} & \small{-.002} & \small{.159} & \small{.162} & \small{.946} & \small{.003} & \small{.169} & \small{.162} & \small{.933} & \small{.006} & \small{.157} & \small{.162} & \small{.963}  \\
	       & \small{ODE-LT} & \small{-.001} & \small{.159} & \small{.163} & \small{.947} & \small{.003} & \small{.169} & \small{.163} & \small{.933} & \small{.007} & \small{.157} & \small{.163} & \small{.961}  \\
	        \midrule
	        \multirow{2}{*}{4000} & \small{NPMLE} & \small{.004} & \small{.117} & \small{.115} & \small{.949} & \small{-.001} & \small{.114} & \small{.115} & \small{.951}  & \small{.003} & \small{.113} & \small{.115} & \small{.960} \\
	        & \small{ODE-LT} & \small{.005} & \small{.117} & \small{.115} & \small{.950} & \small{-.000} & \small{.114} & \small{.115} & \small{.951} & \small{.003} & \small{.113} & \small{.115} & \small{.961}\\ 
	        \midrule
	        \multirow{2}{*}{8000} & \small{NPMLE} & \small{-.005} & \small{.079} & \small{.081} & \small{.956} & \small{.000} & \small{.078} & \small{.081} & \small{.963} & \small{-.001} & \small{.079} & \small{.081} & \small{.950} \\
	       & \small{ODE-LT} & \small{-.004} & \small{.079} & \small{.081} & \small{.957} & \small{.001} & \small{.078} & \small{.081} & \small{.963} & \small{-.000} & \small{.079} & \small{.081} & \small{.951}  \\
	        \bottomrule
	    \end{tabular}
	    \begin{tablenotes}
	 	 \item  \footnotesize{Bias is the difference between the mean of estimates and the true value, SE is the sample standard error of the estimates, and Mean is the mean of IMSE. ESE is the mean of the standard error estimators, and CP is the corresponding coverage proportion of 95\% confidence intervals. 
	  	}
	  	\end{tablenotes}
	  	\end{threeparttable}
	  \end{center}
	\end{table*}

\subsection{\new{Comparison with the rank-based method under the AFT model}}
	\new{In setting 3), we compare the proposed sieve MLE for the ODE-AFT model, where the function $\alpha$ is set to 1, with the rank-based estimation approach implemented using the R package {\it aftgee}. For the ODE-AFT model, we fit $\log q(t)$ by cubic B-splines with $\lfloor N^{\frac{1}{7}} \rfloor$ interior knots. Note that the argument of the function $q(\cdot)$ is the cumulative hazard. Unlike fitting the function $\alpha(\cdot)$ whose argument is the event time in the ODE-Cox model, we do not observe the corresponding cumulative hazard directly. Therefore, we use the estimated cumulative hazards under the Cox model as a remedy. Let $\hat{\Lambda}_i^{Cox}$ denote the estimated cumulative hazard for individual $i$ under the Cox model. The interior knots are located at the quantiles of $\{\hat{\Lambda}_i^{Cox}\}_{i=1}^n$.} 

\begin{figure}[!h]
    \centering
    \includegraphics[width=0.8\textwidth]{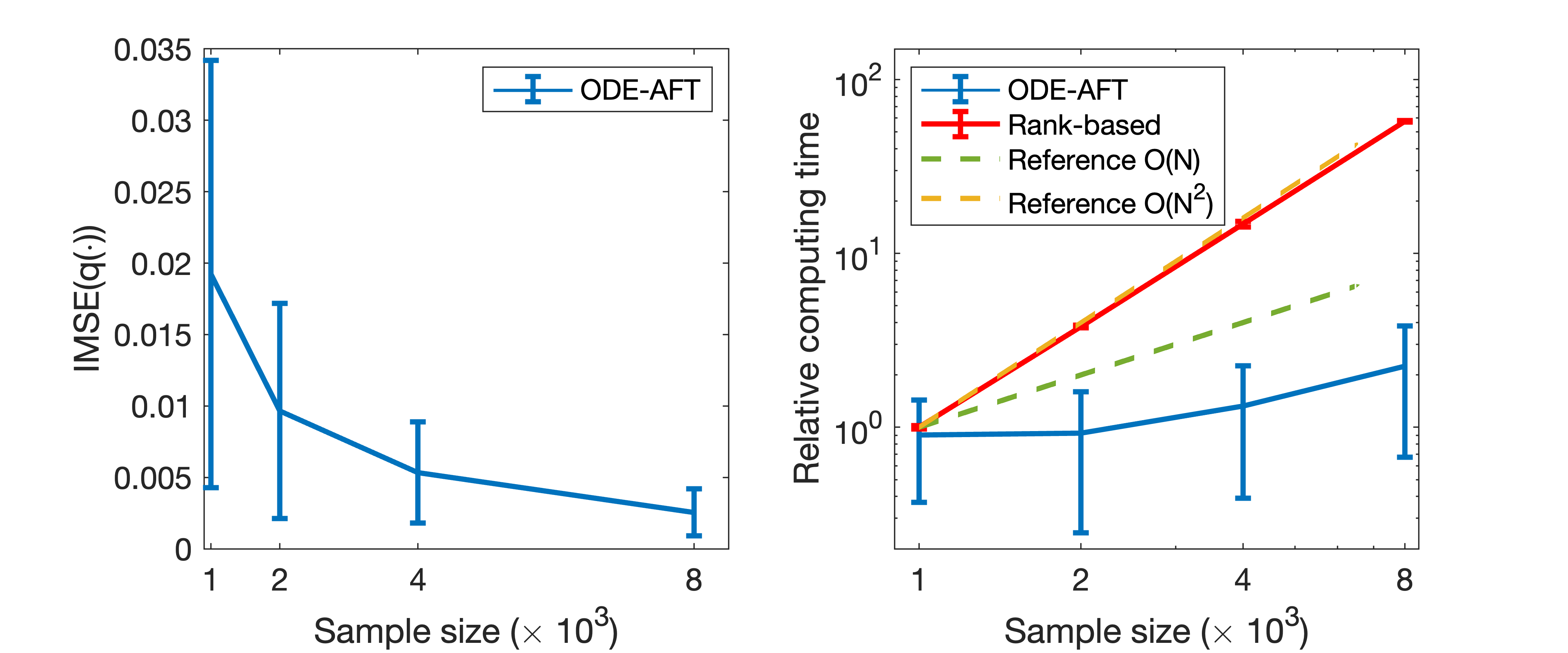}
    \caption{Integrated mean square error (IMSE) of estimated baseline cumulative hazard functions and the log-log plot of mean relative computing time with respect to the sample size under the AFT model are provided from left to right.}
    \label{fig:aft_ode_summary}
	\end{figure}

	\new{Table~\ref{tab: aft} summarizes the estimates of regression coefficients $\beta$ with varying sample sizes. Although the bias of the ODE approach is relatively greater than that of the rank-based method when the sample size is small, the bias of the estimates becomes negligible as the sample size increases. As shown in Figure~\ref{fig:aft_ode_summary}, the relative computing time of the proposed ODE approach increases in a slower rate than that of the rank-based method for the semi-parametric ODE-AFT model. Remarkably, the proposed ODE approach takes just $6$ seconds for estimating the ODE-AFT model but the rank-based method takes $349$ seconds when the sample size is $8,000$.}

	\begin{table*}[!h]
	    \begin{center}
	    \caption{Simulation results under the AFT model.}
	    \label{tab: aft}
	    \begin{threeparttable}
	    \begin{tabular}{cc|cccc|cccc|cccc}
	    \toprule
	        N & Method & \multicolumn{4}{c}{$\beta_1=1$} & \multicolumn{4}{c}{$\beta_2=1$} & \multicolumn{4}{c}{$\beta_3=1$}\\
	        & & \small{Bias} & \small{SE} & \small{ESE} & \small{CP} & \small{Bias} & \small{SE} & \small{ESE} & \small{CP} &  \small{Bias} & \small{SE} & \small{ESE} & \small{CP} \\
	        \midrule
	        \multirow{2}{*}{1000} & \small{Rank-based} & \small{-.000} & \small{.204} & \small{.206} & \small{.952} & \small{-.009} & \small{.213} & \small{.205} & \small{.925} & \small{-.013} & \small{.200} & \small{.206} & \small{.942}  \\
	         & \small{ODE-AFT} & \small{-.014} & \small{.197} & \small{.191} & \small{.944} & \small{-.024} & \small{.209} & \small{.192} & \small{.931}  & \small{-.032} & \small{.199} & \small{.192} & \small{.932} \\
	         \midrule
	         \multirow{2}{*}{2000} & \small{Rank-based} & \small{-.002} & \small{.147} & \small{.145} & \small{.938} & \small{.005} & \small{.147} & \small{.145} & \small{.951} & \small{.004} & \small{.146} & \small{.146} & \small{.945}  \\
	         & \small{ODE-AFT} & \small{-.010} & \small{.144} & \small{.137} & \small{.932} & \small{-.006} & \small{.144} & \small{.137} & \small{.937}  & \small{-.005} & \small{.142} & \small{.137} & \small{.943} \\
	         \midrule
	         \multirow{2}{*}{4000} & \small{Rank-based} & \small{.004} & \small{.105} & \small{.102} & \small{.944} & \small{-.001} & \small{.102} & \small{.102} & \small{.950}  & \small{.002} & \small{.100} & \small{.103} & \small{.954} \\
	         & \small{ODE-AFT} & \small{.000} & \small{.102} & \small{.097} & \small{.944} & \small{-.005} & \small{.100} & \small{.097} & \small{.944}  & \small{-.002} & \small{.097} & \small{.097} & \small{.950} \\
	         \midrule
	         \multirow{2}{*}{8000} & \small{Rank-based} & \small{-.003} & \small{.071} & \small{.073} & \small{.956} & \small{.001} & \small{.071} & \small{.073} & \small{.962} & \small{.000} & \small{.072} & \small{.073} & \small{.949} \\
	         & \small{ODE-AFT} & \small{-.006} & \small{.070} & \small{.069} & \small{.950} & \small{-.003} & \small{.068} & \small{.069} & \small{.967}  & \small{-.004} & \small{.071} & \small{.069} & \small{.945} \\ 
	        \bottomrule
	    \end{tabular}
	    \begin{tablenotes}
	 	 \item  \footnotesize{Bias is the difference between the mean of estimates and the true value, and SE is the sample standard error of the estimates. ESE is the mean of the standard error estimators, and CP is the corresponding coverage proportion of 95\% confidence intervals.
	  	}
	  		\end{tablenotes}
	  		\end{threeparttable}
	 	 \end{center}
	\end{table*}
	
\subsection{\new{Comparison with the smoothed partial rank method under the general linear transformation model}}

\new{In settings 1)-4), we compare the sieve MLE for the general linear transformation model (ODE-Flex), where both $q(\cdot)$ and $\alpha(\cdot)$ are unspecified, with the smoothed partial rank (SPR) estimation method in~\citet{Song_2006}, which is a rank-based estimation method for censored data. As the original code of SPR is not available, we implement the SPR estimation and inference methods by our own, and we verify that our implementations are able to reproduce the simulation results in~\citet{Song_2006}. Note that SPR introduces an additional parameter $c$ in the objective function to improve the estimation accuracy. We evaluate SPR with various values of the parameter $c$ and the sample size $N$ under our data settings 1)-4). We observe that SPR may return extreme estimates, so we count estimates with more than $5$ deviation from the truth as failed replications.}

\begin{figure}%[!h]
		    \centering
		    \includegraphics[width=\textwidth]{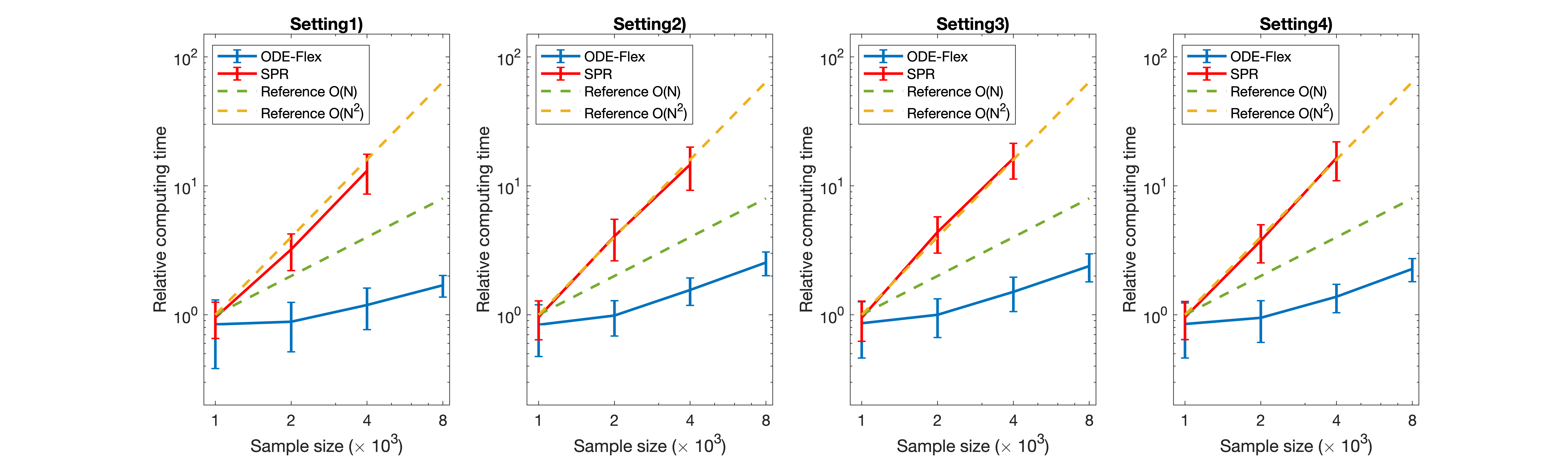}
		    \caption{The log-log plot of mean relative computing time with respect to the sample size under the nonparametric linear transformation model.}
		    \label{fig:spr_ode_computing_summary}
		\end{figure}

\new{Tables~\ref{tab: partial rank - setting 1 & 2}-\ref{tab: partial rank - setting 3 & 4} summarize the estimates of $\beta_2$ under settings 1)-4) over $1,000$ replications. (We observe similar performance for $\beta_3$  and so we omit its results here.) 
In terms of estimation accuracy, both the SPR estimator and ODE-Flex estimator show negligible biases when the sample size is large. However, two inference methods in~\citet{Song_2006} are sensitive to the choices of the parameter $c$~: the sandwich estimator  seriously underestimates the standard deviation for various values of the parameter $c$ and the corresponding coverage proportion is far below the nominal level; the weighted bootstrap estimator overestimates the standard deviation for small values of $c$ and underestimates it for relatively large values of $c$. In contrast, the proposed ODE-Flex method performs well across various sample sizes: the standard error estimators approximate the empirical standard deviations well and the coverage proportions are close to the nominal level. 
 	In terms of numerical stability, the proposed ODE-Flex method can stably return good estimates over $1,000$ replications, especially for large sample sizes: only less than 1\% replications meet with numerical errors when $N = 4,000$ and  100\% replications successfully return accurate estimates when $N=8,000$. We note that this result is reported under a universal precision for ODE solvers and we find that these failed replications can be easily fixed by adjusting the precision of the ODE solver.  However, the SPR method fails to return a reasonable point estimator for more than 12\% realized resampling on average when computing the standard error estimator by the weighted bootstrap. We also observe that it is difficult to obtain the SPR point estimator for larger sample size such as $N=8,000$ or larger parameter $c$ such as $10^{-1}$ and $1$ (success rate less than 10\%).
 	In terms of computation efficiency, as shown in Figure~\ref{fig:spr_ode_computing_summary}, the computing time of ODE-Flex increases in a much smaller rate than that of SPR as the sample size grows, which implies that the proposed estimation method is computationally more efficient for large sample size.}

\begin{table*}%[!h]
	    \begin{center}
	    \caption{Simulation results of $\beta_2$ under the general linear transformation model with both $q(\cdot)$ and $\alpha(\cdot)$ unspecified in settings 1) and 2).}
	    \label{tab: partial rank - setting 1 & 2}
	    \begin{threeparttable}
\begin{tabular}{c|c|cc|cccccc|cc}
\toprule
& Method & N & c & & & \multicolumn{2}{c}{Sandwich}  & \multicolumn{2}{c}{Bootstrap} & & Bootstrap \\
  & & & & Bias &     SE & ESE & CP & ESE & CP  & Succ. \% & Succ. \% \\
   \midrule
\multirow{14}{*}{1)} & \multirow{9}{*}{SPR} &  \multirow{3}{*}{1000}& $10^{-4}$ &  .030 &  .331 &        .000 &       .000 &          .697 &        .974 &        98.3 &                  87.3 \\
		& & & $10^{-3}$ & .034 &  .250 &        .000 &       .003 &         .478 &        .960 &      97.4 &                  84.0 \\
		& & & $10^{-2}$ & .048 &  .295 &        .002 &       .020 &         .103 &        .432 &       80.1 &                  75.5 \\
		    \cmidrule{3-12}
		& &  \multirow{3}{*}{2000} &  $10^{-4}$ &    -.003 &  .313 &        .000 &       .000 &         .668 &        .989 &     98.4 &                  85.0 \\
		& & &  $10^{-3}$ &    .013 &  .210 &        .000 &       .003 &         .314 &        .906 &       94.5 &                  80.4 \\
		& & & $10^{-2}$ &    .007 &  .159 &        .002 &       .022 &         .033 &        .279 &       71.8 &                  70.9 \\
		    \cmidrule{3-12}
		&  & \multirow{3}{*}{4000} & $10^{-4}$ &    .007 &  .153 &        .000 &       .001 &         .552 &        .994 &    97.9 &                  83.1 \\
		& &  & $10^{-3}$ &     .008 &  .120 &        .000 &       .000 &         .136 &        .762 &    95.2 &                  77.7 \\
		& &  & $10^{-2}$ &    .005 &  .105 &        .002 &       .022 &         .016 &        .222 &    67.7 &                  67.8 \\
\cmidrule{2-12}
& & \multicolumn{2}{c|}{N} & Bias &     SE & ESE & CP  & & & Succ. \% &  \\
 \cmidrule{2-12}
& \multirow{4}{*}{ODE-Flex} & \multicolumn{2}{c|}{1000} & .067 & .248 & .243 & .958 & & & 93.6\\
& & \multicolumn{2}{c|}{2000} & .024 & .162 & .158 & .950 & & & 98.4\\
& & \multicolumn{2}{c|}{4000} & .008 & .106 & .107 & .947 & & & 99.5\\
& & \multicolumn{2}{c|}{8000} & .012 & .076 & .075 & .946 & & & 100.0\\
\midrule
& Method & N & c & & & \multicolumn{2}{c}{Sandwich}  & \multicolumn{2}{c}{Bootstrap} & & Bootstrap \\
  & & & & Bias &     SE & ESE & CP & ESE & CP  & Succ. \% & Succ. \% \\
   \midrule
\multirow{14}{*}{2)} & \multirow{9}{*}{SPR} &  \multirow{3}{*}{1000}& $10^{-4}$ &  .082 &  .522 &        .000 &       .000 &          .739 &        .949 &        97.8 &                  87.2 \\
		& & & $10^{-3}$ & .091 &  .449 &        .000 &       .002 &         .538 &        .910 &      96.5 &                  84.2 \\
		& & & $10^{-2}$ & .104 &  .464 &        .003 &       .015 &         .166 &        .457 &       81.8 &                  74.1 \\
		    \cmidrule{3-12}
		&  & \multirow{3}{*}{2000} &  $10^{-4}$ &    .020 &  .347 &        .000 &       .000 &         .702 &        .988 &     98.3 &                  85.5 \\
		& & &  $10^{-3}$ &    .015 &  .320 &        .000 &       .000 &         .393 &        .895 &       95.5 &                  80.3 \\
		& & & $10^{-2}$ &    .044 &  .337 &        .002 &       .005 &         .052 &        .262 &       75.7 &                  69.3 \\
		    \cmidrule{3-12}
		& & \multirow{3}{*}{4000} & $10^{-4}$ &    .014 &  .244 &        .000 &       .000 &         .585 &        .995 &    98.5 &                  83.9 \\
		& & &  $10^{-3}$ &     .019 &  .191 &        .000 &       .001 &         .183 &        .709 &    93.6 &                  77.4 \\
		& & & $10^{-2}$ &    .022 &  .171 &        .002 &       .010 &         .019 &        .158 &    67.1 &                  65.2 \\
\cmidrule{2-12}
 & & \multicolumn{2}{c|}{N} & Bias &     SE & ESE & CP  & & & Succ. \% &  \\
 \cmidrule{2-12}
& \multirow{4}{*}{ODE-Flex} & \multicolumn{2}{c|}{1000} & .024 & .357 & .312 & .918 & & & 98.5\\
& & \multicolumn{2}{c|}{2000} & .009 & .246 & .218 & .931 & & & 99.5\\
& & \multicolumn{2}{c|}{4000} & -.019 & .161 & .151 & .927 & & & 100.0\\
& & \multicolumn{2}{c|}{8000} & -.020 & .113 & .107 & .939 & & & 100.0\\
\bottomrule
\end{tabular}
\begin{tablenotes}
	 	 \item  \footnotesize{Bias is the difference between the mean of estimates and the true value, and SE is the sample standard error of the estimates. ESE is the mean of the standard error estimators, and CP is the corresponding coverage proportion of 95\% confidence intervals.
	  	}
	  		\end{tablenotes}
	  		\end{threeparttable}
\end{center}
\vspace{-3mm}
\end{table*}

\begin{table*}%[!h]
	    \begin{center}
	    \caption{Simulation results of $\beta_2$ under the general linear transformation model with both $q(\cdot)$ and $\alpha(\cdot)$ unspecified in settings 3) and 4).}
	    \label{tab: partial rank - setting 3 & 4}
	    \begin{threeparttable}
\begin{tabular}{c|c|cc|cccccc|cc}
\toprule
& Method & N & c & & & \multicolumn{2}{c}{Sandwich}  & \multicolumn{2}{c}{Bootstrap} & & Bootstrap \\
  & & & & Bias &     SE & ESE & CP & ESE & CP  & Succ. \% & Succ. \% \\
   \midrule
\multirow{14}{*}{3)} & \multirow{9}{*}{SPR} &  \multirow{3}{*}{1000}& $10^{-4}$ &  .053 &  .369 &        .000 &       .000 &          .779 &        .980 &        97.3 &                  86.1 \\
		& & & $10^{-3}$ & .056 &  .386 &        .000 &       .000 &         .529 &        .945 &      95.7 &                  82.9 \\
		& & & $10^{-2}$ & .079 &  .372 &        .004 &       .010 &         .128 &        .454 &       79.9 &                  73.4 \\
		    \cmidrule{3-12}
		& &  \multirow{3}{*}{2000} &  $10^{-4}$ &    .004 &  .304 &        .000 &       .000 &         .721 &        .992 &     97.8 &                  84.4 \\
		& & &  $10^{-3}$ &    .010 &  .308 &        .000 &       .000 &         .357 &        .888 &       96.0 &                  79.2 \\
		& & & $10^{-2}$ &    .010 &  .222 &        .002 &       .016 &         .040 &        .251 &       74.8 &                  68.3 \\
		    \cmidrule{3-12}
		&  & \multirow{3}{*}{4000} & $10^{-4}$ &    .005 &  .194 &        .000 &       .000 &         .602 &        .996 &    97.5 &                  82.4 \\
		& &  & $10^{-3}$ &     .007 &  .146 &        .000 &       .001 &         .154 &        .732 &    92.4 &                  76.0 \\
		& &  & $10^{-2}$ &    .011 &  .141 &        .002 &       .025 &         .020 &        .194 &    68.1 &                  63.4 \\
\cmidrule{2-12}
& & \multicolumn{2}{c|}{N} & Bias &     SE & ESE & CP  & & & Succ. \% &  \\
 \cmidrule{2-12}
& \multirow{4}{*}{ODE-Flex} & \multicolumn{2}{c|}{1000} & .016 & .293 & .270 & .940 & & & 95.9\\
& &\multicolumn{2}{c|}{2000} & .014 & .197 & .191 & .948 & & & 99.0\\
& &\multicolumn{2}{c|}{4000} & -.014 & .134 & .131 & .941 & & & 99.7\\
& &\multicolumn{2}{c|}{8000} & -.019 & .088 & .092 & .957 & & & 100.0\\
\midrule
& Method & N & c & & & \multicolumn{2}{c}{Sandwich}  & \multicolumn{2}{c}{Bootstrap} & & Bootstrap \\
  & & & & Bias &     SE & ESE & CP & ESE & CP  & Succ. \% & Succ. \% \\
   \midrule
\multirow{14}{*}{4)} & \multirow{9}{*}{SPR} &  \multirow{3}{*}{1000}& $10^{-4}$ &  .023 &  .349 &        .000 &       .000 &          .756 &        .987 &        97.1 &                  84.3 \\
		& & & $10^{-3}$ & .030 &  .226 &        .000 &       .003 &         .473 &        .963 &      95.4 &                  80.5 \\
		& & & $10^{-2}$ & .032 &  .227 &        .003 &       .022 &         .083 &        .417 &       77.9 &                  71.8 \\
		    \cmidrule{3-12}
		&  & \multirow{3}{*}{2000} &  $10^{-4}$ &    -.006 &  .253 &        .000 &       .000 &         .719 &        .993 &     97.5 &                  82.0 \\
		& & &  $10^{-3}$ &    .006 &  .147 &        .000 &       .002 &         .274 &        .902 &       95.2 &                  76.8 \\
		& & & $10^{-2}$ &    .007 &  .136 &        .004 &       .034 &         .027 &        .275 &       73.8 &                  66.9 \\
		    \cmidrule{3-12}
		& & \multirow{3}{*}{4000} & $10^{-4}$ &    .001 &  .146 &        .000 &       .000 &         .574 &        .995 &    96.9 &                  79.5 \\
		& & &  $10^{-3}$ &     .004 &  .089 &        .000 &       .004 &         .108 &        .781 &    94.2 &                  73.9 \\
		& & & $10^{-2}$ &    .000 &  .086 &        .002 &       .029 &         .019 &        .240 &    66.3 &                  64.1 \\
\cmidrule{2-12}
 & & \multicolumn{2}{c|}{N} & Bias &     SE & ESE & CP  & & & Succ. \% &  \\
 \cmidrule{2-12}
& \multirow{4}{*}{ODE-Flex} & \multicolumn{2}{c|}{1000} & .020 & .182 & .191 & .954 & & & 96.7\\
& &\multicolumn{2}{c|}{2000} & .016 & .132 & .131 & .958 & & & 98.8\\
& &\multicolumn{2}{c|}{4000} & .001 & .092 & .090 & .938 & & & 99.9\\
& &\multicolumn{2}{c|}{8000} & .008 & .062 & .064 & .960 & & & 100.0\\
\bottomrule
\end{tabular}
\begin{tablenotes}
	 	 \item  \footnotesize{Bias is the difference between the mean of estimates and the true value, and SE is the sample standard error of the estimates. ESE is the mean of the standard error estimators, and CP is the corresponding coverage proportion of 95\% confidence intervals.
	  	}
	  		\end{tablenotes}
	  		\end{threeparttable}
\end{center}
\vspace{-3mm}
\end{table*}

\subsection{\new{Dependence on knots selection}}
	\new{To investigate how the numerical performance of the proposed method depends on the knot selection, we have done several simulation studies to compare two natural placements of knots for the ODE-Cox model, the ODE-AFT model, and the general linear transformation model. Specifically,  
	\begin{itemize}
	\item For the ODE-Cox model, we compare the following two placements of knots when using the B-spline to fit the function $\log \alpha(\cdot)$: (K1) the interior knots are located at the $K_n = \lfloor N^{\frac{1}{5}} \rfloor$ quantiles of the distinct observation time points; (K2) the interior knots equally separate the time interval from $0$ to the maximum of observed times.
	\item For the ODE-AFT model, we compare the following two placements of knots when using the B-spline to fit the function $\log q(\cdot)$: (K1) the interior knots are located at the $K_n = \lfloor N^{\frac{1}{7}} \rfloor$ quantiles of the estimated cumulative hazards $\{\hat{\Lambda}_i^{Cox}\}_{i=1}^n$ under the Cox model; (K2) the interior knots equally separate the interval from $0$ to $2\max_{1\le i\le n} \{\hat{\Lambda}_i^{Cox}\}$.
	\item For the general linear transformation model, we compare combinations of the above knots placements when using the B-spline to fit both functions $\log \alpha(\cdot)$ and $\log q(\cdot)$: (K1) the interior knots for both functions are located at the corresponding quantiles; (K2) the interior knots for both functions equally separate the corresponding intervals. 
	\end{itemize}
	}
	
\begin{figure*}%[!h]
    \centering
    \includegraphics[width=0.8\textwidth]{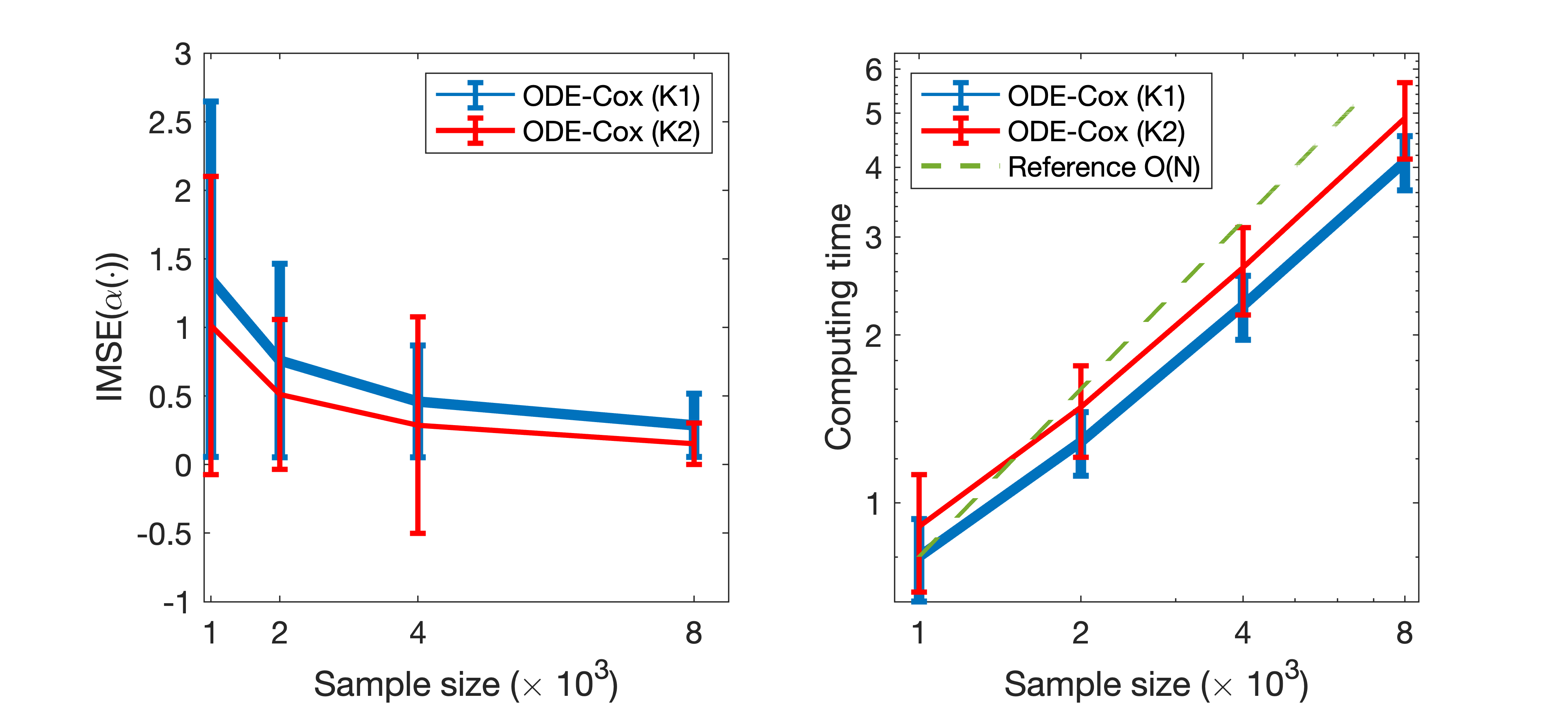}
    \caption{Integrated mean square error (IMSE) of estimated $\alpha(\cdot)$ and the log-log plot of the computing time with respect to the sample size under the Cox model are provided from left to right.}
    \label{fig:cox_knots_comparison}
	\end{figure*}

\begin{figure*}%[!h]
    \centering
    \includegraphics[width=0.8\textwidth]{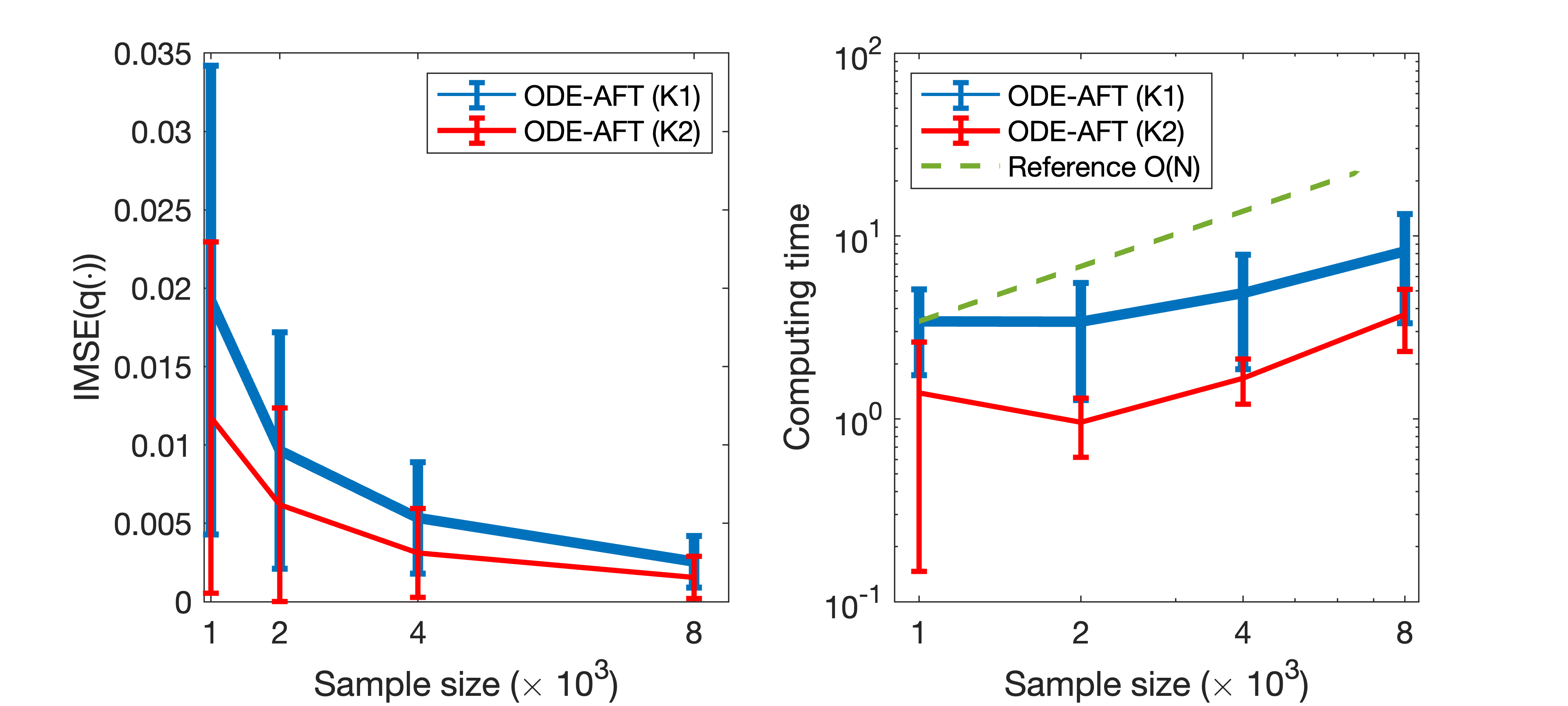}
    \caption{Integrated mean square error (IMSE) of estimated $\alpha(\cdot)$ and the log-log plot of the computing time with respect to the sample size under the AFT model are provided from left to right.}
    \label{fig:aft_knots_comparison}
	\end{figure*}

	\new{Tables~\ref{tab: cox_knots_selection}-\ref{tab: glt_knots_selection} compare the estimates of regression coefficients $\beta$ with two natural placements of knots for the ODE-Cox model, the ODE-AFT model, and the general linear transformation model respectively. Figures~\ref{fig:cox_knots_comparison}-\ref{fig:aft_knots_comparison} compare the integrated mean square errors (IMSE) of estimated functions, and the computing time associated with $K1$ and $K2$ from left to right for the ODE-Cox model and the ODE-AFT model. We can see that both two types of knot locations $K1$ and $K2$ return good estimates of parameters and standard errors. Overall, our numerical results suggest that knot selection does not appear critical for the proposed method in various simulation settings.}
	
	\begin{table*}%[!h]
    \begin{center}
    \caption{Simulation results for two placements of knots under the Cox model.}
	\label{tab: cox_knots_selection}
    \begin{threeparttable}
    \begin{tabular}{cc|cccc|cccc|cccc}
    \toprule
        N & Knots & \multicolumn{4}{c}{$\beta_1=1$} & \multicolumn{4}{c}{$\beta_2=1$} & \multicolumn{4}{c}{$\beta_3=1$}\\
        & & \small{Bias} & \small{SE} & \small{ESE} & \small{CP} & \small{Bias} & \small{SE} & \small{ESE} & \small{CP} &  \small{Bias} & \small{SE} & \small{ESE} & \small{CP} \\
        \midrule
        \multirow{2}{*}{1000} & \small{K1} & \small{.009} & \small{.153} & \small{.157} & \small{.952} & \small{.013} & \small{.157} & \small{.157} & \small{.952} & \small{.007} & \small{.152} & \small{.158} & \small{.961} \\
         & \small{K2} & \small{.009} & \small{.153} & \small{.157} & \small{.953} & \small{.013} & \small{.157} & \small{.157} & \small{.951} & \small{.007} & \small{.152} & \small{.158} & \small{.960} \\ 
         \midrule
        \multirow{2}{*}{2000} & \small{K1} & \small{.007} & \small{.106} & \small{.109} & \small{.956} & \small{-.001} & \small{.107} & \small{.109} & \small{.955} & \small{.007} & \small{.105} & \small{.109} & \small{.961} \\
         & \small{K2} & \small{.006} & \small{.106} & \small{.110} & \small{.958} & \small{-.000} & \small{.107} & \small{.109} & \small{.956} & \small{.007} & \small{.105} & \small{.109} & \small{.960} \\
         \midrule
        \multirow{2}{*}{4000} & \small{K1} & \small{.003} & \small{.076} & \small{.076} & \small{.936} & \small{-.002} & \small{.075} & \small{.076} & \small{.942} & \small{.000} & \small{.074} & \small{.076} & \small{.955} \\
         & \small{K2} & \small{.002} & \small{.076} & \small{.076} & \small{.937} & \small{-.002} & \small{.075} & \small{.077} & \small{.944} & \small{-.000} & \small{.074} & \small{.077} & \small{.955} \\ 
         \midrule
        \multirow{2}{*}{8000} & \small{K1} & \small{-.002} & \small{.053} & \small{.054} & \small{.953} & \small{-.000} & \small{.052} & \small{.054} & \small{.957} & \small{-.002} & \small{.054} & \small{.054} & \small{.947} \\
         & \small{K2} & \small{-.001} & \small{.053} & \small{.054} & \small{.957} & \small{.001} & \small{.053} & \small{.054} & \small{.955} & \small{-.001} & \small{.053} & \small{.054} & \small{.949} \\ 
        \bottomrule
    \end{tabular}
    \begin{tablenotes}
 	 \item  \footnotesize{Bias is the difference between the mean of estimates and the true value, and SE is the sample standard error of the estimates. ESE is the mean of the standard error estimators, and CP is the corresponding coverage proportion of 95\% confidence intervals. In (K1), the interior knots are located at the $K_n = \lfloor N^{\frac{1}{5}} \rfloor$ quantiles of the distinct observation time points. In (K2), the interior knots equally separate the time interval from $0$ to the maximum of observed times.
  	}
  		\end{tablenotes}
  		\end{threeparttable}
 	 \end{center}
	\end{table*}
	
	\begin{table*}%[!h]
    \begin{center}
    \caption{Simulation results for two placements of knots under the AFT model.}
	\label{tab: aft_knots_selection}
    \begin{threeparttable}
    \begin{tabular}{cc|cccc|cccc|cccc}
    \toprule
        N & Knots & \multicolumn{4}{c}{$\beta_1=1$} & \multicolumn{4}{c}{$\beta_2=1$} & \multicolumn{4}{c}{$\beta_3=1$}\\
        & & \small{Bias} & \small{SE} & \small{ESE} & \small{CP} & \small{Bias} & \small{SE} & \small{ESE} & \small{CP} &  \small{Bias} & \small{SE} & \small{ESE} & \small{CP} \\
        \midrule
        \multirow{2}{*}{1000} & \small{K1} & \small{-.014} & \small{.197} & \small{.191} & \small{.944} & \small{-.024} & \small{.209} & \small{.192} & \small{.931}  & \small{-.032} & \small{.199} & \small{.192} & \small{.932} \\
         & \small{K2} & \small{-.001} & \small{.194} & \small{.197} & \small{.954} & \small{-.010} & \small{.203} & \small{.197} & \small{.943}  & \small{-.017} & \small{.195} & \small{.197} & \small{.945} \\ 
         \midrule
        \multirow{2}{*}{2000} & \small{K1} & \small{-.010} & \small{.144} & \small{.137} & \small{.932} & \small{-.006} & \small{.144} & \small{.137} & \small{.937}  & \small{-.005} & \small{.142} & \small{.137} & \small{.943} \\
         & \small{K2} & \small{-.005} & \small{.143} & \small{.139} & \small{.941} & \small{.000} & \small{.143} & \small{.139} & \small{.942}  & \small{-.001} & \small{.141} & \small{.139} & \small{.953} \\ 
         \midrule
        \multirow{2}{*}{4000} & \small{K1} & \small{.000} & \small{.102} & \small{.097} & \small{.944} & \small{-.005} & \small{.100} & \small{.097} & \small{.944}  & \small{-.002} & \small{.097} & \small{.097} & \small{.950} \\
         & \small{K2} & \small{.002} & \small{.102} & \small{.098} & \small{.936} & \small{-.002} & \small{.100} & \small{.098} & \small{.938}  & \small{.001} & \small{.097} & \small{.098} & \small{.950} \\
         \midrule
        \multirow{2}{*}{8000} & \small{K1} & \small{-.006} & \small{.070} & \small{.069} & \small{.950} & \small{-.003} & \small{.068} & \small{.069} & \small{.967}  & \small{-.004} & \small{.071} & \small{.069} & \small{.945} \\
         & \small{K2} & \small{-.005} & \small{.070} & \small{.069} & \small{.951} & \small{-.001} & \small{.068} & \small{.069} & \small{.958}  & \small{-.004} & \small{.071} & \small{.069} & \small{.942} \\ 
        \bottomrule
    \end{tabular}
    \begin{tablenotes}
 	 \item  \footnotesize{Bias is the difference between the mean of estimates and the true value, and SE is the sample standard error of the estimates. ESE is the mean of the standard error estimators, and CP is the corresponding coverage proportion of 95\% confidence intervals. In (K1), the interior knots are located at the $K_n = \lfloor N^{\frac{1}{7}} \rfloor$ quantiles of the estimated cumulative hazards under the Cox model. In (K2), the interior knots equally separate the interval from $0$ to two times the maximum of the estimated cumulative hazards.
  	}
  		\end{tablenotes}
  		\end{threeparttable}
 	 \end{center}
	\end{table*}

	\begin{table*}%[!h]
    \begin{center}
    \caption{Simulation results for two placements of knots under the general linear transformation model where both functions $\alpha(\cdot)$ and $q(\cdot)$ are unknown.}
	\label{tab: glt_knots_selection}
    \begin{threeparttable}
    \begin{tabular}{cc|cccc|cccc}
    \toprule
        Setting & Knots &  \multicolumn{4}{c}{$\beta_2=1$} & \multicolumn{4}{c}{$\beta_3=1$}\\
        & & \small{Bias} & \small{SE} & \small{ESE} & \small{CP} &  \small{Bias} & \small{SE} & \small{ESE} & \small{CP} \\
        \midrule
        \multirow{2}{*}{1)} & \small{K1} & \small{.008} & \small{.106} & \small{.107} & \small{.947}  & \small{.012} & \small{.104} & \small{.107} & \small{.959} \\
         & \small{K2} & \small{-.002} & \small{.098} & \small{.097} & \small{.946}  & \small{.000} & \small{.095} & \small{.097} & \small{.955} \\ 
         \midrule
        \multirow{2}{*}{2)} & \small{K1} & \small{-.019} & \small{.161} & \small{.151} & \small{.927} & \small{-.016} & \small{.159} & \small{.151} & \small{.938} \\
         & \small{K2} & \small{.005} & \small{.152} & \small{.142} & \small{.936} & \small{.009} & \small{.155} & \small{.142} & \small{.931} \\ 
         \midrule
        \multirow{2}{*}{3)} & \small{K1} & \small{-.014} & \small{.134} & \small{.131} & \small{.941} & \small{-.012} & \small{.131} & \small{.132} & \small{.945} \\
         & \small{K2} & \small{.002} & \small{.131} & \small{.124} & \small{.936} & \small{.004} & \small{.131} & \small{.128} & \small{.939} \\ 
         \midrule
        \multirow{2}{*}{4)} & \small{K1} & \small{.001} & \small{.092} & \small{.090} & \small{.939}  & \small{.005} & \small{.091} & \small{.090} & \small{.954} \\
         & \small{K2} & \small{-.002} & \small{.087} & \small{.084} & \small{.940}  & \small{.002} & \small{.085} & \small{.084} & \small{.957} \\ 
        \bottomrule
    \end{tabular}
    \begin{tablenotes}
 	 \item  \footnotesize{Bias is the difference between the mean of estimates and the true value, and SE is the sample standard error of the estimates. ESE is the mean of the standard error estimators, and CP is the corresponding coverage proportion of 95\% confidence intervals.
  	}
  		\end{tablenotes}
  		\end{threeparttable}
 	 \end{center}
	\end{table*}

\subsection{\new{Model diagnostics}}
\new{In this section, we use the linear transformation model as an example to illustrate how the unification of the proposed ODE framework along with the proposed estimation and inference procedure can be applied to model diagnostics and provide preliminary numerical results.}

\new{Recall that, under certain regularity conditions in Proposition~\ref{degeneration}, the linear transformation model, i.e.,
\begin{equation*}
\left\{
\begin{array}{lr}
\Lambda'(t) = \exp(x^T\beta + \gamma(t)+g(\Lambda(t))) \\
\Lambda(0) =0
\end{array}
\right.,
\end{equation*}
reduces to the Cox model if and only if there exist positive constants $c_1$ and $c_2$ such that $g(t) = \log c_2 + (1-c_1)\log t$, 
and it reduces to the AFT model if and only if there exist positive constants $c_1$ and $c_2$ such that $\alpha(t)=\log c_2 + (c_1-1)\log t$ for $t>0$. Therefore, to check whether the Cox or the AFT model is correctly specified, we can artificially create an additional basis function, $B(t)$, that does not belong to the linear span of $\{1, \log t\}$ and make inference about its coefficient.}
  
\new{Specifically, for checking the Cox model, we consider the following linear transformation model
\begin{equation}
	\label{eq: cox model diagnostics}
	\Lambda'_x(t) = \exp(a_1 \log(\Lambda_x) + a_2 B(\Lambda_x)+x^T\beta + \gamma(t)),
\end{equation}
with unspecified $\gamma(\cdot)$.
Then a local test of the null hypothesis $H_0: a_2 =0$ is a test for checking the Cox model specification. Correspondingly, a local test of the null hypothesis $H_0: b_2 =0$ under the model with unspecified $g(\cdot)$:
\begin{equation}
	\label{eq: aft model diagnostics}
	\Lambda'_x(t) = \exp(g(\Lambda_x)+x^T\beta + b_1 \log(t)+b_2 B(t))
\end{equation}
is a test for checking the AFT model specification. We note that, under $H_0$, the models (\ref{eq: cox model diagnostics}) and (\ref{eq: aft model diagnostics}) are identifiable up to a constant respectively, which is a direct result of Proposition~\ref{degeneration}. Thus, to guarantee the identifiability, we constrain $a_1=0$ and $b_1=0$ in the models (\ref{eq: cox model diagnostics}) and (\ref{eq: aft model diagnostics}) respectively. The proposed estimation and inference procedure can be applied to obtain the estimates of $(a_2, \beta, \gamma(\cdot))$ or $(b_2, \beta, g(\cdot))$ along with the  local test of the corresponding $H_0$.}

\new{Next, we examine the above method under the simulation settings (1) and (3) in the main text, where the Cox and the AFT model are correctly specified respectively. We consider two choices of the known basis function: $B(t)=t$ and $B(t)=\log(1+t)$. And we fit the unknown functions $\gamma(\cdot)$ and $g(\cdot)$ by cubic B-splines with the same placements of knots as described in the main text. The sample size varies from $1000$ to $8000$.}

\new{Table~\ref{tab: model diagnostics} summarizes the estimates of the coefficients of interests based on $1000$ replications. We can see that the bias of the estimator is nearly negligible in all settings. When the sample size is large, the coverage proportion of 95\% confidence intervals, where the standard error estimator is obtained by inverting the estimated information matrix of all parameters including the coefficients of spline bases, is slightly greater than the nominal level. The corresponding t-statistics would lead to a conservative local test for $H_0$. We also find that the sample standard errors of the estimates vary with the choice of the basis $B(\cdot)$, and the ability to detect the model specification depends on $B(\cdot)$ as well. It may be preferable to make both functions $\gamma(\cdot)$ and $g(\cdot)$ unknown in the nonparametric linear transformation model for model diagnostics, which requires the asymptotic distributional theory for the functional parameters. We leave this interesting direction for future work.}

\begin{table*}%[!h]
    \begin{center}
    \caption{Simulation results for checking the Cox and the AFT model specification.}
	\label{tab: model diagnostics}
    \begin{threeparttable}
    \begin{tabular}{cc|cccc|cccc}
    \toprule
        Setting &  &  \multicolumn{4}{c}{$B(t)=t$} & \multicolumn{4}{c}{$B(t)=\log(1+t)$}\\
        & N & \small{1000} & \small{2000} & \small{4000} & \small{8000} &  \small{1000} & \small{2000} & \small{4000} & \small{8000} \\
        \midrule
        \multirow{4}{*}{\parbox{3cm}{\small Cox is correctly specified: $a_2=0$}} & \small{Bias} & \small{.023} & \small{.007} & \small{.006} & \small{.003} & \small{.041} & \small{.000} & \small{.009} & \small{.007} \\
         & \small{SE} & \small{.133} & \small{.087} & \small{.056} & \small{.033} & \small{.459} & \small{.314} & \small{.206} & \small{.112} \\ 
          & \small{ESE} & \small{.136} & \small{.092} & \small{.062} & \small{.043}  & \small{.467} & \small{.325} & \small{.226} & \small{.159} \\
         & \small{CP} & \small{.949} & \small{.954} & \small{.956} & \small{.968}  & \small{.943} & \small{.945} & \small{.956} & \small{.977} \\ 
         \midrule
        \multirow{4}{*}{\parbox{3cm}{\small AFT is correctly specified: $b_2=0$}} & \small{Bias} & \small{-.003} & \small{.000} & \small{.003} & \small{.000} & \small{-.011} & \small{.003} & \small{.002} & \small{-.002} \\
         & \small{SE} & \small{.182} & \small{.130} & \small{.083} & \small{.067} & \small{.423} & \small{.295} & \small{.198} & \small{.156} \\ 
          & \small{ESE} & \small{.214} & \small{.155} & \small{.111} & \small{.079}  & \small{.515} & \small{.376} & \small{.271} & \small{.195} \\
         & \small{CP} & \small{.968} & \small{.960} & \small{.978} & \small{.961}  & \small{.975} & \small{.963} & \small{.979} & \small{.964} \\ 
        \bottomrule
    \end{tabular}
    \begin{tablenotes}
 	 \item  \footnotesize{Bias is the difference between the mean of estimates and the true value, and SE is the sample standard error of the estimates. ESE is the mean of the standard error estimators, and CP is the corresponding coverage proportion of 95\% confidence intervals.
  	}
  		\end{tablenotes}
  		\end{threeparttable}
 	 \end{center}
	\end{table*}

\end{document}